\documentclass[11pt,a4paper]{article}
\usepackage[margin=2cm]{geometry}
\usepackage{comment}
\usepackage{amsmath,amssymb,extarrows,graphicx,subfigure,setspace}
\usepackage{cite}
\usepackage{slashed}
\usepackage{color}
\usepackage{braket}
\usepackage{float}
\makeatother
\usepackage{pdfpages}
\usepackage{blindtext}
\usepackage{amsfonts}
\usepackage{tensor}
\usepackage{graphicx}
\usepackage{pict2e}
\usepackage[useregional]{datetime2}
\usepackage{amssymb, amsmath,environ}
\usepackage[
colorlinks=true,
linkcolor=blue,
urlcolor=blue,
filecolor=blue,
citecolor=red,
pdfstartview=FitV,
pdftitle={},
pdfauthor={},
pdfsubject={},
pdfkeywords={},
pdfpagemode=None,
bookmarksopen=true
]{hyperref}

\usepackage{epsfig}

\usepackage{hyperref}

\newcommand{\be}{\begin{equation}}
\newcommand{\bea}{\begin{eqnarray}}
\newcommand{\eea}{\end{eqnarray}}
\newcommand{\ba}{\begin{array}}
\newcommand{\ea}{\end{array}}
\newcommand{\ee}{\end{equation}}
\newcommand{\bes}{\begin{equation*}}
\newcommand{\beas}{\begin{eqnarray*}}
\newcommand{\eeas}{\end{eqnarray*}}
\newcommand{\bas}{\begin{array*}}
\newcommand{\eas}{\end{array*}}
\newcommand{\ees}{\end{equation*}}

\numberwithin{equation}{section}
\begin{document}	
	\onehalfspacing
	\noindent
	
	\begin{titlepage}
		\vspace{10mm}
		\begin{flushright}
			
		\end{flushright}
		
		\vspace*{20mm}
		\begin{center}
			
			{\Large {\bf  Complexity for Charged Thermofield Double States}\\
			}
			
			\vspace*{15mm}
			\vspace*{1mm}
			{Mehregan Doroudiani${}^{a,b}$, Ali Naseh${}^{a}$, Reza Pirmoradian${}^{a,c}$}
			
			\vspace*{1cm}
			
			{\it 
				${}^a$ School of Particles and Accelerators, 
				Institute for Research in Fundamental Sciences (IPM)\\
				P.O. Box 19395-5531, Tehran, Iran
				\\  
				${}^b$ Department of Physics, Sharif University of Technology,\\
				P.O. Box 11365-9161, Tehran, Iran
				\\  
				${}^c$ Department of physics, Islamic Azad University Central Tehran Branch, Tehran, Iran 
			}
			
			\vspace*{0.5cm}
			{E-mails: {\tt doroudiani,naseh,rezapirmoradian@ipm.ir}}%
			
			\vspace{.5cm}
			
			\vspace*{1cm}
		\end{center}
		
		\begin{abstract}
			We study Nielsen's circuit complexity for a charged thermofield double state (cTFD) of free complex scalar quantum field theory in the presence of background electric field. We show that the ratio of the complexity of formation for cTFD state to the thermodynamic entropy is finite and it depends just on the temperature and chemical potential. Moreover, this ratio smoothly approaches the value for real scalar theory. We compare our field theory calculations with holographic complexity of charged black holes and confirm that the same cost function which is used for neutral case continues to work in presence of $U(1)$ background field. For $t>0$, the complexity of cTFD state evolves in time and contrasts with holographic results, it saturates after a time of the order of inverse temperature. This discrepancy can be understood by the fact that holographic QFTs are actually strong interacting theories, not free ones.
		\end{abstract}
		
	\end{titlepage}

	\section{Introduction}
	One of the most intriguing discoveries in the past decade is finding interesting connections between different quantum notions of information theory with quantum gravity and quantum field theory. These connections have been provided through Gauge/Gravity duality which the first one of them is reproducing entanglement entropy of conformal field theories (CFTs) using minimal co-dimension two surfaces in geometry of asymptotically locally AdS (AlAdS) spacetimes \cite{Ryu:2006bv,Ryu:2006ef}. Intriguingly, this surface, which is known as the Ryu-Takayanagi surface, is also observed in a completely different area i.e. multi-scale entanglement renormalization ansatz (MERA) for a ground state of critical systems. It is observed by Swingle \cite{Swingle:2009bg} that the Ryu-Takayanagi curve is similar to a special curve with minimum length in MERA tensor network and this similarity implies that gravity is an emergent quantity, indeed it emerges from entanglement entropy. Albeit this is a genuine idea, but it is discussed by Susskind \cite{Susskind:2014moa} that entanglement entropy is not enough since it can not give information about the interior of black hole. This latter claim was a starting point to use another notion of quantum information theory in high energy physics community, i.e. computational complexity or more precisely state complexity.
	
	According to complexity action (CA) proposal of Susskind \cite{Susskind:2014rva,Brown:2015bva,Brown:2015lvg}, the complexity of a boundary state is dual to a value of gravitational action in a special part of spacetime, known as WDW  patch, which interestingly contains also the black hole interior. It is shown by Lloyd \cite{Lloyd} that the complexity growth rate for a given state is smaller than twice the average energy of the system at that state and intriguingly the CA proposal gives the same value for neutral black holes at late times \cite{Brown:2015bva,Carmi:2017jqz,Carmi:2016wjl}.  The ability of CA proposal to probe the interior of black holes not only gives us the opportunity to understand thermalization process in quantum many-body systems better, but also, by using that, one can hopefully resolve the black hole information paradox. Moreover, this proposal can play a great role in improving our understanding about scrambling and quantum chaos \cite{Shenker:2013pqa}, ER$=$EPR \cite{Maldacena:2013xja} and firewalls \cite{Almheiri:2013hfa,Papadodimas:2012aq}. It is worth noting that the main idea for this holographic proposal comes from rigorous computations for quantum mechanical systems. Accordingly, to understand the holographic results better, before anything, one needs to develop state complexity for QFTs with infinite degrees of freedom. 
	
	Let us remind that the complexity of a state $\mid \hspace{-1mm}\Psi_{\text{T}} \rangle$ in a system of qubits is defined as the minimum number of gates which are needed to approximately produce target state from a reference state, $\mid\hspace{-1mm} \Psi_{R}\rangle$. The reference state is chosen as a state with no spatial entanglement entropy, i.e. ground state of a Hamiltonian with a very large mass in comparison with the kinetic term. Moreover, the gates are unitary operators and one calls the sequence of gates a circuit. Two main extensions of this idea have been recently made to quantify state complexity in QFTs. The first extension \cite{Jefferson:2017sdb} is based on the idea of Nielsen  \cite{Nielsen:2005,Nielsen:2006,Nielsen:2007} in which, one associates a geometry to the space of unitaries based on the algebra of gates. Then finding the optimal circuit is mapped to finding a minimal geodesic in the space of unitaries and the state complexity becomes the length of that geodesic. Up to now, this procedure has been well established for the free bosonic and fermionic QFTs \cite{Guo:2018kzl,Hackl:2018ptj,Caceres:2019pgf,Jiang:2018nzg,Jiang:2018gft,Khan:2018rzm} and preliminary steps for weakly interacting theories has been done in \cite{Bhattacharyya:2018bbv}. In the second extension, the optimal circuit is mapped to an optimized path integral definition of the target state \cite{Caputa:2017urj,Caputa:2017yrh}. The intuition behind this extension comes from noting the relation between MERA and discretized path integral representation of a state. Accordingly, the cost of computing path integral on a Weyl-rescaled geometry is less than the cost of computing it on a flat two-dimensional lattice. This cost is quantified by Liouville action and very interestingly it is shown \cite{Camargo:2019isp} that this action has a deep connection with the length of minimal geodesics in the Nielsen approach i.e. in the first extension. Remarkably, both of the above extensions give rise to UV divergences similar to the ones in holographic computations. We explore the first extension in this paper and we discuss path integral complexity no more. 
	
	Recently, Chapman, Eisert, Hackl, Heller, Jefferson, Marrochio and Myers \cite{Chapman:2018hou} evaluated the complexity of TFD state of a free scalar field theory. Their study led to discovery a cost function which is completely consistent with the properties of holographic complexity in the absence of any background charge. To discover these similarities, one special cost function, known as $F_{1}$ which will be explained below, is used for counting gates. By this cost function, the length of the path which connects reference state to the target state depends on the basis chosen for generators. Intriguingly, the advantage of this basis dependence has been taken in evaluating the complexity of formation and it is shown that it can produce the complexity of formation that compares well with holographic results. For $U(1)$ charged black holes, a generalization of Lloyd's bound has been proposed \cite{Brown:2015bva},
	\bea
	\label{CA}
	\frac{d\mathcal{C}_{A}}{dt} \leq \frac{2}{\pi} \left[(M-\mu Q)-(M-\mu Q)_{gs}\right],
	\eea
	where the subscript $gs$ indicates a state which for a given chemical potential gives the lowest $(M-\mu Q)$.  For charged small black holes, i.e. for the ones where the size of the horizon is much smaller than
	AdS radius, the rate of the change of complexity at late times saturates the bound
	\bea\label{cdotcTFD}
	\frac{d\mathcal{C}_{A}}{dt} = \frac{2}{\pi} (M-\mu Q),
	\eea
	where for a given $\mu$, the smallest value of $(M-\mu Q)$ is zero. In this paper, we would like to extend the analysis of \cite{Chapman:2018hou} to free complex scalar QFT in the presence of chemical potential. To be more precise, we would like to check whether that special cost function can also provide similar results with the holographic proposal (\ref{CA}), i.e. a linear growth of complexity at late times, its dependency to the detailed structure of null boundary terms, its UV divergence as spatial volume
	and its IR divergence for vanishing temperature at a fixed chemical potential. We will show that for the first case they clearly differ, since the holographic results are applicable for strong interacting theories but our results are obtained for a free theory. Moreover, without any change in the holographic proposal, for third and fourth cases the QFT results based on using $F_{1}$ cost function in Left-Right (LR) basis are compatible with holographic results but for the second one, it needs some modification.  
	
	We organize the paper as follows: In section \ref{cTFD}, we reconstruct the cTFD state of free complex scalar theory by putting it on a lattice with a UV cut off. In the same section, the time-dependent cTFD state is also studied. In section \ref{covmatrix}, after a brief review of covariance matrix formalism for manipulating the state complexity,  we proceed to evaluate the complexity and its time dependency for cTFD state. In section \ref{conti} we extend the same analysis for continuous system and confirm that the prefer cost function introduced in \cite{Chapman:2018hou} continues to work also in the presence of $U(1)$ global charge. In appendix \ref{def} some definitions are provided. Moreover, in appendix \ref{diagonal} the similar results for another basis are presented, i.e. diagonal one and in appendix \ref{compareholo} more concrete comparison with holographic complexity of charged black holes is provided.
	
	\textbf{Note Added:} When our paper was in the final stage to be submitted to arxiv, the paper \cite{Chapman:2019clq} appeared which studied the same problem. The main difference between our approach with theirs is that we just have particles on the left side and as a result, just anti-particles on the right side. But their cTFD is made of particles and anti-particles on both sides, so their cTFD is a tensor product of our cTFD and a state with anti-particles on the left side and particles on the right. Our approach divides the phase space in half. Furthermore, it let us reach the conformal limit of the theory.
	
	\section{cTFD states}\label{cTFD}
	In this section, we construct the cTFD state in the Hilbert space of free complex scalar theory. Just for simplicity we consider this theory in $(1+1)$ dimensions and in section.\ref{conti}, we easily extend the study to general dimensions. The Lagrangian density of a free complex scalar field is given by
	\bea
	\label{Lcomplex1}
	\mathcal{L} = (D_{\mu}\Phi)(D_{\mu}\Phi)^* - m^2 \Phi \Phi^*,
	\eea
	where the covariant derivative is defined by $D_{\mu} = \partial_{\mu} - iqA_{\mu}$. For the background electric field,  $A_{\mu}\hspace{-.1cm}=\hspace{-.1cm}(-\mu, 0)$, so this Lagrangian density is simplified to
	\bea
	\mathcal{L} = \dot{\Phi}\dot{\Phi}^* - \partial_{\vec{x}}\Phi\partial_{\vec{x}}\Phi^*-i\mu q(\dot{\Phi}\Phi^* - \Phi\dot{\Phi}^*)-(m^2-q^2\mu^2)\Phi\Phi^*,
	\eea
	with the dot means derivative with respect to the time. The conjugate momentums are given by
	\begin{align}
	\Pi = \frac{\partial\mathcal{L}}{\partial{\dot{\Phi}}} = \dot{\Phi}^*-i\mu q\Phi^*,
	\hspace{1cm}\Pi^* = \frac{\partial\mathcal{L}}{\partial{\dot{\Phi}^*}} = \dot{\Phi}+i\mu q\Phi,
	\end{align}
	and the Hamiltonian on a circle with circumference $L$  becomes
	\begin{align}
	\label{Hcomplex1}
	H = \int_{-\frac{L}{2}}^{\frac{L}{2}} dx\hspace{.5mm}\Big( \Pi\hspace{.5mm}\Pi^*-i\mu q\left(\Pi\hspace{.5mm}\Phi-\Pi^*\hspace{.5mm}\Phi^*\right)+\partial_{x}\hspace{.5mm}\Phi\partial_{x}\Phi^*+m^2\Phi\hspace{.5mm}\Phi^*\Big).
	\end{align}
	To regulate this theory in the UV, we consider that circle as a lattice with $N$ sites and lattice spacing $\delta$ which it implies that 
	\bea\label{delta}
	\delta = \frac{L}{N},\hspace{1cm}\mathbf{x}_a = \Phi(x_a)\delta, \hspace{1cm} \mathbf{p}_a = \Pi(x_a),
	\eea
	with $\mathbf{x}_a$, $\mathbf{p}_a$ are redefined canonical variables and $\mathbf{x}_{N+1} = \mathbf{x}_1$ and $\mathbf{p}_{N+1} = \mathbf{p}_N$.
	Following by these variables, the Hamiltonian (\ref{Hcomplex1}) changes to
	\begin{align}
	\label{Hcomplex2}
	H = \sum_{a=1}^{N}\bigg(\delta\hspace{.5mm} \mathbf{p}_a\hspace{.5mm}\mathbf{p}_a^*-i\mu q\left(\mathbf{p}_a\hspace{.5mm}\mathbf{x}_a-\mathbf{p}_a^*\hspace{.5mm}\mathbf{x}_a^*\right)+\frac{m^2}{\delta}\hspace{.5mm}\mathbf{x}_a\hspace{.5mm}\mathbf{x}_a^* +\frac{1}{\delta^3}\left(\mathbf{x}_{a+1}-\mathbf{x}_a\right)\left(\mathbf{x}_{a+1}^*-\mathbf{x}_a^*\right)\bigg).
	\end{align}
	By defining the following canonical transformation in momentum variable
	\bea
	\mathbf{p}_a' = \mathbf{p}_a+i\frac{\mu q}{\delta}\hspace{.5mm}\mathbf{x}^{*}_a,
	\eea
	the Hamiltonian (\ref{Hcomplex2}) becomes
	\begin{align}
	\label{Hcomplex3.3}
	H = \sum_{a=1}^{N}\bigg(\delta\hspace{.5mm} \mathbf{p}_a'\hspace{.5mm}\mathbf{p}_a'^*+\frac{m^2-\mu^2 q^2}{\delta}\hspace{.5mm}\mathbf{x}_a\hspace{.5mm}\mathbf{x}_a^* +\frac{1}{\delta^3}\left(\mathbf{x}_{a+1}-\mathbf{x}_a\right)\left(\mathbf{x}_{a+1}^*-\mathbf{x}_a^*\right)\bigg).
	\end{align}
	To decompose the contribution of different modes, one can use discrete Fourier transformation of variables as follows
	\begin{align}
	\label{Fourier}
	\tilde{\mathbf{x}}_k = \frac{1}{\sqrt{N}}\sum_{a=1}^{N}e^{\frac{2\pi ika}{N}}\hspace{.5mm}\mathbf{x}_a,\hspace{1cm}\tilde{\mathbf{p}}^{\prime}_k = \frac{1}{\sqrt{N}}\sum_{a=1}^{N}e^{\frac{-2\pi ika}{N}}\hspace{.5mm}\mathbf{p}_a,
	\end{align}
	where $\tilde{\mathbf{x}}^{*}_{k} =\tilde{\mathbf{x}}_{N-k}$ and similarly for the momentum variable. According to (\ref{Fourier}), the Hamiltonian (\ref{Hcomplex3.3}) reduces to
	\begin{align}
	\label{Hcomplex3.4}
	H = \sum_{k=0}^{N-1}\bigg(\delta\hspace{.5mm} |\tilde{\mathbf{p}}^{\prime}_k|^2+\frac{m^2-\mu^2 q^2+\frac{4}{\delta^2}\sin^2(\frac{\pi k}{N})}{\delta}\hspace{.5mm}|\tilde{\mathbf{x}}_k|^2\bigg).
	\end{align}
	Now using the Fourier transformation $
	\tilde{\mathbf{p}}'_k = \tilde{\mathbf{p}}_k+i\frac{\mu q}{\delta}\hspace{.5mm}\tilde{\mathbf{x}}^{*}_{N-k},$
	the Hamiltonian (\ref{Hcomplex3.4}) changes to 
	\begin{align}
	\label{Hcomplex3}
	H = \sum_{k=0}^{N-1}\bigg(\delta\hspace{.5mm} |\tilde{\mathbf{p}}_k|^2+\frac{\omega_k^2}{\delta}\hspace{.5mm}|\tilde{\mathbf{x}}_k|^2-i\mu q\left(\tilde{\mathbf{p}}_k\hspace{.5mm}\tilde{\mathbf{x}}_{N-k}-\tilde{\mathbf{p}}_k^*\hspace{.5mm}\tilde{\mathbf{x}}_{N-k}^*\right)\bigg),
	\end{align}
	where $\omega_k$ is given by  
	\bea
	\label{omegak}
	\omega_k^2 = m^2+\frac{4}{\delta^2}\sin^2(\frac{\pi k}{N}).
	\eea 
	To construct the cTFD state of theory (\ref{Hcomplex3}), one needs to quantize its Hamiltonian and it can be achieved by defining two sets of creation and annihilation operators
	\begin{align}
	\label{aad}
	\tilde{\mathbf{x}}_k = \sqrt{\frac{\delta}{2\omega_k}}\left(\hat{a}_{N-k}+\hat{b}_k^{\dagger}\right),\hspace{1cm}\tilde{\mathbf{p}}_k = i \sqrt{\frac{\omega_k}{2\delta}}\left(\hat{a}_k^{\dagger}-\hat{b}_{N-k}\right),
	\end{align}
	with $[\hat{a}_k,\hat{a}^{\dagger}_k]=[\hat{b}_k,\hat{b}^{\dagger}_k]=1$ and other commutators are zero. Let remind that $(\hat{a}_k,\hat{a}^{\dagger}_k)$ and $(\hat{b}_k,\hat{b}^{\dagger}_k)$ are respectively the "annihilation, creation" operators for particles and anti-particles.  Substituting (\ref{aad}) in (\ref{Hcomplex3}) gives the following quantized Hamiltonian,
	\begin{align}
	\label{Hcomplex4}
	\hat{H}=\sum_{k=0}^{N-1} \omega_k \left(\hat{a}_k^{\dagger}\hspace{.5mm}\hat{a}_k+\hat{b}_k^{\dagger}\hspace{.5mm}\hat{b}_k+1\right) + \mu q\left(\hat{a}_k^{\dagger}\hat{a}_k-\hat{b}_k^{\dagger}\hat{b}_k\right).
	\end{align}
	The Hamiltonian (\ref{Hcomplex4}) contains two different parts. One part describes two standard harmonic oscillators for each mode (constant momentum $k$)
	\begin{align}
	\label{Hharmonic}
	\hat{H}_{0}=\sum_{k=0}^{N-1} \omega_k \left(\hat{a}_k^{\dagger}\hspace{.5mm}\hat{a}_k+\hat{b}_k^{\dagger}\hspace{.5mm}\hat{b}_k+1\right).
	\end{align}
	By noting that the $U(1)$ electric Noether charge associated to the complex scalar field reads as
	\bea
	Q = q \int dx \left(\Phi^{*}\dot{\Phi}-\Phi \dot{\Phi}^{*}\right),
	\eea
	the second part is the product of this charge operator with chemical potential 
	\bea\label{Q}
	\mu \hat{Q} = \sum_{k=0}^{N-1} \mu q\left(\hat{a}_k^{\dagger}\hspace{.5mm}\hat{a}_k-\hat{b}_k^{\dagger}\hspace{.5mm}\hat{b}_k\right).
	\eea
	Accordingly, the Hamiltonian (\ref{Hcomplex3}) constitutes  $N$ decoupled harmonic oscillators with equal masses  $M=\delta^{-1}$ (not to
	be confused with the physical mass $m$) and $k$-dependent frequencies $\omega_{k}$. If we omit the constant part, the total energy of level "$n$" for each of these oscillators with fixed momentum $k$ is given by $ n \left(\omega_{k}+\mu q\right)$. Since $\omega_{0}$ vanishes when $m=0$, the zero mode Hamiltonian does not have a normalizable ground state. But it can be regularized by introducing a very small dimensionless mass, $m L \ll 1$. Moreover, since each mode $k$ in (\ref{Hcomplex3}) is decoupled from the other modes,
	the respective cTFD state will be the product of cTFD states for each oscillator. Accordingly, in the following, we focus on a single mode (fixed momentum $k$)  and construct its time-dependent cTFD state. To attain that, let us remind that in thermal equilibrium, the ensemble average of any operator is given by
	\begin{align}
	\label{expO}
	\langle \hat{\mathcal{O}}\rangle_{\beta}= \frac{1}{Z}\hspace{.5mm}\text{Tr}[e^{-\beta \hat{H}}\hspace{.5mm}\hat{\mathcal{O}}] = \frac{1}{Z}\sum_{n}e^{-\beta\hspace{.5mm} \mathcal{E}_{n,q}}\langle \mathcal{E}_{n,q}\hspace{-1mm} \mid \hat{\mathcal{O}}\mid\hspace{-1mm}  \mathcal{E}_{n,q}\rangle,
	\end{align}
	where $Z$ is the partition function of the system and $\mathcal{E}_{n,q}= n\left(\omega+ \mu q\right)$. One can express the above ensemble average of any operator as an expectation value in a thermal vacuum by defining a state $\mid\hspace{-1mm}0\rangle_\beta$ such that
	\begin{align}
	\label{expO1}
	\langle \hat{\mathcal{O}}\rangle_{\beta} =\hspace{.5mm}_{\beta} \langle 0\hspace{-1mm}\mid\hspace{-1mm}\hat{\mathcal{O}}\hspace{-1mm}\mid\hspace{-1mm} 0\rangle_\beta=\frac{1}{Z}\sum_{n}e^{-\beta\hspace{.5mm} \mathcal{E}_{n,q}}\langle \mathcal{E}_{n,q}\hspace{-1mm} \mid \hat{\mathcal{O}}\mid\hspace{-1mm}  \mathcal{E}_{n,q}\rangle.
	\end{align}
	Assuming thermal vaccum $\mid\hspace{-1mm} 0\rangle_\beta$ can be written as a linear superposition of the states of Hilbert space $\mid\hspace{-1mm}  \mathcal{E}_{n,q}\rangle$,
	\begin{align}
	\label{expO2}
	\mid\hspace{-1mm} 0\rangle_\beta =\sum_{n} \mid\hspace{-1mm}  \mathcal{E}_{n,q}\rangle \langle \mathcal{E}_{n,q}\hspace{-1mm} \mid\hspace{-1mm} 0\rangle_\beta =\sum_{n} \alpha_{n,q}(\beta) \mid\hspace{-1mm}  \mathcal{E}_{n,q}\rangle,
	\end{align}
	implies that
	\begin{align}
	\hspace{.5mm}_{\beta} \langle 0\hspace{-1mm}\mid\hspace{-1mm}\hat{\mathcal{O}}\hspace{-1mm}\mid\hspace{-1mm} 0\rangle_\beta =\sum_{n,m} \alpha^{*}_{m,q}(\beta)\hspace{.5mm}\alpha_{n,q}(\beta) \langle \mathcal{E}_{m,q}\hspace{-1mm} \mid\hspace{-1mm}\hat{\mathcal{O}}\hspace{-1mm}\mid\hspace{-1mm}  \mathcal{E}_{n,q}\rangle.
	\end{align}
	This agrees with (\ref{expO1}) provided
	\bea\label{expO3}
	\alpha^{*}_{m,q}(\beta)\hspace{.5mm}\alpha_{n,q}(\beta) = Z^{-1} e^{-\beta \mathcal{E}_{n,q}}\hspace{1mm}\delta_{mn}.
	\eea
	However, as it is clear from (\ref{expO2}), $\alpha_{n,q}$'s are ordinary numbers and therefore it is not possible to satisfy (\ref{expO3}). This means that we can not define the thermal vacuum as long as we restrict ourselves to the original Hilbert space. Interestingly, the condition (\ref{expO3}) is quite analogous to orthonormality condition for state vectors. Let us introduce a fictitious system which is an identical copy of the original system and denote it with tilde system. Now, expanding the thermal vacuum as following
	\begin{align}
	\label{thermalvac}
	\mid\hspace{-1mm} 0\rangle_\beta = \sum_{n} \alpha_{n,q}(\beta) \mid\hspace{-1mm} n,\tilde{n}\rangle = \sum_{n} \alpha_{n,q}(\beta)\mid\hspace{-1mm} \mathcal{E}_{n,q}\rangle\hspace{1mm}\otimes \mid\mathcal{E}_{\tilde{n},-q}\rangle,
	\end{align}
	implies that
	\bea
	\label{expO4}
	&&\hspace{.5mm}_{\beta} \langle 0\hspace{-1mm}\mid\hspace{-1mm}\hat{\mathcal{O}}\hspace{-1mm}\mid\hspace{-1mm} 0\rangle_\beta 
	=\sum_{n}\alpha^{*}_{n,q}(\beta)\hspace{.5mm}\alpha_{n,q}(\beta)\hspace{.5mm} \langle \mathcal{E}_{n,q}\hspace{-1mm} \mid\hspace{-1mm}\hat{\mathcal{O}}\hspace{-1mm}\mid\hspace{-1mm}  \mathcal{E}_{n,q}\rangle.
	\eea
	Comparing (\ref{expO4}) with (\ref{expO1}) implies that
	\bea
	\alpha^{*}_{n,q}(\beta)\hspace{.5mm}\alpha_{n,q}(\beta)\hspace{.5mm} =Z^{-1}\hspace{.5mm}e^{-\beta \mathcal{E}_{n,q}},
	\eea
	which has a simple real solution
	\bea
	\label{expO5}
	\alpha_{n,q}(\beta)=\alpha^{*}_{n,q}(\beta)= Z^{-1/2}\hspace{.5mm}e^{-\frac{\beta}{2}\mathcal{E}_{n,q}}.
	\eea
	By Substituting $\alpha_{n,q}$ from (\ref{expO5}) in (\ref{thermalvac}), the charged thermal vacuum (cTFD state) becomes
	\bea\label{TFD.1}  
	\mid\hspace{-1mm}  \text{cTFD}\rangle=Z^{-\frac{1}{2}}\displaystyle\sum_{n=0}^{\infty}e^{-\frac{\beta}{2}\mathcal{E}_{n,q}}\mid\hspace{-1mm}  \mathcal{E}_{n,q}\rangle_L\hspace{1mm}\otimes\mid\hspace{-1mm}  \mathcal{E}_{n,-q}\rangle_R,
	\eea
	where we denoted the original system by subscript $R$ and the tilde system with subscript $L$, just in analogy with the right and left copies of CFT in the penrose diagram of eternal black hole. It is easy to see that the normalization factor becomes $Z^{-1}= 1-e^{-\beta(\omega+\mu q)}$. To proceed furthermore, we note that the level "n" energy eigenstate can be expressed as
	\bea\label{staten}
	\mid\hspace{.5mm}\mathcal{E}_{n,q}\rangle = \frac{1}{\sqrt{n!}} (\hat{a}^\dagger)^n \mid \hspace{-1mm}0 \rangle,
	\eea
	where 
	\bea
	\hat{a}^\dagger \hspace{-1mm}\mid\hspace{-1mm}n\rangle = \sqrt{n+1}\mid\hspace{-1mm}n+1\rangle,\hspace{.5cm}\hat{a}\hspace{-1mm} \mid\hspace{-1mm}n\rangle = \sqrt{n}\mid\hspace{-1mm}n-1\rangle,
	\eea
	with 
	\bea\label{a}
	\hat{a}^\dagger=\sqrt{\frac{m\omega}{2}}\left(\hat{x}-i\frac{\hat p}{m\omega}\right),  \hspace{.5cm}
	\hat{a}=\sqrt{\frac{m\omega}{2}}\left(\hat{x}+i\frac{\hat p}{m\omega}\right),
	\eea
	and $[\hat{a},\hat{a}^\dagger]=1$.
	In comparison with (\ref{Hcomplex4}) and noting to (\ref{Hharmonic}) and (\ref{Q}), it is clear that to construct the state (\ref{TFD.1}) one can use the $(\hat{a}_k,\hat{a}^{\dagger}_k)$ for the left side and $(\hat{b}_k,\hat{b}^{\dagger}_k)$ for the right side. This observation implies that (\ref{TFD.1}) can be simplified to
	\bea
	\label{CTFD.2.1}
	&&\ket{\text{cTFD}}= 
	\sqrt{1-e^{-\beta(\omega+\mu q)}}\hspace{1mm}\exp{\left(e^{-\frac{\beta }{2}(\omega+\mu q)}\hspace{1mm}\hat{a}^\dagger_{L}\hat{b}^\dagger_{R}\right)}\mid\hspace{-1mm}0\rangle_L\hspace{.5mm}\otimes\mid \hspace{-1mm}0\rangle_R.
	\eea
	We would like to re-express the  thermofield double state (\ref{CTFD.2.1}) by acting a unitary operator on the vacuum state $\mid\hspace{-1mm}0\rangle_L\hspace{.5mm}\otimes\mid \hspace{-1mm}0\rangle_R$. To achieve this, one can define the following operator
	\bea
	\label{O}
	\mathcal{\hat{O}}=\exp\left(\alpha_+\hat{\mathcal{K}}_+\hspace{.5mm}+\hspace{.5mm}\alpha_-\hat{\mathcal{K}}_-\hspace{.5mm}+\hspace{.5mm}\alpha_{0} \hat{\mathcal{K}}_0\right),
	\eea
	with 
	\bea\label{Ks} \hat{\mathcal{K}}_{-}=\hat{a}_{L}\hspace{.5mm}\hat{b}_{R}\hspace{1cm}\hat{\mathcal{K}}_{+}=\hat{a}^\dagger_{L}\hspace{.5mm}\hat{b}^\dagger_{R}
	\hspace{1cm}\hat{\mathcal{K}}_{0}=\frac{1}{2}\left(\hat{a}^\dagger_{L}\hspace{.5mm}\hat{a}_{L}+\hat{b}^\dagger_{R}\hspace{.5mm}\hat{b}_{R}+1\right),
	\eea
	and
	\bea
	[\hat{\mathcal{K}}_-,\hat{\mathcal{K}}_+]=2\hat{\mathcal{K}}_0,  \hspace{1cm} [\hat{\mathcal{K}}_0,\hat{\mathcal{K}}_\pm]=\pm \hat{\mathcal{K}}_\pm.
	\eea
	The operator $\hat{\mathcal{O}}$ is unitary for $\alpha_+=-\alpha_-$ and $\alpha_{0}\in R$ and it can be decomposed as \cite{Klimov}  
	\bea
	\mathcal{\hat{O}}= e^{\gamma_+\hat{\mathcal{K}}_+}\hspace{1mm} e^{(\log{\gamma_0})\hat{\mathcal{K}}_0}\hspace{1mm}e^{\gamma_-\hat{\mathcal{K}}_-},
	\eea
	where
	\bea
	\gamma _0=\left(\cosh \Theta-\frac{\alpha_{0}}{2\Theta}\sinh \Theta\right)^{-2},
	\hspace{.5cm}\gamma _\pm=\frac{2\alpha_\pm\sinh \Theta}{2\Theta\cosh \Theta-\alpha_{0}\sinh \Theta}\hspace{.5mm},
	\hspace{0.5cm}\Theta^2=\frac{\alpha_{0}^2}{4}-\alpha_+\alpha_-.
	\eea
	For the special case $\alpha_{0}=0$ and $\alpha_+=-\alpha_-=\alpha\in \mathbb{R}$, the operator $\mathcal{\hat{O}}$ can be simplified to
	\bea
	\mathcal{\hat{O}}=e^{\alpha(\hat{\mathcal{K}}_+-\hat{\mathcal{K}}_-)}  = e^{(\tanh\alpha)\hat{\mathcal{K}}_+}\hspace{1mm}e^{-2(\log\cosh\alpha)\hat{ \mathcal{K}}_0}\hspace{1mm}e^{-(\tanh\alpha)\hat{\mathcal{K}}_-}.
	\eea
	By noting that
	\bea
	\hat{\mathcal{K}}_- \mid\hspace{-1mm}0\rangle_L\hspace{.5mm}\otimes\mid\hspace{-1mm} 0\rangle_R=0,\hspace{1cm} \hspace{3mm}\hat{\mathcal{K}}_0\mid\hspace{-1mm}0\rangle_L\hspace{.5mm}\otimes\mid\hspace{-1mm} 0\rangle_R=\frac{1}{2}\mid\hspace{-1mm}0\rangle_L\otimes\mid\hspace{-1mm} 0\rangle_R,
	\eea 
	we have
	\bea
	\label{OO}
	e^{\alpha(\hat{\mathcal{K}}_+-\hat{\mathcal{K}}_-)}\mid\hspace{-1mm}0\rangle_L\hspace{.5mm}\otimes\hspace{-.5mm}\mid \hspace{-1mm}0\rangle_R=\frac{1}{\cosh\alpha}\hspace{1mm}e^{(\tanh\alpha)\hat{ \mathcal{K}}_+}\mid\hspace{-1mm}0\rangle_L\otimes\hspace{-.5mm}\mid \hspace{-1mm}0\rangle_R.
	\eea
	Now, by comparing the right hand side of (\ref{OO}) with (\ref{CTFD.2.1}), one can write the cTFD state (\ref{CTFD.2.1}) in desired form
	\bea
	\label{TFD.3}
	\mid \hspace{-1mm}\text{cTFD}\rangle = e^{\alpha\left(\hat{a}_L^{\dagger}\hat{b}_R^{\dagger}-\hat{a}_L\hat{b}_R\right)}\ket{0}_L\otimes\ket{0}_R \hspace{1cm}\text{with}\hspace{.5cm}\tanh{\alpha} = e^{-\frac{\beta}{2}(\omega+\mu q)}.
	\eea
	The time evolution of state (\ref{TFD.3}) is given by 
	\bea
	\label{TTFD.2.1}
	\mid\hspace{-.5mm}\text{cTFD}(t)\rangle = e^{-i(\hat{H}_{L}+\mu \hat{Q}_{L})t_{L}} e^{-i(\hat{H}_{R}-\mu \hat{Q}_{R})t_{R}} \mid\hspace{-.5mm}\text{cTFD} \rangle,
	\eea
	with 
	\bea
	\hat{H}_{L} = \hat{a}^{\dagger}_{L}\hspace{.5mm}\hat{a}_{L}+\frac{1}{2},\hspace{.5cm}\hat{H}_{R} = \hat{b}^{\dagger}_{R}\hspace{.5mm}\hat{b}_{R}+\frac{1}{2},
	\hspace{.5cm}\mu \hat{Q}_{L} = \mu q\hspace{.5mm} \hat{a}^{\dagger}_{L}\hspace{.5mm}\hat{a}_{L},\hspace{.5cm}\mu \hat{Q}_{R} = -\mu q\hspace{.5mm} \hat{b}^{\dagger}_{R}\hspace{.5mm}\hat{b}_{R}.
	\eea
	By choosing $t_{L}=t_{R}=t/2$ following the common convention in holography\footnote{According to the boost symmetry, the evaluation of the holographic complexity will depend on $t = t_L + t_R$ and not on each of the boundary times separately. So, without loss of generality, on can choose symmetric times $t_L = t_R = t/2$.}, the time-dependent cTFD state (\ref{TTFD.2.1}) after short computations becomes 
	\bea
	\label{cTFDt}
	\mid\hspace{-.5mm}\text{cTFD}(t)\rangle 
	=e^{-\frac{i}{2}(\omega+\mu q)t}\hspace{.5mm}\sqrt{1-e^{-\beta(\omega+\mu q)}}\hspace{1mm}\exp\hspace{-.5mm}\bigg[e^{-\frac{\beta}{2}(\omega+\mu q)}e^{-i(\omega+\mu q)t}\hspace{1mm}a_L^{\dagger}b_R^{\dagger}\bigg]\ket{0}_L\otimes\ket{0}_R.
	\eea
	To re-express the state (\ref{cTFDt}) as acting a unitary operator on the $\ket{0}_L\otimes\ket{0}_R$, one can define
	\bea
	z = \alpha\hspace{.5mm}e^{-i(\omega+\mu q) t},
	\eea
	which by that and doing similar procedure as (\ref{O})-(\ref{OO}), this state becomes\footnote{We have dropped the above global time-dependent phase, since this does not change the physical state.}
	\bea
	\label{TTFD.2.2}
	\mid\hspace{-1mm} \text{cTFD}(t)\rangle=\exp\hspace{-.5mm}\bigg[z\hspace{.5mm}\hat{a}^\dagger_{L}\hspace{.5mm}\hat{b}^\dagger_{R}-z^*\hspace{.5mm}\hat{a}_{L}\hspace{.5mm}\hat{b}_{R}\bigg]\hspace{-1mm}\mid\hspace{-1mm}0\rangle_L\otimes\mid\hspace{-1mm}0\rangle_R.
	\eea
	For future application, let us find the wave function representation of cTFD state (\ref{TFD.3}). By defining the proper normal coordinates 
	\bea\label{xpm}
	\hat{x} _\pm=\frac{1}{\sqrt{2}}\bigg(\hat{x}_L\pm \hat{x}_R\bigg),\hspace{1cm}
	\hat{p}_\pm=\frac{1}{\sqrt{2}}\bigg(\hat{p}_L\pm \hat{p}_R\bigg),
	\eea
	one can see
	\bea
	\label{pm}
	\left(\hat{a}^\dagger_{L}\hat{b}^\dagger_{R}-\hat{a}_{L}\hat{b}_{R}\right)
	= - i \left(\hat{x}_+ \hat{p}_+ -\hat{x}_- \hat{p}_-\right),
	\eea
	which this implies that (\ref{CTFD.2.1}) changes to
	\bea
	\label{TFDpm}
	\mid \hspace{-1mm}\text{cTFD}\rangle = e^{-i\alpha(\hat{x}_+\hat{p}_+ -\hat{x}_- \hat{p}_-)}\mid\hspace{-1mm}0\rangle_+\hspace{.5mm}\otimes\mid\hspace{-1mm} 0\rangle_- =  e^{-i\alpha\hat{x}_+\hat{p}_+}\mid\hspace{-1mm}0\rangle_+\otimes e^{i\alpha\hat{x}_-\hat{p}_-}\hspace{-1mm}\mid\hspace{-1mm}0\rangle_-.
	\eea
	By noting that the ground-state wave functions are given by 
	\bea
	\langle x\hspace{-1mm}\mid\hspace{-.8mm}0\rangle_{\pm} \equiv \Psi_0=\frac{(m\omega)^\frac{1}{4}}{\pi^\frac{1}{4}}e^{-\frac{m\omega}{2}{x_{\pm}^2}},
	\eea
	and
	\bea
	&& e^{-\frac{i\alpha}{2}(\hat{x}_+\hat{p}_+ \hspace{.5mm}+\hspace{.5mm}\hat{p}_+\hat{x}_+)}\Psi_0(x_+)  \simeq \Psi_0(e^{-\alpha}x_+)  \simeq e^{-\frac{m\omega}{2}e^{-2\alpha}{x^2_+}},
	\cr \nonumber\\&&
	e^{\frac{i\alpha}{2}(\hat{x}_-\hat{p}_-+\hat{p}_-\hat{x}_-)}\Psi_0(x_-)  \simeq \Psi_0(e^{\alpha}x_-) \simeq e^{-\frac{m\omega}{2}e^{2\alpha}{x^2_-}},
	\eea
	the wavefunctional cTFD state (\ref{TFDpm}) becomes
	\bea
	\label{WFCTFDpm}
	\langle x_+,x_{-}\hspace{-1mm}\mid \hspace{-1mm}\text{cTFD}\rangle\equiv\Psi_{\text{cTFD}\hspace{1mm}}\simeq\exp\hspace{-.5mm}\bigg[ -\frac{m\omega}{2}\bigg(e^{-2\alpha} x^2_+ + e^{2\alpha} x^2_-\bigg)\bigg].
	\eea
	In the standard coordinates ($x_{L},x_R$) (\ref{xpm}), the above wave function changes to
	\bea
	\label{WFCTFD}
	\Psi_{\text{cTFD}}\simeq  \exp\hspace{-.5mm}\bigg[-\frac{m\omega}{2}\bigg(\cosh2\alpha\hspace{.5mm}\big(x_L^2+x_R^2\big)-2\sinh2\alpha\left(x_L x_R\right)\bigg)\bigg].
	\eea
	Moreover, for later convenience let us write the time-dependent cTFD state (\ref{TTFD.2.2}),  by using (\ref{pm}), as following
	\bea
	\label{TCTFD.3}
	\mid\hspace{-1mm} \text{cTFD}(t)\rangle= e^{-i\alpha\hspace{.5mm}\hat{{O}}_{+}(t)} \hspace{-.5mm}\mid\hspace{-1mm}0\rangle_+ \otimes e^{i\alpha\hspace{.5mm}\hat{\mathcal{O}}_{-}(t)} \hspace{-.5mm}\mid\hspace{-1mm}0\rangle_-,
	\eea
	with
	\bea\label{opm}
	&&\hat{\mathcal{O}}_{\pm}(t)= \frac{1}{2}\bigg(\cos\left((\omega+\mu q) t\right)\hspace{.5mm}(\hat{x}_\pm \hat{p}_\pm + \hat{p}_\pm \hat{x}_\pm)+\sin\left((\omega+\mu q) t\right)\hspace{.5mm}(m\omega{\hat{x}}^2_\pm-\frac{1}{m\omega}\hat{p}^2_\pm)\bigg).
	\eea
	By using this representation, one can easily find the time-dependent cTFD wavefunctional state similar to the procedure which is done in (\ref{TFDpm})-(\ref{WFCTFD}). Now, by having the cTFD state (and its time-dependent counterpart) one can calculate its complexity. The way we have chosen to calculate this quantity is based on the Nielsen geometric approach which is developed for QFTs in \cite{Jefferson:2017sdb,Guo:2018kzl}. In the next section, we give a brief review on this approach and after that, we present the results for complexity with different cost functions.
	\section{Nielsen's complexity of cTFD state}\label{covmatrix}
	As we discussed above, since each mode $k$ is decoupled from the other modes in (\ref{Hcomplex3}), 
	the respective cTFD state will be the product of TFD states for each of the oscillators. This means that the complexity of ground state of Hamiltonian (\ref{Hcomplex3}) on a lattice with $N$ site is easily a sum over the complexity for each mode $k$. Therefore, it is sufficient to find the complexity for a single mode (\ref{WFCTFDpm}) and its time-dependent counterpart, (\ref{TCTFD.3}), by using covariance matrix approach and then sum over all modes. To do that, let us provide firstly a
	brief review of Nielsen's geometric approach for evaluating circuit complexity. This geometric approach is a base for a group theoretic perspective to calculate the complexity of a state that is named as covariance matrix approach. As said before, the state complexity is defined as the minimal number of unitary gates required to prepare a certain target state $\mid\hspace{-1.5mm}\Psi_{T}\rangle$ from a specific reference state $\mid\hspace{-1.5mm}\Psi_{R}\rangle$, up to an error of $\epsilon$. The reference state $\mid\hspace{-1mm}\Psi_{R}\rangle$ is actually a factorizable state in position space which means that it is an unentangled state. Practically, the reference state is the ground state of the Hamiltonian
	\bea
	\label{HR}
	H_{R} = \int_{-\frac{L}{2}}^{\frac{L}{2}}\hspace{.5mm} dx \left(\frac{1}{2}\Pi^{2}(x)+\frac{1}{2}\omega_{R}^{2}\hspace{.5mm}\Phi^{2}(x) \right) = \displaystyle\sum_{k=0}^{N-1}\left(\frac{\delta}{2}\hspace{.5mm}{|\tilde{\mathbf{p}}_k|}^2+\frac{\omega_R^2}{2\delta}\hspace{.5mm} {|\tilde{\mathbf{x}}_k|}^2\right).
	\eea
	Nielsen and collaborators  \cite{Nielsen:2005,Nielsen:2006,Nielsen:2007} introduced a geometric approach to identify the optimal unitary transformation $\hat{U}$ between reference and target state
	\bea\label{u}
	\mid\hspace{-1mm}\Psi_{T}\rangle=\hat{U} \mid\hspace{-1mm}\Psi_{R}\rangle,
	\eea
	as a string of continuous unitary operators
	\bea\label{U}
	\hat{U} = \mathcal{P}\hspace{1mm} e^{-i \int_{0}^{1} dw\hspace{1mm} \hat{H}(w)},
	\eea
	where the path-dependent Hamiltonian $H(w)$ is expanded in terms of a basis of Hermitian operators $\hat{K}_{I}$ as following
	\bea
	\label{YI}
	\hat{H}(w) = \sum _{I} Y^{I}(w)\hspace{.2mm}\hat{K}_{I},
	\eea
	and $\mathcal{P}$ indicates the path ordering operator. It is worth noting that in the case of Gaussian states and with the mentioned gates, one does not need to consider a tolerance $\epsilon$, since the $Y_{I}(w)$ can always be adjusted to produce exactly the desired target state. Moreover, using this framework, one can  consider trajectories in the space of unitaries
	\bea\label{Us}
	\hat{U}(s) = \mathcal{P} \hspace{1mm} e^{-i \int _{0}^{s} dw \hspace{1mm}\hat{H}(w)},
	\eea
	with the boundary conditions $\hat{U}(s=1)=\hat{U}$,  $\hat{U}(s=0)= 1$
	and by that interprets  $Y_{I}(s)$ as
	the tangent vector of the the corresponding trajectory
	\bea
	Y^{I}(s)\hat{\mathcal{O}}_{I} = \left(\partial_{s}\hat{U}(s)\right)\hat{U}^{-1}(s).
	\eea
	Now the state complexity is defined by the value of particular minimized cost $\mathcal{C}(U)$ defined by
	\bea
	\mathcal{C}(U) = \int _{0}^{1} ds \hspace{1mm}F\left(U(s),\overrightarrow{Y}(s)\right),
	\eea 
	where the cost function $F\left(U(s),\overrightarrow{Y}(s)\right)$ is a local functional along the trajectory of $U(s)$. This cost function has different forms, but the one which is the main focus in this paper is \cite{Jefferson:2017sdb}
	\bea
	\label{Fkappa}
	F_{\kappa} (U,\overrightarrow{Y}) = \sum_{I} \mid\hspace{-1mm} Y^{I}\hspace{-1mm}\mid^{\kappa}.
	\eea
	Intriguingly, in applying the above framework to a free scalar QFT in \cite{Jefferson:2017sdb}, a group theoretic structure was found to appear naturally. To see this structure, let us consider a bosonic system with $N$ degrees of freedom. This system can be described by $2N$ observables $\hat{\xi} = (\hat{q}_1, \hat{q}_2,...,\hat{q}_N,\hat{p}_1 ,..., \hat{p}_N)$ with $(\hat{q}_i,\hat{p}_i)$ are canonical operators. The two-point functions of these observables in an arbitrary state $\mid\hspace{-1mm}\Psi\rangle$ can be expressed as
	\bea
	\label{Gab}
	\langle\Psi\hspace{-1mm}\mid \hspace{-1mm}\hat{\xi}^a\hspace{.5mm}\hat{\xi}^b\hspace{-1mm}\mid\hspace{-1mm} \Psi\rangle=\frac{1}{2}\left(G^{a b}+i\Omega^{a b}\right),
	\eea
	where $G^{a b}=G^{(a b)}$ is the symmetric part of the correlation matrix and $\Omega^{ab}=\Omega^{[a b]}$ denotes the antisymmetric part. In fact, for a system with bosonic degrees of freedom, $\Omega^{ab}$ is trivial and simply contains the information about commutation relations of $\hat{q}_{i}$ and $\hat{p}_{i}$. Restricting to the space of Gaussian states implies that the unitary operator $\hat{U}(s)$ can be expressed by Hermitian operators that are quadratic in the canonical operators $\hat{\xi}$,
	\bea
	\label{hatK}
	\hat{U}(s) = e^{-i s \hat{K}},\hspace{1cm}\text{with}\hspace{.5cm}\hat{K} =\frac{1}{2}\hat{\xi}^{a}\hspace{.5mm} k_{(a,b)}\hspace{.5mm}\hat{\xi}^{b} \equiv \frac{1}{2}\hat{\xi}\hspace{.5mm} k\hspace{.5mm} \hat{\xi}^{T}.
	\eea
	To proceed further, one needs the operation of $\hat{U}(s)$ on $\hat{\xi}^{a}$ which can be obtained as follows
	\bea
	\hat{U}^{\dagger}(s)\hspace{.5mm}\hat{\xi}^{a}\hspace{.5mm}\hat{U}(s) = \sum_{n=0}^{\infty} \frac{s^n}{n!} [i \hat{K},\hat{\xi}^{a}]_{(n)},
	\eea
	with $[i \hat{K},\hat{\xi}^{a}]_{(n)}$ is defined recursively by $[i \hat{K},\hat{\xi}^{a}]_{(n)}=[i\hat{K}, [i \hat{K},\hat{\xi}^{a}]_{(n-1)}]$, and $[i \hat{K},\hat{\xi}^{a}]_{(0)}=[i \hat{K},\hat{\xi}^{a}]$. Using (\ref{hatK}) and the commutation relation $[\hat{\xi}^{a},\hat{\xi}^{b}] =i\Omega^{a b}$, it is easy to see that
	\bea
	[i \hat{K},\hat{\xi}^{a}] = \Omega^{a b}\hspace{.5mm} k_{(b, c)}\hspace{.5mm}\hat{\xi}^{c},
	\eea
	which by defining $K^{a}_{b}= \Omega^{a c}\hspace{.5mm} k_{(c,b)}$, it
	can be written as
	\bea\label{Kxi}
	[i \hat{K},\hat{\xi}^{a}] = K^{a}_{b}\hspace{.5mm}\hat{\xi}^{b}.
	\eea
	This latter identity implies that
	\bea
	\label{UdxiU}
	\hat{U}^{\dagger}(s)\hspace{.5mm}\hat{\xi}^{a}\hspace{.5mm}\hat{U}(s) = (e^{s K})^{a}_{b}\hspace{1mm}\hat{\xi}^{b}\equiv U(s)^{a}_{b}\hspace{1mm}\hat{\xi}^{b}.
	\eea
	Now, (\ref{hatK}) together with (\ref{UdxiU}) imply that the covariance matrix for $\mid\hspace{-1mm}\Psi_{\text{G}}(s)\rangle$ becomes
	\bea
	\label{GsG0}
	&& G_{s}^{(a,b)} = 
	U(s)^{a}_{c} \hspace{1mm}G_{0}^{(c,d)} \hspace{1mm}U(s)^{b}_{d}.
	\eea
	Eventually, the complexity of a target state in a basis independent way can be achieved by defining the relative covariance matrix
	\bea\label{Delta}
	\Delta^{a}_{b} = G_{T}^{(a,c)} \hspace{1mm}G^{-1}_{R \hspace{.5mm}(c,b)},
	\eea  
	and make a proper choice for the cost function which by that, the complexity only depends on the eigenvalues of this matrix. One of these choices is the $\kappa =2$ in (\ref{Fkappa}). Indeed one can choose a basis such that the $G_{R}$ becomes identity and then diagonalize $G_{T}$ using only transformations within the stabilizer subgroup $Sta_{G_{R}}$. Then $\kappa=2$ complexity is given by
	\bea
	\label{ck2}
	\mathcal{C}_{\kappa=2}\hspace{.5mm}(G_{T},G_{R}) =\frac{1}{4}\text{Tr} \left[\mid\hspace{-.5mm}\log\Delta\hspace{-.6mm}\mid^2\right].
	\eea 
	It is worth to mention that the $\mathcal{C}_{\kappa=1} = \frac{1}{2}\text{Tr} \left[\mid\hspace{-.5mm}\log\Delta\hspace{-.6mm}\mid\right]$ in general is different from the complexity obtained from 
	$F_1$ in (\ref{Fkappa}). The reason for that is $F_1$ can not be found from any well-defined metric. Despite $C_{\kappa=2}$, choosing the gates matters for $F_1$. One way to estimate the length of shortest path between the two states is to consider orthonormal symplectic group generators (with respect to Frobenius inner product) as our gates and calculate the length of the path that minimizes $F_2$ cost function, i.e. the linear path. Since this path is not necessary the minimal geodesic of $F_1$, then
	\begin{equation}
	\mathcal{C}_1 \leq \sum_{I} |Y^I|.
	\end{equation}
	From now on, we take this upper bound as $\mathcal{C}_1$. Intriguingly, in the case of ground state of free bosonic quantum field theory, this approach gives the same answer as covariance matrix approach \cite{Jefferson:2017sdb}. Last but not least, we need the dimensionless control functions to add them together. This can be achieved by introducing the new dimensionless position and momentum coordinates according to
	\bea
	\label{xpnew}
	x_{\text{new},\pm} = g_{s}\hspace{.5mm} x_{\pm},\hspace{1cm}p_{\text{new},\pm} = \frac{p_{\pm}}{g_s},
	\eea
	where $g_{s}$ is a new gate scale. 
	Accordingly, for each mode we have
	\bea\label{newlambdas}
	\lambda_{R,k} \equiv \lambda_{R} =\frac{\omega_{R}}{\delta\hspace{.5mm}g_{s}^2},\hspace{1cm}\lambda_{k} =\frac{\omega_{k}}{\delta\hspace{.5mm}g_{s}^2}.
	\eea
	Let us write the state (\ref{TCTFD.3}) as following
	\bea
	\label{TCTFD.4}
	\mid\hspace{-1mm} \text{cTFD}(t)\rangle= \hat{U}_+(1)\hspace{-1mm}\mid \hspace{-1mm}G_0\rangle_+\otimes \hat{U}_-(1)\hspace{-1mm}\mid \hspace{-1mm}G_0\rangle_-=e^{-i\hat{K}_+}\mid \hspace{-1mm}G_0\rangle_+ \otimes e^{-i\hat{K}_-}\mid \hspace{-1mm}G_0\rangle_-.
	\eea
	By choosing $\hat{\xi}_{\pm} =(\hat{x}_{\text{new},\pm},\hat{p}_{\text{new},\pm})$ and noting to (\ref{opm}), the unitary matrices $k_{(a,b),+}$ and $k_{(a,b),-}$ are given by 
	\bea 
	k_{(a,b),+}=
	\alpha\begin{pmatrix}
		\lambda\sin\big((\omega+\mu q) t\big) & \cos\big((\omega+\mu q) t\big) \\
		\cos\big((\omega+\mu q) t\big) &  -\frac{1}{ \lambda}\sin\big((\omega+\mu q) t\big).
	\end{pmatrix},\hspace{1cm}k_{(a,b),-} = k_{(a,b),+} (\alpha \rightarrow -\alpha),
	\eea
	with
	\bea\label{lambda}
	\lambda= m\omega/g_s^2.
	\eea 
	The commutator $[\hat{\xi}^{a}_{\pm},\hat{\xi}^{b}_{\pm}] =i\Omega^{a b}_{\pm}$ implies that 
	\bea
	\Omega_+^{a b}=\Omega_-^{a b}=
	\begin{pmatrix}
		0 &1 \\
		-1& 0  \\
	\end{pmatrix},
	\eea
	and by that matrices $K_{\pm}$ become respectively,
	\bea\label{Km}
	{K}_{+}=
	\alpha\begin{pmatrix}
		\cos\big((\omega+\mu q) t\big) & -\frac{1}{ \lambda}\sin\big((\omega+\mu q) t\big)\\
		-\lambda\sin\big((\omega+\mu q) t\big) & -\cos\big((\omega+\mu q) t\big)
	\end{pmatrix},\hspace{1cm} {K}_{-} = {K}_{+} (\alpha \rightarrow -\alpha).
	\eea
	Exponentiation of matrices $K_{\pm}$ gives the unitary matrices $U_{\pm}(1)$,
	\bea\label{Upm}
	U_{+}(1)=
	\begin{pmatrix}
		u_{11}&u_{12}\\
		u_{21} &u_{22}\\
	\end{pmatrix},\hspace{1cm} U_{-}(1) = U_{+}(1) \left(\alpha \rightarrow -\alpha\right),
	\eea
	with
	\bea
	&& u_{11}= \cosh(\alpha)+\cos\big((\omega+\mu q) t\big) \sinh(\alpha),
	\hspace{1cm}u_{22}= u_{11} (\alpha\rightarrow -\alpha),
	\cr\nonumber\\
	&&u_{21}=\lambda^2 u_{12} = -\lambda \sin\big((\omega+\mu q) t\big) \sinh(\alpha).
	\eea
	Now, according to (\ref{Upm}) and noting that 
	\bea
	G_{0,+} = G_{0,-}=
	\begin{pmatrix}
		\frac{1}{\lambda}&0 \\
		0&\lambda
	\end{pmatrix},
	\eea
	the eq.(\ref{GsG0}) implies that
	\bea
	\label{GTpmcomplex}
	{G}_{\text{TFD},+}(t)=
	\begin{pmatrix}
		g_{11,+}& g_{12,+} \\
		g_{21,+}&g_{22,+}   
	\end{pmatrix},\hspace{1cm}
	{G}_{\text{TFD},-}(t) = {G}_{\text{TFD},+}(t) \left(\alpha \rightarrow -\alpha\right),
	\eea
	with
	\bea
	&&g_{11,+}= \frac{1}{\lambda}\bigg( \cosh(2\alpha)+\cos\big((\omega+\mu q) t\big) \sinh(2\alpha)\bigg),\hspace{.5cm} g_{22,+} = \lambda^2 g_{11,+}(\alpha\rightarrow -\alpha),
	\cr \nonumber\\
	&&g_{12,+}=g_{21,+}=-\sin\big((\omega+\mu q) t\big) \sinh(2\alpha) .
	\eea
	The last ingredient to construct the relative covariance matrix (\ref{Delta}) is the covariance matrix for the unentangled states  $\mid\hspace{-1mm}\Psi_{R,\pm}\rangle$. These states are the ground states of Hamiltonian (\ref{HR}) and their covariance matrices are given by 
	\bea
	\label{GRpm} 
	G_{R,+}=G_{R,-}=
	\begin{pmatrix}
		\frac{1}{\lambda_{R}}&0  \\
		0&\lambda_{R}  
	\end{pmatrix},
	\eea
	with 
	\bea
	\label{lambdaR}
	\lambda_{R} = m\omega_{R}/g_s^2.
	\eea
	Having all ingredients, (\ref{GTpmcomplex}) and (\ref{GRpm}), the relative covariance matrix (\ref{Delta}) for right and left moving modes is given by
	\bea
	\label{Deltapmcomplex} 
	\Delta_+(t)=
	\begin{pmatrix}
		\Delta_{11,+}&\Delta_{12,+}  \\
		\Delta_{21,+} &\Delta_{22,+}   
	\end{pmatrix},\hspace{1cm}\Delta_-(t) = \Delta_+(t) \left(\alpha \rightarrow -\alpha\right),
	\eea
	with
	\bea
	&&\Delta_{11_+}= \frac{\lambda_R}{\lambda}\bigg( \cosh(2\alpha)+\cos\big((\omega+\mu q) t\big) \hspace{.5mm}\sinh(2\alpha)\bigg),\hspace{.5cm}\Delta_{22_+} =\frac{\lambda^2}{\lambda_{R}^2} \Delta_{11_+} (\alpha \rightarrow -\alpha),
	\cr \nonumber\\
	&&
	\Delta_{21_+}= \lambda^2_{R} \hspace{1mm}\Delta_{12_+} = -\lambda_R\hspace{.5mm}\sin\big((\omega+\mu q) t\big) \hspace{.5mm}\sinh(2\alpha).
	\eea
	Finally, according to (\ref{ck2})  the $\kappa=2$ complexity for the single mode is given by
	\bea
	\label{ck2app1complex}
	\mathcal{C}_{\kappa=2} (t,q) =\frac{1}{4}\sum_{i=1}^{2}\bigg[\left(\log\Delta^{(i)}_{+}(t)\right)^2\hspace{.5mm}+\left(\log\Delta^{(i)}_{-}(t)\right)^2\bigg],
	\eea
	where $\Delta^{(i)}_{\pm}(t)$ are eigenvalues of $\Delta_{\pm}(t)$ matrices (\ref{Deltapmcomplex}),
	\bea
	\label{EigenDeltap}
	&&\Delta^{(1)}_{+}(t)=\frac{1}{A_1}\Bigg({\lambda}^2 A_2+\lambda^2_R \hspace{1mm}A_3-\bigg(\left(\lambda^2\hspace{1mm}A_2 +\lambda^2_R\hspace{1mm} A_3\right)^2-A^2_1\bigg) ^{\frac{1}{2}}\Bigg),
	\cr\nonumber\\
	&&\Delta^{(2)}_{+}(t)=\frac{1}{A_1}\Bigg({\lambda}^2 A_2+\lambda^2_R\hspace{1mm} A_3+\bigg(\left(\lambda^2\hspace{1mm}A_2 +\lambda^2_R\hspace{1mm} A_3\right)^2-A^2_1\bigg) ^{\frac{1}{2}}\Bigg),
	\cr\nonumber\\
	&&
	\Delta^{(i)}_{-}(t)= \Delta^{(i)}_{+}(t)\hspace{.5mm}\left(\alpha\rightarrow -\alpha\right),
	\eea
	with
	\bea
	&&A_1=2\lambda \hspace{.5mm}\lambda_{R},\hspace{.5cm}
	A_2=\cosh(2\alpha)-\cos\big((\omega+\mu q) t\big)\hspace{.5mm}\sinh(2\alpha),\hspace{.5cm}A_3 = A_2\hspace{.5mm}(\alpha\rightarrow -\alpha).
	\eea
	Having a complexity for single mode, (\ref{ck2app1complex}), and noting that each mode $k$ is decoupled from the other modes, the complexity of the ground state of Hamiltonian (\ref{Hcomplex3}) on a lattice with $N$ site is easily a sum over the complexity for each mode $k$,
	\bea
	\label{ckappa2complex}
	\mathcal{C}_{\kappa=2} (t,q) =\frac{1}{4}\sum_{k=0}^{N-1}\sum_{i=1}^{2}\bigg[\left(\log\Delta^{(i,k)}_{+}(t)\right)^2\hspace{.5mm}+\left(\log\Delta^{(i,k)}_{-}(t)\right)^2\bigg],
	\eea
	where $\Delta^{(i,k)}_{\pm}(t)$ are given by (\ref{EigenDeltap}) upon substituting $\omega$ with $\omega_{k}$. In the subsequent subsections \ref{L1} and \ref{L2} we study the complexity (\ref{ckappa2complex}) for the system with fixed size $L$. The same analysis for the continuum system will be presented in section \ref{conti}.  
	\subsection{Keeping the total size $L$ of the system fixed}\label{L1}
\begin{figure}[H]		
		
		\center{\includegraphics[scale=.245]{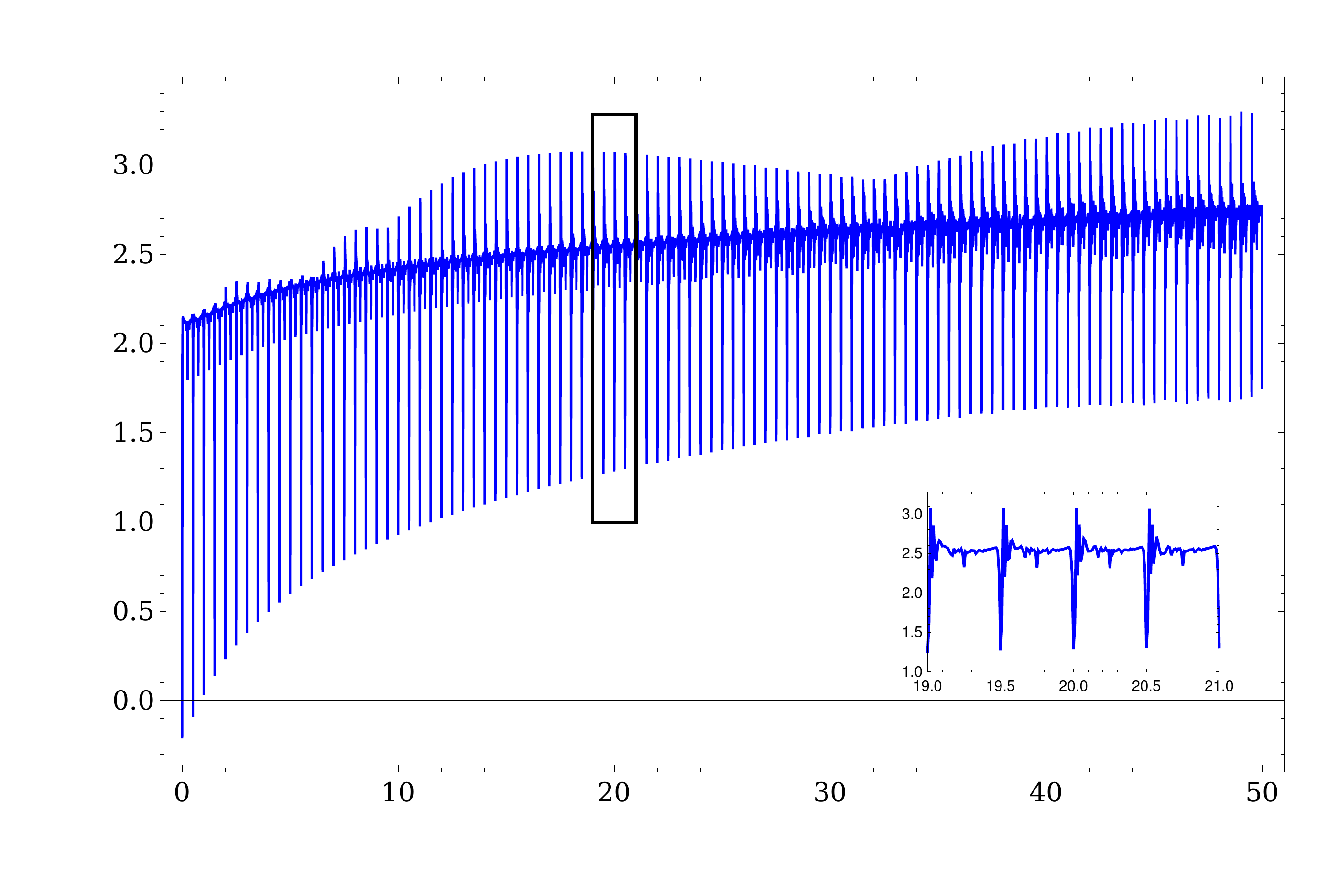}\put(-205,32){\rotatebox{-270}{\fontsize{15}{15}\selectfont $\frac{\mathcal{C}_{\kappa=2}(t)-\mathcal{C}_{\kappa=2}(0)}{S_{th}}$}}		\put(-97,0){{\fontsize{12}{12}\selectfont $t/L$}}
			\put(-105,125){{\fontsize{10}{10}\selectfont $\beta = 10^{-2} L$}}\hspace{1cm}\includegraphics[scale=.241]{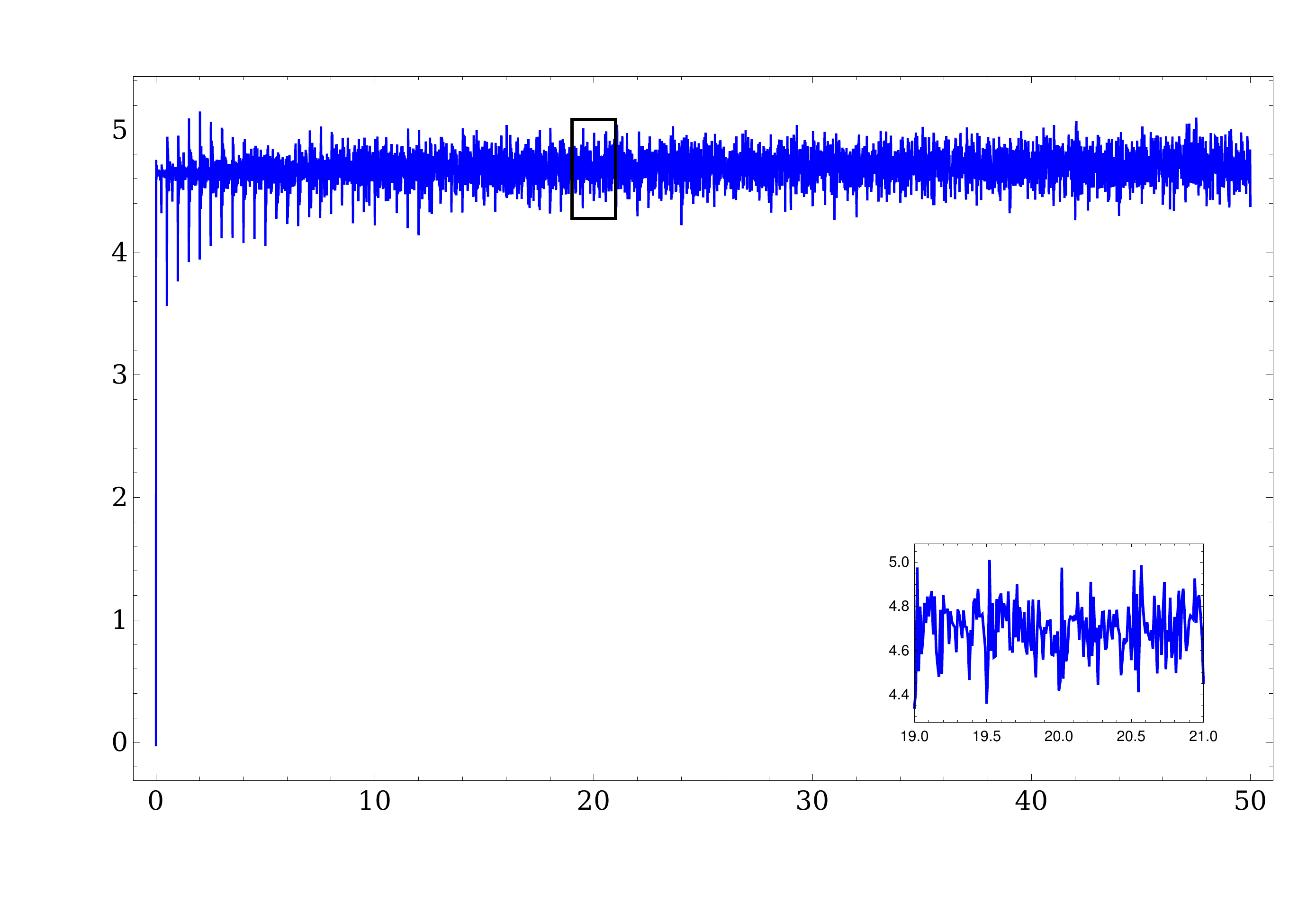}\put(-203,32){\rotatebox{-270}{\fontsize{15}{15}\selectfont $\frac{\mathcal{C}_{\kappa=2}(t)-\mathcal{C}_{\kappa=2}(0)}{S_{th}}$}}		\put(-97,0){{\fontsize{12}{12}\selectfont $t/L$}}
			\put(-110,125){{\fontsize{10}{10}\selectfont $\beta = 10^{-3} L$}}}
		\caption{All mode contribution to time dependence of ${C}_{\kappa=2}$ with the initial value subtracted for the neutral TFD at zero time on a circle with circumference $L$ with $\omega_{R} = 1/L$, $m = 10^{-5}/L$, $\mu q=10^{-5}/L$ and increasing temperatures $T = 1/\beta$. We use 1501 lattice sites on each side.}\label{discrete1}	
	\end{figure}

\begin{figure}[h]	\center{\includegraphics[scale=.25]{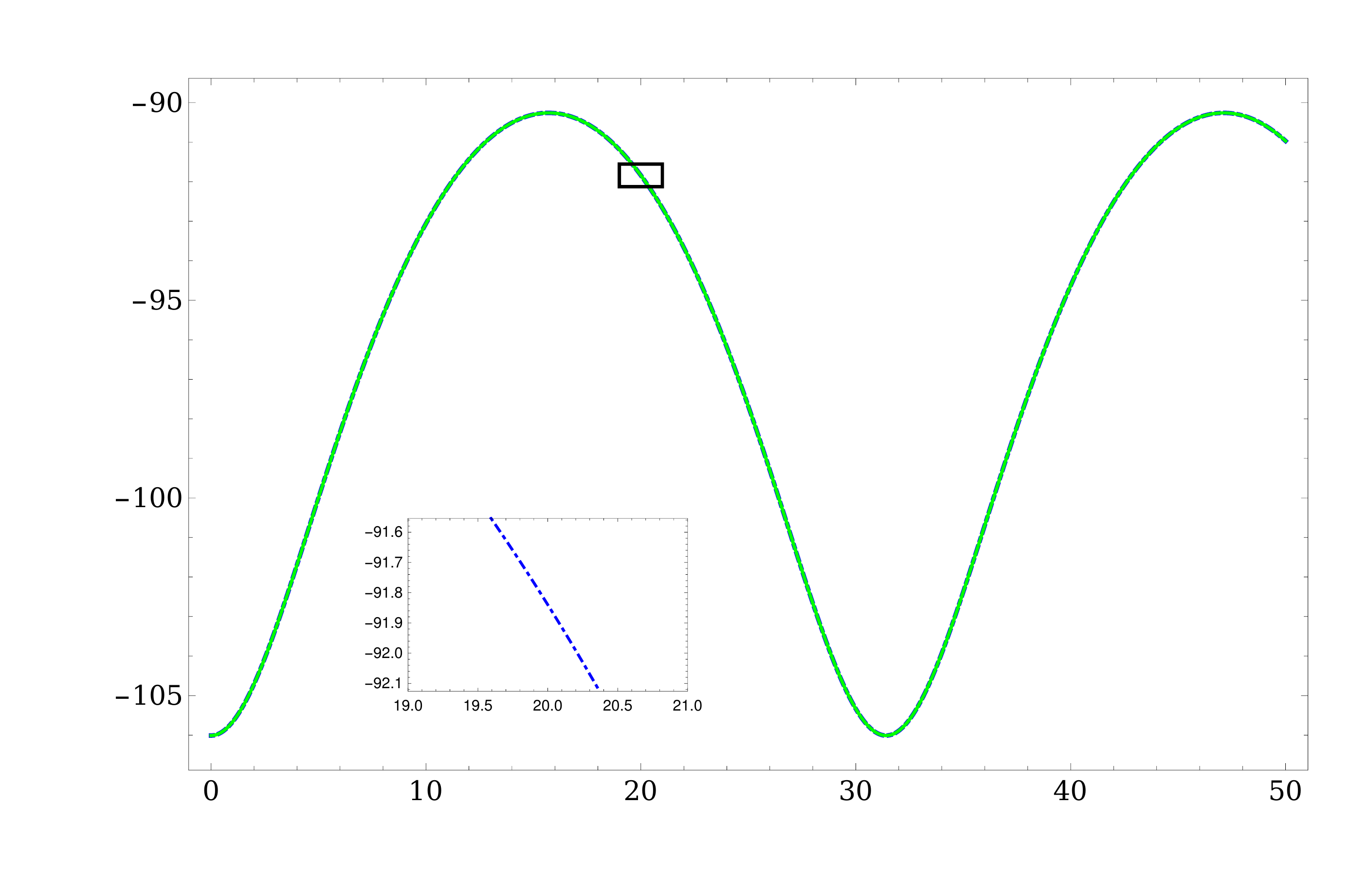}
			\put(-207,36){\rotatebox{-270}{\fontsize{15}{15}\selectfont $\frac{\mathcal{C}_{\kappa=2}(t)-\mathcal{C}_{\kappa=2}(0)}{S_{th}}$}}		\put(-100,0){{\fontsize{12}{12}\selectfont $t/L$}}
			\put(-110,125){{\fontsize{10}{10}\selectfont $\beta = 10 L$}}		
			\hspace{1cm}\includegraphics[scale=.248]{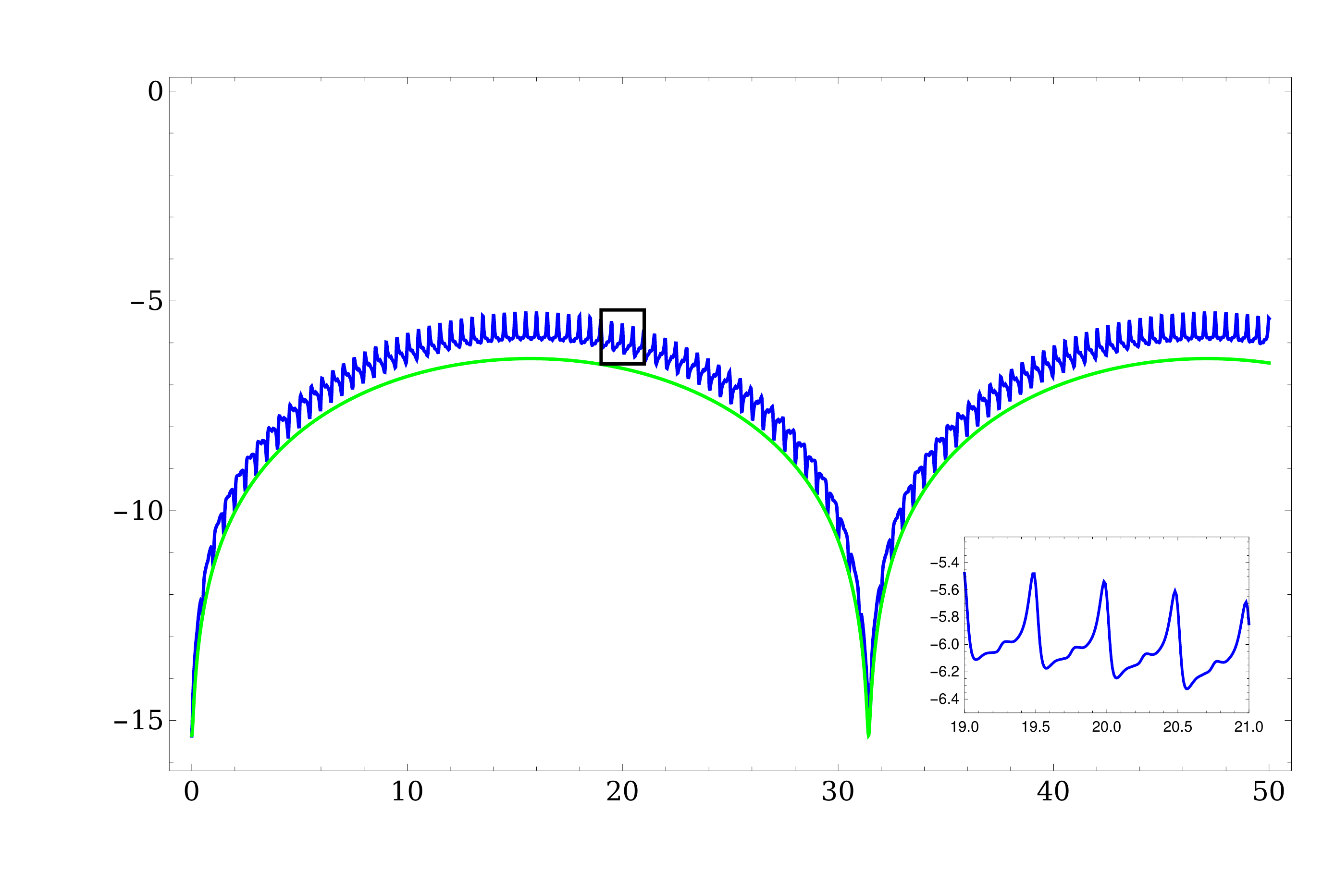}\put(-207,36){\rotatebox{-270}{\fontsize{15}{15}\selectfont $\frac{\mathcal{C}_{\kappa=2}(t)-\mathcal{C}_{\kappa=2}(0)}{S_{th}}$}}		\put(-100,0){{\fontsize{12}{12}\selectfont $t/L$}}
			\put(-110,125){{\fontsize{10}{10}\selectfont $\beta = 10^{-1} L$}}}\vspace{0cm}
		
		\center{\includegraphics[scale=.252]{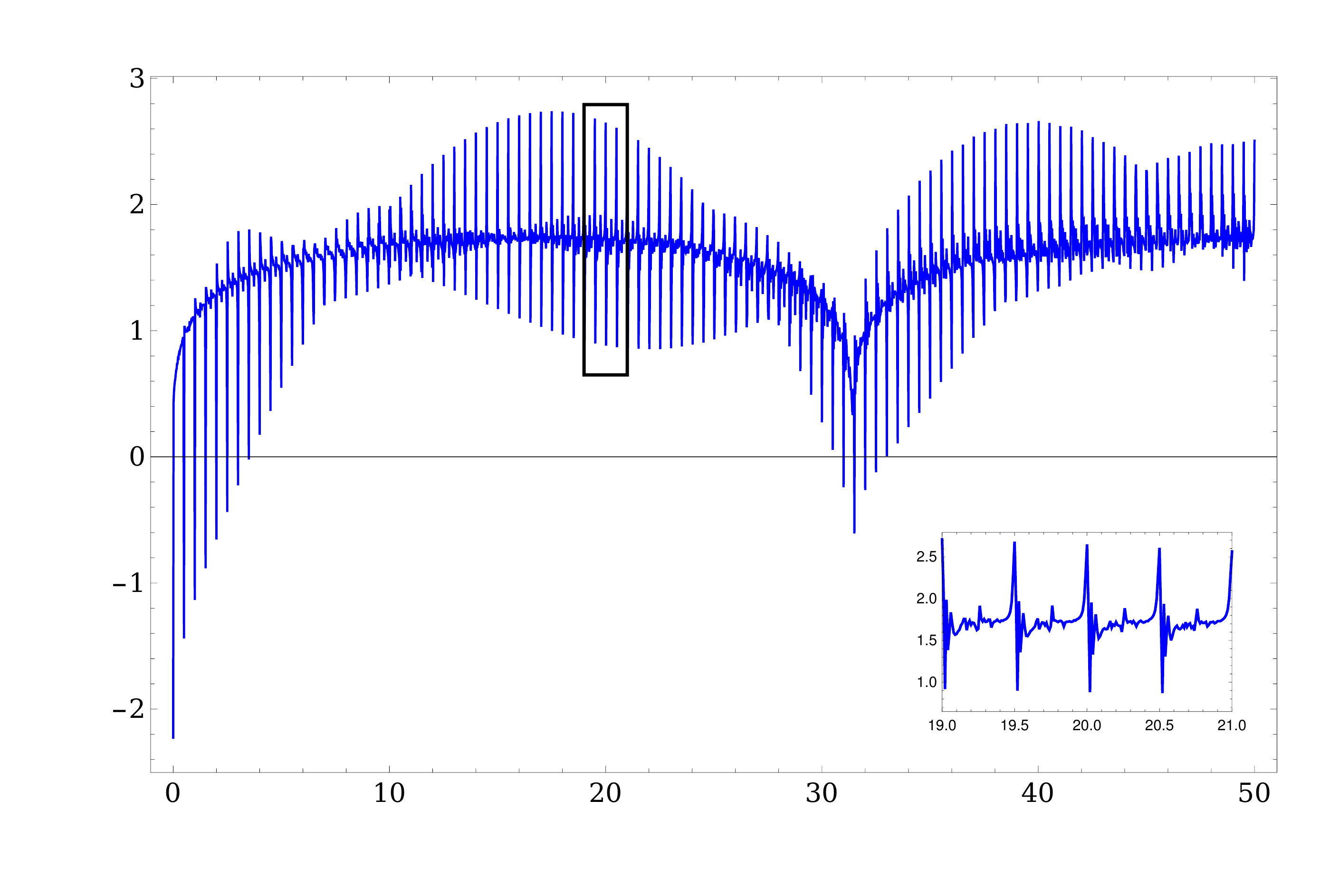}\put(-209,36){\rotatebox{-270}{\fontsize{15}{15}\selectfont $\frac{\mathcal{C}_{\kappa=2}(t)-\mathcal{C}_{\kappa=2}(0)}{S_{th}}$}}		\put(-100,0){{\fontsize{12}{12}\selectfont $t/L$}}
			\put(-110,129){{\fontsize{10}{10}\selectfont $\beta = 10^{-2} L$}}\hspace{1cm}\includegraphics[scale=.25]{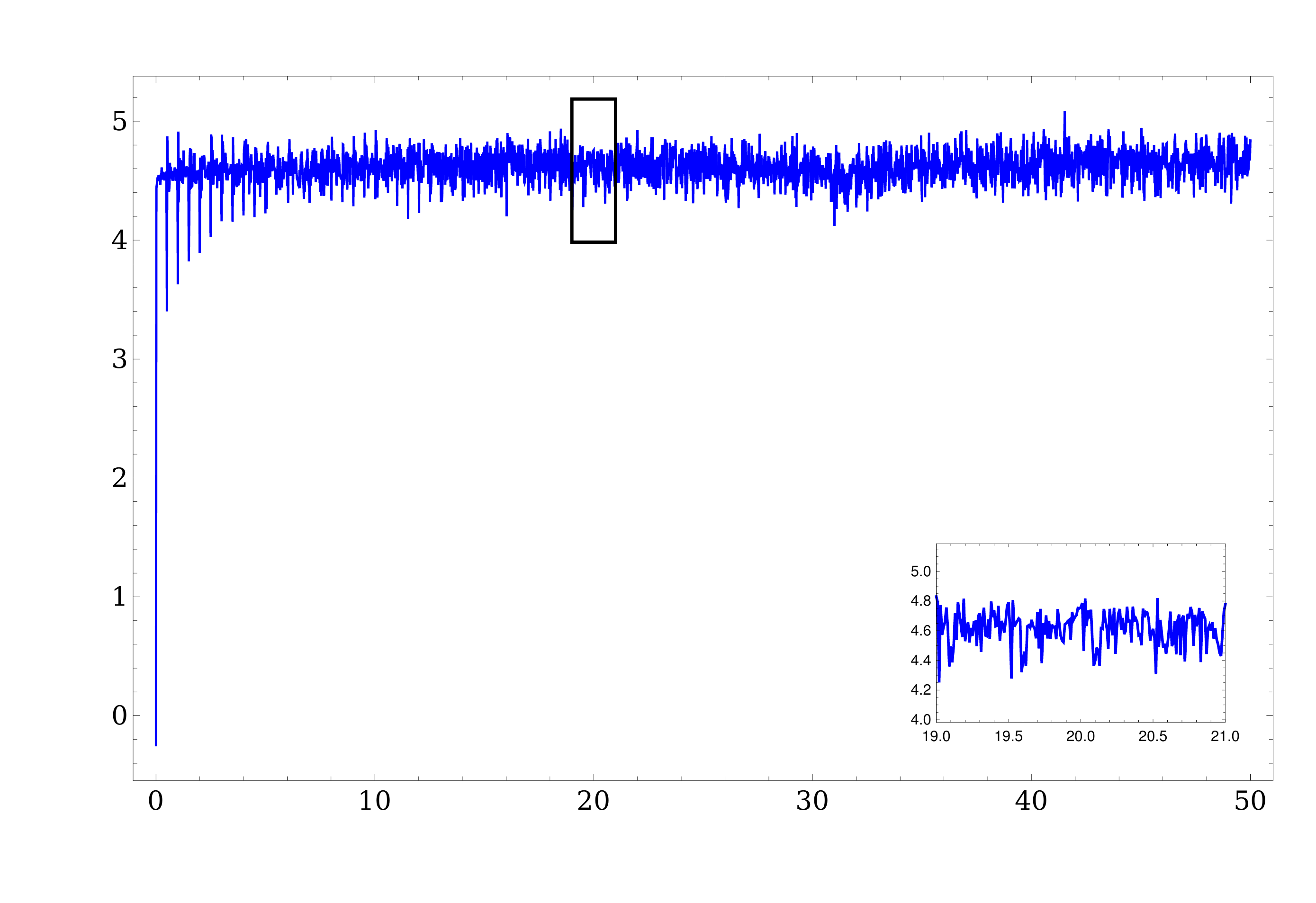}\put(-209,36){\rotatebox{-270}{\fontsize{15}{15}\selectfont $\frac{\mathcal{C}_{\kappa=2}(t)-\mathcal{C}_{\kappa=2}(0)}{S_{th}}$}}		\put(-100,0){{\fontsize{12}{12}\selectfont $t/L$}}
			\put(-110,129){{\fontsize{10}{10}\selectfont $\beta = 10^{-3} L$}}}
		\caption{All mode contribution to time dependence of ${C}_{\kappa=2}$ with the initial value subtracted for the neutral TFD at zero time on a circle with circumference $L$ with $\omega_{R} = 1/L$, $m = 10^{-5}/L$, $\mu q=10^{-1}/L$ and increasing temperatures $T = 1/\beta$. We use 1501 lattice sites on each side.}\label{discrete2}	
	\end{figure}
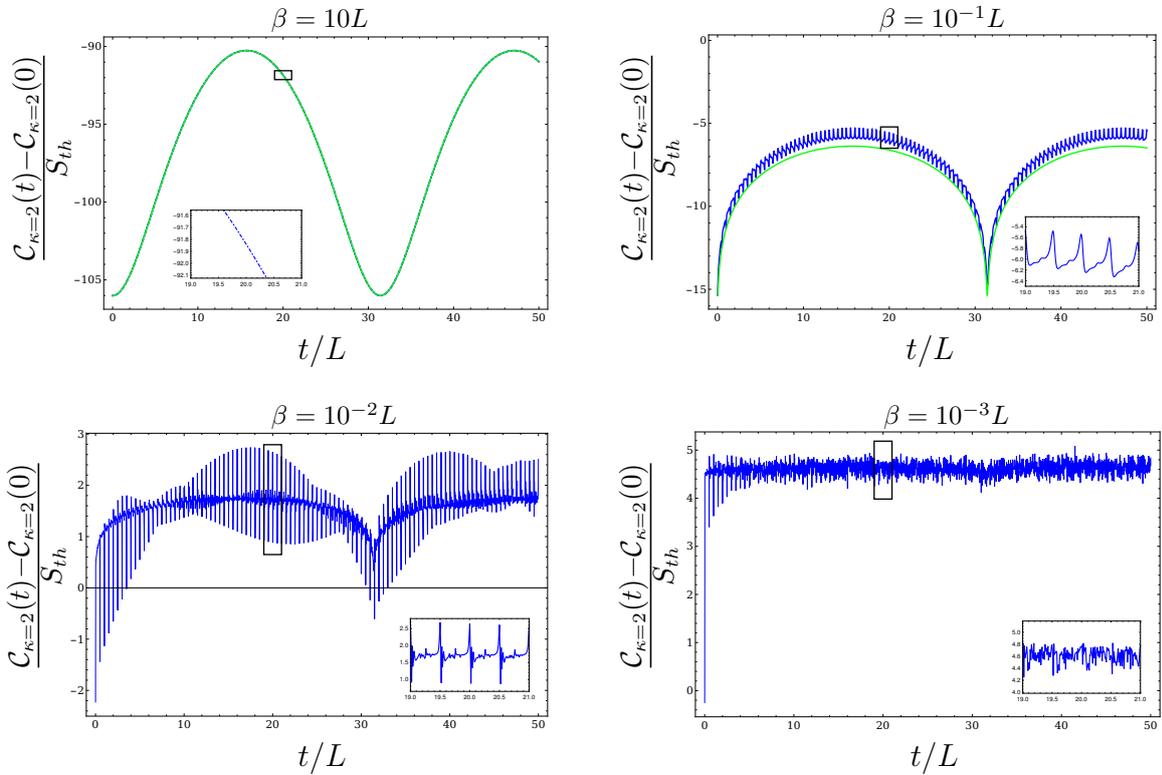

\begin{figure}[H]
		\hspace{1cm}\includegraphics[scale=.36]{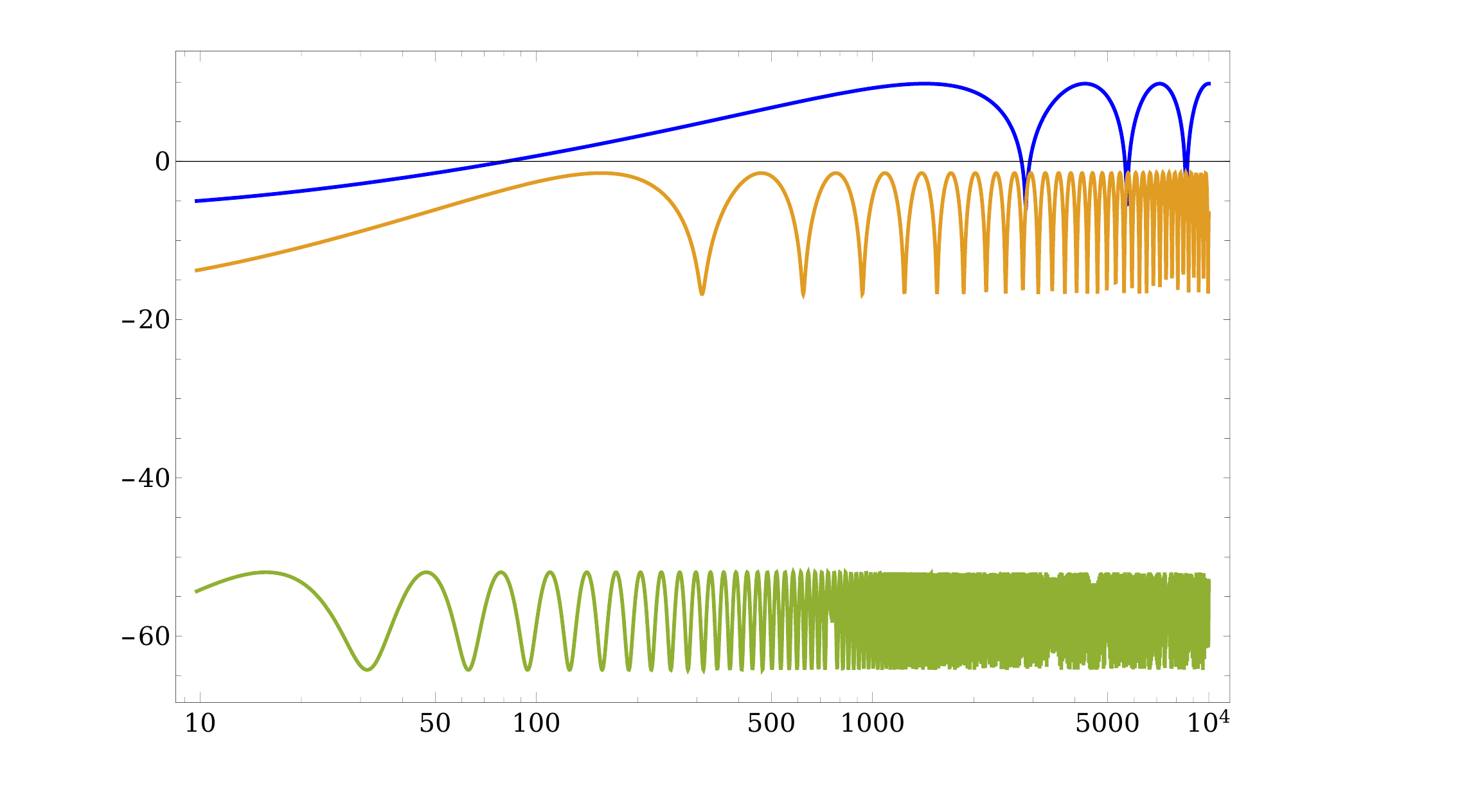}\put(-245,35){\rotatebox{-270}{\fontsize{16}{16}\selectfont $\frac{\mathcal{C}_{\kappa=2}(t)-\mathcal{C}_{\kappa=2}(0)}{S_{th}}$}}		\put(-130,0){{\fontsize{12}{12}\selectfont $t/L$}}
		\put(-140,125){{\fontsize{10}{10}\selectfont $\beta=10L$}}\hspace{-.8cm}
		\includegraphics[scale=.36]{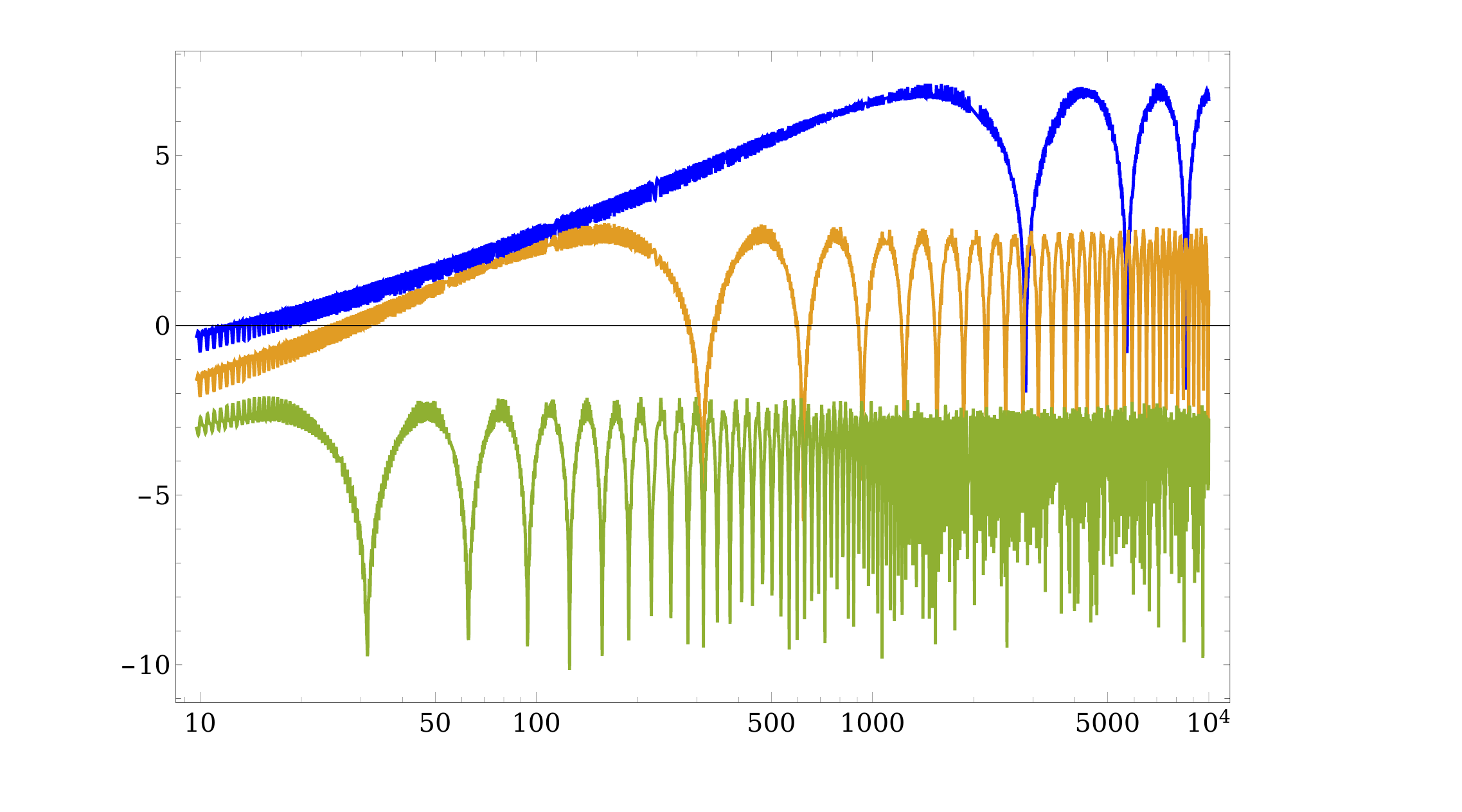}\put(-239,35){\rotatebox{-270}{\fontsize{16}{16}\selectfont $\frac{\mathcal{C}_{\kappa=2}(t)-\mathcal{C}_{\kappa=2}(0)}{S_{th}}$}}		\put(-130,0){{\fontsize{12}{12}\selectfont $t/L$}}
		\put(-140,125){{\fontsize{10}{10}\selectfont $\beta=10^{-1}L$}}
		\vspace{.25cm}
		
		\caption{Time dependence of $\kappa=2$ complexity with the initial value subtracted for TFD state on a circle with circumference $L$ with $\omega_{R} = 1/L$, $m=10^{-4}/L$, $\lambda_{R}=1$, $\beta = 10 L$ (\textbf{left}),
			$\beta = 10^{-1}L$ (\textbf{right}) and different increasing the chemical potential from $10^{-3}/L$ (blue) to $10^{-1}/L$ (green). We use 1501 lattice sites on each side. We see that complexity grows as $\log^2(t/L)$ up to times of the order of $1/(m+\mu q)$ when it starts oscillating around the saturated value. The zero mode is the source for this logarithmic growth but by increasing the temperature this behavior is lost.}\label{discrete3}
	\end{figure}

\begin{figure}[H]
		\hspace{1cm}\includegraphics[scale=.36]{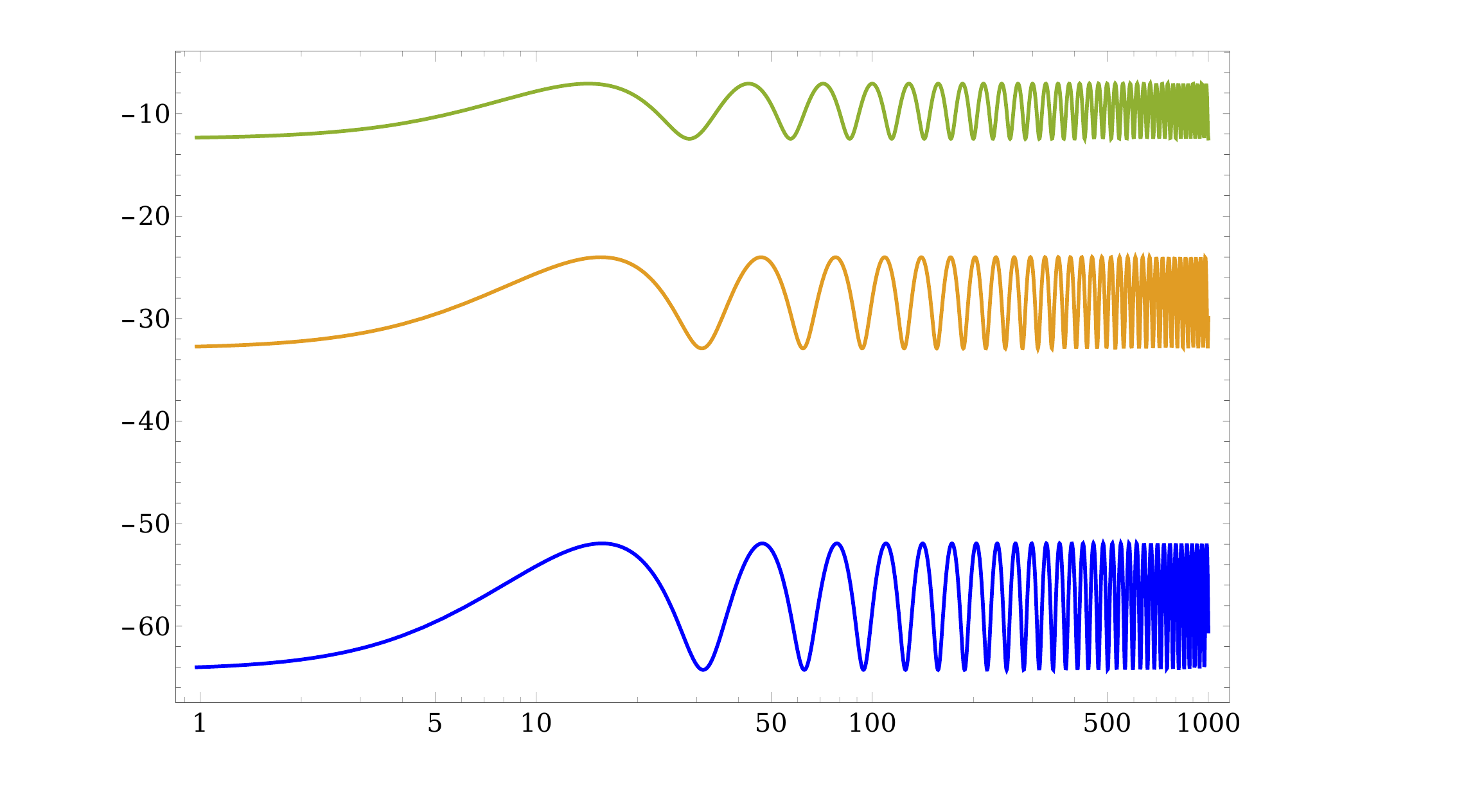}\put(-245,30){\rotatebox{-270}{\fontsize{17}{17}\selectfont $\frac{\mathcal{C}_{\kappa=2}(t)-\mathcal{C}_{\kappa=2}(0)}{S_{th}}$}}		\put(-125,0){{\fontsize{12}{12}\selectfont $t/L$}}
		\put(-140,125){{\fontsize{10}{10}\selectfont $\beta=10L$}}\hspace{-.6cm}
		\includegraphics[scale=.36]{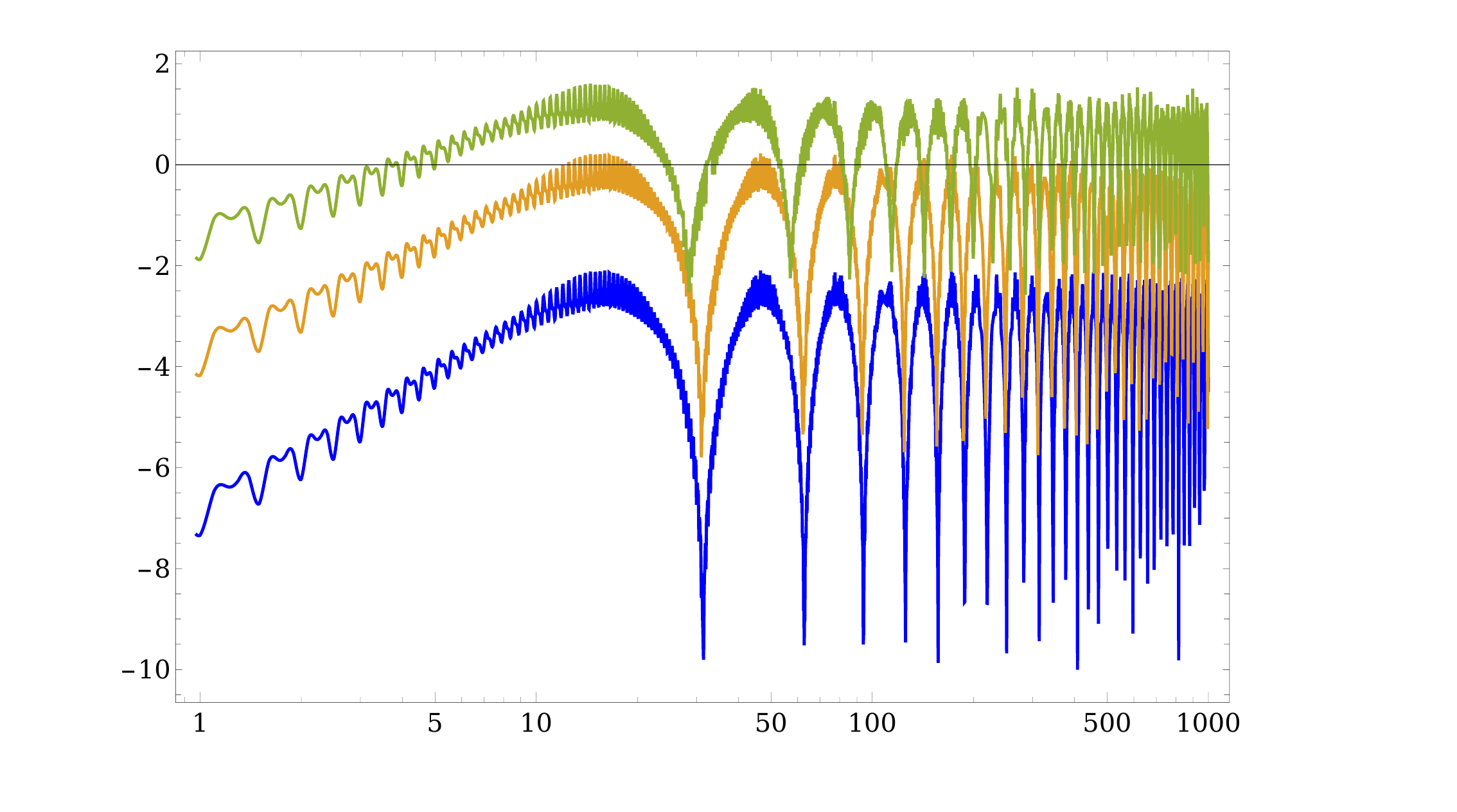}\put(-240,30){\rotatebox{-270}{\fontsize{17}{17}\selectfont $\frac{\mathcal{C}_{\kappa=2}(t)-\mathcal{C}_{\kappa=2}(0)}{S_{th}}$}}		\put(-130,0){{\fontsize{12}{12}\selectfont $t/L$}}
		\put(-140,125){{\fontsize{10}{10}\selectfont $\beta=10^{-1}L$}}
		\vspace{.25cm}
		
		\caption{Time dependence of $\kappa=2$ complexity with the initial value subtracted for TFD state on a circle with circumference $L$ with $\omega_{R} = 1/L$, $\mu q = 10^{-1}/L$, $\beta = 10 L$ (\textbf{left}), $\beta = 10^{-1}/L$ (\textbf{right}) and different increasing the masses, from $10^{-4}/L$ (blue) to $10^{-2}/L$ (green). We use 1501 lattice sites on each side. We see that complexity grows as $\log^{2}(t/L)$ up to times of the order of $1/(m+\mu q)$ when it starts oscillating around the saturated value. The zero mode is the source for this logarithmic growth but by increasing the temperature this behavior is lost.}\label{discrete4}
	\end{figure}

	To explore the consequences of (\ref{ckappa2complex}) for a system with fixed size $L$, we note that the only dimensionless parameters are $L/\beta$ and $\mu q L$. In figs.\ref{discrete1} and \ref{discrete2}, we set $\lambda_{R}=1$ and the parameter $\mu q L$ is fixed but the temperature is changed. It is explained above that we consider the mass parameter $m L$ very small. One may expect that for this case the main contribution to the complexity comes from the zero mode. In the following, we will show that, independent from the value of $\mu q L$, this expectation is correct just for low temperatures.  Indeed the contribution of the zero mode, in comparison with other modes, can be seen effectively in the limit $m \ll \omega_{R}$ where the eigenvalues (\ref{EigenDeltap}) for this mode can be simplified to
	\bea\label{a1}
	\Delta^{(1)}_{\pm}(t) \approx 0,
	\hspace{1.5cm}\Delta^{(2)}_{\pm}(t) \approx \frac{\omega_{R}}{m} \bigg(
	\cosh{2\alpha} \pm \cos\big[{(m+\mu q)t}\big] \sinh{2\alpha}\bigg).
	\eea
	In the limit $\beta (m+\mu q) \ll 1$, the above eigenvalues can be simplified more to  
	\bea\label{a2}
	\Delta^{(2)}_{+}(t) \approx \frac{4\omega_{R}}{\beta m (m+\mu q)} -\frac{\omega_{R}\hspace{.5mm}t^2}{\beta m} (m+\mu q),\hspace{1.5cm} \Delta^{(2)}_{-}(t) \approx \frac{\omega_{R}\hspace{.5mm}t^2}{\beta m}(m+ \mu q),
	\eea
	which by using them and for the times $(m+\mu q)t \ll 1$, the contribution of zero mode to $\kappa =2$ complexity (\ref{ck2app1complex}) becomes
	\bea
	\label{c2complexsimplezeromode}
	\mathcal{C}_{\kappa=2} \approx \frac{1}{4} \log^{2}\left[\frac{\omega_{R}(m+\mu q)}{\beta m} t^2\right]+\frac{1}{4} \log^{2}\left[\frac{4\omega _{R}}{\beta m (m+\mu q)}\right] .
	\eea
	This result is presented with the green curve in figs.\ref{discrete1} and \ref{discrete2} and it exhibits that the contribution of zero mode for small times, $(m+\mu q)t \ll 1$ is proportional to $a_1\log^{2}(a_2\hspace{.5mm}t^2)$. Besides that, it is clear from (\ref{EigenDeltap}) that the biggest contribution of the zero mode happens at the time $t=\pi/2(m +\mu q)$. This means that the maximum contribution of zero mode happens between times $0$ and $\pi/2(m +\mu q)$ where in the limits $(m+\mu q) \ll 1$ and $\beta(m+\mu q) \ll 1$  it becomes
	\bea
	\mathcal{C}_{\kappa=2}^{k=0} \hspace{.5mm}\big(\frac{\pi}{2(m+ \mu q)}\big) - \mathcal{C}_{\kappa=2}^{k=0}\hspace{.5mm}\big(0\big)\approx \frac{1}{2} \log^{2}\left[\frac{2\omega_{R}}{\beta m (m+\mu q)}\right] -\frac{1}{4} \log^{2}\left[\frac{4\omega_{R}}{\beta m(m+\mu q)}\right].
	\eea
	Even though the zero mode contribution diverges at those limits but by paying attention to the figs.\ref{discrete1} and \ref{discrete2}, it is clear that at the decompactification limit i.e. high temperatures and independently from the charge, the contribution of the other modes dominate. More precisely, for higher temperatures, we observe saturation which happens through the presence of many modes which they contribute non-trivially to the sum (\ref{ckappa2complex}). The transition between logarithmic growth regime and saturation regime  is oscillatory, which occurs with a period of half of the circle’s circumference, as if two wave packets were propagating on a circle in opposite directions, see figs.\ref{discrete3} and \ref{discrete4}. For example when $m\ll \omega_{R}$ and $m \ll \mu q$, for the zero mode we have
	\bea\label{aa2}
	\Delta^{(2)}_{\pm}(t) \approx \frac{\omega_{R}}{m}\hspace{1mm}\text{csch}(\frac{\beta \hspace{.5mm}\mu q}{2})\left(\cosh(\frac{\beta \hspace{.5mm}\mu q}{2})\pm \cos [\mu q\hspace{.5mm}t]\right).
	\eea
	\vspace{.5cm}
	
	\begin{figure}[H]	\center{\includegraphics[scale=.24]{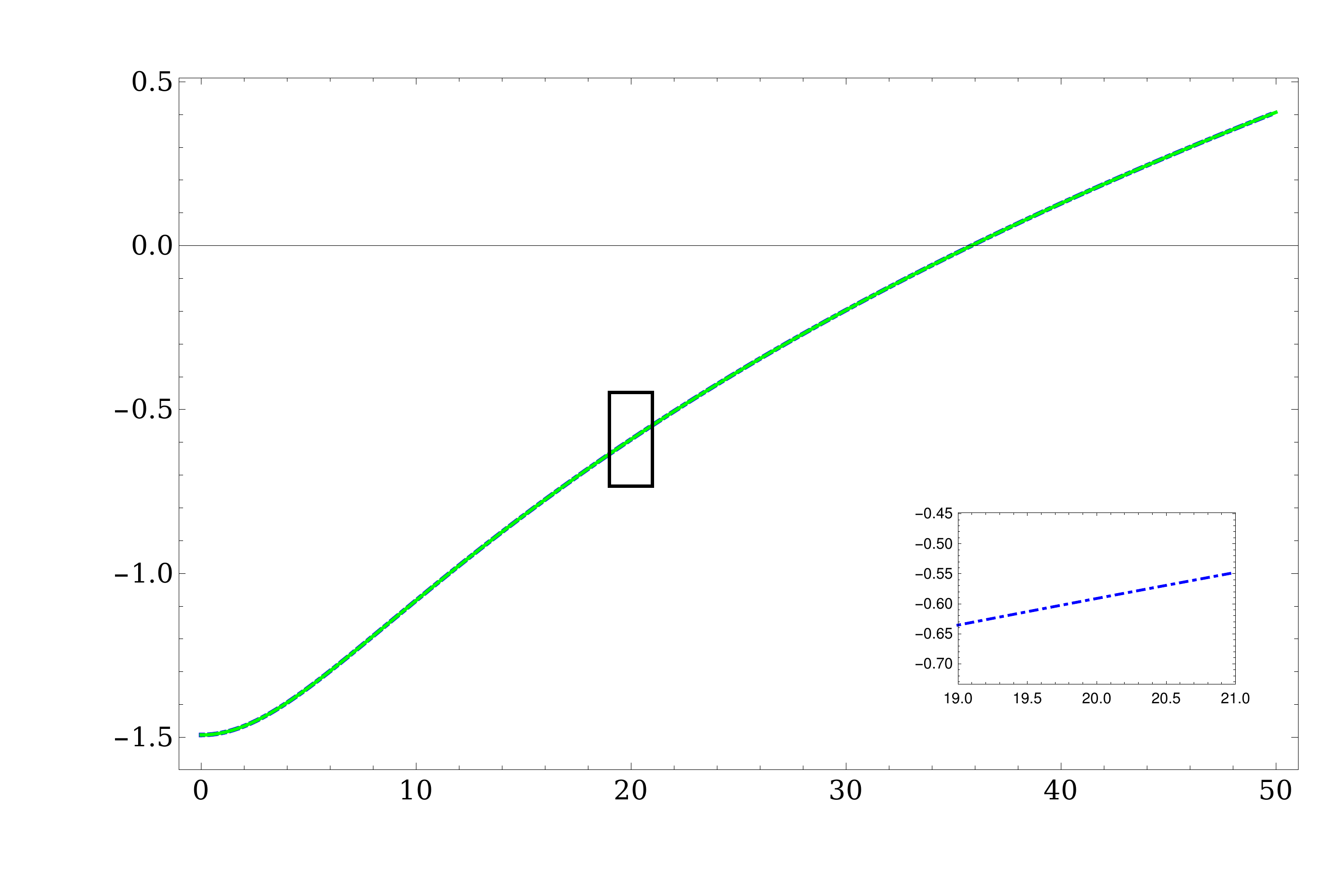}
			\put(-200,32){\rotatebox{-270}{\fontsize{15}{15}\selectfont $\frac{\mathcal{C}_{\kappa=2}(t)-\mathcal{C}_{\kappa=2}(0)}{S_{th}}$}}		\put(-95,0){{\fontsize{12}{12}\selectfont $t/L$}}
			\put(-105,120){{\fontsize{10}{10}\selectfont $\beta = 10 L$}}		
			\hspace{1cm}\includegraphics[scale=.24]{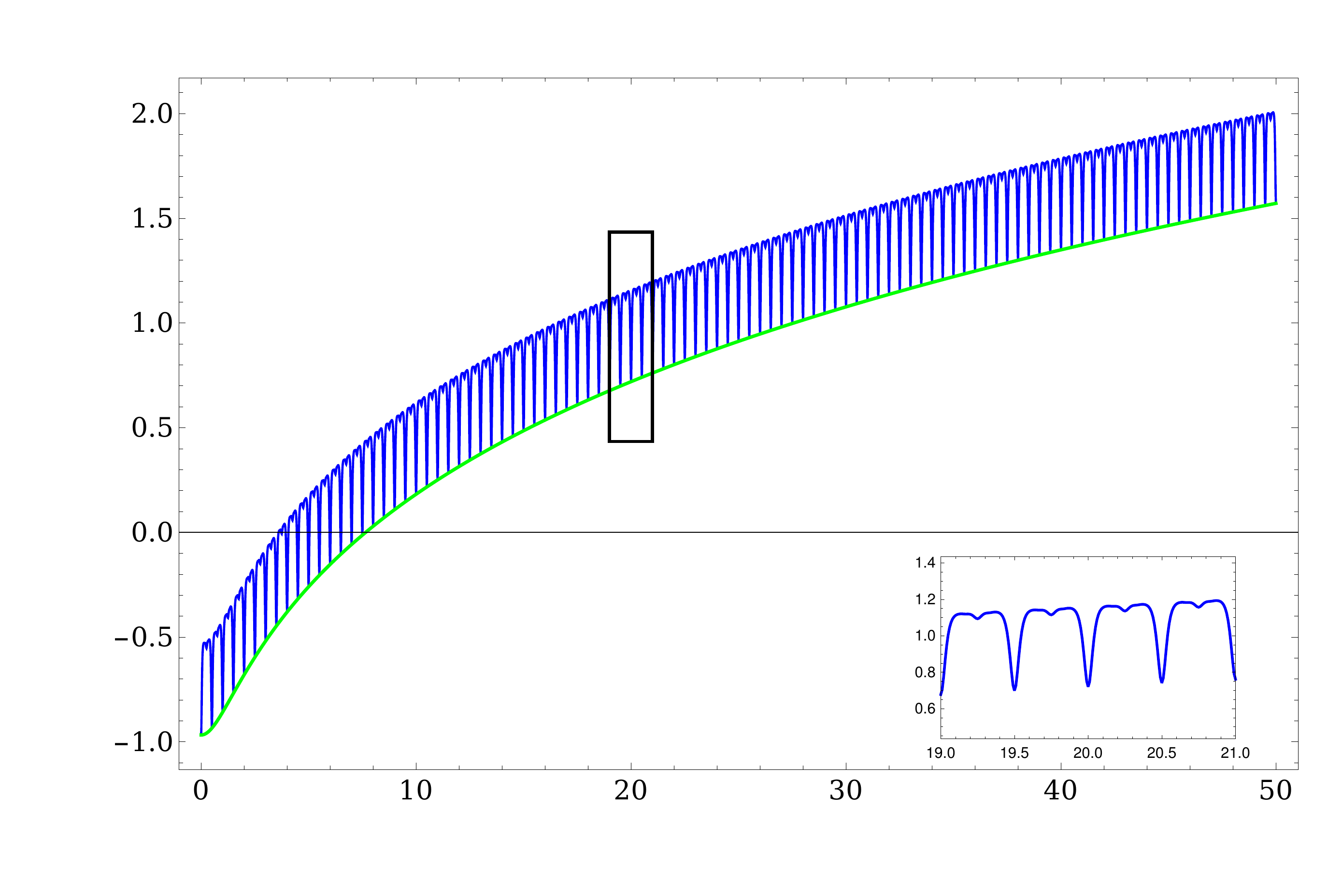}\put(-200,32){\rotatebox{-270}{\fontsize{15}{15}\selectfont $\frac{\mathcal{C}_{\kappa=2}(t)-\mathcal{C}_{\kappa=2}(0)}{S_{th}}$}}		\put(-95,0){{\fontsize{12}{12}\selectfont $t/L$}}
			\put(-105,120){{\fontsize{10}{10}\selectfont $\beta = 10^{-1} L$}}}\vspace{0cm}
	\end{figure}
\begin{figure}[H]
		\center{\includegraphics[scale=.26]{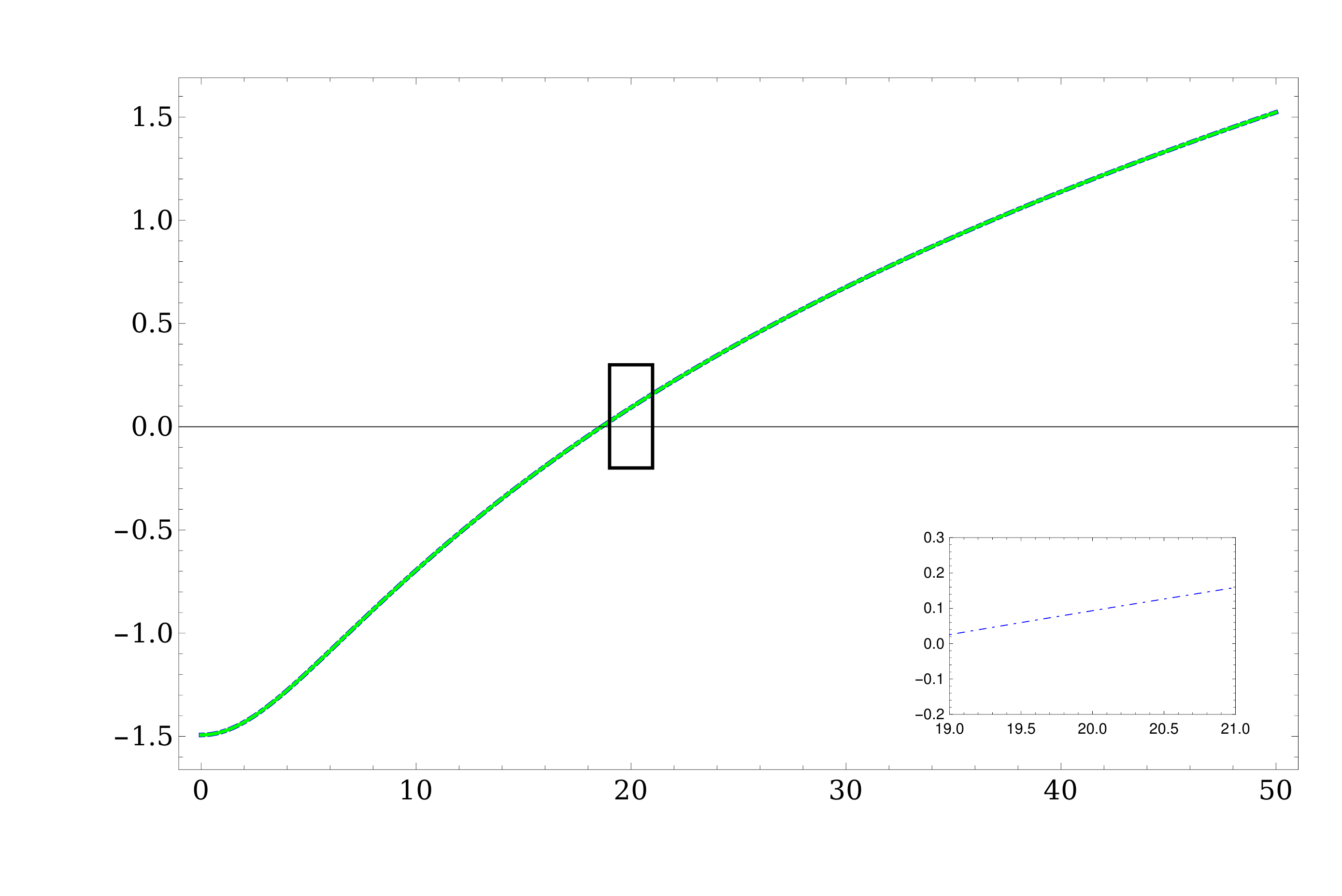}\hspace{.4cm}\put(-215,35){\rotatebox{-270}{\fontsize{15}{15}\selectfont $\frac{\mathcal{C}_{\kappa=2}(t)-\mathcal{C}_{\kappa=2}(0)}{S_{th}}$}}		\put(-105,5){{\fontsize{12}{12}\selectfont $t/L$}}
			\put(-125,132){{\fontsize{10}{10}\selectfont $\beta=10L,\hspace{1mm}\lambda_{R} = 10$}}\hspace{1cm}\includegraphics[scale=.26]{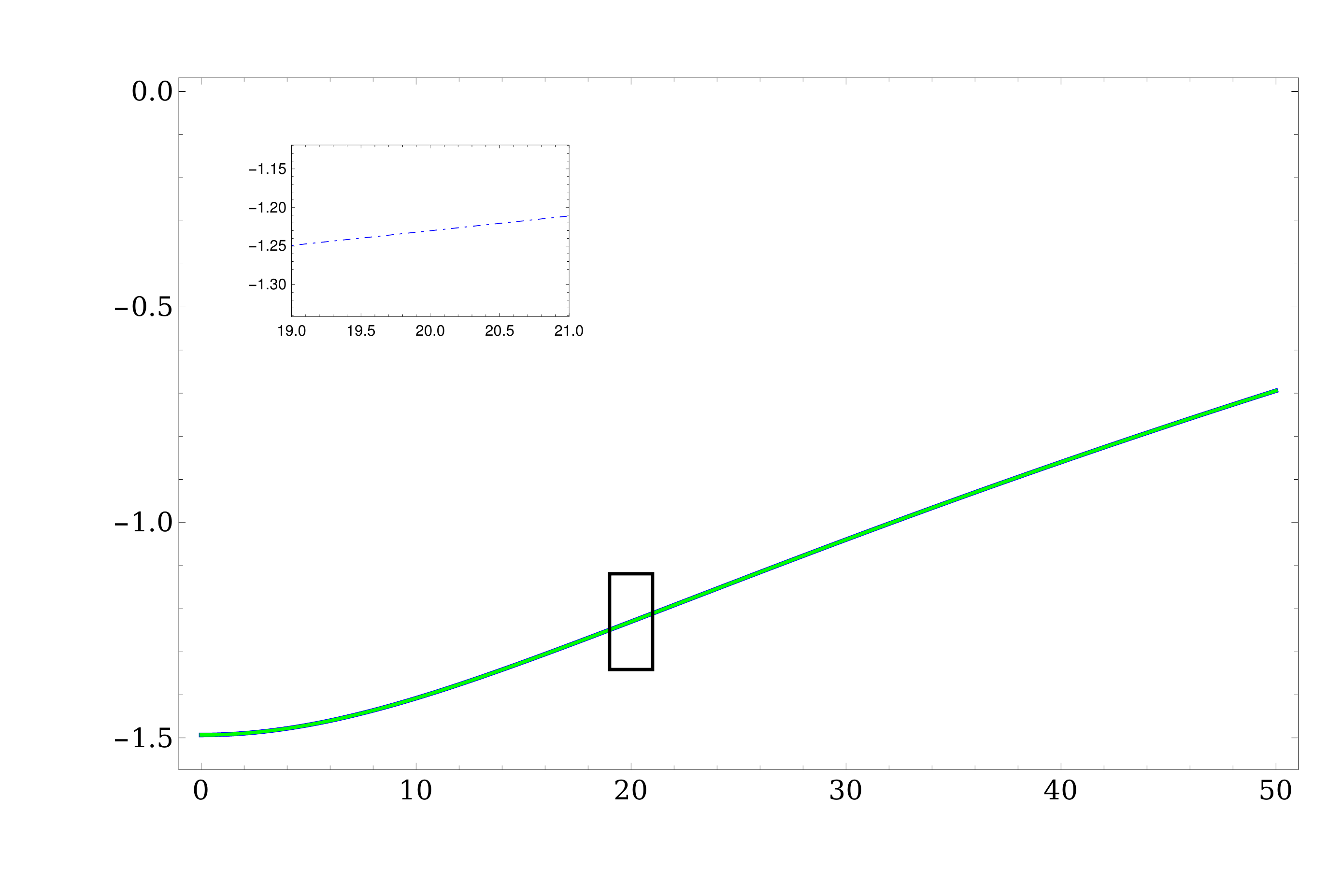}
			\put(-215,35){\rotatebox{-270}{\fontsize{15}{15}\selectfont $\frac{\mathcal{C}_{\kappa=2}(t)-\mathcal{C}_{\kappa=2}(0)}{S_{th}}$}}		\put(-105,5){{\fontsize{12}{12}\selectfont $t/L$}}
			\put(-125,132){{\fontsize{10}{10}\selectfont $\beta=10L,\hspace{1mm}\lambda_{R} = \frac{1}{10}$}}\vspace{.5cm}
			
			\includegraphics[scale=.26]{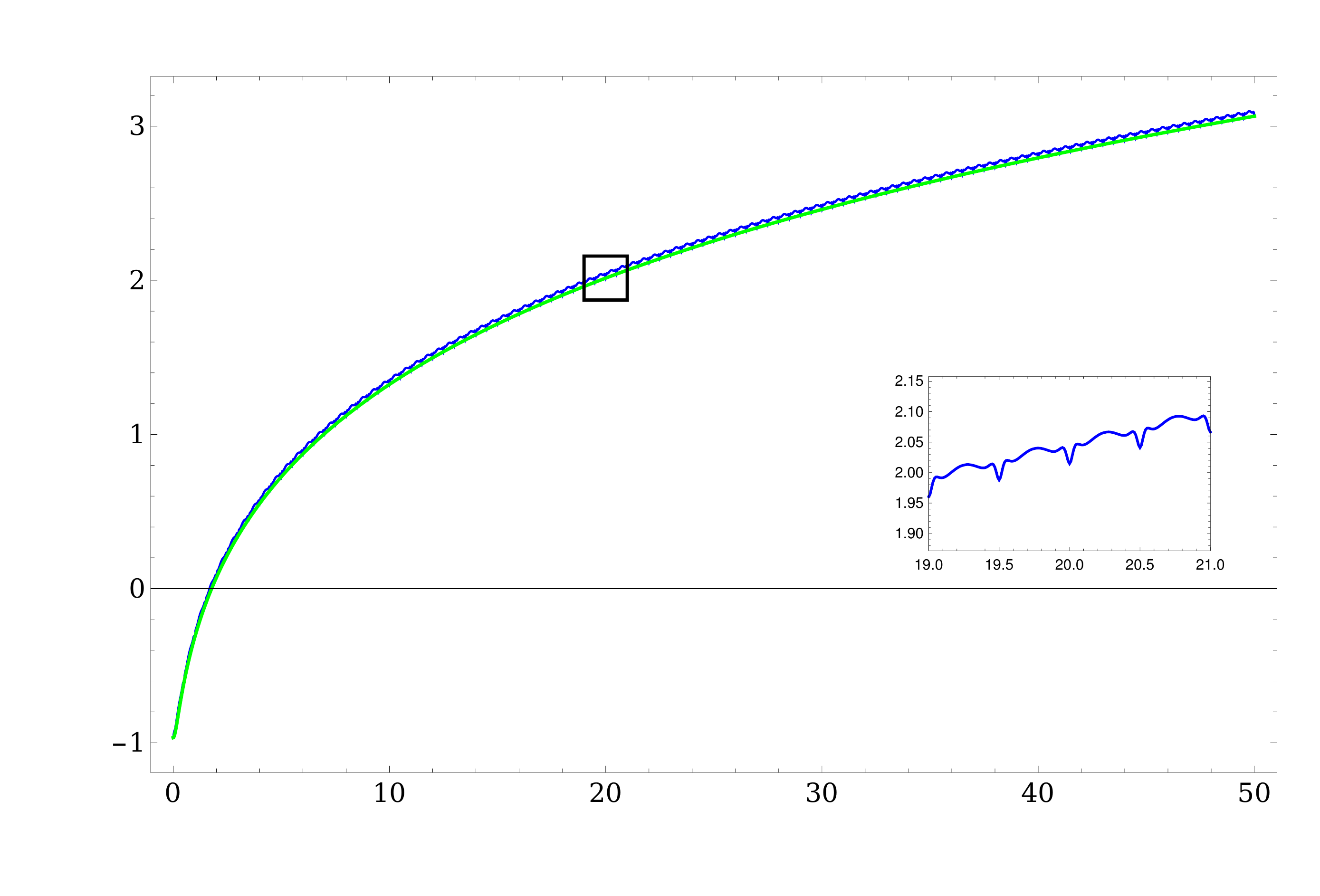}
			\put(-215,35){\rotatebox{-270}{\fontsize{15}{15}\selectfont $\frac{\mathcal{C}_{\kappa=2}(t)-\mathcal{C}_{\kappa=2}(0)}{S_{th}}$}}		\put(-105,5){{\fontsize{12}{12}\selectfont $t/L$}}
			\put(-135,132){{\fontsize{10}{10}\selectfont$\beta=10^{-1}L,\hspace{1mm}\lambda_{R} =10$}}\hspace{1cm}\includegraphics[scale=.27]{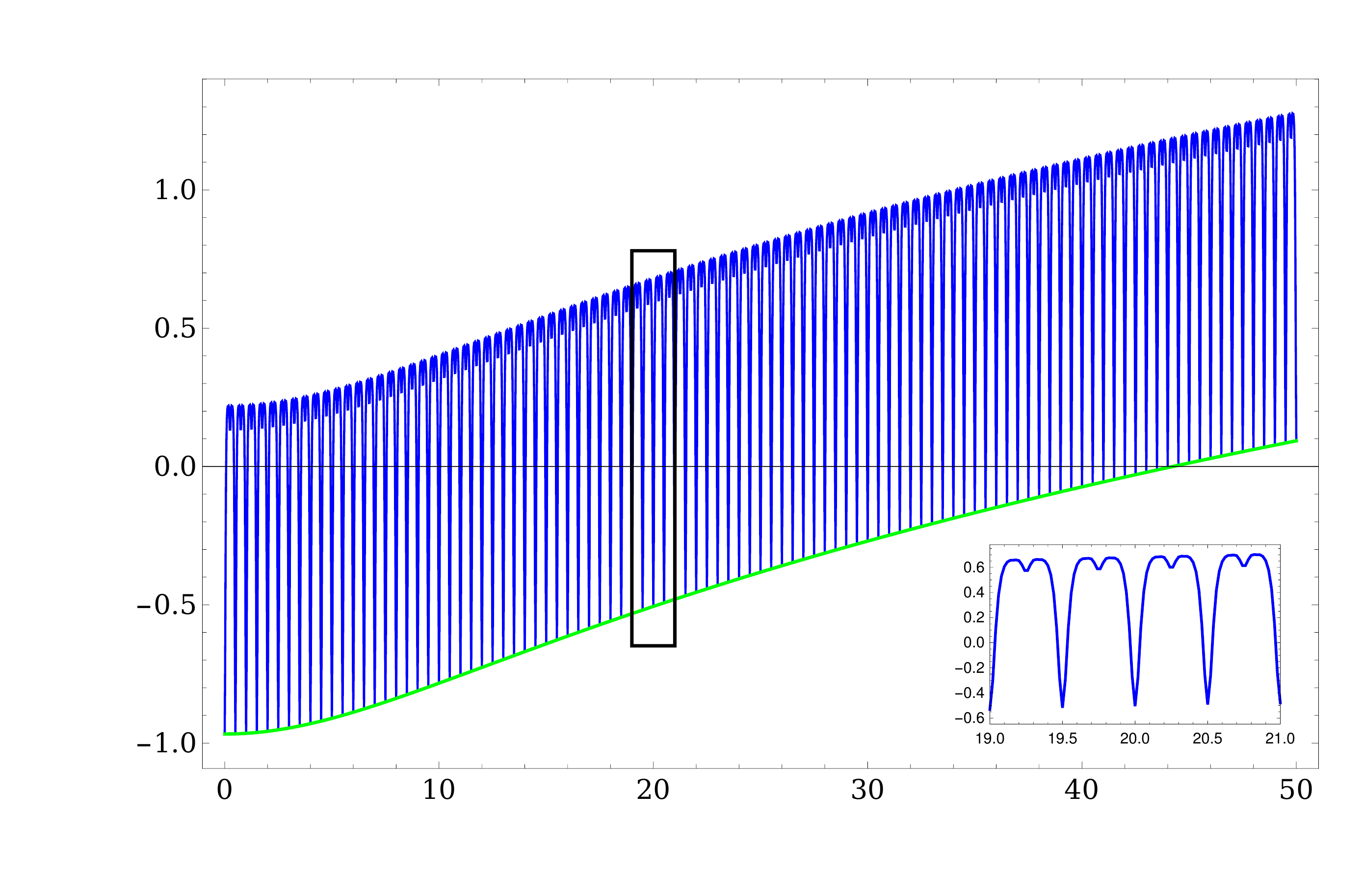}\put(-217,35){\rotatebox{-270}{\fontsize{15}{15}\selectfont $\frac{\mathcal{C}_{\kappa=2}(t)-\mathcal{C}_{\kappa=2}(0)}{S_{th}}$}}		\put(-105,5){{\fontsize{12}{12}\selectfont $t/L$}}
			\put(-135,134){{\fontsize{10}{10}\selectfont$\beta=10^{-1}L,\hspace{1mm}\lambda_{R} =\frac{1}{10}$}}\vspace{.5cm}
			
			\includegraphics[scale=.26]{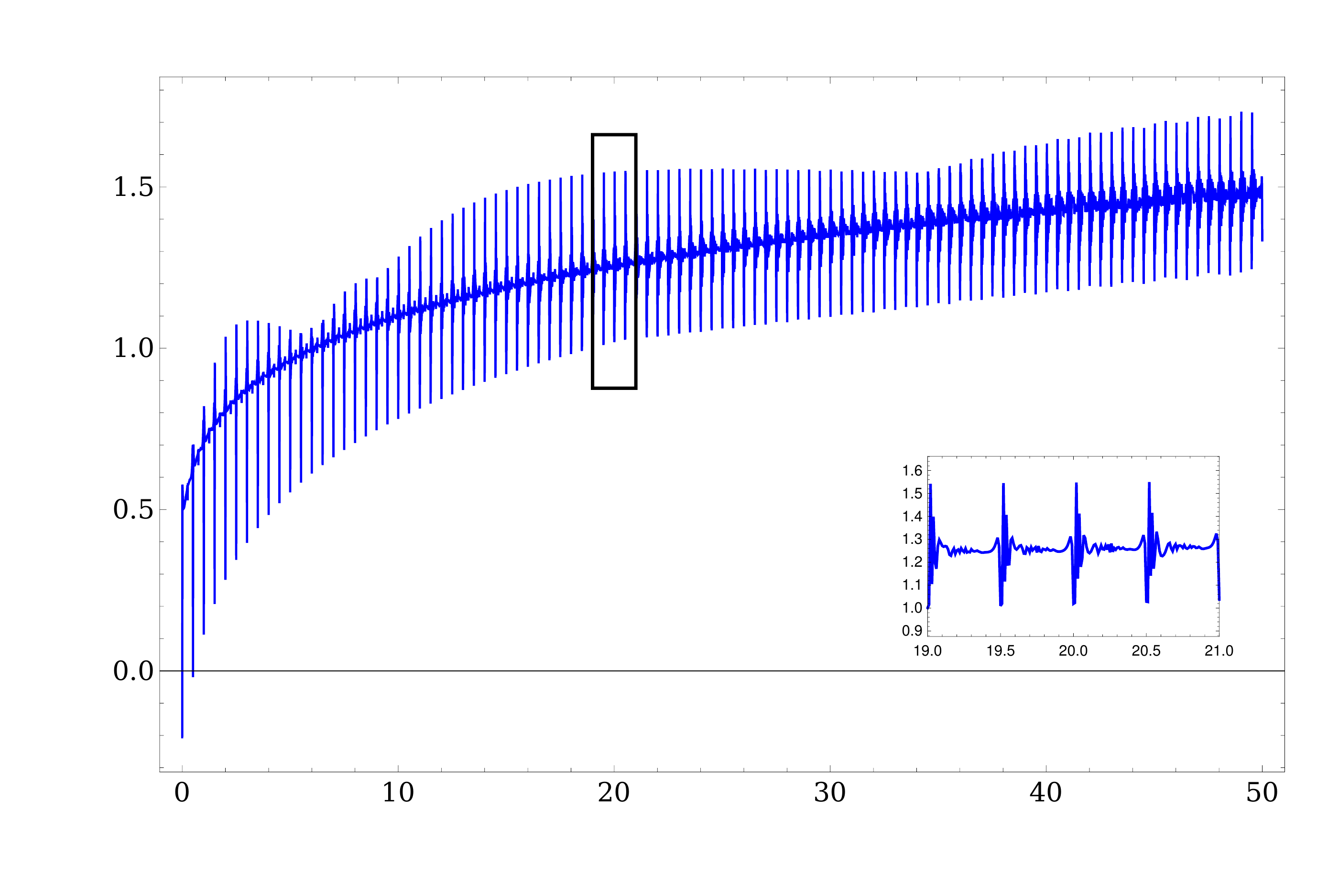}\put(-215,35){\rotatebox{-270}{\fontsize{15}{15}\selectfont $\frac{\mathcal{C}_{\kappa=2}(t)-\mathcal{C}_{\kappa=2}(0)}{S_{th}}$}}		\put(-105,5){{\fontsize{12}{12}\selectfont $t/L$}}
			\put(-135,132){{\fontsize{10}{10}\selectfont$\beta=10^{-2}L,\hspace{1mm}\lambda_{R} =10$}}\hspace{1cm}\includegraphics[scale=.26]{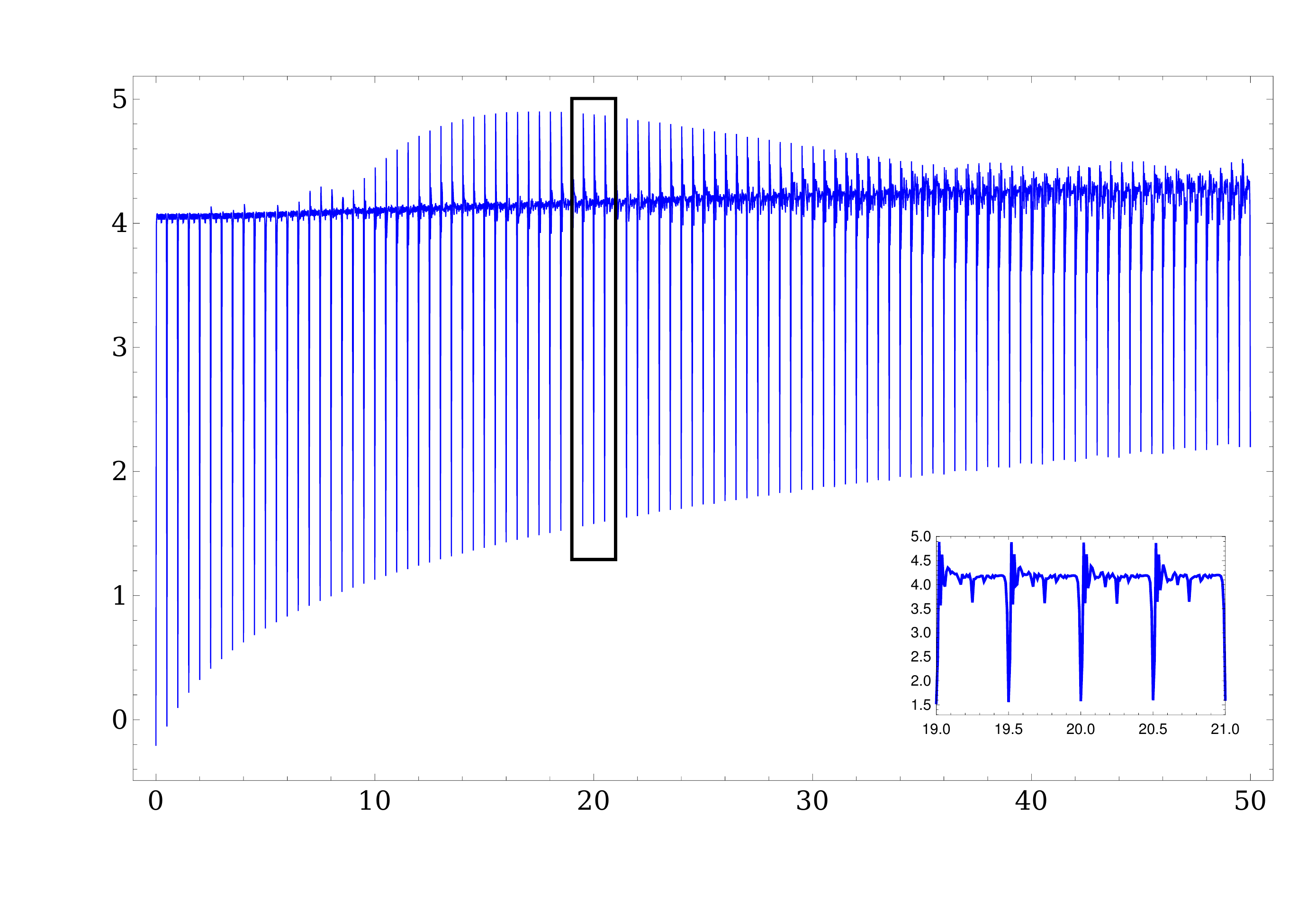}\put(-220,35){\rotatebox{-270}{\fontsize{15}{15}\selectfont $\frac{\mathcal{C}_{\kappa=2}(t)-\mathcal{C}_{\kappa=2}(0)}{S_{th}}$}}		\put(-105,5){{\fontsize{12}{12}\selectfont $t/L$}}
			\put(-135,132){{\fontsize{10}{10}\selectfont$\beta=10^{-2}L,\hspace{1mm}\lambda_{R} =\frac{1}{10}$}}\vspace{.5cm}
			
			\includegraphics[scale=.26]{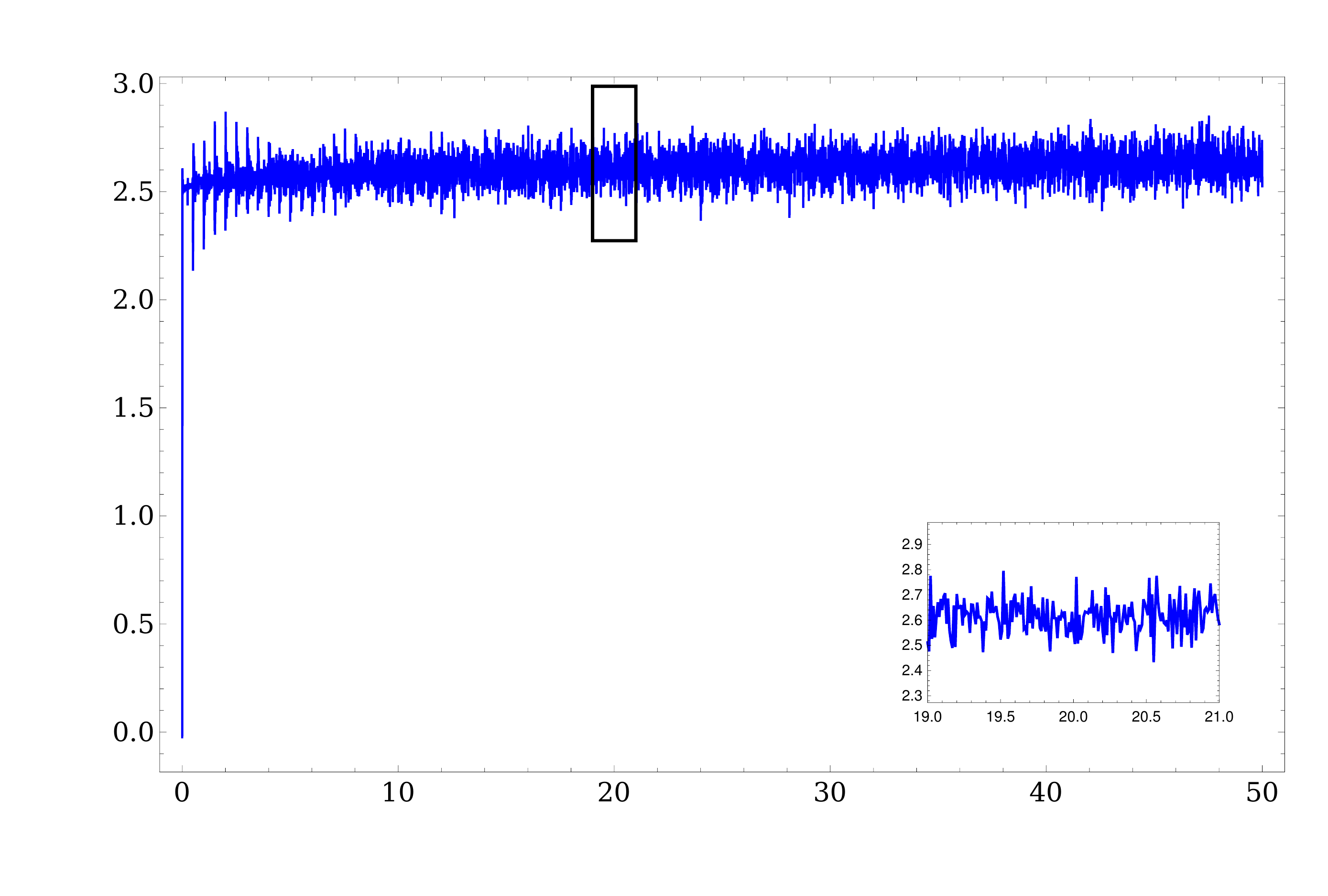}\put(-215,35){\rotatebox{-270}{\fontsize{15}{15}\selectfont $\frac{\mathcal{C}_{\kappa=2}(t)-\mathcal{C}_{\kappa=2}(0)}{S_{th}}$}}		\put(-105,5){{\fontsize{12}{12}\selectfont $t/L$}}
			\put(-135,132){{\fontsize{10}{10}\selectfont$\beta=10^{-3}L,\hspace{1mm}\lambda_{R} =10$}}\hspace{1cm}\includegraphics[scale=.26]{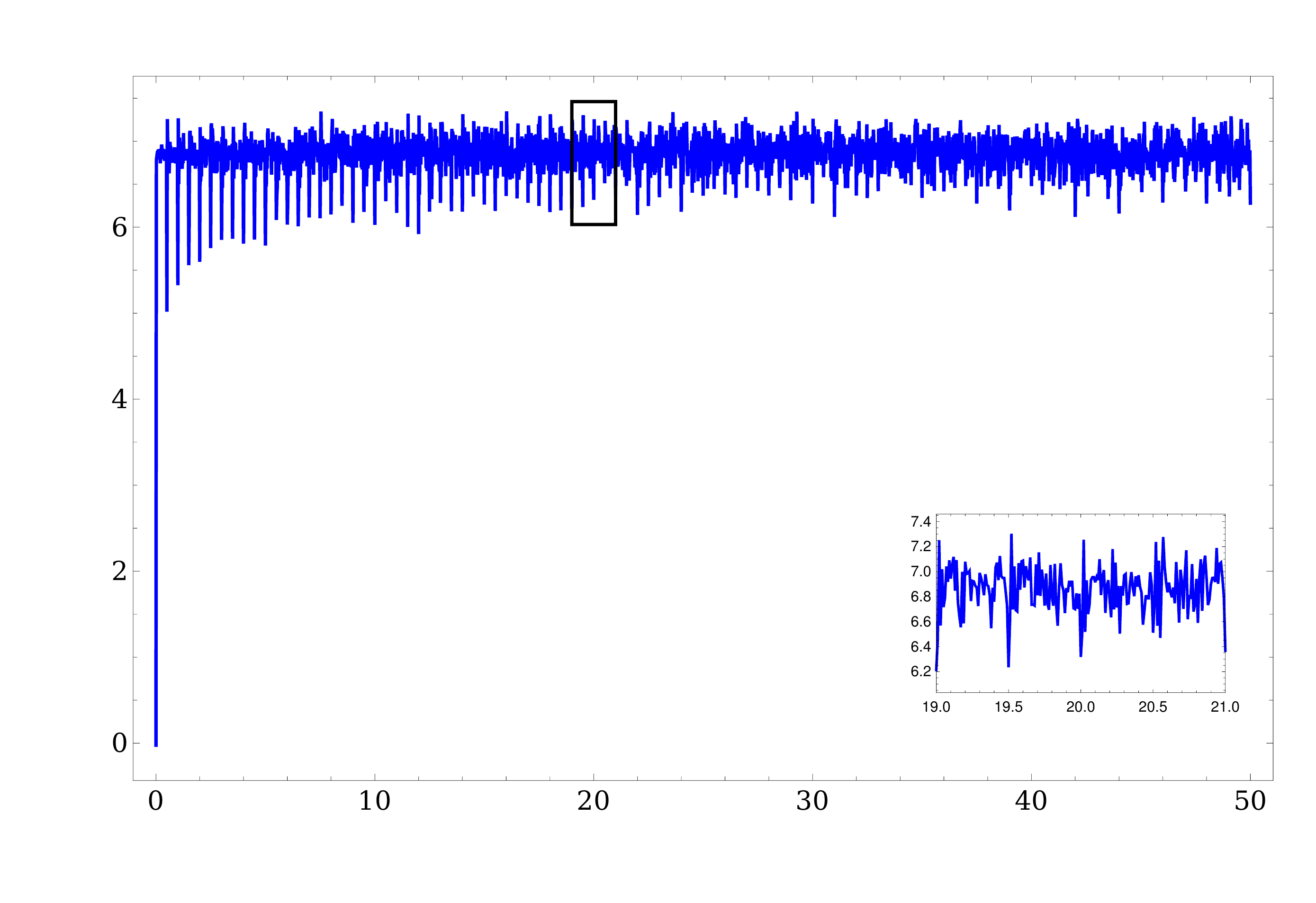}\put(-218,35){\rotatebox{-270}{\fontsize{15}{15}\selectfont $\frac{\mathcal{C}_{\kappa=2}(t)-\mathcal{C}_{\kappa=2}(0)}{S_{th}}$}}		\put(-105,5){{\fontsize{12}{12}\selectfont $t/L$}}
			\put(-135,134){{\fontsize{10}{10}\selectfont$\beta=10^{-3}L,\hspace{1mm}\lambda_{R} =\frac{1}{10}$}}}
		\caption{All mode contribution to time dependence of ${C}_{\kappa=2}$ with the initial value subtracted for the neutral TFD at zero time on a circle with circumference $L$ with $\omega_{R} = 1/L$, $m = 10^{-5}/L$, $\mu q= 10^{-5}/L$. We use 1501 lattice sites on each side.}\label{discrete5}
	\end{figure}

\begin{figure}[H]
		\center{\includegraphics[scale=.26]{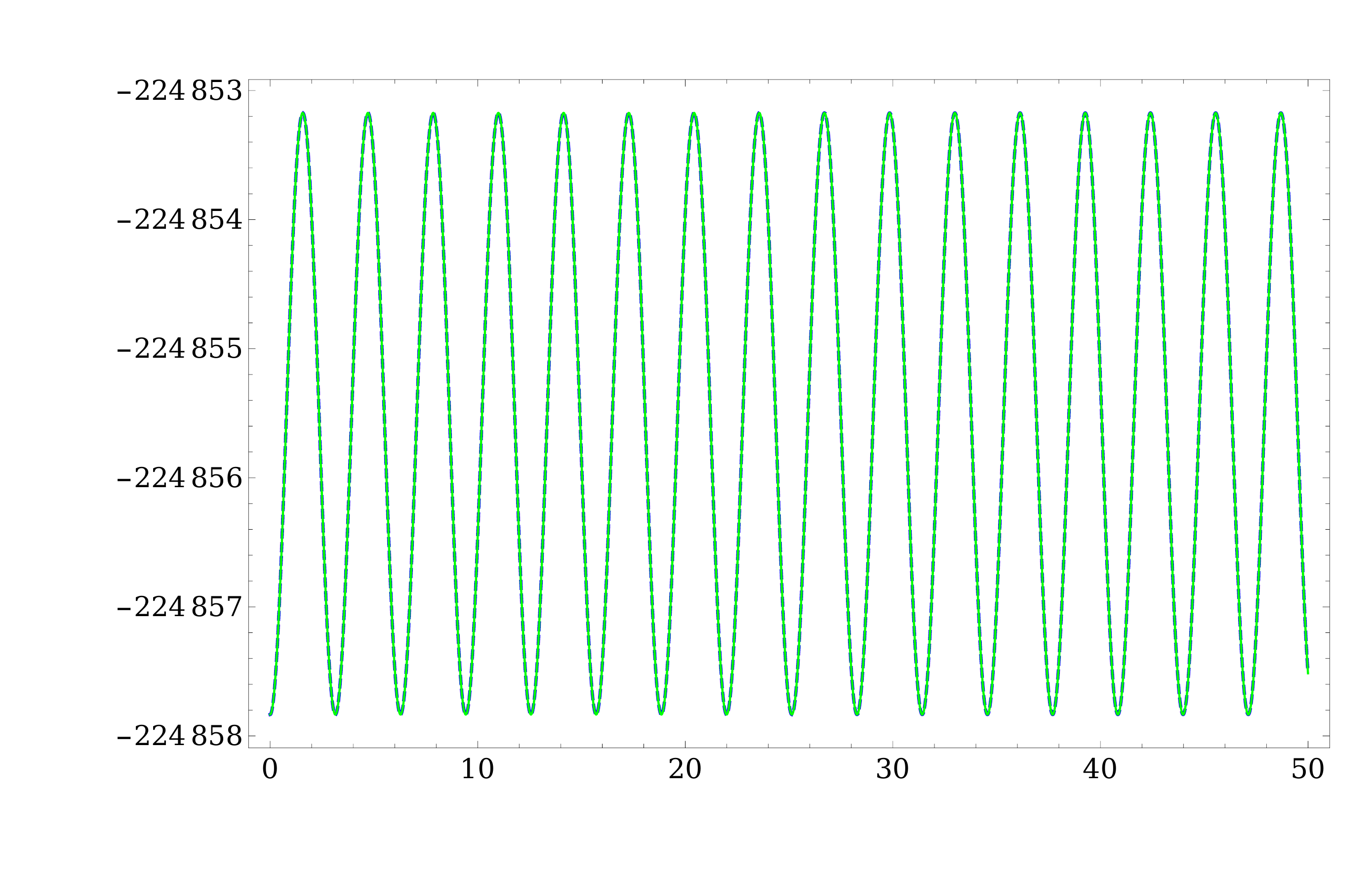}\hspace{.4cm}\put(-215,35){\rotatebox{-270}{\fontsize{15}{15}\selectfont $\frac{\mathcal{C}_{\kappa=2}(t)-\mathcal{C}_{\kappa=2}(0)}{S_{th}}$}}		\put(-105,5){{\fontsize{12}{12}\selectfont $t/L$}}
			\put(-125,125){{\fontsize{10}{10}\selectfont $\beta=10L,\hspace{1mm}\lambda_{R} = 10$}}\hspace{1cm}\includegraphics[scale=.26]{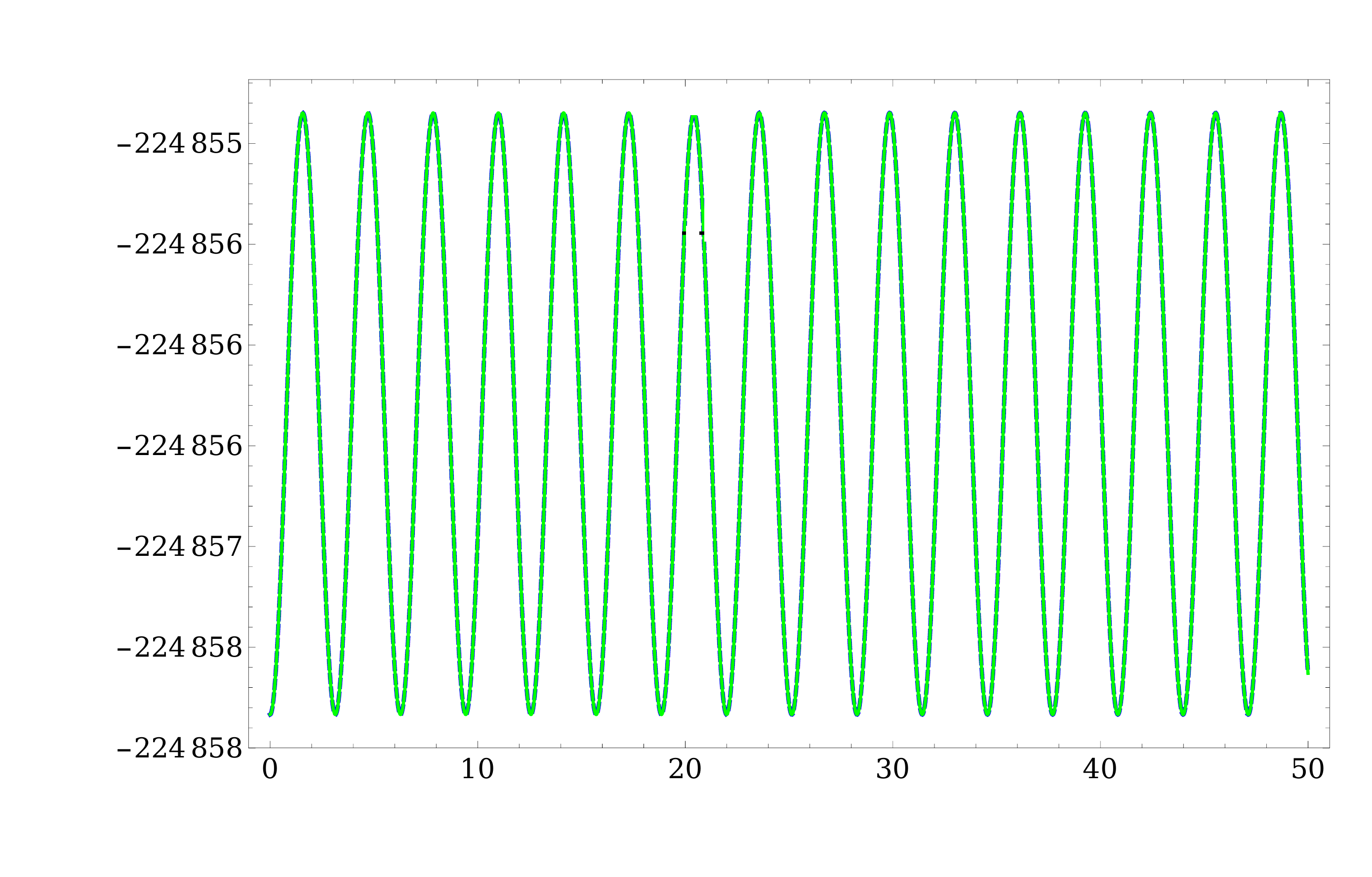}
			\put(-215,35){\rotatebox{-270}{\fontsize{15}{15}\selectfont $\frac{\mathcal{C}_{\kappa=2}(t)-\mathcal{C}_{\kappa=2}(0)}{S_{th}}$}}		\put(-105,5){{\fontsize{12}{12}\selectfont $t/L$}}
			\put(-125,125){{\fontsize{10}{10}\selectfont $\beta=10L,\hspace{1mm}\lambda_{R} = \frac{1}{10}$}}\vspace{.5cm}
			
			\hspace{.5cm}\includegraphics[scale=.26]{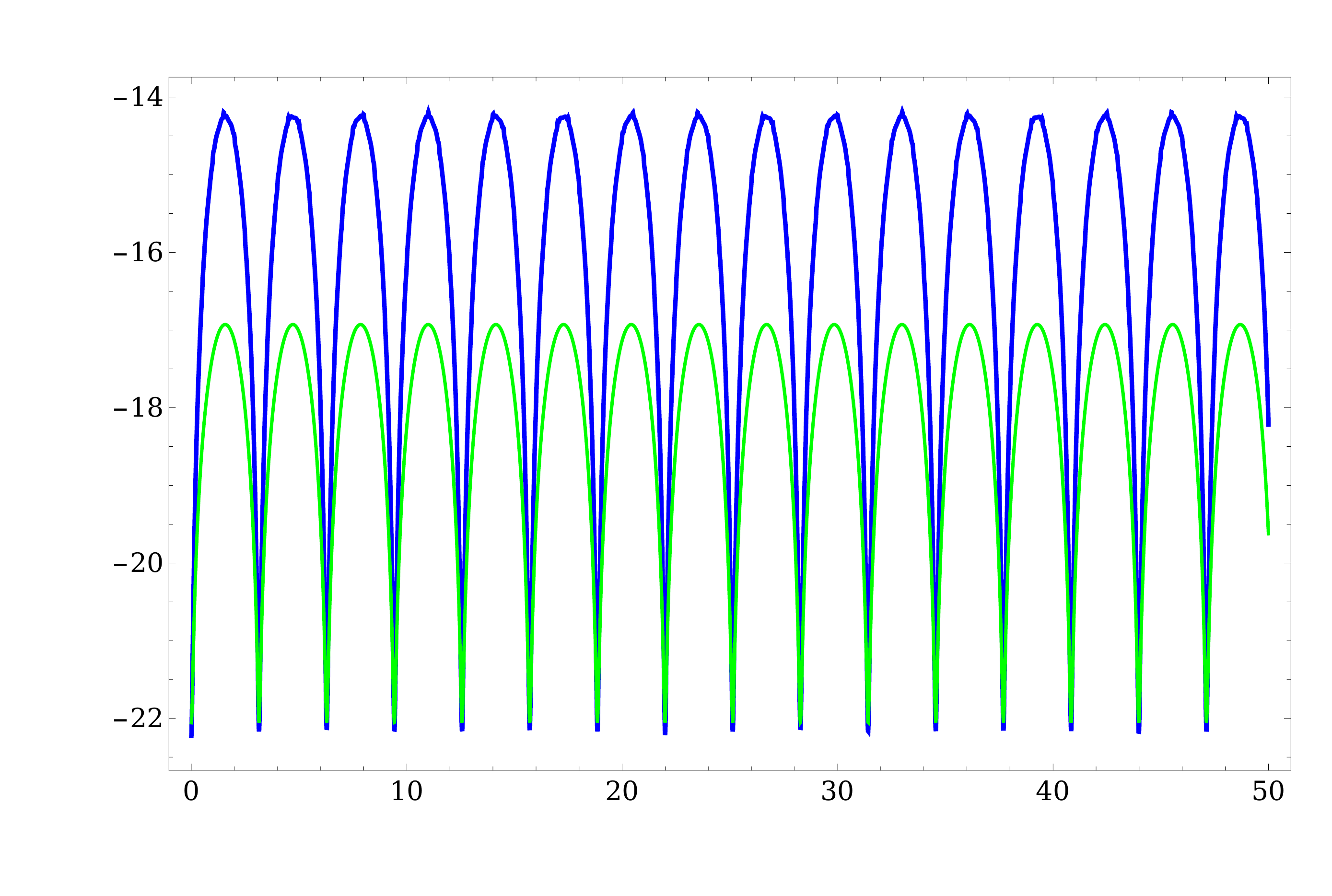}
			\put(-215,35){\rotatebox{-270}{\fontsize{15}{15}\selectfont $\frac{\mathcal{C}_{\kappa=2}(t)-\mathcal{C}_{\kappa=2}(0)}{S_{th}}$}}		\put(-105,5){{\fontsize{12}{12}\selectfont $t/L$}}
			\put(-135,132){{\fontsize{10}{10}\selectfont$\beta=10^{-1}L,\hspace{1mm}\lambda_{R} =10$}}\hspace{1cm}\includegraphics[scale=.26]{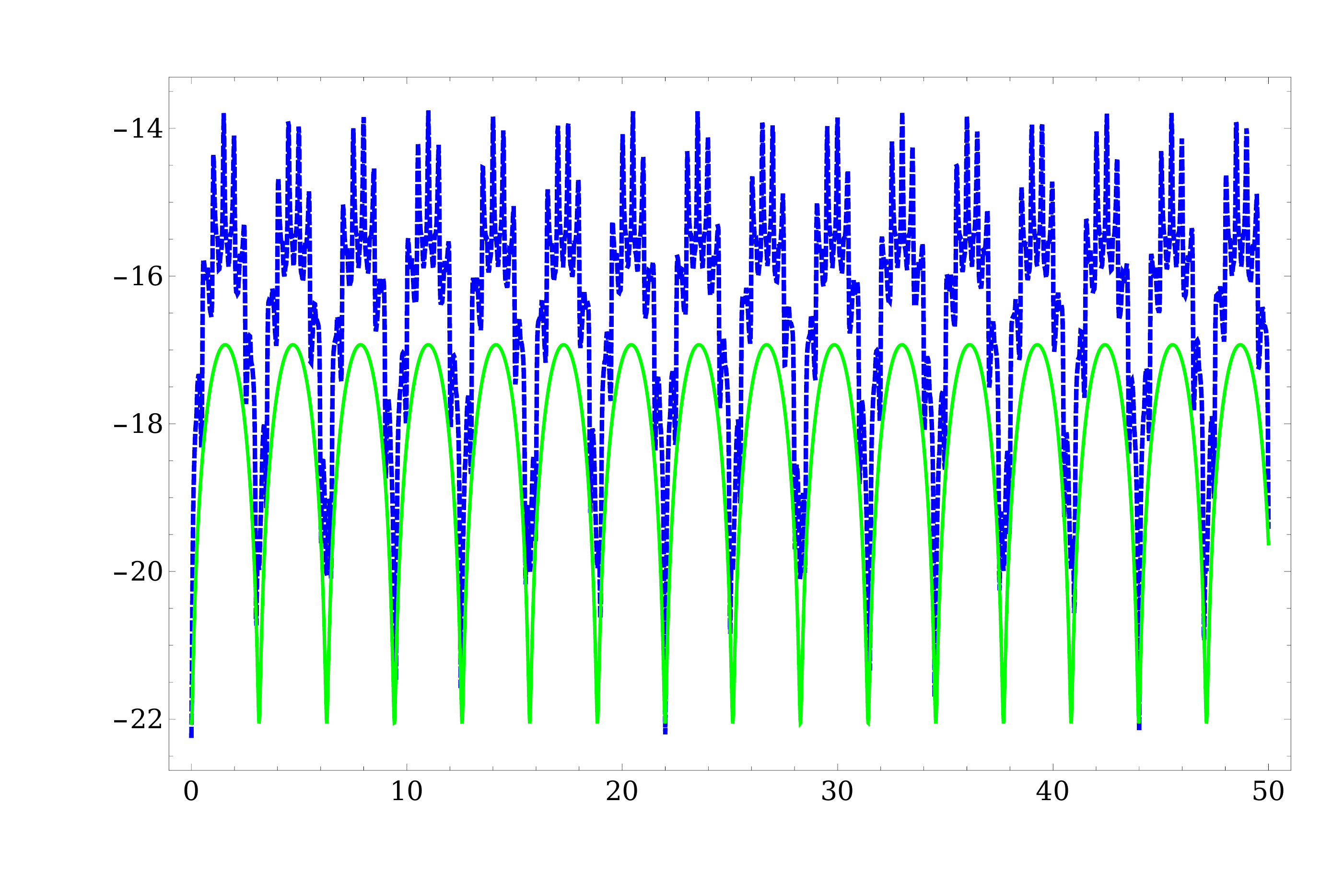}\put(-217,35){\rotatebox{-270}{\fontsize{15}{15}\selectfont $\frac{\mathcal{C}_{\kappa=2}(t)-\mathcal{C}_{\kappa=2}(0)}{S_{th}}$}}		\put(-105,5){{\fontsize{12}{12}\selectfont $t/L$}}
			\put(-135,132){{\fontsize{10}{10}\selectfont$\beta=10^{-1}L,\hspace{1mm}\lambda_{R} =\frac{1}{10}$}}\vspace{.5cm}
			
			\hspace{.3cm}\includegraphics[scale=.25]{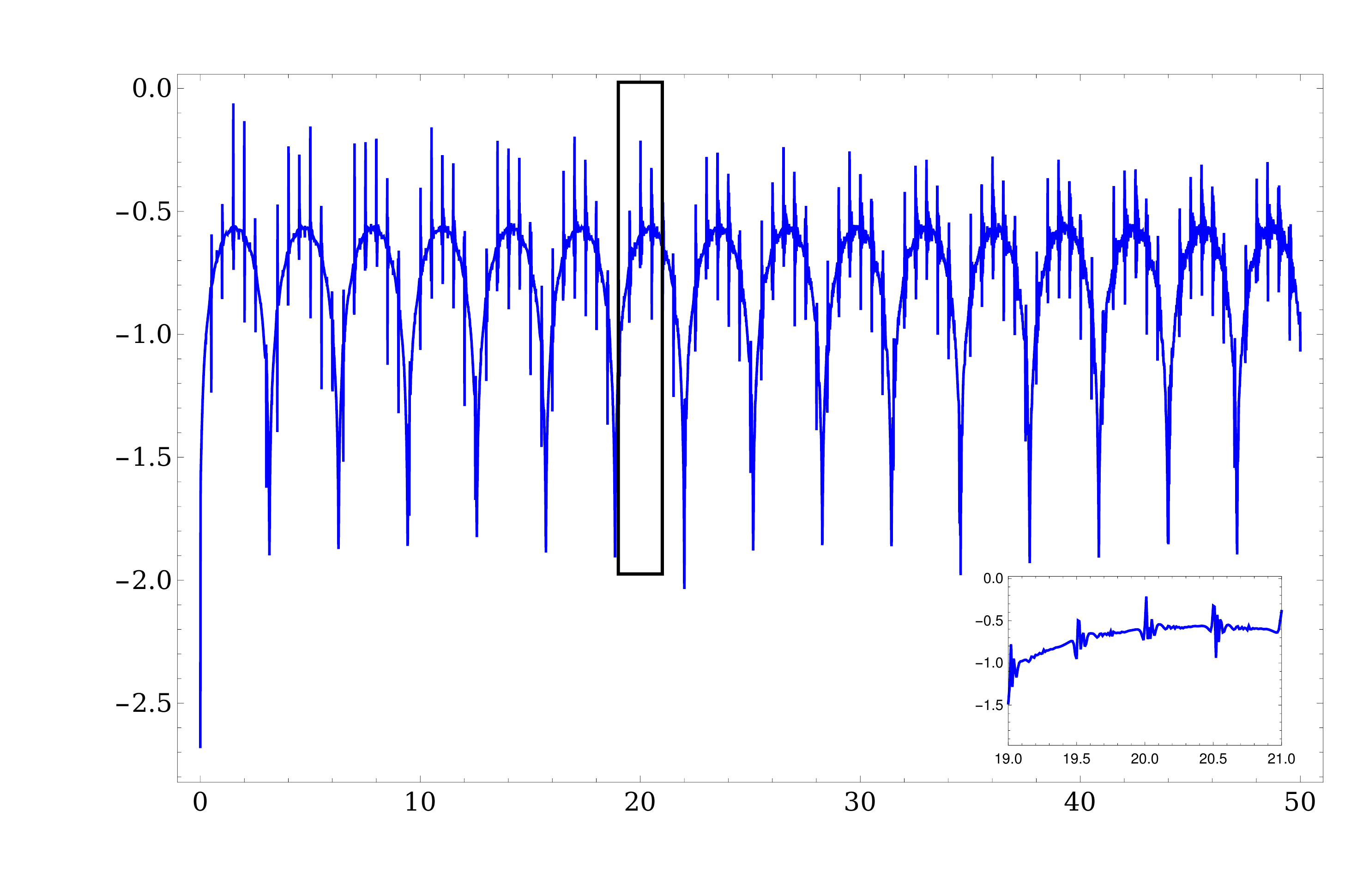}\put(-218,35){\rotatebox{-270}{\fontsize{15}{15}\selectfont $\frac{\mathcal{C}_{\kappa=2}(t)-\mathcal{C}_{\kappa=2}(0)}{S_{th}}$}}		\put(-105,5){{\fontsize{12}{12}\selectfont $t/L$}}
			\put(-135,132){{\fontsize{10}{10}\selectfont$\beta=10^{-2}L,\hspace{1mm}\lambda_{R} =10$}}\hspace{1.2cm}\includegraphics[scale=.26]{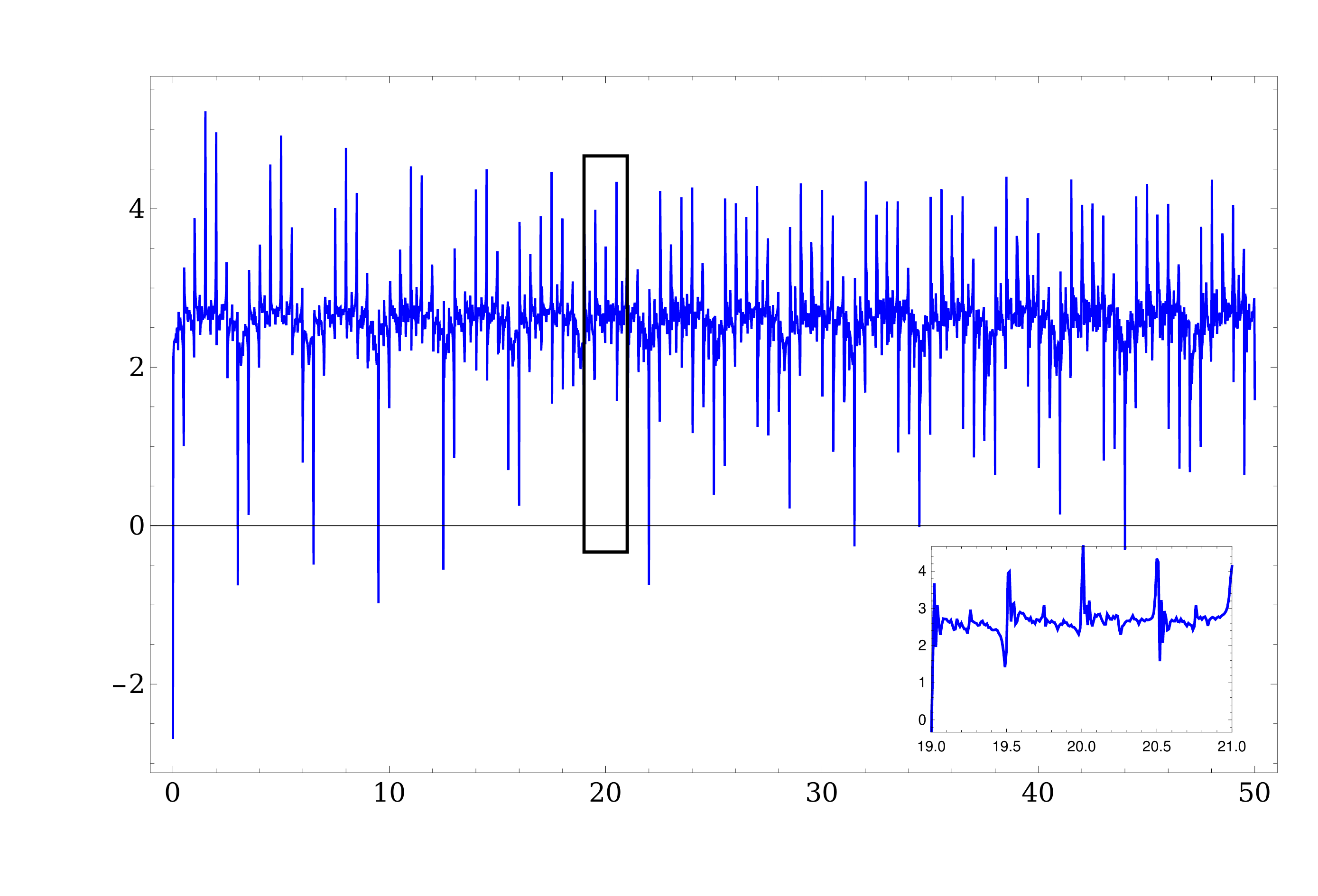}\put(-216,35){\rotatebox{-270}{\fontsize{15}{15}\selectfont $\frac{\mathcal{C}_{\kappa=2}(t)-\mathcal{C}_{\kappa=2}(0)}{S_{th}}$}}		\put(-105,5){{\fontsize{12}{12}\selectfont $t/L$}}
			\put(-135,134){{\fontsize{10}{10}\selectfont$\beta=10^{-2}L,\hspace{1mm}\lambda_{R} =\frac{1}{10}$}}\vspace{.5cm}
			
			\includegraphics[scale=.26]{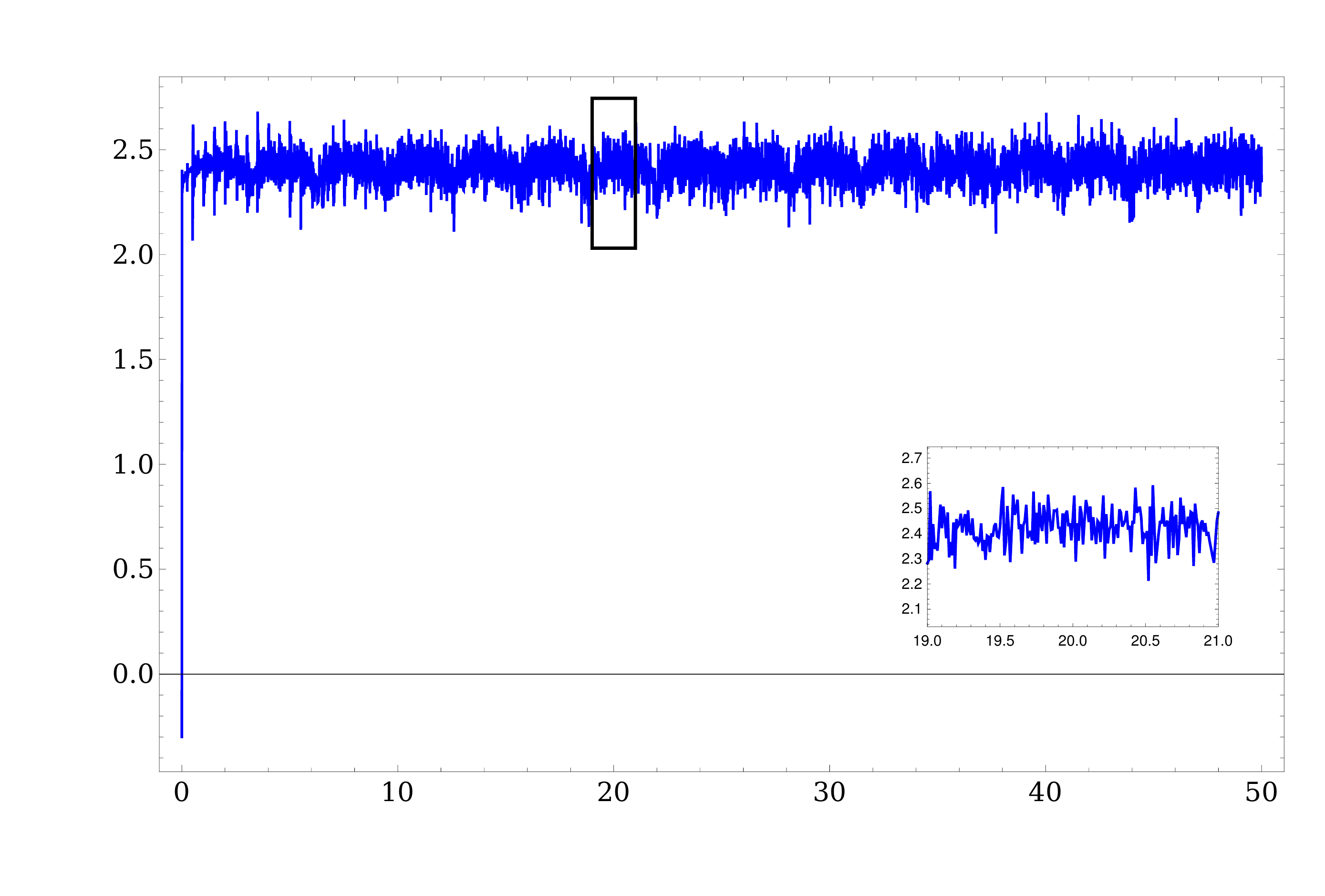}\put(-215,35){\rotatebox{-270}{\fontsize{15}{15}\selectfont $\frac{\mathcal{C}_{\kappa=2}(t)-\mathcal{C}_{\kappa=2}(0)}{S_{th}}$}}		\put(-105,5){{\fontsize{12}{12}\selectfont $t/L$}}
			\put(-135,132){{\fontsize{10}{10}\selectfont$\beta=10^{-3}L,\hspace{1mm}\lambda_{R} =10$}}\hspace{1cm}\includegraphics[scale=.26]{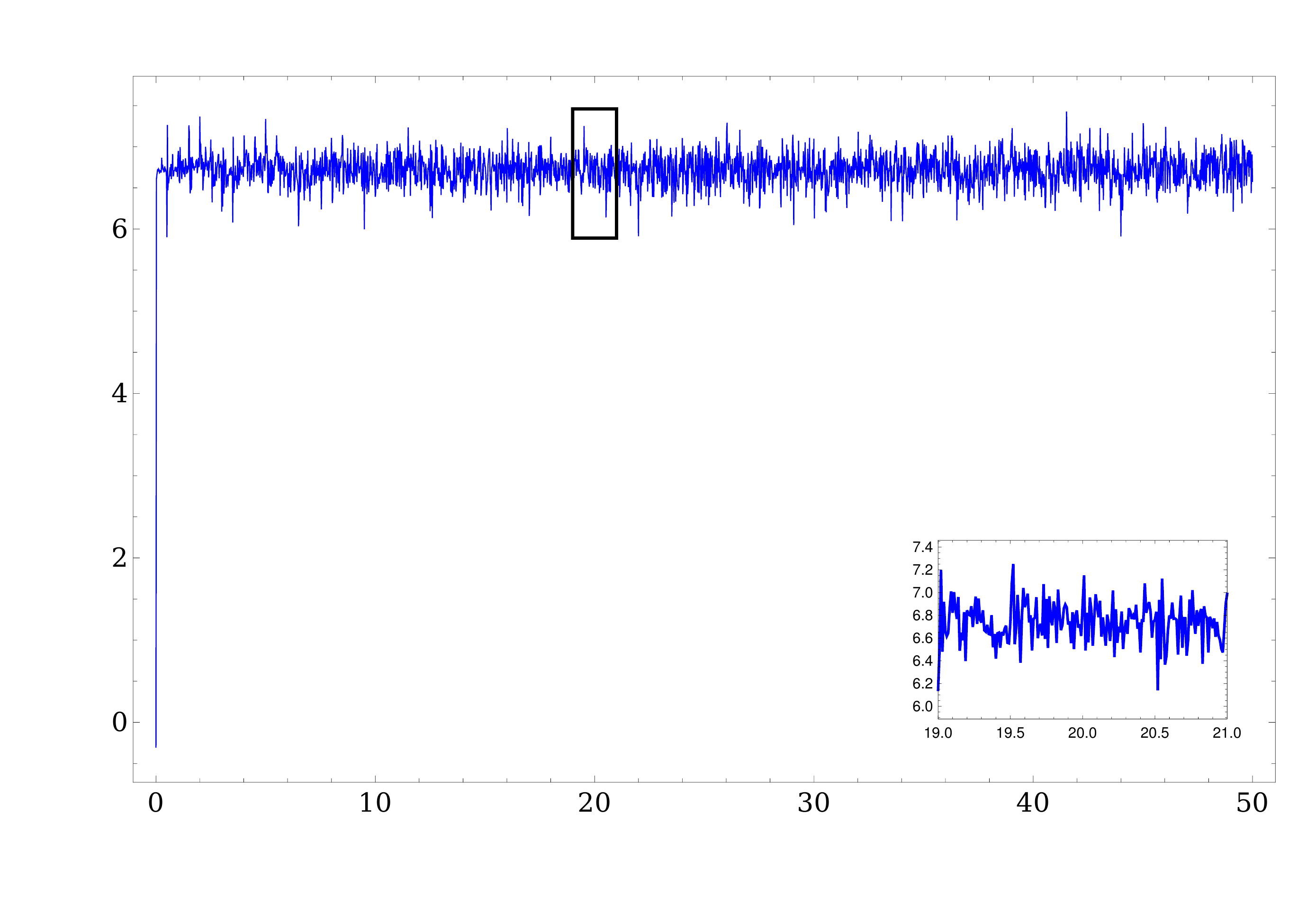}\put(-218,35){\rotatebox{-270}{\fontsize{15}{15}\selectfont $\frac{\mathcal{C}_{\kappa=2}(t)-\mathcal{C}_{\kappa=2}(0)}{S_{th}}$}}		\put(-105,5){{\fontsize{12}{12}\selectfont $t/L$}}
			\put(-135,136){{\fontsize{10}{10}\selectfont$\beta=10^{-3}L,\hspace{1mm}\lambda_{R} =\frac{1}{10}$}}}
		\caption{All mode contribution to time dependence of ${C}_{\kappa=2}$ with the initial value subtracted for the neutral TFD at zero time on a circle with circumference $L$ with $\omega_{R} = 1/L$, $m = 10^{-5}/L$, $\mu q= 1/L$. We use 1501 lattice sites on each side.}\label{discrete6}
	\end{figure}
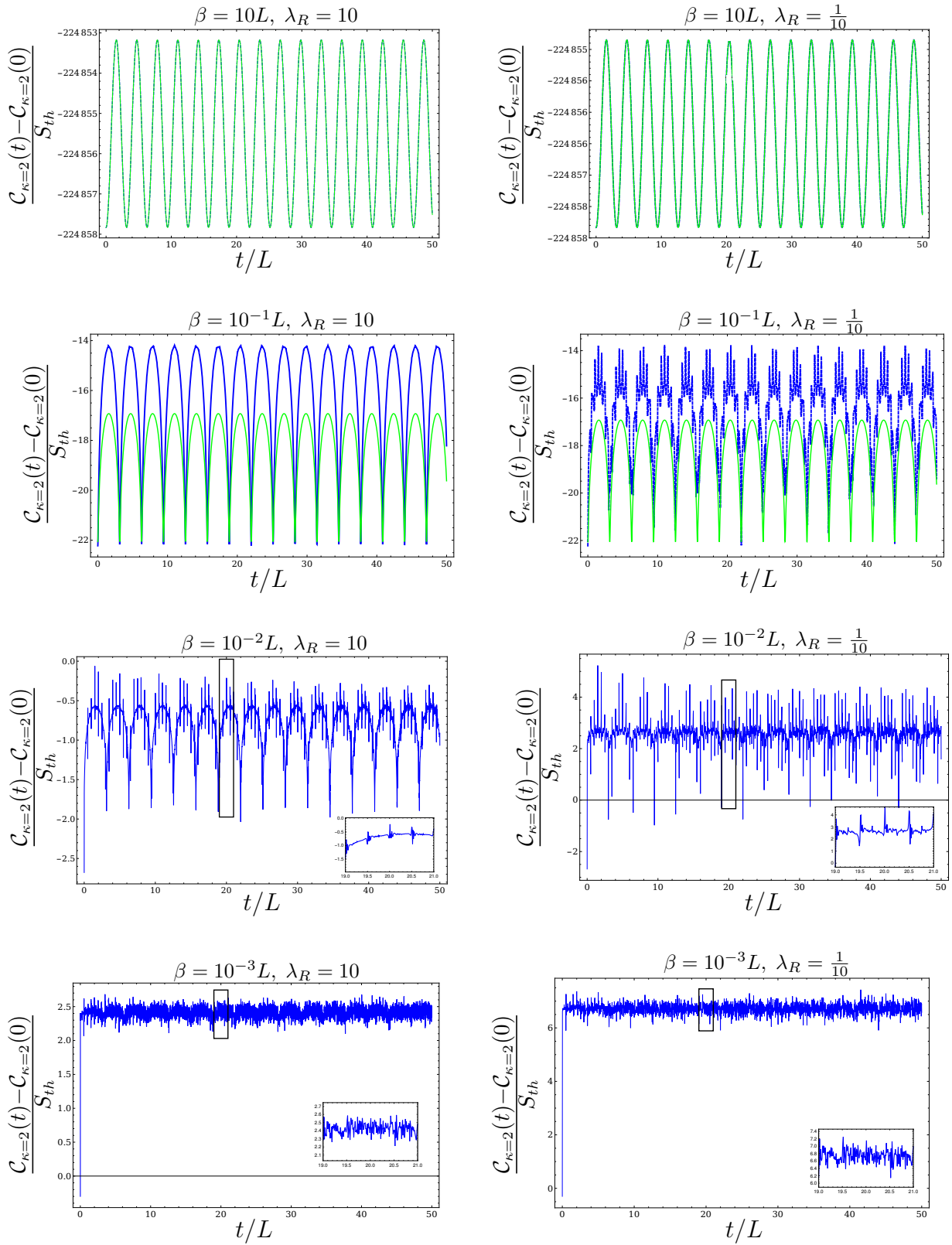

	We would like to emphasize again that above, $\lambda_{R}=1$ is considered since for this special value, it is proven that the optimal circuit does not mix the normal modes. Even though this simplicity of the optimal circuits might be lost 
	when $\lambda_{R} \neq 1$,
	just to make the numerical problem tractable, we still assume that there is
	no mixing between the normal modes along the circuit. The same analysis as before for $\lambda_{R} \neq 1$ is presented in figs.\ref{discrete5} and \ref{discrete6} which they show the validity of the results for $\lambda_{R}=1$ also in this case.
	\vspace{10cm}
	
	\vspace{1cm}
	
\begin{figure}[h]
		\hspace{1.3cm}\includegraphics[scale=.34]{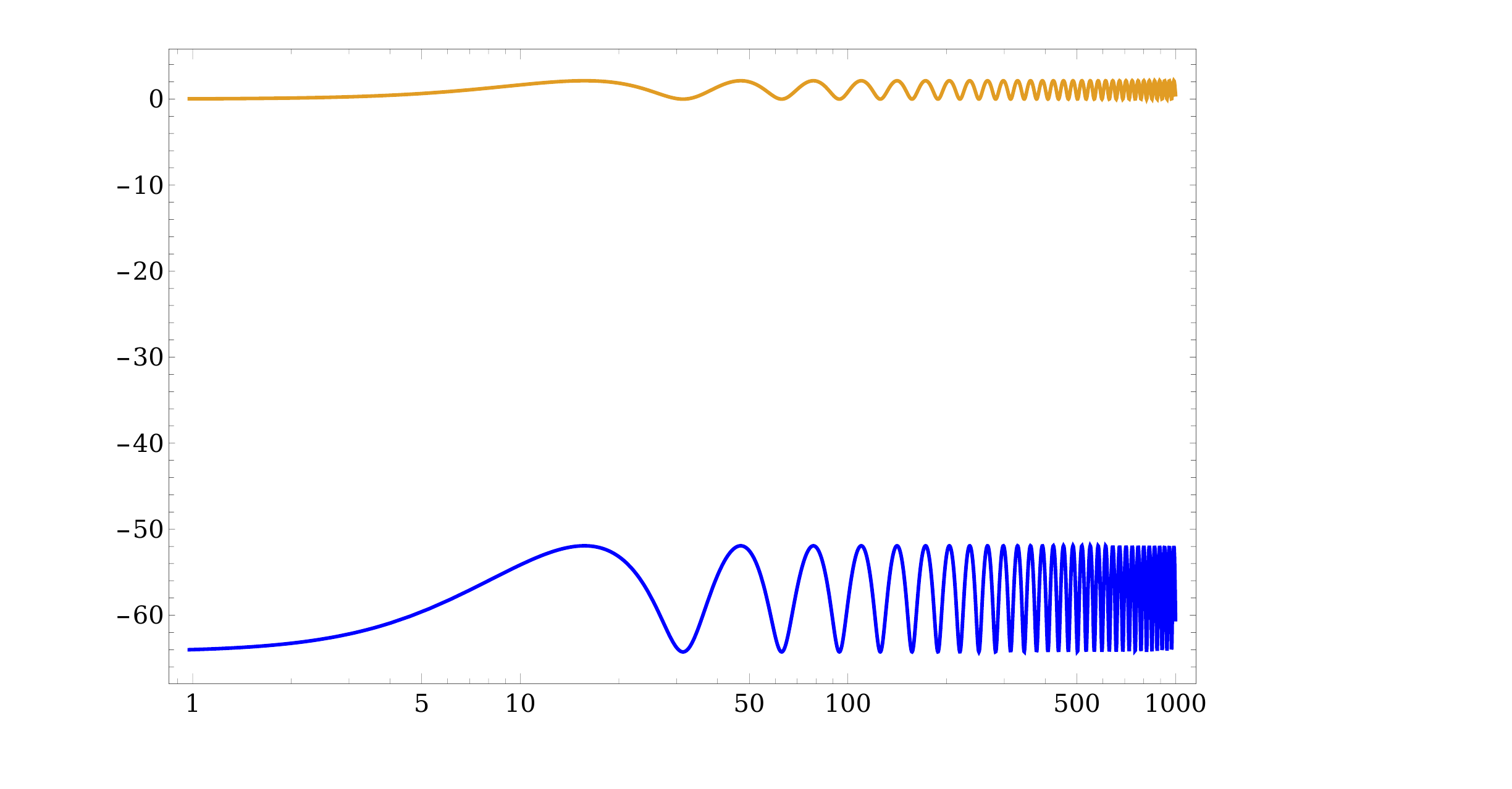}\put(-248,32){\rotatebox{-270}{\fontsize{15}{15}\selectfont $\frac{\mathcal{C}_{\kappa=2}(t)-\mathcal{C}_{\kappa=2}(0)}{S_{th}}$}}		\put(-140,0){{\fontsize{12}{12}\selectfont $t/L$}}
		\put(-150,120){{\fontsize{10}{10}\selectfont $\beta=10L$}}\hspace{-1cm}
		\includegraphics[scale=.34]{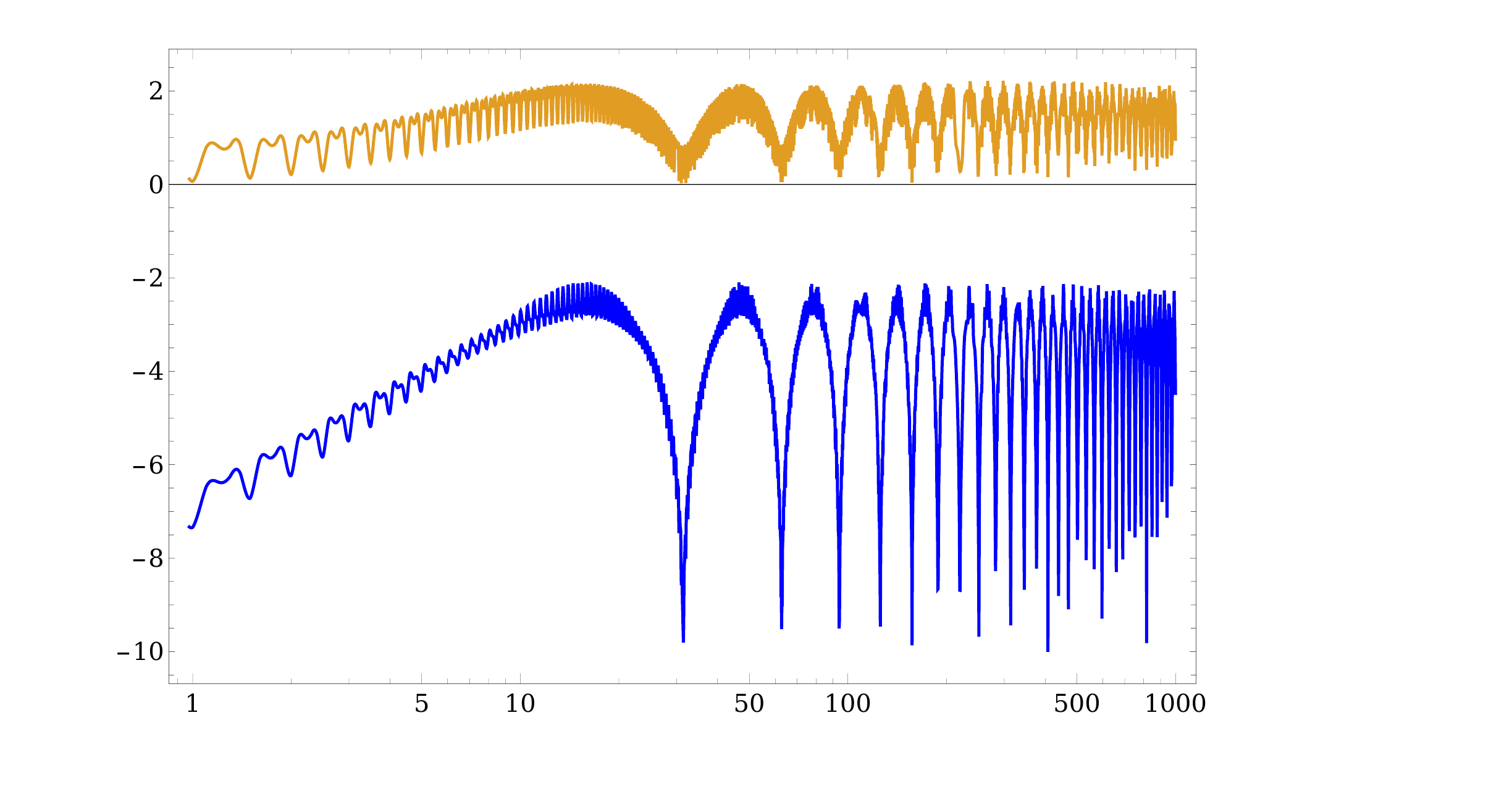}\put(-245,32){\rotatebox{-270}{\fontsize{15}{15}\selectfont $\frac{\mathcal{C}_{\kappa=2}(t)-\mathcal{C}_{\kappa=2}(0)}{S_{th}}$}}		\put(-140,0){{\fontsize{12}{12}\selectfont $t/L$}}
		\put(-150,120){{\fontsize{10}{10}\selectfont $\beta=10^{-1}L$}}
		\caption{Time dependence of $\kappa=2$ complexity with the initial value subtracted for TFD state on a circle with circumference $L$ with $\omega_{R} = 1/L$, ($m=10^{-4}/L, \mu q = 10^{-1}/L$) (blue), ($m = 10^{-1}/L, \mu q=10^{-4}/L$) (brown). \textbf{Left}: $\beta=10L$ and \textbf{Right}: $\beta=10^{-1}L$. We use 1501 lattice sites on each side.}\label{discrete7}
	\end{figure}
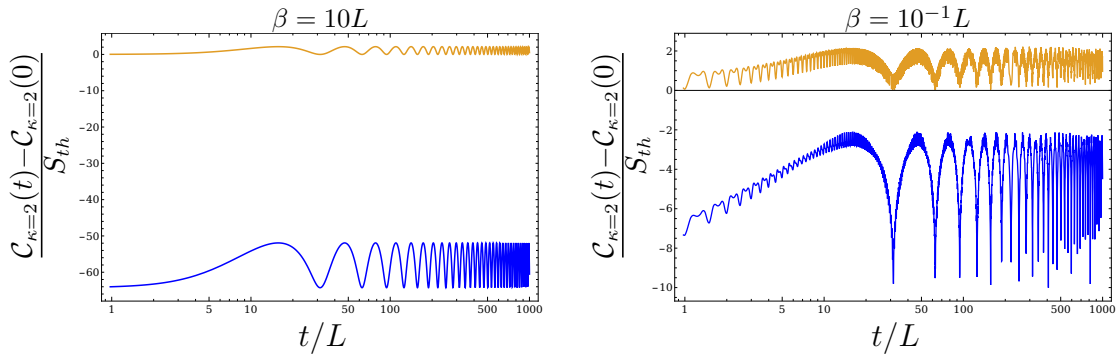

	To close this section, we show the asymmetry between the effect of mass and charge times chemical potential in fig.\ref{discrete7}
	\subsection{Keeping the lattice spacing fixed with increasing $N$}\label{L2}
\begin{figure}[H]
		\hspace{1.5cm}\includegraphics[scale=.33]{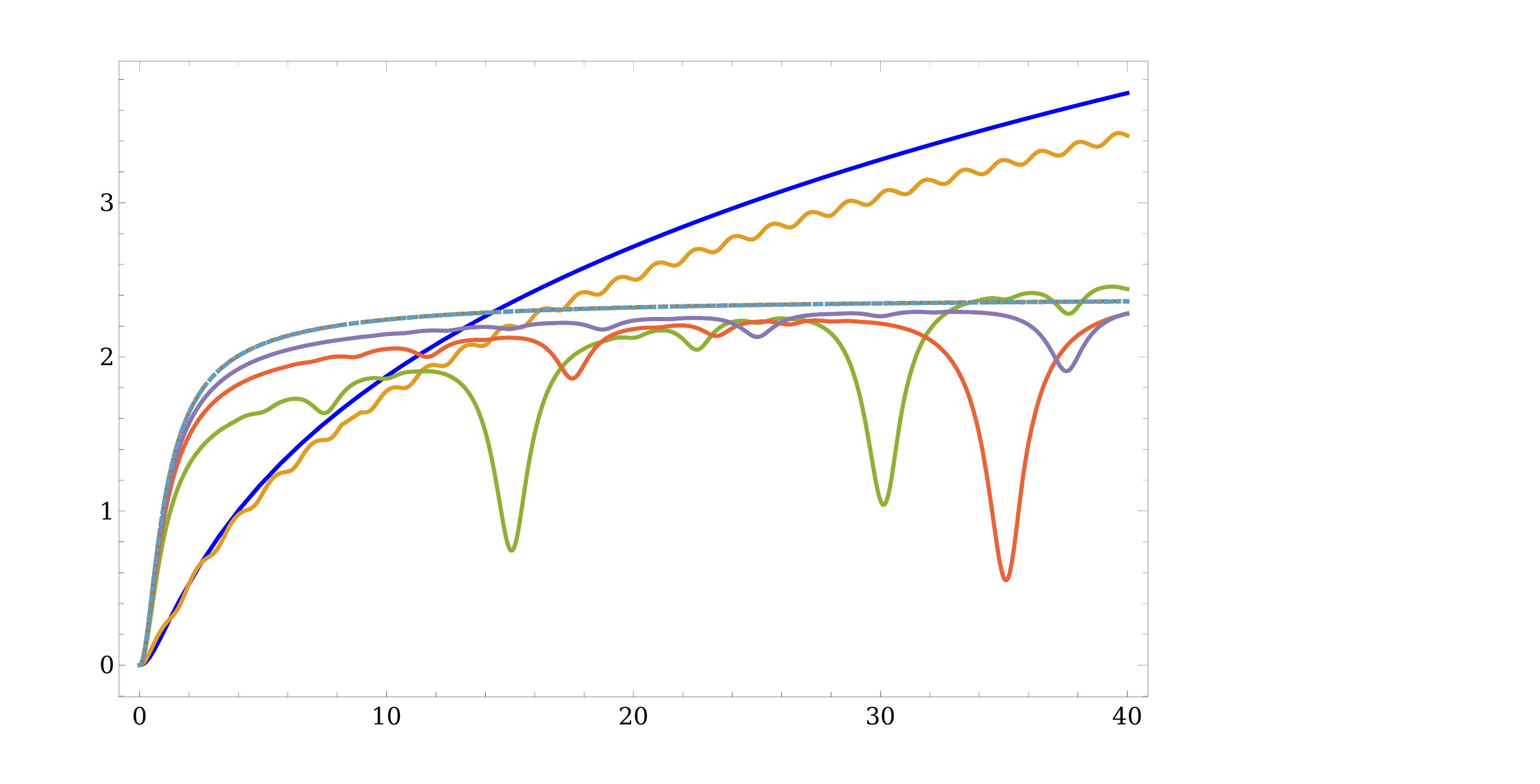}\put(-256,30){\rotatebox{-270}{\fontsize{15}{15}\selectfont $\frac{\mathcal{C}_{\kappa=2}(t)-\mathcal{C}_{\kappa=2}(0)}{S_{th}}$}}		\put(-155,-5){{\fontsize{14}{14}\selectfont $t T$}}\hspace{-1.cm}\includegraphics[scale=.33]{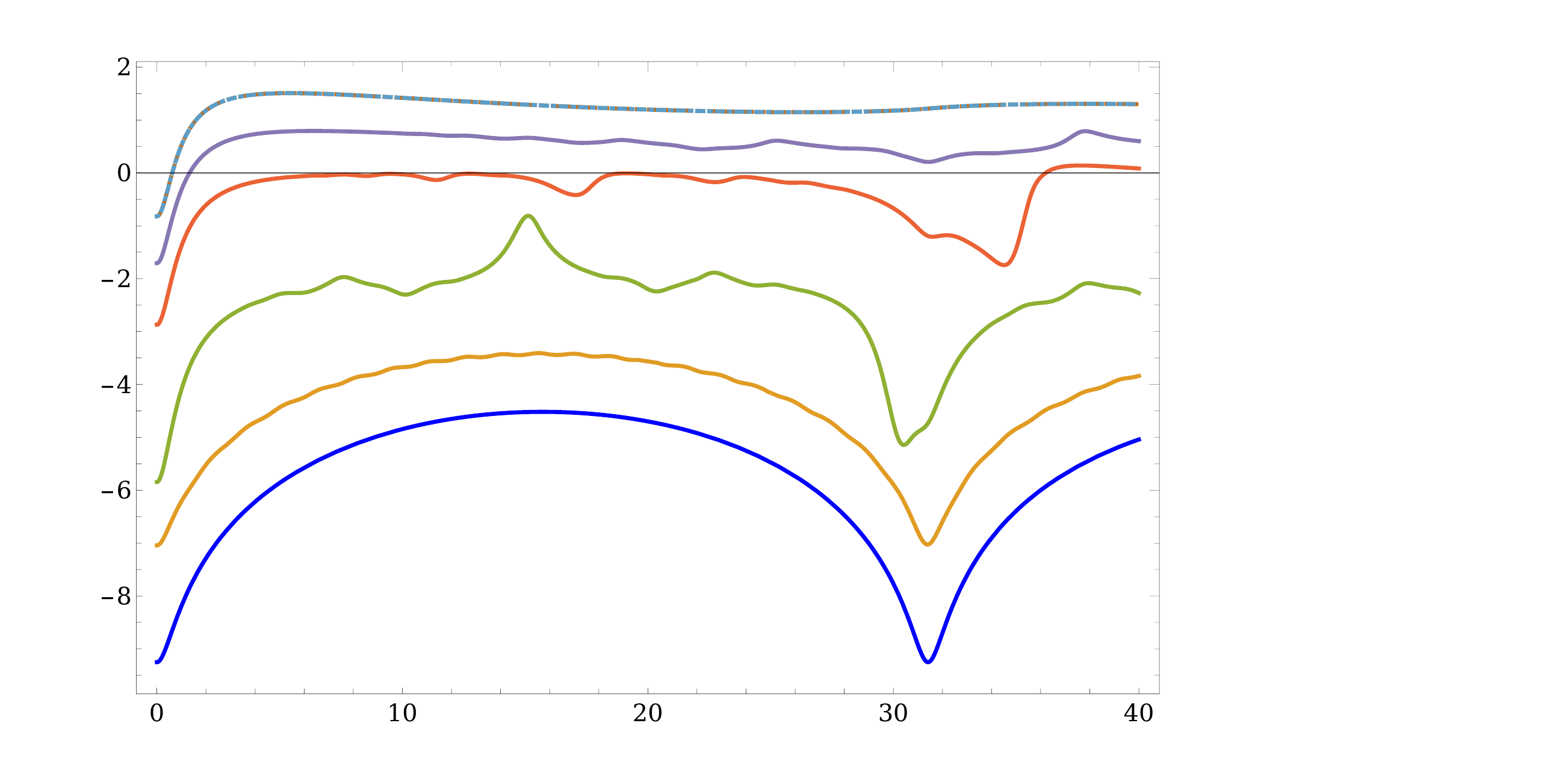}\put(-256,30){\rotatebox{-270}{\fontsize{15}{15}\selectfont $\frac{\mathcal{C}_{\kappa=2}(t)-\mathcal{C}_{\kappa=2}(0)}{S_{th}}$}}		\put(-155,-5){{\fontsize{14}{14}\selectfont $t T$}}
		\caption{Time dependence of $\kappa=2$ complexity with the initial value subtracted for TFD state with fixed lattice spacing $\delta$, $m = 10^{-6}/\delta$, $\beta=10\delta$ and $\omega_{R} = 1/\delta$, $\mu q =0$ (\textbf{left}) and $\mu q =10^{-2}/\delta$ (\textbf{right}). The solid curves represents the theory on a circle with the total number of cites $N=1$ (blue), $N=31$ (brown), $N=301$ (green), $N=701$ (red), $N=1501$ (purple). The solid blue and brown curves are scaled with $1/5$ for the cTFD state. The dashed brown and dotted dashed blue curve represent the result respectively for the cMERA inspired technique with UV cutoff $\Lambda =1/\delta$ and the one obtained directly for a theory on an infinite line.}\label{diffN}
	\end{figure}

	What we have done in previous subsection was introducing a UV-regularization and mode decomposition in which the Hamiltonian of a continuous quantum many-body system became
	a sum over independent harmonic oscillators (bosonic modes). The latter one implies the complexity for QFT is found by simply adding up contributions for each bosonic mode.
	One can use an interesting different regularization which is based on the continuous multiscale entanglement renormalization ansatz (cMERA). In cMERA approach, a UV regulator is introduced such that the ground state for large momenta $|p| > \Lambda$ behaves as a product state i.e. the ground state of ultralocal Hamiltonian (\ref{HR}). We extend this regularization also for the cTFD state. In next section, we will study the continuous system on the line with details. For the moment, the fig.\ref{diffN} demonstrates that the decompactification limit of the circle ($\delta$ is fixed and $N\rightarrow \infty$) quantitatively reproduces the results on an infinite line, and provides an example of the use of cMERA-inspired techniques in the context of complexity.
	\section{Working with the infinite system}\label{conti}
	To find the complexity in the continuum limit, we use the cMERA inspired regularization. Let us introduce a continuous label $p \hspace{.5mm}\in\hspace{.5mm} [-\frac{\pi}{\delta},\frac{\pi}{\delta}]$ defined as
	\bea
	p = \frac{2\pi}{N}\frac{k}{\delta}.
	\eea
	This definition for continuous momentum $p$ implies that $\omega_{p}$, (\ref{omegak}), becomes
	\bea\label{omegap}
	\omega_{p}^{2} = m^2+ \frac{4}{\delta^2} \sin^2\left(\frac{p\hspace{.5mm}\delta}{2}\right).
	\eea
	The continuity of momentum $p$ implies that we have taken the limit of large chain, $L \gg \delta$ (i.e. $N\gg 1$ while keeping $\delta$ fixed). Since energy of all modes is less than the UV cut off $\Lambda$ and $\Lambda$ itself is much less than $\pi/\delta$, therefore $\omega_{p}$ (\ref{omegap}) is simplified to\footnote{One may also demand that the frequency of the oscillator is continuous at the transition point, i.e. $\omega_{p=\Lambda} = \omega_{R}$ where $\omega_{p}$ was defined in (\ref{omegap}).}
	\bea
	\omega_{p} = \sqrt{m^2+p^2}.
	\eea
	In the following, we consider the
	complex scalar theory (\ref{Lcomplex1}) in d-dimensional flat spacetime and with mass m.
	In this case, one may also modify the gate sale. For  d-dimensional spacetime, the relations in (\ref{delta}) become
	\bea\label{}
	\mathbf{x}_a = \Phi(x_a)\hspace{.5mm}\delta^{d/2}, \hspace{1cm} \mathbf{p}_a = \Pi(x_a)\hspace{.5mm}\delta^{d/2 -1},
	\eea
	which they imply that in going from the lattice to the continuum expressions, we should absorb a factor of $1/\delta^{d-1}$ into control functions and also define the gate scale as follows
	\bea
	\omega_{R,g} \equiv \delta \hspace{.5mm}g_s^2.
	\eea 
	Accordingly, the dimensionless ratios in (\ref{newlambdas}) change to
	\bea
	\lambda_{R} = \frac{\omega_{R}}{\omega_{R,g}},\hspace{1cm} \lambda= \frac{\omega_{k}}{\omega_{R,g}}.
	\eea
	Furthermore, in order to make contact with holography, we are primarily interested in CFTs (i.e.
	massless models). However as we have pointed above, the massless limit of (\ref{EigenDeltap}) is ill-defined since the zero mode gives a divergent contribution. We regulate this divergence by instead
	working with a small but non-zero mass. Moreover, in order to UV regulate our results for the complexity, one can subtract its value at the initial time $t = 0$. Accordingly, based on covariance matrix approach (\ref{ck2}), the desired complexities  in the continuum limit reads
	\bea
	\label{Cscovariance}
	&& \mathcal{C}_{\kappa=2}(t) = \text{vol} \int_{p\leq \Lambda} \frac{d^{d-1}p}{(2\pi)^{d-1}} \hspace{1mm}\frac{1}{4}\sum_{i=1}^{2}\bigg[\left(\log\Delta^{(i)}_{+}(t,p)\right)^2\hspace{.5mm}+\left(\log\Delta^{(i)}_{-}(t,p)\right)^2\bigg],
	\cr\nonumber\\
	&&\mathcal{C}_{2}(t) = \text{vol} \int_{p\leq \Lambda} \frac{d^{d-1}p}{(2\pi)^{d-1}} \hspace{1mm}\bigg(\frac{1}{4}\sum_{i=1}^{2}\bigg[\left(\log\Delta^{(i)}_{+}(t,p)\right)^2\hspace{.5mm}+\left(\log\Delta^{(i)}_{-}(t,p)\right)^2\bigg]\bigg)^{\frac{1}{2}},
	\cr \nonumber\\&&
	\mathcal{C}_{1}(t) = \text{vol} \int_{p\leq \Lambda} \frac{d^{d-1}p}{(2\pi)^{d-1}} \hspace{1mm}\frac{1}{2}\sum_{i=1}^{2}\bigg[\hspace{1mm}\bigg{|}\log\Delta^{(i)}_{+}(t,p)\bigg{|}\hspace{.5mm}+\bigg{|}\log\Delta^{(i)}_{-}(t,p)\bigg{|}\hspace{1mm}\bigg],
	\eea
	where $\Delta^{(i)}_{\pm}(t,p)$ are obtained by substituting 
	\bea
	\lambda \rightarrow \lambda_{p}
	,\hspace{1cm}\omega \rightarrow \omega_{p},
	\eea
	in eigenvalues (\ref{EigenDeltap}). In the first view, one may suspect that these integrals will produce UV divergences. But these potential
	divergences are eliminated, because we have powers
	of $e^{-\beta \Lambda}$ competing against (positive) powers of $\Lambda$ in these contributions, and so they actually vanish in the limit of $\Lambda \rightarrow \infty$. According to this fact, 
	we remove the UV regulator and integrating all the way to
	infinite momenta. But the above integrals have IR divergences which they are encoded in the "\text{vol}" coefficients. These IR divergences can be remove by dividing the change of complexity with thermal entropy 
	\bea
	\label{Sthq}
	S_{\text{th}} = \frac{\text{vol}}{\beta^{d-1}}\frac{\Omega_{d-2}}{(2\pi)^{d-1}} \int_{0}^{\infty}du\hspace{1mm}u^{d-2}\hspace{1mm}\bigg[\frac{\sqrt{u^{2}+\beta^2 m^2}+\beta \mu q}{e^{\sqrt{u^2+\beta^2 m^2}+\beta\mu q}-1}-\log\bigg(1-e^{-\sqrt{u^2+\beta^2 m^2}-\beta \mu q})\bigg)\bigg],
	\eea
	where $\Omega_{d-2}$ is the volume of a $(d-2)$-sphere, i.e. $\Omega_{d-2}=2\pi^{\frac{d-1}{2}}/\Gamma(\frac{d-1}{2})$. Last but not least, to take into account the reference scale together with massless limit, it will be useful to define dimensionless variables
	\bea
	\label{deftildes}
	\tilde{t} = \frac{t}{\beta},\hspace{1cm}\tilde{k} = \beta k,\hspace{.5cm}\tilde{\gamma} =\frac{1}{\beta \omega_{R}},\hspace{1cm} \tilde{Q} = \mu q \beta.
	\eea
	\subsection{Complexity of formation: A prob for prefer cost function}\label{prefereC}
\begin{figure}[h]
		\hspace{1.1cm}\includegraphics[scale=.33]{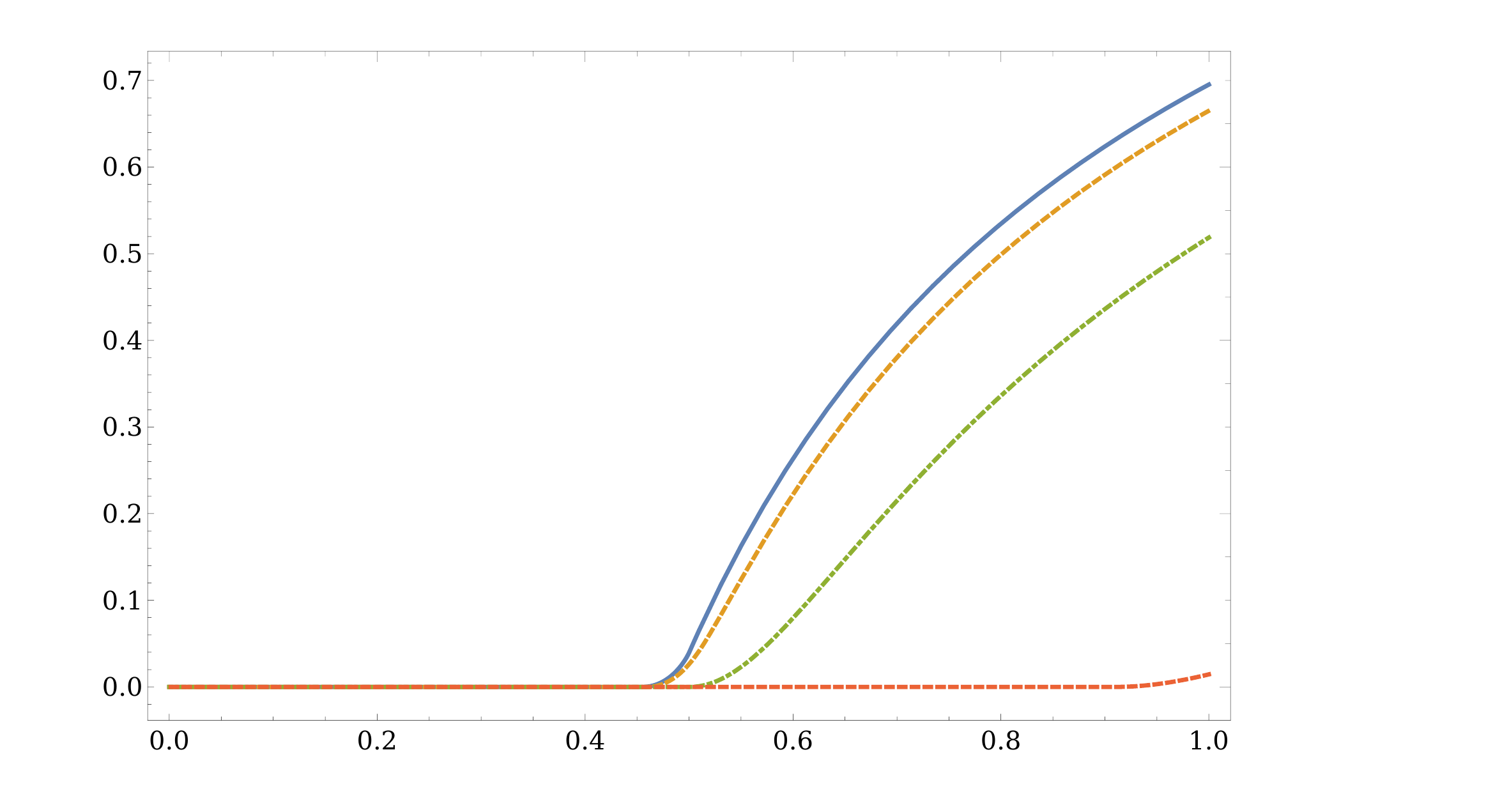}\put(-235,35){\rotatebox{-270}{\fontsize{15}{15}\selectfont $\frac{\mathcal{C}_{1}(0)-\mathcal{C}_{1}(\text{vac})}{S_{th}}$}}		\put(-135,-5){{\fontsize{12}{12}\selectfont $T/\Lambda$}}\hspace{-.4cm}\includegraphics[scale=.33]{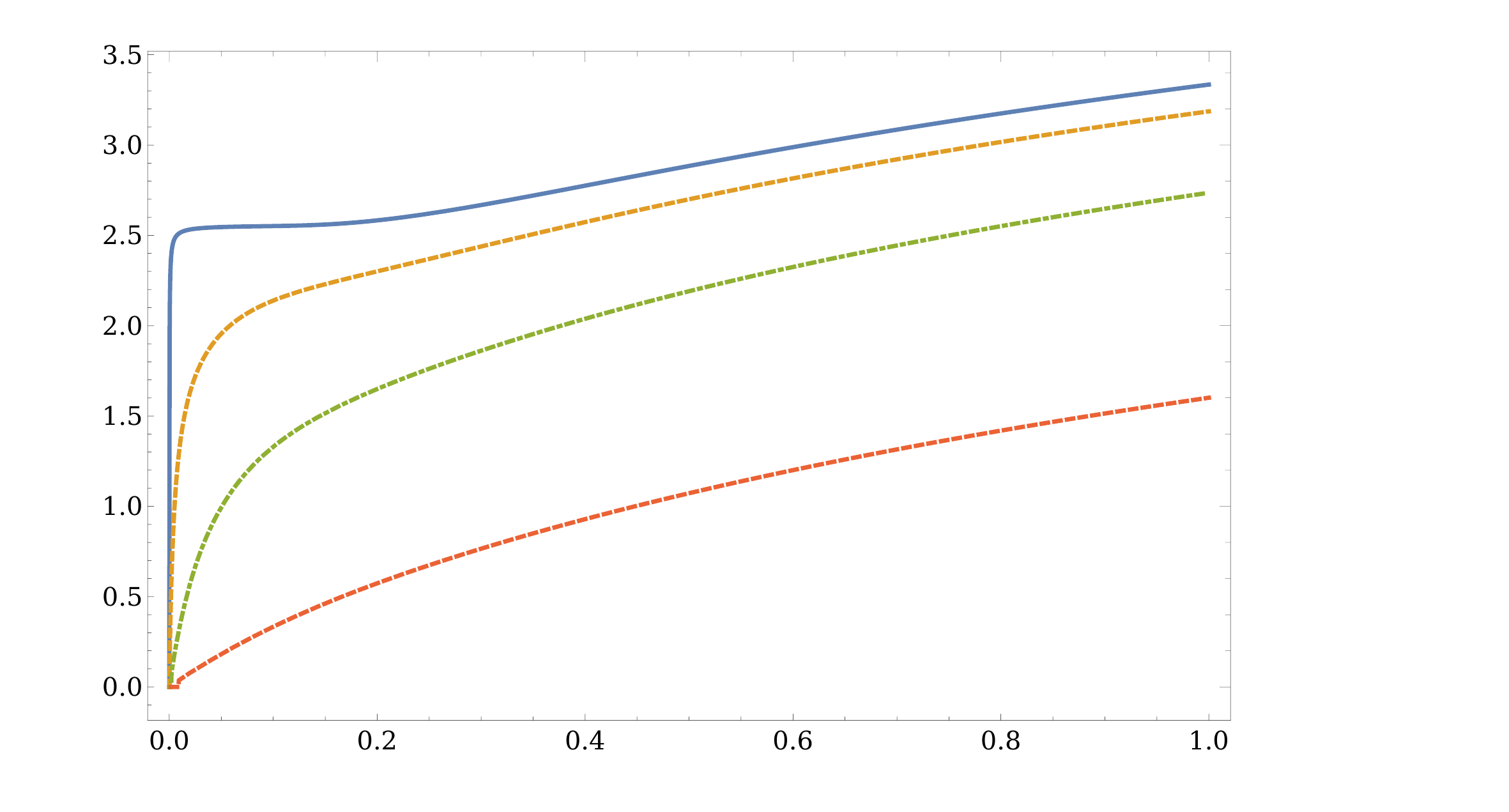}\put(-235,25){\rotatebox{-270}{\fontsize{14}{14}\selectfont $\frac{\mathcal{C}_{\kappa=2}(0)-\mathcal{C}_{\kappa=2}(\text{vac})}{S_{th}}$}}	\put(-135,-5){{\fontsize{12}{12}\selectfont$T/\Lambda$}}\vspace{.2cm}
		
		\hspace{4.5cm}\includegraphics[scale=.33]{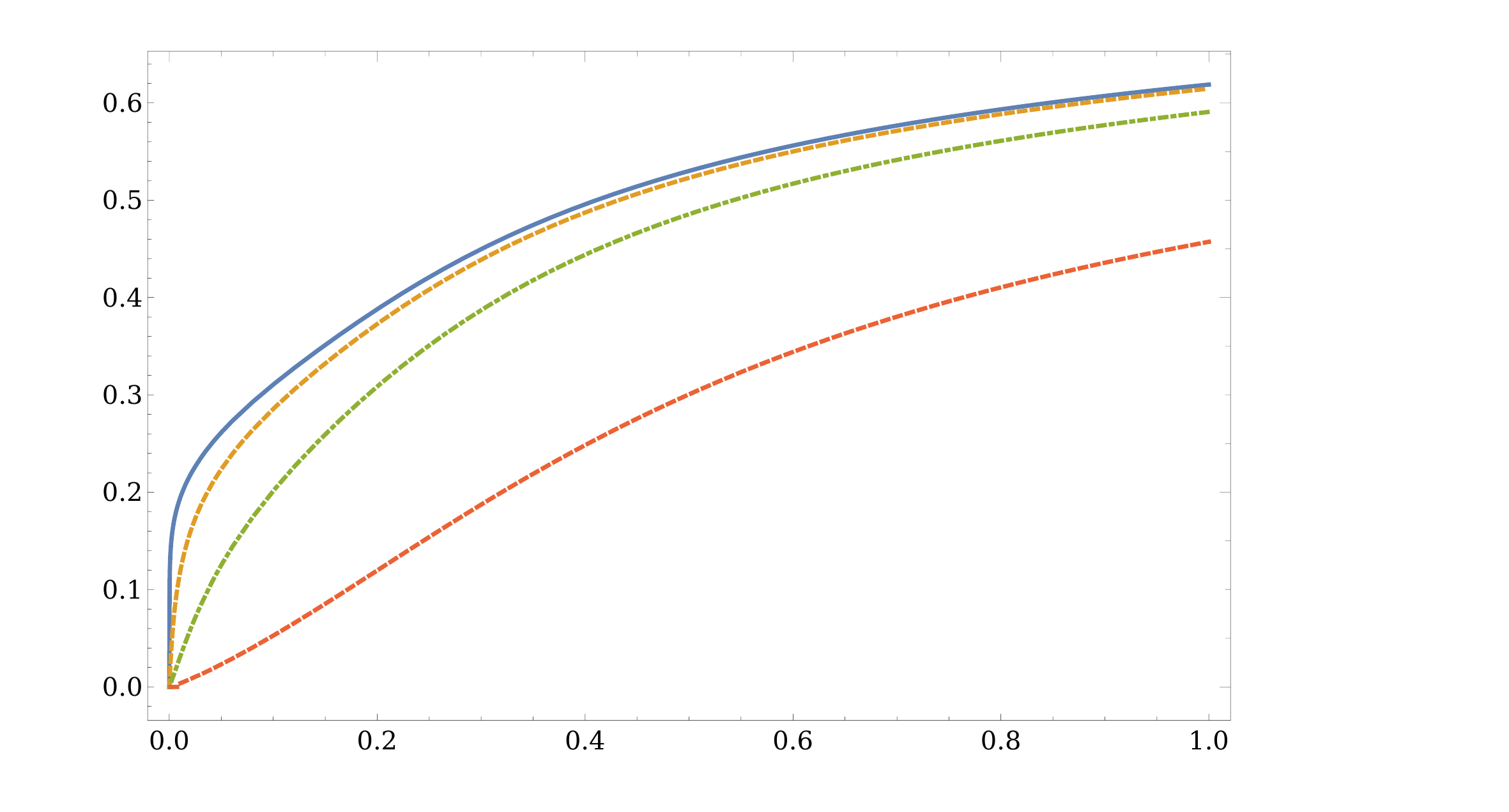}\put(-236,35){\rotatebox{-270}{\fontsize{15}{15}\selectfont $\frac{\mathcal{C}_{2}(0)-\mathcal{C}_{2}(\text{vac})}{S_{th}}$}}		\put(-135,-5){{\fontsize{12}{12}\selectfont $T/\Lambda$}}
		\caption{ $(\mathcal{C}_{1},\mathcal{C}_{\kappa=2}, \mathcal{C}_{2})$ complexities of formation normalized by the entropy for $\mu q=10^{-5}\Lambda$ (blue), $10^{-2}\Lambda$ (dashed orange), $10^{-1}\Lambda$ (dotted dashed green), $\Lambda$ (dashed red) and $d=1+1, m=10^{-5}\Lambda, \omega_{R} = 2\Lambda$. For small temperatures, the profile of the ratio of the complexity of formation over the entropy becomes small. For higher temperatures, it develops a dependence on the temperature and the cutoff scale $\Lambda$. The nontrivial profile of the ratio of the complexity of formation to the entropy contrasts with the holographic results of ref. \cite{Carmi:2017jqz}.}\label{compforms}	
	\end{figure}
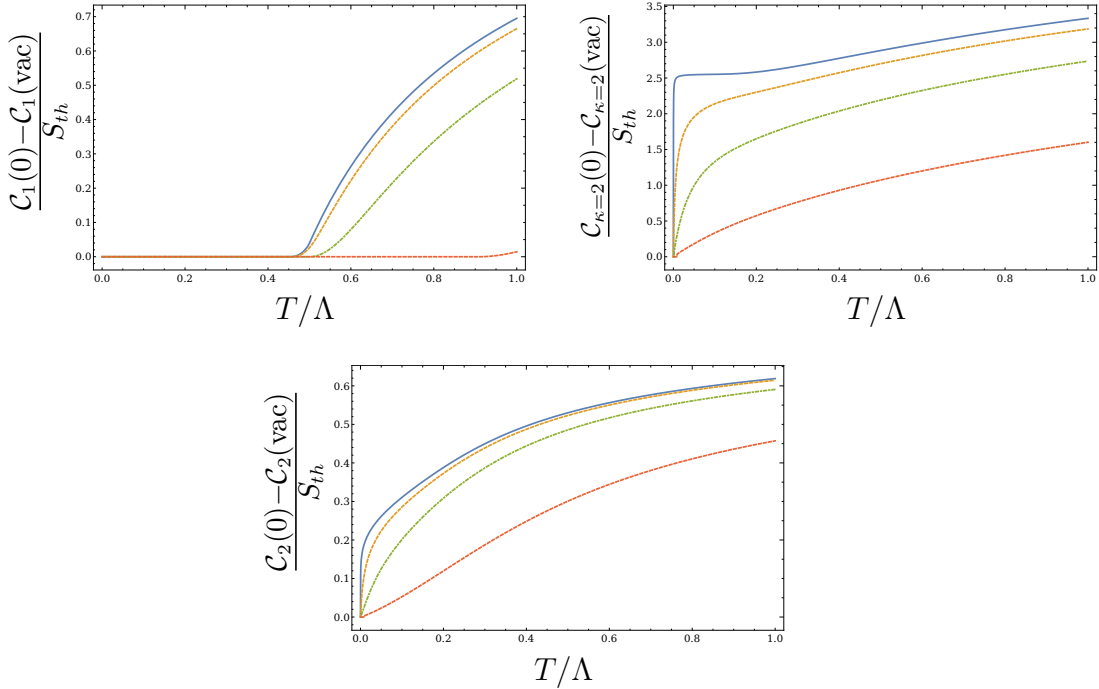

	In this subsection we study the complexity of formation for cTFD state (\ref{TFD.3}) carefully to see which cost function is consistent with the UV structure and third law of holographic complexity \cite{Chapman:2018hou}. The complexity of formation is indeed the 
	extra complexity required to prepare the two copies of a complex scalar field theory in the cTFD
	state compared to simply preparing each of the copies in the vacuum state, i.e. $\Delta \mathcal{C} = \mathcal{C}(0) -\mathcal{C}(0)\big{|}_{\beta \rightarrow \infty}
	$. For the case where the frequency of all modes $\omega_{p}$ is much less than the reference scale $\omega_{R}$, the eigenvalues (\ref{EigenDeltap}) can be simplified to
	\bea
	\label{b1}
	\Delta^{(1)}_{\pm}(t) \approx 0,
	\hspace{1.5cm}\Delta^{(2)}_{\pm}(t) \approx \frac{\omega_{R}}{\omega_{p}} \bigg(
	\cosh{2\alpha_{p}} \pm \cos\big[{(\omega_{p}+\mu q)t}\big] \sinh{2\alpha_{p}}\bigg).
	\eea
	It is worth noting that very high
	frequency contributions are exponentially suppressed and so in fact, we only need to consider
	$\omega_{R} \gg T$. In this regime, the complexity of formation for the ones in (\ref{Cscovariance}) becomes 
	\bea
	\label{diagonalcomplex}
	&& \Delta \mathcal{C}_{\kappa=2} \hspace{1mm}
	= \text{vol} \int_{p\leq \Lambda} \frac{d^{d-1}p}{(2\pi)^{d-1}}\hspace{1mm} 2\alpha_{p}^{2},
	\cr\nonumber\\&&
	\Delta \mathcal{C}_{2} \hspace{1mm}= \text{vol} \int_{p\leq \Lambda} \frac{d^{d-1}p}{(2\pi)^{d-1}}\hspace{2mm}\sqrt{\frac{1}{2}\log^{2}\left(\frac{\omega_{P}}{\omega_{R}}\right)+2\alpha_{p}^2}-\frac{1}{\sqrt{2}}\bigg{|}\log\frac{\omega_{p}}{\omega_{R}}\bigg{|},
	\cr\nonumber\\
	&&\Delta \mathcal{C}_{1} \hspace{1mm}= \text{vol} \int_{p\leq \Lambda} \frac{d^{d-1}p}{(2\pi)^{d-1}}\left(\bigg{|}\frac{1}{2}\log\frac{\omega_{p}}{\omega_{R}}-\alpha_{p} \bigg{|}+\bigg{|}\frac{1}{2}\log\frac{\omega_{p}}{\omega_{R}}+\alpha_{p} \bigg{|}-\bigg{|}\log\frac{\omega_{p}}{\omega_{R}}\bigg{|}\hspace{1mm}\right).
	\eea
	These complexities are presented in fig.\ref{compforms}. To probe which of them is consistent with UV divergences in holographic results, let us focus for the moment on the vacuum state ($\alpha_{p}\rightarrow 0$). According to (\ref{b1}) for this state we have
	\bea
	\label{Cs}
	\mathcal{C}_{\kappa=2} =\frac{1}{4} \sum_{[p_i]=0}^{N-1}\left(\log\frac{\omega_{\vec{p}}}{\omega_{R}}\right)^2,\hspace{.5cm}\mathcal{C}_{2} =\frac{1}{2}\sqrt{\sum_{[p_i]=0}^{N-1}\left(\log\frac{\omega_{\vec{p}}}{\omega_{R}}\right)^{2}},\hspace{.5cm}\mathcal{C}_{1}=\frac{1}{2} \sum_{[p_i]=0}^{N-1}\bigg{|}\log\frac{\omega_{\vec{p}}}{\omega_{R}}\hspace{.5mm}\bigg{|}
	\eea
	with $p_i$'s are the components of the momentum vector $\vec{p}=(p_1,p_2,...,p_{d-1})$ and the total number of oscillators is $
	N^{d-1} =V/\delta^{d-1}$. Therefore, contribution of UV modes i.e. $\omega_{\vec{p}}\sim 1/\delta$ to (\ref{Cs}) become   
	\bea
	\label{Cs2}
	\mathcal{C}_{\kappa=2} \sim  \frac{V}{\delta^{d-1}}\left(\log\frac{1}{\delta\hspace{.5mm}\omega_{R}}\right)^2,\hspace{.5cm}\mathcal{C}_{2} \sim (\frac{V}{\delta^{d-1}})^{1/2}\log\left(\frac{1}{\delta\hspace{1mm}\omega_{R}}\right),\hspace{.5cm}\mathcal{C}_{1}\sim \frac{V}{\delta^{d-1}}\bigg{|}\log\frac{1}{\delta\hspace{.5mm}\omega_{R}}\hspace{.5mm}\bigg{|}
	\eea
	The leading divergence appearing in CA proposal takes the 
	\bea
	\label{Cholog}
	\mathcal{C}_{\text{holography}} \sim \frac{V}{\delta^{d-1}}\log\left(\frac{l}{\tilde{\alpha} \delta}\right),
	\eea
	where $\delta$ is the short-distance cut-off scale in the boundary CFT,
	$l$ is the AdS curvature
	scale of the bulk spacetime, and $\tilde{\alpha}$
	is an arbitrary (dimensionless) coefficient which fixes the
	normalization of the null normals on the boundary of the WDW patch.  Since $\mathcal{C}_{\text{holography}}$ is a quantity
	which is to be defined in the boundary CFT, it should not depend on the bulk AdS scale. One can eliminate this factor with the freedom in choosing $\tilde{\alpha}=\tilde{\omega}_{R}\hspace{.5mm}l$
	where $\tilde{\omega}_{R}$
	is some arbitrary frequency. In this case (\ref{Cholog}) is simplified to
	\bea
	\mathcal{C}_{\text{holography}} \sim \frac{V}{\delta^{d-1}}\log\left(\frac{1}{\delta\hspace{.5mm}\tilde{\omega}_{R}}\right),
	\eea
	which interestingly agrees with $\mathcal{C}_{1}$ in (\ref{Cs2}). It is discussed \cite{Lehner:2016vdi} that the CA proposal which leds to (\ref{Cholog}) does not have a reparameterization invariance of null boundary coordinates. To recover this symmetry
	one can consider a new boundary term which intriguingly adding it also removes that UV logarithmic divergence. The $\mathcal{C}_{1}$ complexity (\ref{Cs2}) agrees with holography also in the presence of this new boundary term if, 
	for example $\omega_{R}$ is set by the cut-off scale i.e. $\omega_{R} \sim 1/\delta$.
	
\begin{figure}[H]
		\hspace{1.1cm}\includegraphics[scale=.35]{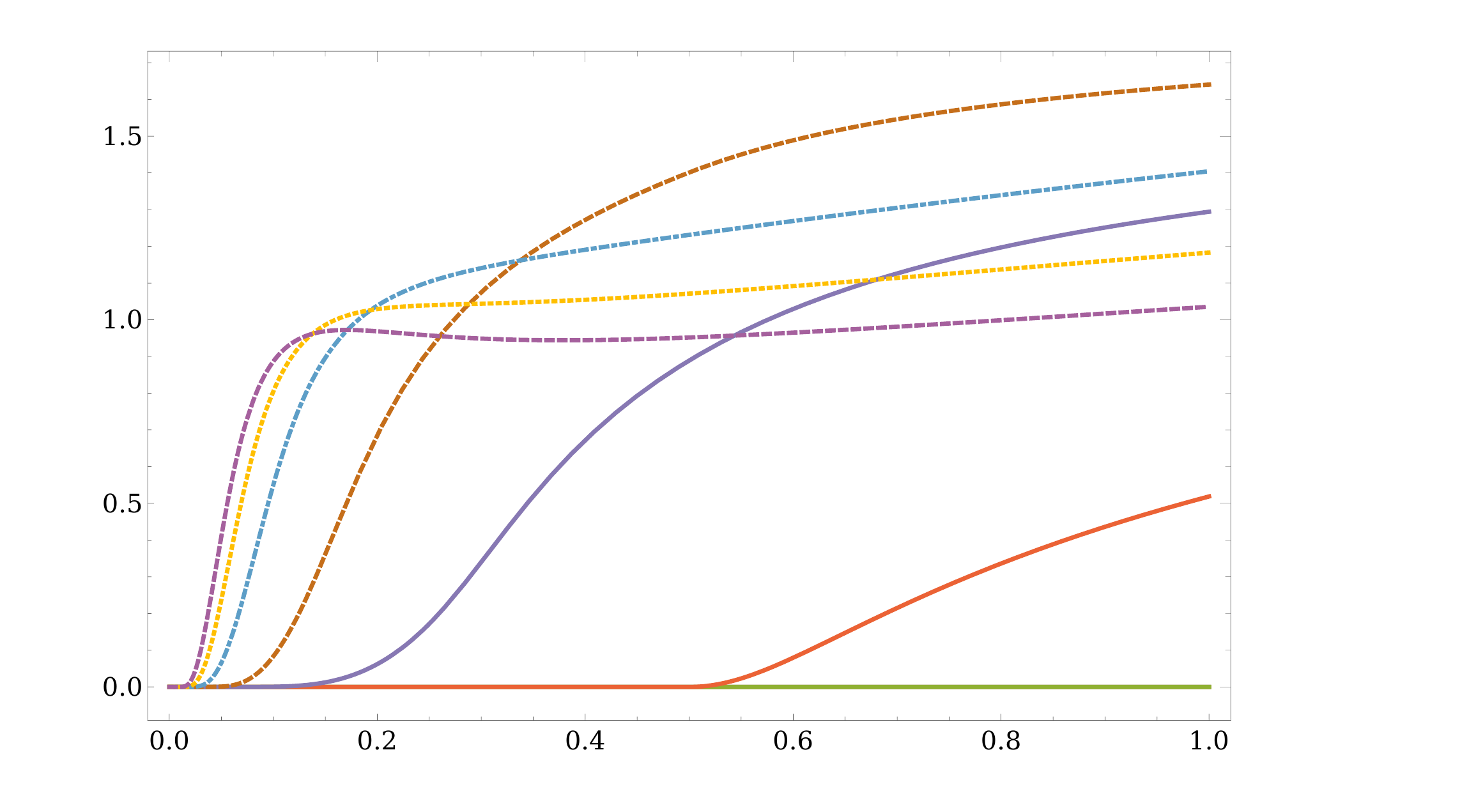}\put(-243,40){\rotatebox{-270}{\fontsize{15}{15}\selectfont $\frac{\mathcal{C}_{1}(0)-\mathcal{C}_{1}(\text{vac})}{S_{th}}$}}	\put(-135,-5){{\fontsize{12}{12}\selectfont$T/\Lambda$}}\hspace{-.1cm}\includegraphics[scale=.35]{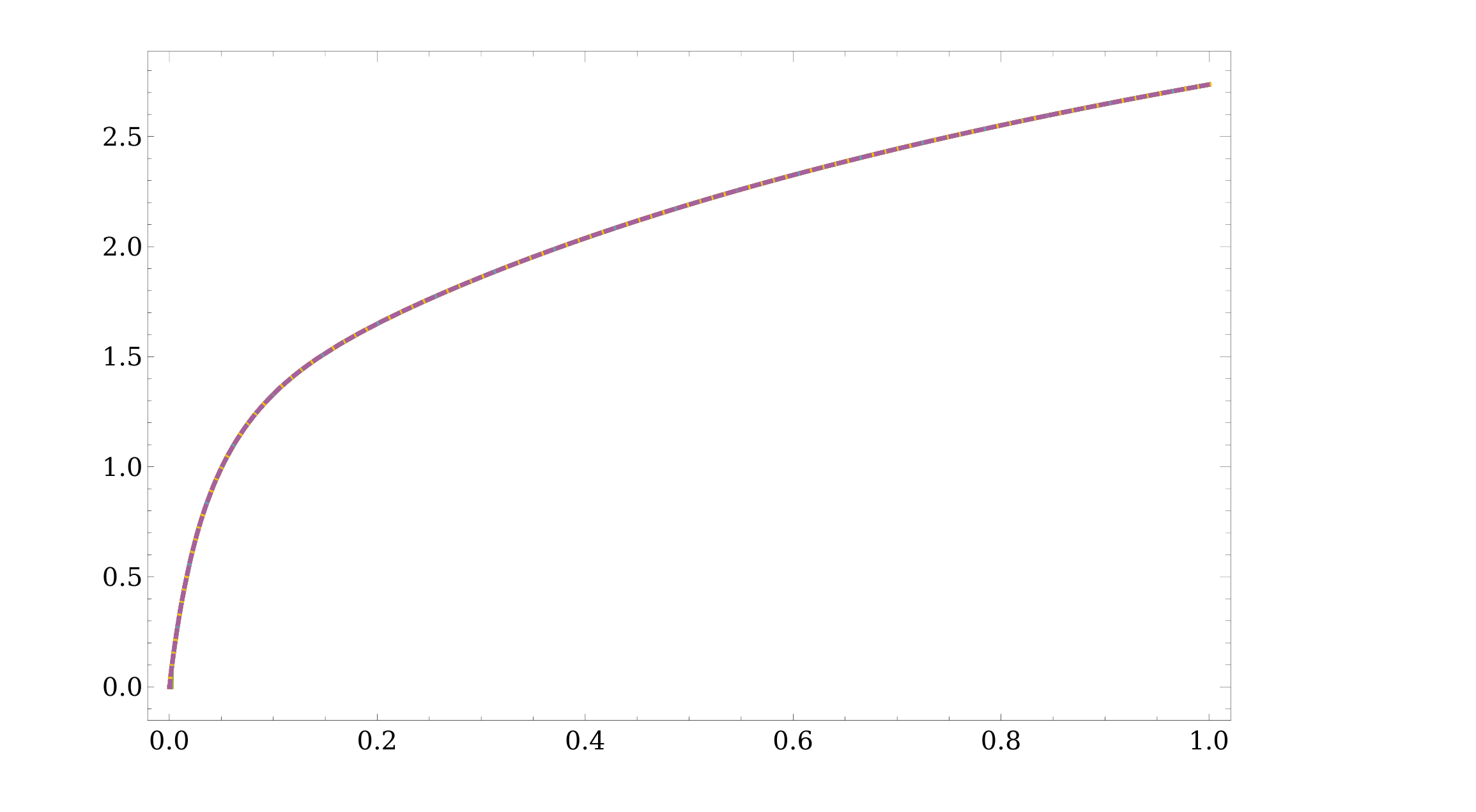}\put(-246,30){\rotatebox{-270}{\fontsize{14}{14}\selectfont $\frac{\mathcal{C}_{\kappa=2}(0)-\mathcal{C}_{\kappa=2}(\text{vac})}{S_{th}}$}}	\put(-135,-5){{\fontsize{12}{12}\selectfont$T/\Lambda$}}
		\vspace{.3cm}
		
		\hspace{4.7cm}\includegraphics[scale=.36]{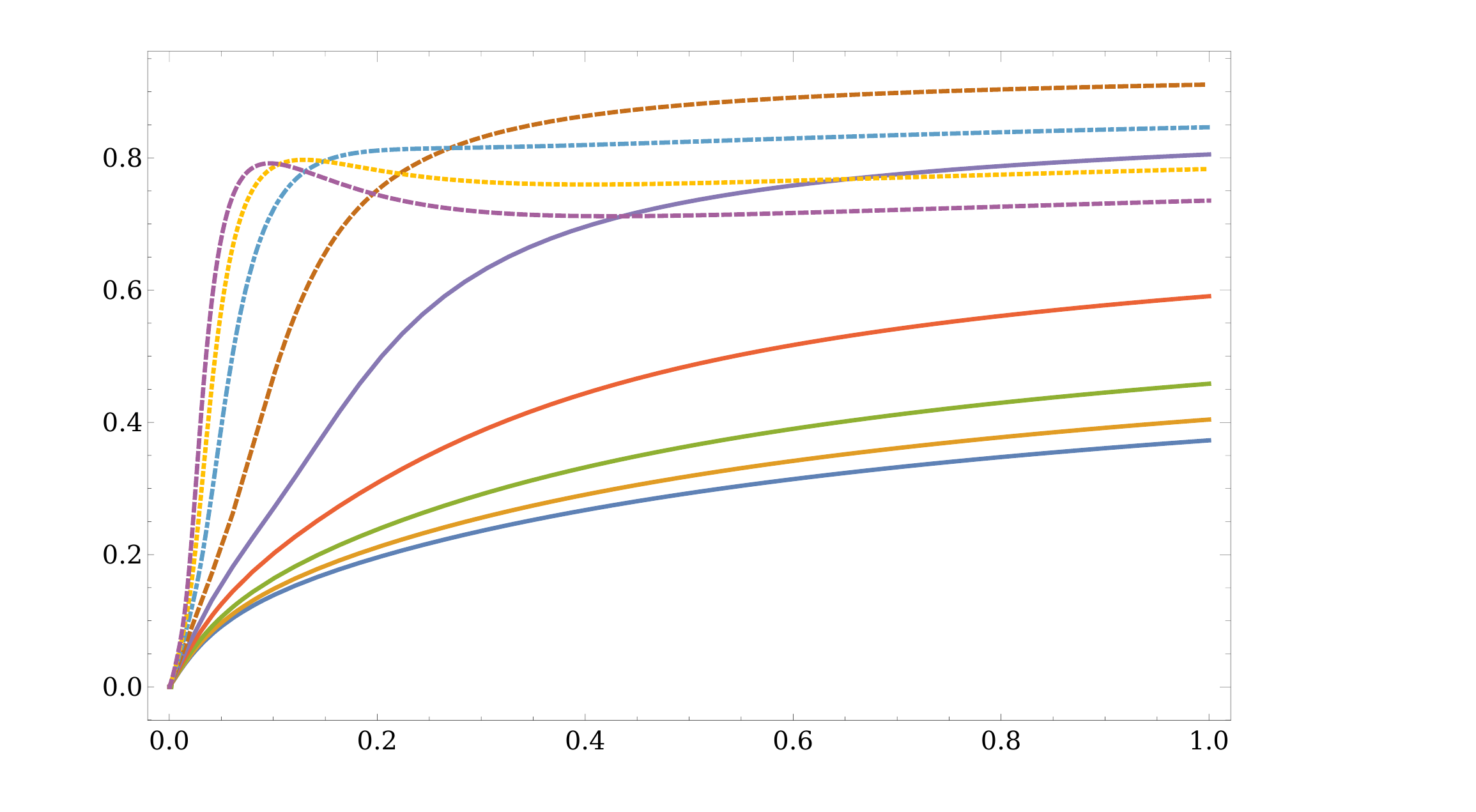}\put(-249,37){\rotatebox{-270}{\fontsize{15}{15}\selectfont $\frac{\mathcal{C}_{2}(0)-\mathcal{C}_{2}(\text{vac})}{S_{th}}$}}	\put(-135,-5){{\fontsize{13}{13}\selectfont$T/\Lambda$}}\put(0,8){\includegraphics[scale=.6]{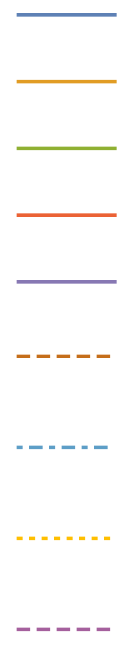}}\put(25,117){{\fontsize{9.5}{9.5}\selectfont$\omega_{R}=8\Lambda$}}\put(25,105){{\fontsize{9.5}{9.5}\selectfont$\omega_{R}=6\Lambda$}}\put(25,94){{\fontsize{9.5}{9.5}\selectfont$\omega_{R}=4\Lambda$}}\put(25,82){{\fontsize{9.5}{9.5}\selectfont$\omega_{R}=2\Lambda$}}\put(25,70){{\fontsize{9.5}{9.5}\selectfont$\omega_{R}=\Lambda$}}\put(25,57){{\fontsize{9.5}{9.5}\selectfont$\omega_{R}=\Lambda/2$}}\put(25,42){{\fontsize{9.5}{9.5}\selectfont$\omega_{R}=\Lambda/4$}}\put(25,26){{\fontsize{9.5}{9.5}\selectfont$\omega_{R}=\Lambda/6$}}\put(25,10){{\fontsize{9.5}{9.5}\selectfont$\omega_{R}=\Lambda/8$}}
		\caption{$\mathcal{C}_{1},\mathcal{C}_{\kappa=2}, \mathcal{C}_{2}$ complexities of formation normalized by the entropy for different values of $\omega_{R}$
			and $d=1+1, m=10^{-5}\Lambda, \mu q=10^{-1}\Lambda$. For small temperatures, the profile of the ratio of the complexity of formation over the entropy becomes small. For higher temperatures, it develops a dependence on the temperature and the cutoff scale $\Lambda$. The nontrivial profile of the ratio of the complexity of formation to the entropy contrasts with the holographic results of ref. \cite{Carmi:2017jqz}.}\label{compforms2}	
	\end{figure}

\begin{figure}[h]
		\hspace{1.2cm}\includegraphics[scale=.35]{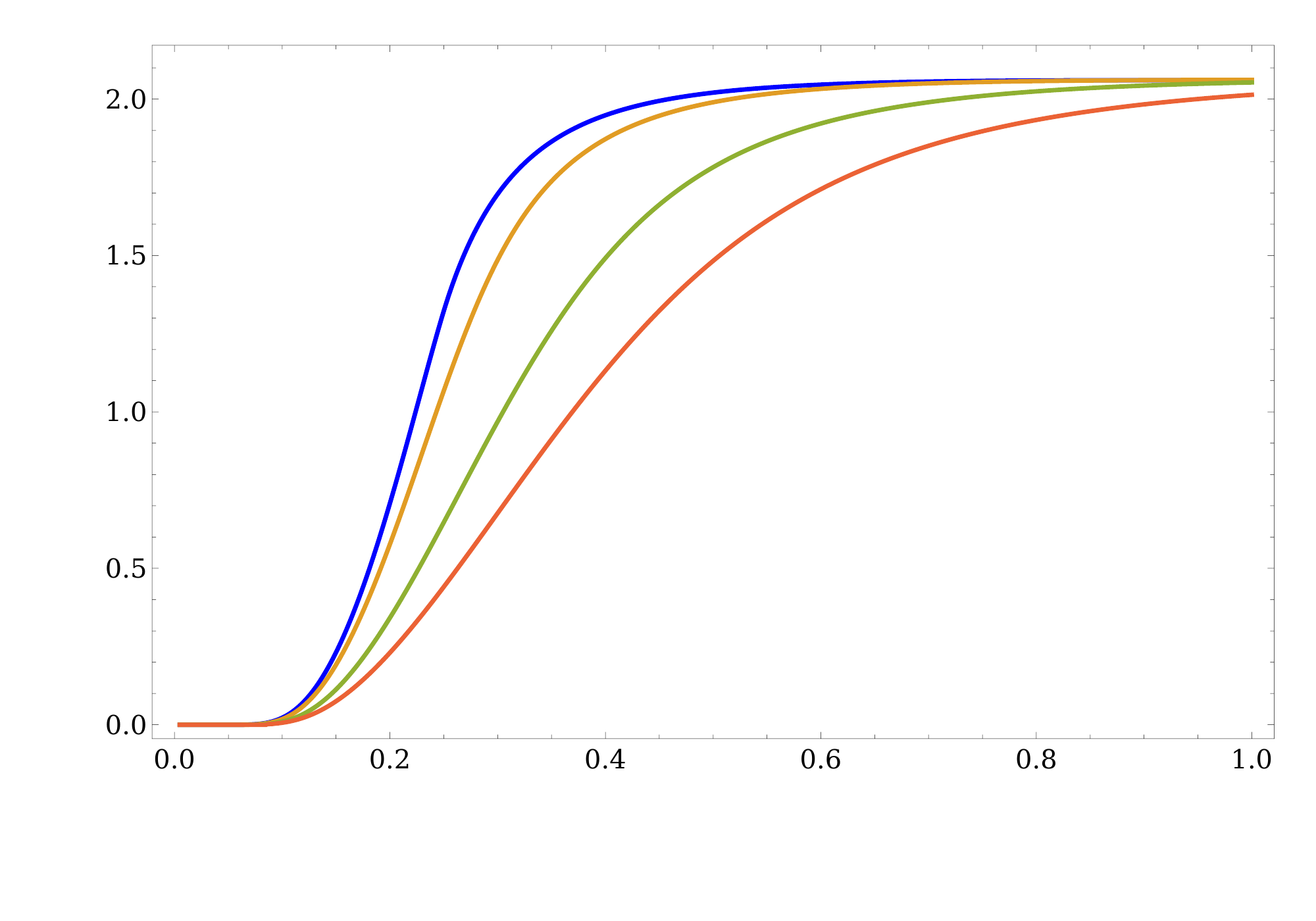}\put(-215,50){\rotatebox{-271}{\fontsize{14}{14}\selectfont $\frac{\mathcal{C}_{1}(0)-\mathcal{C}_{1}(\text{vac})}{S_{th}}$}}	\put(-105,12){{\fontsize{12}{12}\selectfont$T/\Lambda$}}\put(-135,140){{\fontsize{11}{11}\selectfont$d=3+1, \omega_{R}=\Lambda$}}\hspace{2cm}\put(+15,-5){\includegraphics[scale=.35]{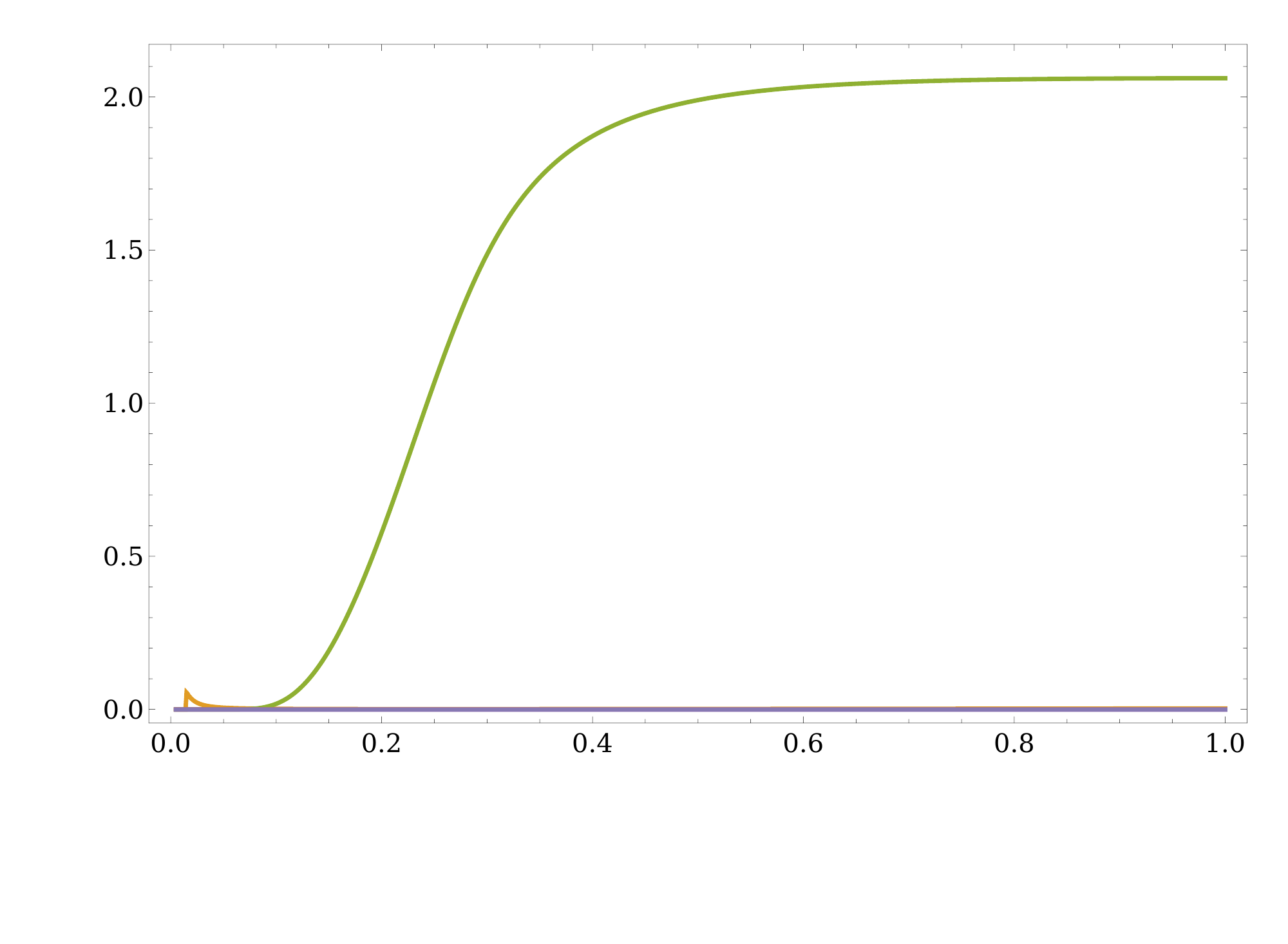}}\put(5,52){\rotatebox{-271}{\fontsize{14}{14}\selectfont $\frac{\mathcal{C}_{1}(0)-\mathcal{C}_{1}(\text{vac})}{S_{th}}$}}	\put(110,12){{\fontsize{12}{12}\selectfont$T/\Lambda$}}\put(70,140){{\fontsize{11}{11}\selectfont$d=3+1, \mu q = 10^{-1}\Lambda$}}\vspace{-.5cm}
		
		\caption{\textbf{Left:} $\mathcal{C}_{1}$ complexity of formation normalized by the entropy for $\mu q = 0$ (blue), $10^{-1}\Lambda$ (brown), $\Lambda/2$ (green), $\Lambda$ (red) and $d=3+1, m=10^{-5}\Lambda, \omega_{R} =\Lambda$. For small temperatures, the profile of the ratio of the complexity of formation over the entropy becomes small. For higher temperatures, it develops a dependence on the temperature and the cutoff scale $\Lambda$. The nontrivial profile of the ratio of the complexity of formation to the entropy contrasts with the holographic results of \cite{Carmi:2017jqz}. \textbf{Right:} $\mathcal{C}_{1}$ complexity of formation normalized by the entropy for $\omega_{R} =10^{-2}\Lambda$ (brown), $\Lambda$ (green), $10^{4}\Lambda$ (purple) and $d=3+1, m=10^{-5}\Lambda, \mu q= 10^{-1} \Lambda$. For small and large values of reference scale, the profile of the ratio of the complexity of formation over the entropy is almost very small. For high temperatures, for reference scale in order of UV cutoff it develops a dependence on the temperature and the cutoff scale $\Lambda$.}\label{c13p1}	
	\end{figure}
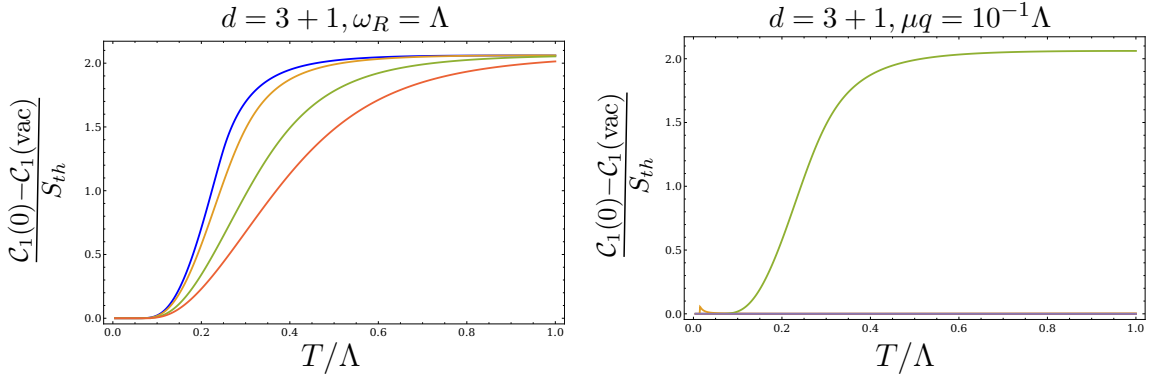

	We would like to mention here that according to the above discussion, it seems $\mathcal{C}_{1}$ complexity (\ref{Cs}) for ground state agrees with the holographic proposal. But by carefully noting to fig.\ref{compforms}, we see that $\mathcal{C}_{1}$ complexity is in contrast with the third law of holographic complexity \cite{Carmi:2017jqz}. The third law of holographic complexity expresses that the complexity of formation for a charged black hole diverges in extremal limit. The origin of this mismatch can be understood by evaluating carefully that the sign changes
	inside each argument of the logarithms in (\ref{Cs}). Let us concentrate on the massless theory for the moment.  In this case, 
	$\frac{1}{2}\log\frac{p}{\omega_{R}}+\alpha_{p}$ changes sign at a value $p_{c}$ given by
	\bea
	\label{pc}
	p_{c}\coth\left(\frac{p_{c}+\mu q}{4T}\right) =\omega_{R}
	\eea
	There are two important limits to the above  equation: when $p_c+\mu q$ is very small
	and when $p_c$ is close to the cutoff scale $\Lambda$. Solving for the temperature in these two regimes, we find
	\bea
	T_{\text{c}_1}= \frac{(\Lambda+\mu q)}{2} \frac{1}{\log\left(\frac{\omega_{R}+\Lambda}{\omega_{R}-\Lambda}\right)},\hspace{1cm}T_{\text{c}_2} = \frac{\omega_{R}}{4}\left(1+\frac{\mu q}{p_c}\right).
	\eea
	For $T < T_{\text{c}_1}$ , the arguments of all of the absolute values in (\ref{diagonalcomplex}) are
	negative which implies that the complexity of formation is identically zero, $\Delta\mathcal{C}_{1}(T < T_{\text{cl}}) = 0$. For temperature within the range $ T_{\text{c}_1} < T < T_{\text{c}_2}$, there is a single solution  $p_c$ to the (\ref{pc}) in the range $[0, \Lambda]$. We find that for $p < p_c$ the argument of
	the second absolute value is negative, and for $p > p_c$ the argument is positive. The complexity
	of formation in this situation is found by integrating only over modes larger than $p_c$,
	\bea
	\Delta\mathcal{C}_{1}(T_{\text{c}_1}< T < T_{\text{c}_2}) =\text{vol}\frac{\Omega_{d-2}}{(2\pi)^{d-1}} \int_{p_c}^{\Lambda} dp\hspace{.5mm} p^{d-2}\left(2\alpha_{p}-\log\frac{p}{\omega_{R}}\right).
	\eea
	Finally, if the temperature is bigger than $T_{\text{c}_2}$, we find that $\frac{1}{2}\log\frac{p}{\omega_{R}}+\alpha_{p}$ is always positive in
	the range of momenta $[0,\Lambda]$. Therefore, we have
	\bea
	\Delta\mathcal{C}_{1}( T > T_{\text{c}_2}) =\text{vol}\frac{\Omega_{d-2}}{(2\pi)^{d-1}} \int_{0}^{\Lambda} dp\hspace{.5mm} p^{d-2}\left(2\alpha_{p}-\log\frac{p}{\omega_{R}}\right).
	\eea
	Therefore for small temperatures with respect to the cutoff scale, we find that the complexity of formation is exactly zero. For higher temperatures, there are some nontrivial cancellations between the circuits that introduce some dependence on T and $\Lambda$ that contrasts
	with the holographic results of ref.\cite{Carmi:2017jqz}. To complete this discussion, we also present the same
	analysis for different $\omega_{R}$ in fig.\ref{compforms2} and for $d=3+1$ in fig.\ref{c13p1}. 
	
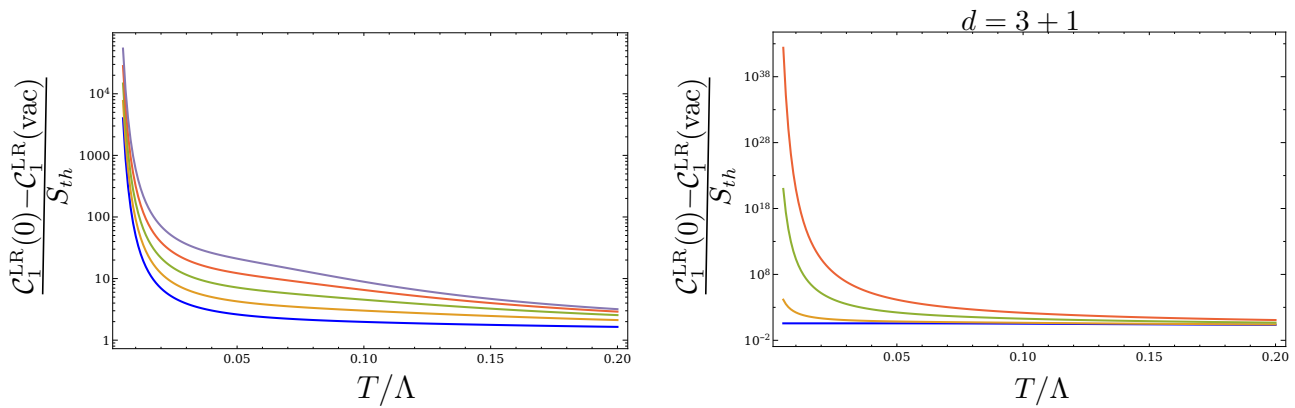
\begin{figure}[h]
		\hspace{.2cm}\includegraphics[scale=.4]{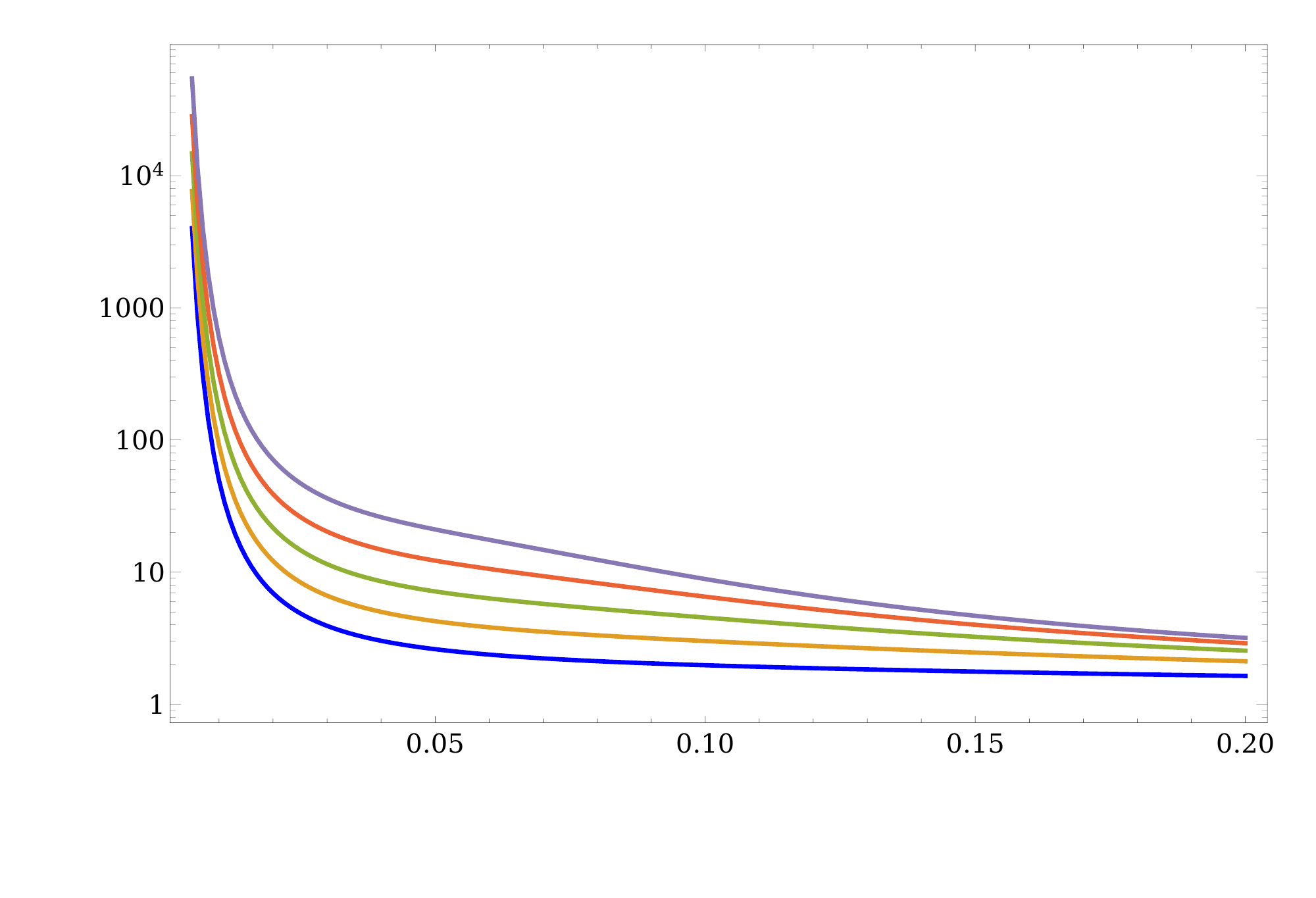}\put(-240,50){\rotatebox{-271}{\fontsize{15}{15}\selectfont $\frac{\mathcal{C}_{1}^{\text{LR}}(0)-\mathcal{C}_{1}^{\text{LR}}(\text{vac})}{S_{th}}$}}	\put(-110,12){{\fontsize{12}{12}\selectfont$T/\Lambda$}}\hspace{.5cm}
		\centering\includegraphics[scale=.395]{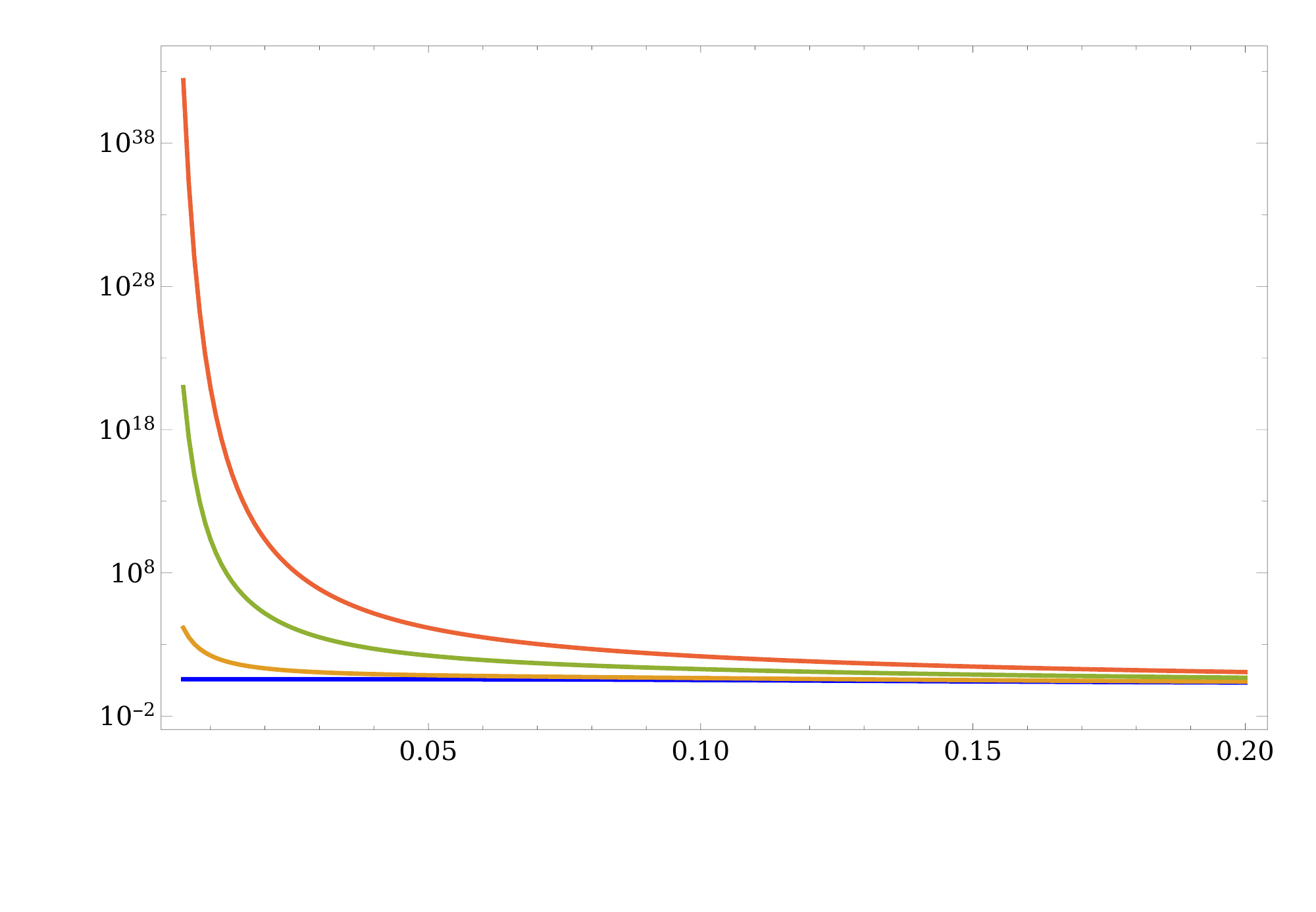}\put(-238,50){\rotatebox{-271}{\fontsize{15}{15}\selectfont $\frac{\mathcal{C}_{1}^{\text{LR}}(0)-\mathcal{C}_{1}^{\text{LR}}(\text{vac})}{S_{th}}$}}	\put(-110,12){{\fontsize{11}{11}\selectfont$T/\Lambda$}}\put(-130,151){{\fontsize{11}{11}\selectfont$d=3+1$}}\vspace{-.5cm}
		\caption{\textbf{Left:}\hspace{1mm}$ \mathcal{C}_{1}$ complexity of  formation normalized by the entropy in $LR$ basis at constant chemical potential $\mu q =10^{-1}\Lambda$ and very small mass $m=10^{-5}\Lambda$ for $d = 1+1$ (blue), $2+1$ (brown), $3+1$ (green), $4+1$ (red) and $5+1$ (purple). For small temperatures, the profile of the ratio of the complexity of formation over the entropy diverges. For higher temperatures, it decreases to a constant value which it depends to  dimension. \textbf{Right:}\hspace{1mm}$\mathcal{C}_{1}$ complexity of formation normalized by the entropy in $LR$ basis at different constant chemical potential $\mu q =0$ (blue), $\mu q=10^{-1}\Lambda$ (brown), $\mu q =\Lambda/2$ (green) and $\mu q= \Lambda$ (red) with very small mass $m=10^{-5}\Lambda$. For small temperatures, the profile of the ratio of the complexity of formation over the entropy diverges. For higher temperatures, it decreases to a constant value.}\label{c1lrt0}	
	\end{figure}

\begin{figure}[H]
		\hspace{2cm}\includegraphics[scale=.17]{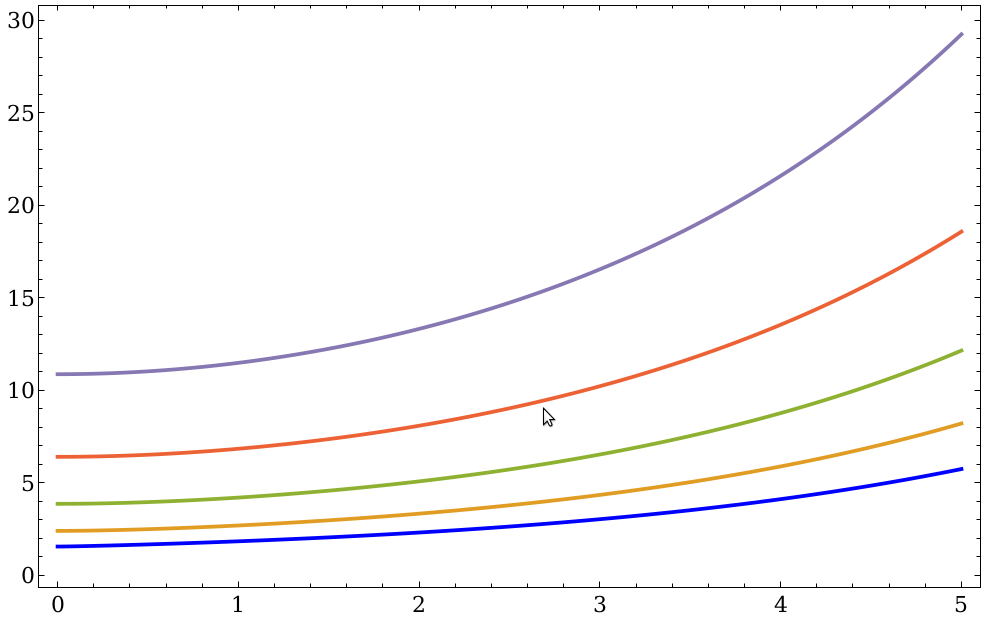}\put(-195,20){\rotatebox{-271}{\fontsize{14}{14}\selectfont $\frac{\mathcal{C}_{1}^{\text{LR}}(0)-\mathcal{C}_{1}^{\text{LR}}(\text{vac})}{S_{th}}$}}\put(-100,-15){{\fontsize{13}{13}\selectfont$m/T$}}\hspace{1.5cm}\includegraphics[scale=.2]{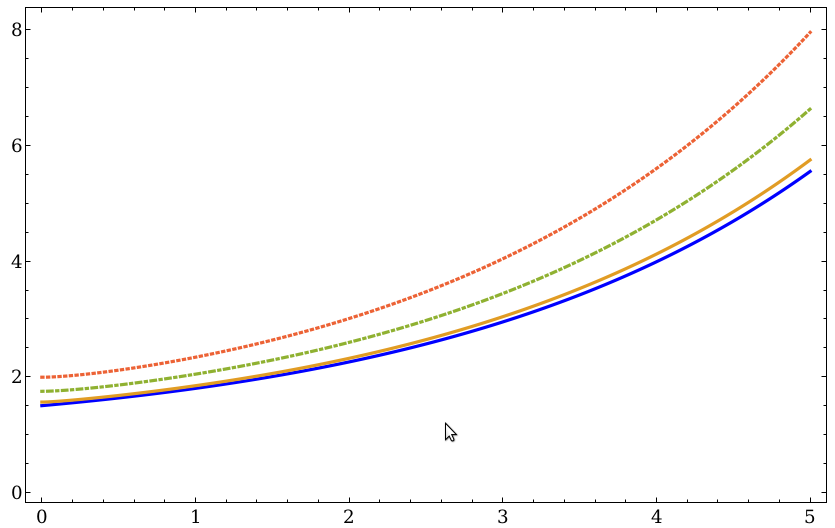}\put(-189,20){\rotatebox{-271}{\fontsize{13}{13}\selectfont $\frac{\mathcal{C}_{1}^{\text{LR}}(0)-\mathcal{C}_{1}^{\text{LR}}(\text{vac})}{S_{th}}$}}	\put(-100,-15){{\fontsize{13}{13}\selectfont$m/T$}}
		\caption{\textbf{Left:} The ratio of complexity of formation with thermal entropy in the LR basis for the cTFD state of a free complex scalar with a mass $m$ and $\tilde{Q} = 10^{-1}$. We show the dependence on the dimension from $d = 2$ (bottom) to $d = 6$ (top).  \textbf{Right:} The complexity of formation in the LR basis for the cTFD state of a free complex scalar with a mass $m$ and different $\tilde{Q}=10^{-6}$ (blue), $10^{-1}$ (brown), $1/2$ (dot-dashed green) and $1$ (dotted red) in $d=1+1$.}\label{c1lrmass}	
	\end{figure}

	Based on the above contradiction with holography, we should use another cost function which also gives the same UV divergence as $\mathcal{C}_{1}$. It is argued that \cite{Chapman:2018hou} the basis-dependent $L^1$ norm in (\ref{Fkappa}) is a preferred choice. According to (\ref{TTFD.2.2}) and (\ref{TFDpm}), in the $LR$ basis, two copies of the physical degrees of freedom
	entangled in the cTFD state and in the diagonal or $\pm$ basis the cTFD state factorizes. For a single mode, these basis are related by a simple rotation
	\bea 
	\label{rotation}
	\begin{pmatrix}
		\hat{x}_{+} \\
		\hat{p}_{+} \\
		\hat{x}_{-} \\
		\hat{p}_{-}
	\end{pmatrix} = R_{4} \begin{pmatrix}
		\hat{x}_{L} \\
		\hat{x}_{R}  \\
		\hat{p}_{L} \\
		\hat{p}_{R}
	\end{pmatrix},\hspace{1.5cm}
	\text{with}\hspace{.5cm}
	R_{4} = \frac{1}{\sqrt{2}}\begin{pmatrix}
		1& 1 & 0 & 0\\
		0& 0 & 1 &1 \\
		1& -1& 0& 0 \\
		0 &0 & 1& -1
	\end{pmatrix}.
	\eea
	Accordingly, the two-point function transforms as follows
	\bea
	G^{(\pm)} = R_{4}\hspace{1mm}G^{(LR)} \hspace{1mm}R_{4}^{\text{T}},
	\eea
	which it implies that if we have a circuit acting in the diagonal basis as $G^{(\pm)}_{T} = U^{(\pm)} \hspace{1mm}G^{(\pm)}_{R}\hspace{1mm} U^{(\pm),\text{T}}$, then 
	the same circuit in the $LR$ basis is
	\bea
	G^{(LR)}_{\text{T}} = U^{(LR)} \hspace{1mm} G^{(LR)}_{R}\hspace{1mm} (U^{(LR)}\hspace{.5mm})^{\text{T}} \hspace{1cm}\text{with}\hspace{.5cm}U^{(LR)} = R_{4}^{\text{T}}\hspace{1mm}U^{(\pm)}\hspace{1mm}R_{4}.
	\eea
	For the $F_1$ cost function, we also have considerable freedom in choosing
	the basis of generators $K_I$ in (\ref{YI}). One can impose that  the generators be orthonormal under the
	inner product inducing the $F_2$ cost function. For the single degree of freedom the generator group is
	$Sp(2, R) = SL(2, R)$, whose algebra is given by the traceless matrices $K_I \in T_1, T_2, T_3$, 
	\bea
	\label{T123}
	T_{1} = \begin{pmatrix}
		1 & 0 \\
		0 & -1 
	\end{pmatrix},\hspace{1cm}
	T_{2} = \begin{pmatrix}
		0 & 0 \\
		-\sqrt{2} & 0 
	\end{pmatrix},\hspace{1cm}
	T_{1} = \begin{pmatrix}
		0 & \sqrt{2} \\
		0 & 0 
	\end{pmatrix},
	\eea
	with
	\bea
	[T_{2},T_{1}]= 2T_{2},\hspace{1cm}[T_{1},T_{3}] = 2T_{3},\hspace{1cm} [T_{2},T_{3}] = 2T_{1}.
	\eea
	Exponentiating these generators yields the group elements that will serve as the elementary gates used in the construction of quantum circuits
	\bea
	U_{T_{1}} = e^{\epsilon T_{1}} = \begin{pmatrix}
		e^{\epsilon} & 0 \\
		0 & e^{-\epsilon} 
	\end{pmatrix},\hspace{.75cm}U_{T_{2}} = e^{\epsilon T_{2}} =\begin{pmatrix}
		1 & 0 \\
		-\sqrt{2}\epsilon & 1 
	\end{pmatrix},\hspace{.75cm}U_{T_{3}} = e^{\epsilon T_{3}} = \begin{pmatrix}
		1 & \sqrt{2}\epsilon \\
		0 & 1 
	\end{pmatrix},
	\eea
	with $\epsilon$ is a real parameter with $\epsilon \ll 1$. According to (\ref{GTpmcomplex}) for $
	G_{T}^{(\pm)} = U^{(\pm)} G_{R}^{(\pm)} \left(U^{(\pm)}\right)^{T}$, it is easy to see
	\bea
	U_{+} =  \left(e^{T_{1,+}}\right)^{(\alpha-\frac{1}{2}\log\frac{\lambda}{\lambda_{R}})},\hspace{.5cm}
	U_{-} = \big(e^{T_{1,-}}\big)^{(-\alpha-\frac{1}{2}\log\frac{\lambda}{\lambda_{R}})}.
	\eea
	The corresponding matrix $U$ in $LR$ bases is $U^{(LR)} = R_{4}^{T} \hspace{1mm}U^{(\pm)}\hspace{1mm} R_{4}
	$ where
	\bea
	U^{(\pm)} = e^{K^{(\pm)}}, \hspace{1cm} K^{(\pm)} = \begin{pmatrix}
		\alpha-\frac{1}{2}\log\frac{\lambda}{\lambda_{R}}&0&0&0 \\
		0&-\alpha+\frac{1}{2}\log\frac{\lambda}{\lambda_{R}}&0&0 \\
		0&0&-\alpha-\frac{1}{2}\log\frac{\lambda}{\lambda_{R}}&0 \\
		0&0&0&\alpha+\frac{1}{2}\log\frac{\lambda}{\lambda_{R}}
	\end{pmatrix}  
	\eea
	After a short computation, one can find $U^{(LR)}$ which by knowing that, the generator $K^{(LR)}$ in LR basis become
	\bea
	\label{KLR}
	K^{(LR)} =
	\begin{pmatrix}
		-\frac{1}{2}\log\frac{\lambda}{\lambda_{R}}&\alpha&0&0 \\
		\alpha&-\frac{1}{2}\log\frac{\lambda}{\lambda_{R}}&0&0 \\
		0&0&\frac{1}{2}\log\frac{\lambda}{\lambda_{R}}&-\alpha \\
		0&0&-\alpha&\frac{1}{2}\log\frac{\lambda}{\lambda_{R}}
	\end{pmatrix}.
	\eea
	To compute the complexity, we need to decompose the generator $K^{(LR)}$ (\ref{KLR}) in terms of
	the generators of $Sp(4,R)$. The matrix generators for the $Sp(4, R)$ can be split into the $Sp(2, R)$
	subalgebra acting on the left oscillator only
	\bea\label{gen1}
	T_{L,L}^{(1)}= 
	\begin{pmatrix}
		1&0&0&0 \\
		0&0&0&0 \\
		0&0&-1&0 \\
		0&0&0&0
	\end{pmatrix},\hspace{1cm}
	T_{L,L}^{(2)}= 
	\begin{pmatrix}
		0&0&0&0 \\
		0&0&0&0 \\
		-\sqrt{2}&0&0&0 \\
		0&0&0&0
	\end{pmatrix},\hspace{1cm}T_{L,L}^{(3)}= 
	\begin{pmatrix}
		0&0&\sqrt{2}&0 \\
		0&0&0&0 \\
		0&0&0&0 \\
		0&0&0&0
	\end{pmatrix}
	\eea
	the Sp(2, R) subalgebra acting on the right oscillator only
	\bea\label{gen2}
	T_{R,R}^{(1)}= \begin{pmatrix}
		0&0&0&0 \\
		0&1&0&0 \\
		0&0&0&0 \\
		0&0&0&-1
	\end{pmatrix},\hspace{1cm}
	T_{R,R}^{(2)}= \begin{pmatrix}
		0&0&0&0 \\
		0&0&0&0 \\
		0&0&0&0 \\
		0&-\sqrt{2}&0&0
	\end{pmatrix}
	,\hspace{1cm}
	T_{R,R}^{(3)}= \begin{pmatrix}
		0&0&0&0 \\
		0&0&0&\sqrt{2} \\
		0&0&0&0 \\
		0&0&0&0
	\end{pmatrix},
	\eea
	and the remaining generators which entangle the two oscillators
	\bea\label{gen3}
	&& T_{L,R}^{(1)}=  \begin{pmatrix}
		0&0&0&0 \\
		1&0&0&0 \\
		0&0&0&-1 \\
		0&0&0&0
	\end{pmatrix},\hspace{1cm}
	T_{R,L}^{(1)}= \begin{pmatrix}
		0&1&0&0 \\
		0&0&0&0 \\
		0&0&0&0 \\
		0&0&-1&0
	\end{pmatrix},
	\cr\nonumber\\
	&&T_{L,R}^{(2)}=  \begin{pmatrix}
		0&0&0&0 \\
		0&0&0&0 \\
		0&-1&0&0 \\
		-1&0&0&0
	\end{pmatrix},\hspace{1cm}
	T_{L,R}^{(3)}=  \begin{pmatrix}
		0&0&0&1 \\
		0&0&1&0 \\
		0&0&0&0 \\
		0&0&0&0
	\end{pmatrix}.
	\eea
	According to the above generators, $K^{(LR)}$ (\ref{KLR}) can be written as follows
	\bea
	K^{(LR)} = -\frac{1}{2} \log\frac{\lambda}{\lambda_{R}} \left(T_{L,L}^{(1)}+T_{R,R}^{(1)}\right)+\alpha \left(T_{L,R}^{(1)}+T_{R,L}^{(1)}\right),
	\eea
	which it means that only four components of the tangent vector $Y^I$ are non-vanishing. The above result implies that the $F_{1}$  complexity becomes
	\bea
	\label{C1lr}
	\mathcal{C}_{1}^{(LR)} = |Y_{L,L}^{(1)}|+|Y_{L,R}^{(1)}|+|Y_{R,L}^{(1)}|+|Y_{R,R}^{(1)}| = 2|\alpha|+|\log\frac{\lambda}{\lambda_{R}}|,
	\eea
	and by that, the complexity of formation is given by
	\bea
	\label{C1LR}
	\mathcal{C}_{1}^{(LR)}-\mathcal{C}_{1}^{(LR)}(\text{vac}) = 2|\alpha|,
	\eea
	where it only
	depends on $\beta(\omega+\mu q)$ and it contains no information about the reference state\footnote{It is worth noting that in (\ref{C1lr}) all gates are uniformly weighted.}. Accordingly, the contribution of all modes to the complexity of formation becomes
	\bea
	\label{c1lr2}
	\mathcal{C}_1^{(LR)} -\mathcal{C}_1^{(LR)}(\text{vac}) = \frac{\text{vol}\hspace{1mm}\Omega_{d-2}}{\beta^{d-1}(2\pi)^{d-1}}\int_{0}^{\infty} du\hspace{1mm}u^{d-2} \log\left(\frac{1+e^{-\frac{1}{2}(\sqrt{u^{2}+s^{2}}+\tilde{Q})}}{1-e^{-\frac{1}{2}(\sqrt{u^{2}+s^{2}}+\tilde{Q})}}\right).
	\eea
	The ratio of the complexity of formation to thermal entropy is provided in fig.\ref{c1lrt0} for different dimensions and different values of chemical potential. Intriguingly, this ratio is free of UV divergence but has a new IR divergence for $T \rightarrow 0$ case.  Similarly, in the holographic counterpart, for the
	complexity of formation there is still a cancellation of the UV divergences associated
	with the asymptotic boundary and instead exists an IR divergence associated with the infinitely long throat of the extremal black holes. The above results suggests that preparing the ‘extremal’ thermofield double states at zero temperature and finite chemical  potential  are  infinitely hard compared to the  finite temperature states in complete agreement with third law of holographic complexity, \cite{Carmi:2017jqz}.
	To complete the study of complexity of formation in this basis, let us present the result for   
	massive complex scalar. The effect of changing dimension and chemical potential for this case are presented fig.\ref{c1lrmass}. Both the complexity of formation and the entropy go to zero as the
	parameter $\beta m$ increases, but the ratio increases exponentially as a function of $\beta m$.
	\subsection{Time dependency of Complexity}
\begin{figure}[H]	\hspace{.7cm}\includegraphics[scale=.26]{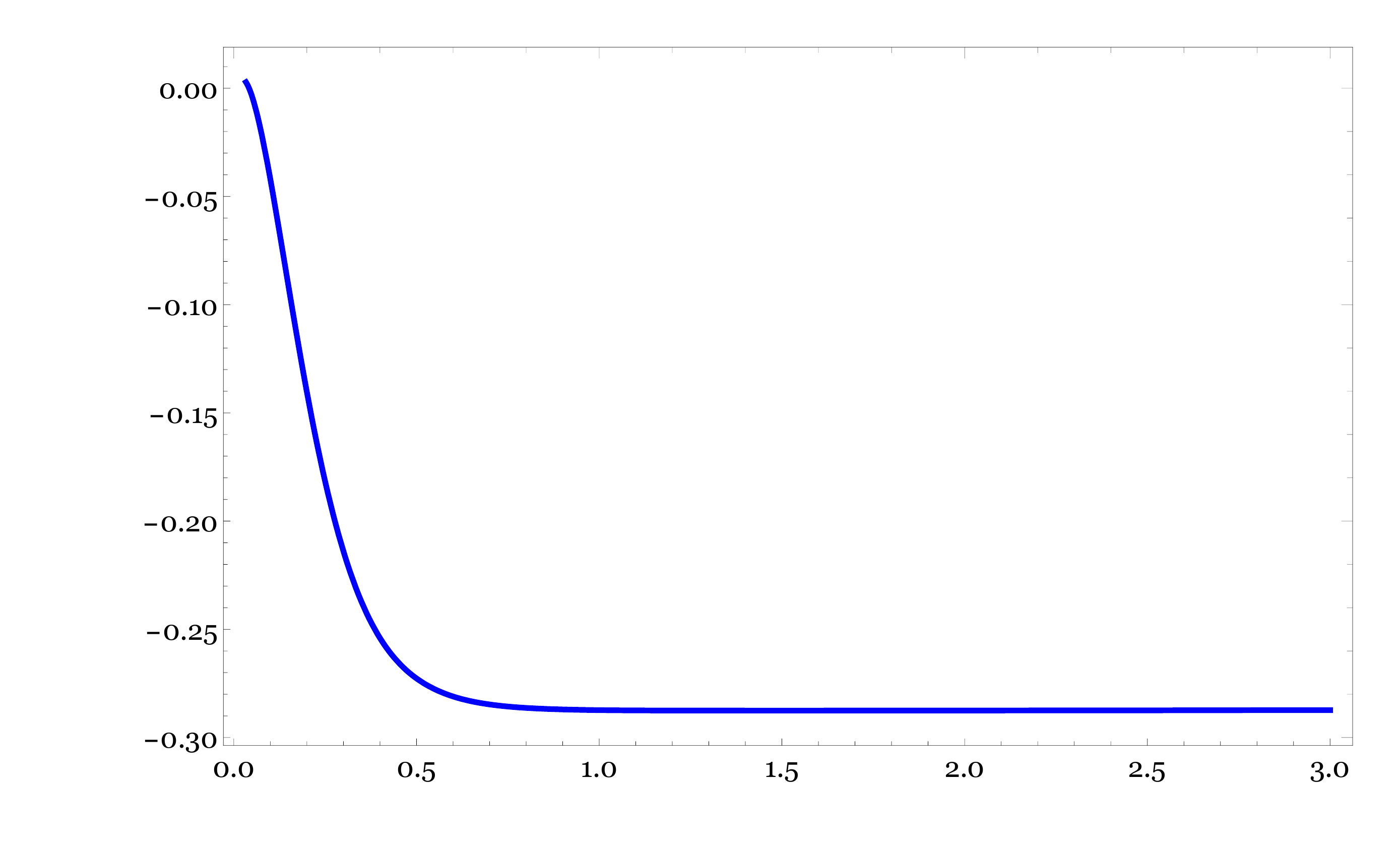}\put(-210,30){\rotatebox{-271}{\fontsize{14}{14}\selectfont $\frac{\Delta\mathcal{C}_{1}^{\text{LR}}(t)-\Delta\mathcal{C}_{1}^{\text{LR}}(0)}{S_{th}}$}}	\put(-100,-5){{\fontsize{13}{13}\selectfont$t\hspace{1mm} T$}}
		\put(-120,120){{\fontsize{11}{11}\selectfont$d=1+1$}}
		\hspace{.5cm}\includegraphics[scale=.25]{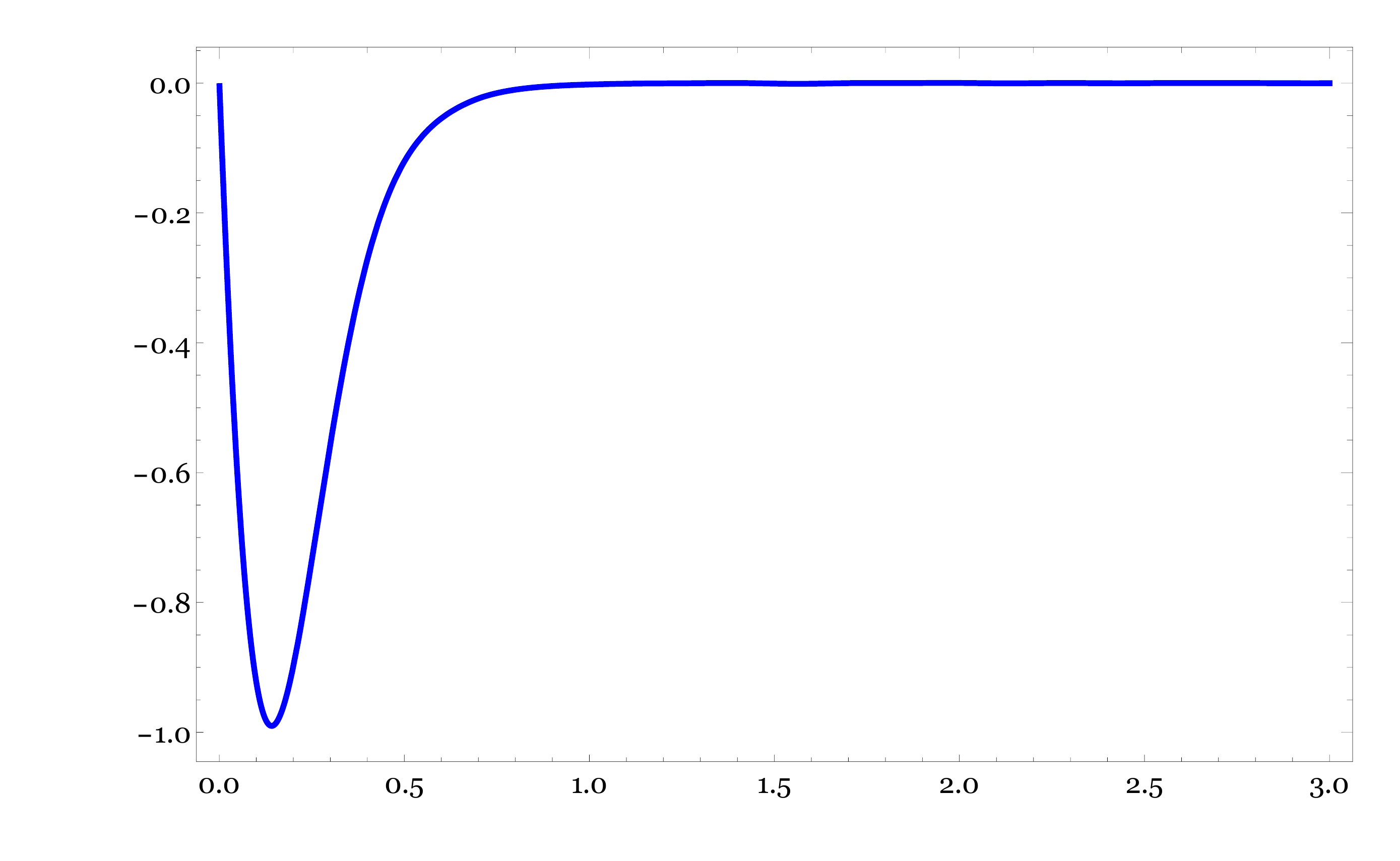}\put(-210,35){\rotatebox{-271}{\fontsize{15}{15}\selectfont $\frac{1}{T\hspace{.5mm}S_{th}}\frac{d\mathcal{C}_{1}^{\text{LR}}(t)}{dt}$}}	\put(-100,-5){{\fontsize{13}{13}\selectfont$t\hspace{1mm} T$}}
		\put(-120,118){{\fontsize{11}{11}\selectfont$d=1+1$}}\vspace{.7cm}
		
		\hspace{.7cm}\includegraphics[scale=.26]{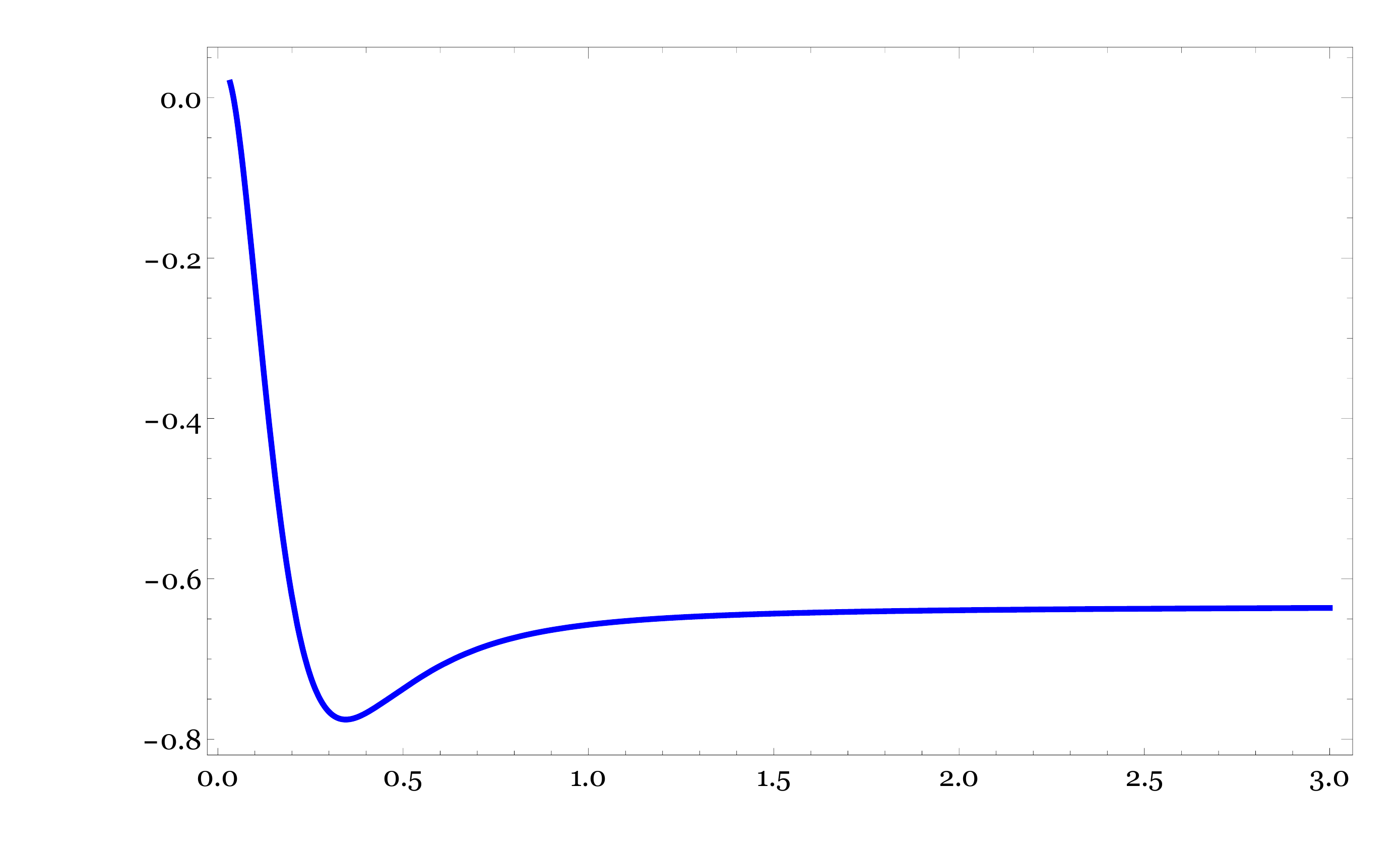}\put(-210,30){\rotatebox{-271}{\fontsize{14}{14}\selectfont $\frac{\Delta\mathcal{C}_{1}^{\text{LR}}(t)-\Delta\mathcal{C}_{1}^{\text{LR}}(0)}{S_{th}}$}}	\put(-100,-5){{\fontsize{13}{13}\selectfont$t\hspace{1mm} T$}}
		\put(-120,125){{\fontsize{11}{11}\selectfont$d=2+1$}}\hspace{.5cm}\includegraphics[scale=.25]{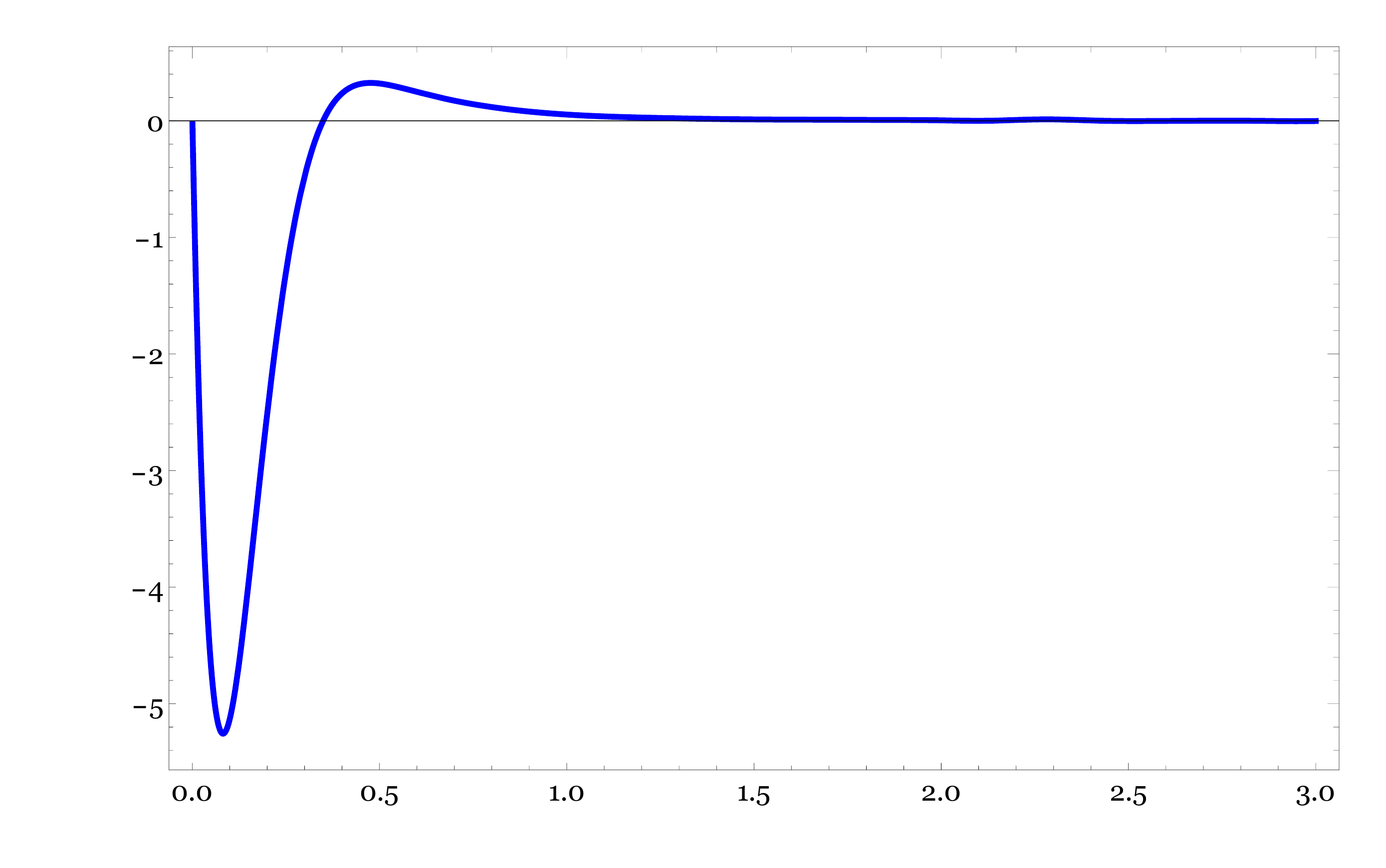}\put(-210,40){\rotatebox{-271}{\fontsize{17}{17}\selectfont $\frac{1}{T\hspace{.5mm}S_{th}}\frac{d\mathcal{C}_{1}^{\text{LR}}(t)}{dt}$}}	\put(-100,-5){{\fontsize{13}{13}\selectfont$t\hspace{1mm} T$}}
		\put(-120,120){{\fontsize{11}{11}\selectfont$d=2+1$}}
		\caption{The $L^{1}$ norm complexity and its time derivative for a cTFD state of a massless complex scalar field with $\tilde{Q} = 1/10$ in the LR basis and simple limit ($\lambda_{p}\rightarrow 0$). $d=1+1$ (\textbf{up}) and $d=2+1$ (\textbf{down}). In contrast to holography, the
			rate of change is negative at first, then saturates to zero at times of the order of the inverse
			temperature.}\label{c1lrt2-3}	
	\end{figure}

\begin{figure}[H]
		\hspace{.7cm}\includegraphics[scale=.26]{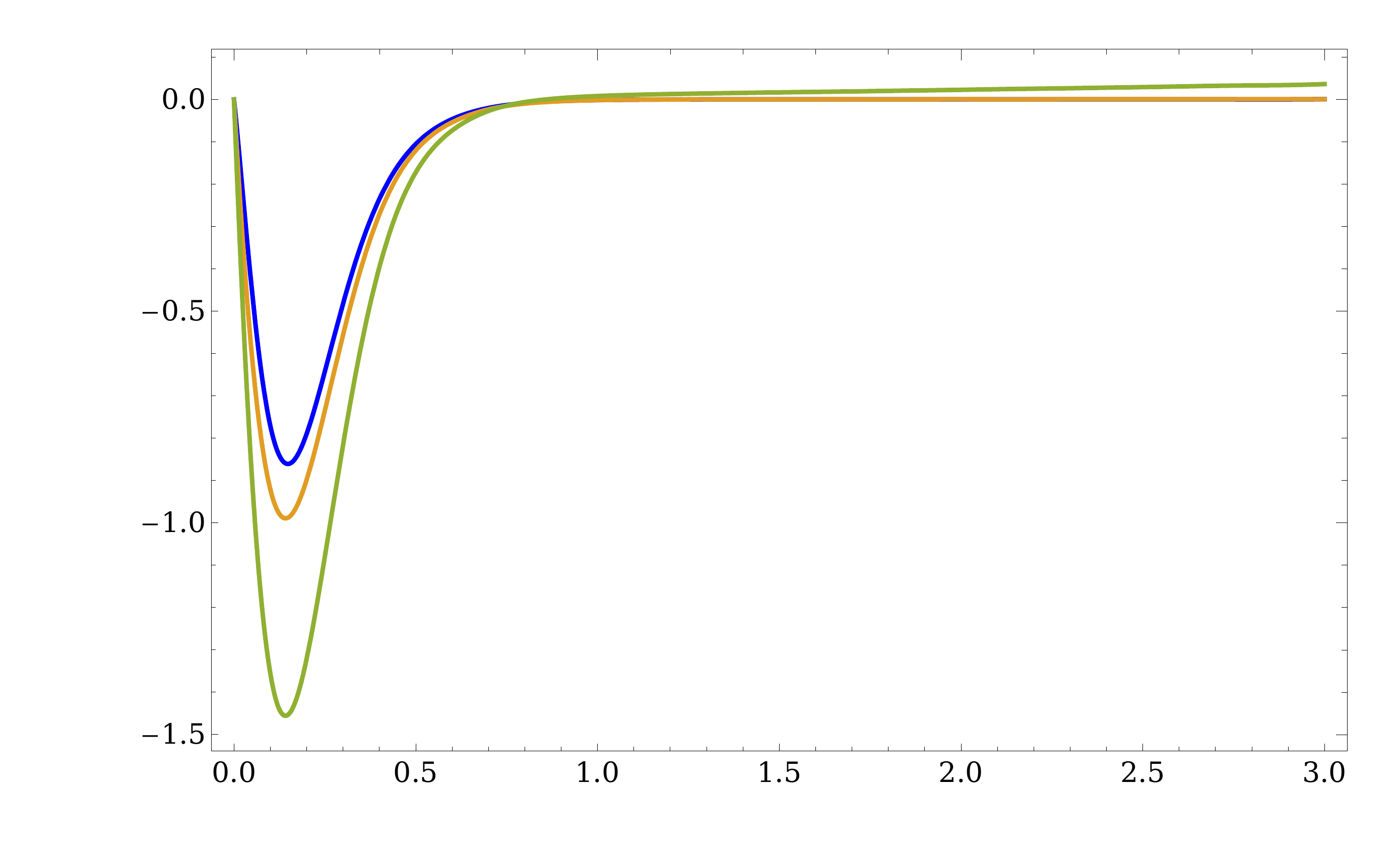}\put(-215,40){\rotatebox{-271}{\fontsize{17}{17}\selectfont $\frac{1}{T\hspace{.5mm}S_{th}}\frac{d\mathcal{C}_{1}^{\text{LR}}(t)}{dt}$}}	\put(-100,-2){{\fontsize{13}{13}\selectfont$t\hspace{1mm} T$}}
		\put(-120,121){{\fontsize{11}{11}\selectfont$d=1+1$}}\hspace{.5cm}\includegraphics[scale=.26]{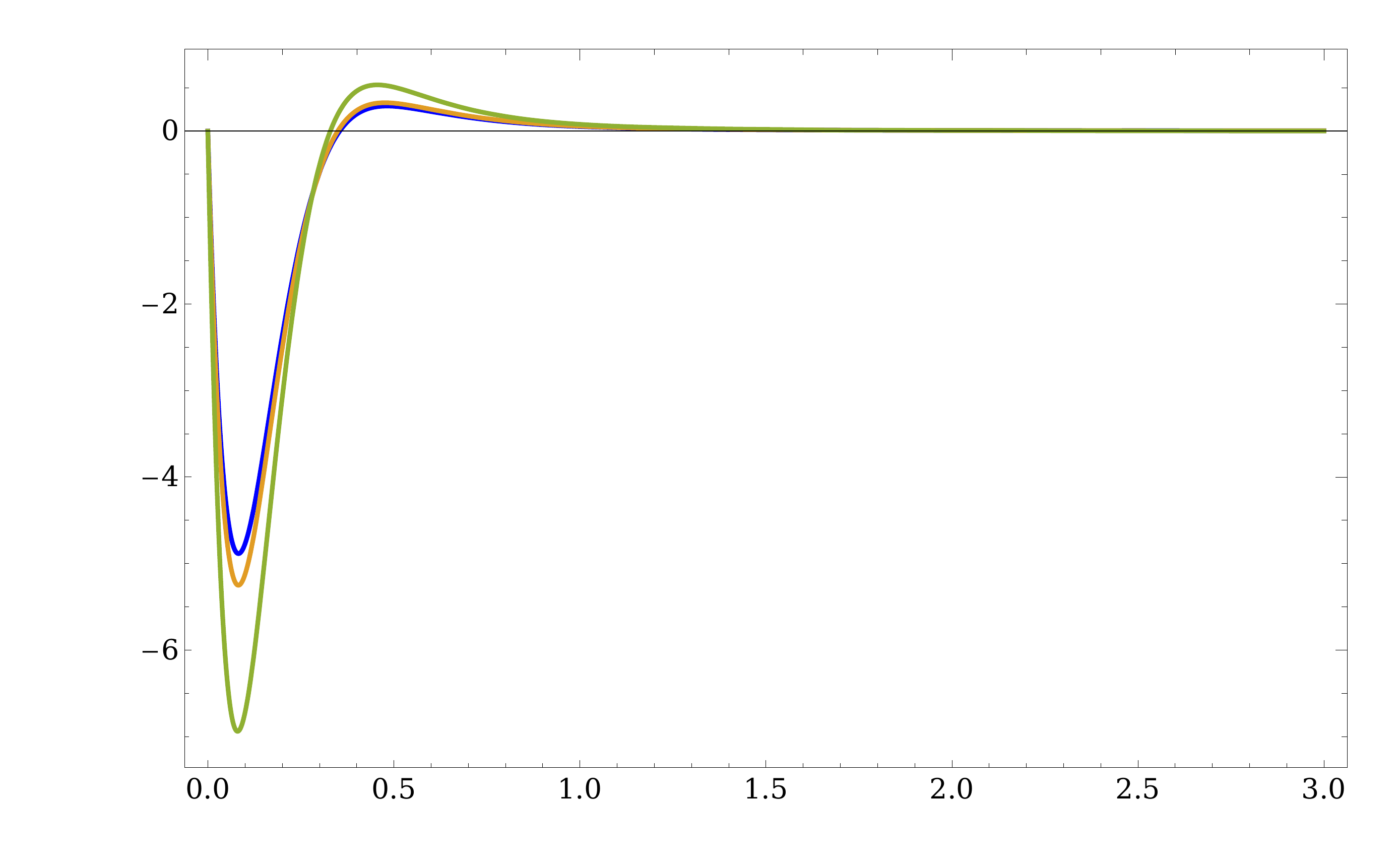}\put(-215,40){\rotatebox{-271}{\fontsize{17}{17}\selectfont $\frac{1}{T\hspace{.5mm}S_{th}}\frac{d\mathcal{C}_{1}^{\text{LR}}(t)}{dt}$}}	\put(-100,-2){{\fontsize{13}{13}\selectfont$t\hspace{1mm} T$}}
		\put(-120,123){{\fontsize{11}{11}\selectfont$d=2+1$}}
		\caption{The $F_{1}$ complexity growth rate over entropy times temperature for cTFD state of a masless complex scalar with different chemical potential in $d=1+1$ (\textbf{left}) and $d=2+1$ (\textbf{right}). $\tilde{Q}=10^{-6}$ (blue), $\tilde{Q}=10^{-1}$ (brown) and $\tilde{Q}=1/2$ (green).}\label{c1lrt2-4}
	\end{figure}
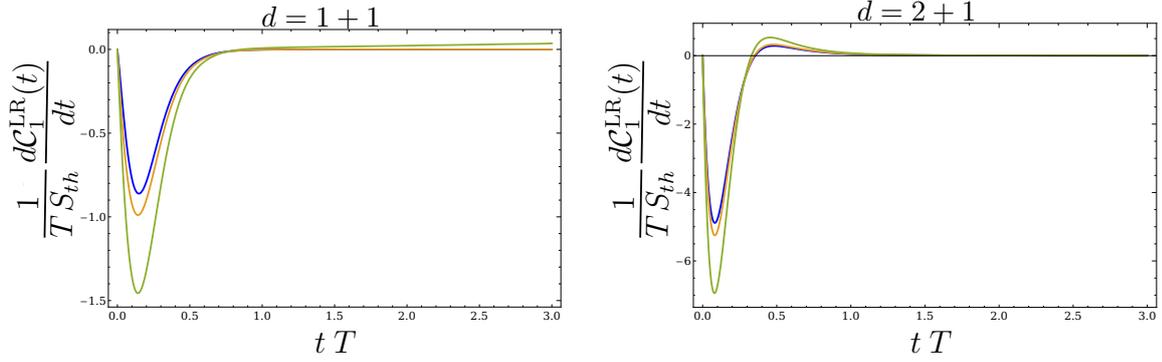

	To evaluate the complexity growth rate of cTFD state in LR basis we need the full time-dependent relative covariance matrix. Similar to previous analysis, let us firstly concentrate on a single mode. To obtain that matrix, we start with the covariance
	matrix itself in diagonal basis
	\bea
	G^{(\pm)}(t) = G^{+}_{\text{TFD}}(t) \oplus G^{-}_{\text{TFD}}(t),
	\eea
	where the direct sum inputs the $+$ and $-$ components in a $4$ by $4$ combined matrix and
	where the $G^+_
	{\text{TFD}}(t)$ was defined in (\ref{GTpmcomplex}) by noting to $\alpha\rightarrow \alpha_{p}$ and $G^-_{\text{TFD}}(t)$ is a same matrix just with $\alpha_{p}\rightarrow -\alpha_{p}$.
	According to the following definition
	\bea
	G_{\text{T},+}(t) = U_{+}(t) \hspace{1mm}G_{R,+}\hspace{1mm} U_{+}^{\text{T}}(t),
	\eea
	for $\lambda_{R} =1$, it is easy to see that the transfer matrix $U_{+}(t)$ can be written as
	\bea\label{Upt}
	U_{+}(t) = 
	\begin{pmatrix}
		\cosh s_{1,+}-\sin s_{2,+}\sinh s_{1,+} & \cos s_{2,+}\sinh s_{1,+} \\
		\cos s_{2,+}\sinh s_{1,+} & 
		\cosh s_{1,+}+\sin s_{2,+}\sinh s_{1,+}
	\end{pmatrix},
	\eea
	with
	\bea
	\label{upt}
	&& s_{1,+} = \frac{1}{2} \cosh^{-1}\left(\frac{1+\lambda^2_{p}}{2\lambda_{p}}\cosh 2\alpha_{p}+\frac{1-\lambda^{2}_{p}}{2\lambda_{p}}\sinh 2\alpha_{p} \cos[(\omega_{p}+\mu q)t]\right),
	\cr\nonumber\\
	&& s_{2,+} = \tan^{-1} \left(\frac{1+\lambda^{2}_{p}}{2\lambda_{p}} \cot[(\omega_{p}+\mu q)t] +\frac{1-\lambda^{2}_p}{2\lambda_{p}}\coth 2\alpha_p \csc [(\omega_{p}+\mu q)t]\right).
	\eea
	Clearly, the identity $U_{-}(t) = U_{+}(t) (\alpha_{p}\rightarrow -\alpha_{p})$, implies that
	\bea
	s_{i,-} = s_{i,+} (\alpha_{p} \rightarrow -\alpha_{p}).
	\eea
	The transfer matrix (\ref{upt})
	by using the $SL(2,p)$ generators (\ref{T123}) can be decomposed as following
	\bea
	U_{+}(t) = \exp\left[\left(
	-s_{1,+}\sin s_{2,+}\right) T_{1,+} +\left(\frac{s_{1,+}\cos s_{2,+}}{\sqrt{2}}\right)\left(T_{2,+}-T_{3,+}\right)\right].
	\eea
	Therefore, the full circuit $U_{(\pm)}$ becomes
	\bea
	U_{\pm}(t)\equiv \exp[M_{\pm}(t)] = \exp\left[
	\begin{pmatrix}
		-s_{1,+} \sin s_{2,+}& s_{1,+}\cos s_{2,+} & 0 & 0\\
		s_{1,+}\cos s_{2,+}& s_{1,+}\sin s_{2,+}&0 &0 \\
		0& 0& -s_{1,-}\sin{s_{2,-}}& s_{1,-}\cos s_{2,-} \\	0 &0 & s_{1,-}\cos s_{2,-}& s_{1,-}\sin s_{2,-}
	\end{pmatrix}\right].
	\eea
	Now, we can find relevant generator $M_{(LR)}(t) = R_{4}^{\text{T}} M_{(\pm)} R_{4}$ in the LR basis by apply the transformation (\ref{rotation}). It is easy to see that
	\bea
	M_{(LR)}(t) = \frac{1}{2}
	\begin{pmatrix}
		\label{MLRt}
		M_{11}(t)& M_{12}(t) & M_{13}(t) & M_{14}(t)\\
		M_{12}(t)& M_{11}(t)&M_{14}(t) &M_{13}(t)\\
		M_{13}(t)& M_{14}(t)& -M_{11}(t)& M_{34}(t)\\	
		M_{14}(t) &M_{13}(t) & M_{34}(t)& -M_{11}(t)
	\end{pmatrix},
	\eea
	with
	\bea
	&& M_{11}(t) = -s_{1,-}\sin s_{2,-} -s_{1,+}\sin s_{2,+},\hspace{1cm}
	M_{12}(t) = s_{1,-}\sin s_{2,-} -s_{1,+}\sin s_{2,+},\cr\nonumber\\
	&&M_{13}(t) = s_{1,-}\cos s_{2,-} +s_{1,+}\cos s_{2,+},\hspace{1cm}M_{14}(t)= -s_{1,-}\cos s_{2,-} +s_{1,+}\cos s_{2,+},\cr\nonumber\\
	&&\hspace{4cm}M_{34}(t)= -s_{1,-}\sin s_{2,-} +s_{1,+}\sin s_{2,+}.
	\eea
	The matrix $M_{(LR)}(t)$ (\ref{MLRt}) can be decomposed in terms of
	the generators of $Sp(4, R)$, (\ref{gen1})-(\ref{gen3}) as follows
	\bea
	&& M_{LR}(t) = a_{L,L}^{(1)} \left(T_{L,L}^{(1)}+T_{R,R}^{(1)}\right) + a_{L,R}^{(1)}\left(T_{L,R}^{(1)}+T_{R,L}^{(1)}\right)+ a_{L,R}^{(2)}\left(T_{L,R}^{(2)}-T_{L,R}^{(3)}\right)
	\cr\nonumber\\
	&& \hspace{2.2cm}+a_{L,L}^{(2)}\left(T_{L,L}^{(2)}+T_{R,R}^{(2)}-T_{L,L}^{(3)} -T_{R,R}^{(3)}\right),
	\eea
	where
	\bea
	\label{as}
	&&a_{L,L}^{(1)} = -\frac{1}{2}\bigg(s_{1,-}\sin s_{2,-}+s_{1,+}\sin s_{2,+}\bigg),\hspace{.5cm}a_{L,L}^{(2)} = -\frac{1}{2\sqrt{2}}\bigg(s_{1,-}\cos s_{2,-}+s_{1,+}\cos s_{2,+}\bigg),
	\cr\nonumber\\
	&&a_{L,R}^{(1)} = \frac{1}{2}\bigg(s_{1,-}\sin s_{2,-}-s_{1,+}\sin s_{2,+}\bigg),
	\hspace{.5cm}a_{L,R}^{(2)} = \frac{1}{2} \bigg(s_{1,-}\cos s_{2,-} -s_{1,+}\cos s_{2,+}\bigg),
	\eea
	and
	\bea
	&& s_{1,\pm} = \frac{1}{2} \cosh^{-1}\left(\frac{1}{2\lambda_{p}}\cosh 2\alpha_{p}\pm\frac{1}{2\lambda_{p}}\sinh 2\alpha_{p} \cos[(\omega_{p}+\mu q)t]\right),
	\cr\nonumber\\
	&& s_{2,\pm} = \tan^{-1} \left(\frac{1}{2\lambda_{p}} \cot[(\omega_{p}+\mu q)t] \pm\frac{1}{2\lambda_{p}}\coth 2\alpha_p \csc [(\omega_{p}+\mu q)t]\right).
	\eea
	Now, using the $F_{1}$ cost function , (\ref{Fkappa}), we arrive at
	\bea
	\label{c1LRt}
	\mathcal{C}_{1}(t) = 2\hspace{1mm} |a_{L,L}^{(1)}| + 2\hspace{1mm} |a_{L,R}^{(1)}| + 2\hspace{1mm}|a_{L,R}^{(2)}| + 4 \hspace{1mm}|a_{L,L}^{(2)}|.
	\eea
	Firstly, let us study (\ref{c1LRt}) in simple limit where $\omega_{R}$ is much bigger than any other scale, i.e. $\lambda_{p} \rightarrow  0$. In this limit we have
	\bea
	\label{deltaC1LR}
	\mathcal{C}_1^{(LR)}(t) -\mathcal{C}_1^{(LR)}(\text{vac}) = \frac{\text{vol}}{(2\pi)^{d-1}}\int_{0}^{\infty} d^{d-1}p\hspace{1mm} \log\bigg(\cosh2\alpha_p+\bigg{|}\cos(\omega_{p}+\mu q)t\bigg{|}\sinh2\alpha_p\hspace{1mm}\bigg).
	\eea
	From the above equation, one may conclude that the complexity oscillates in time. But as it is clear from fig.\ref{c1lrt2-3}, a saturation happens at the order of inverse temperature\footnote{Our result is in contradiction with previous attempts \cite{Sinamuli:2019utz} to study complexity of cTFD state of free complex scalar theory, since the authors find that the complexity grows linearly for a long time.}. This fast saturation happens since oscillatory behavior for all modes quickly dephase and as a result, summing over them averages out at a time of the inverse temperature. Of course, this is very different than what we see for holographic complexity, i.e. where
	we see a linear growth at late times. The late time saturation for holographic CFTs is indeed related to their 
	chaotic/fast-scrambling characteristic. The complexity growth rate differs more with  holography since it is negative at first and it becomes more negative by increasing the $U(1)$ global charge, see fig.\ref{c1lrt2-4}. The reason for this negativity can be easily understood from (\ref{deltaC1LR}) where at $t=0$ the contribution of all of individual modes take their maximum value and after this time all oscillatory terms become misaligned and therefore the complexity begins to decrease.
	\vspace{.5cm}
	
\begin{figure}[H]	\hspace{.5cm}\includegraphics[scale=.26]{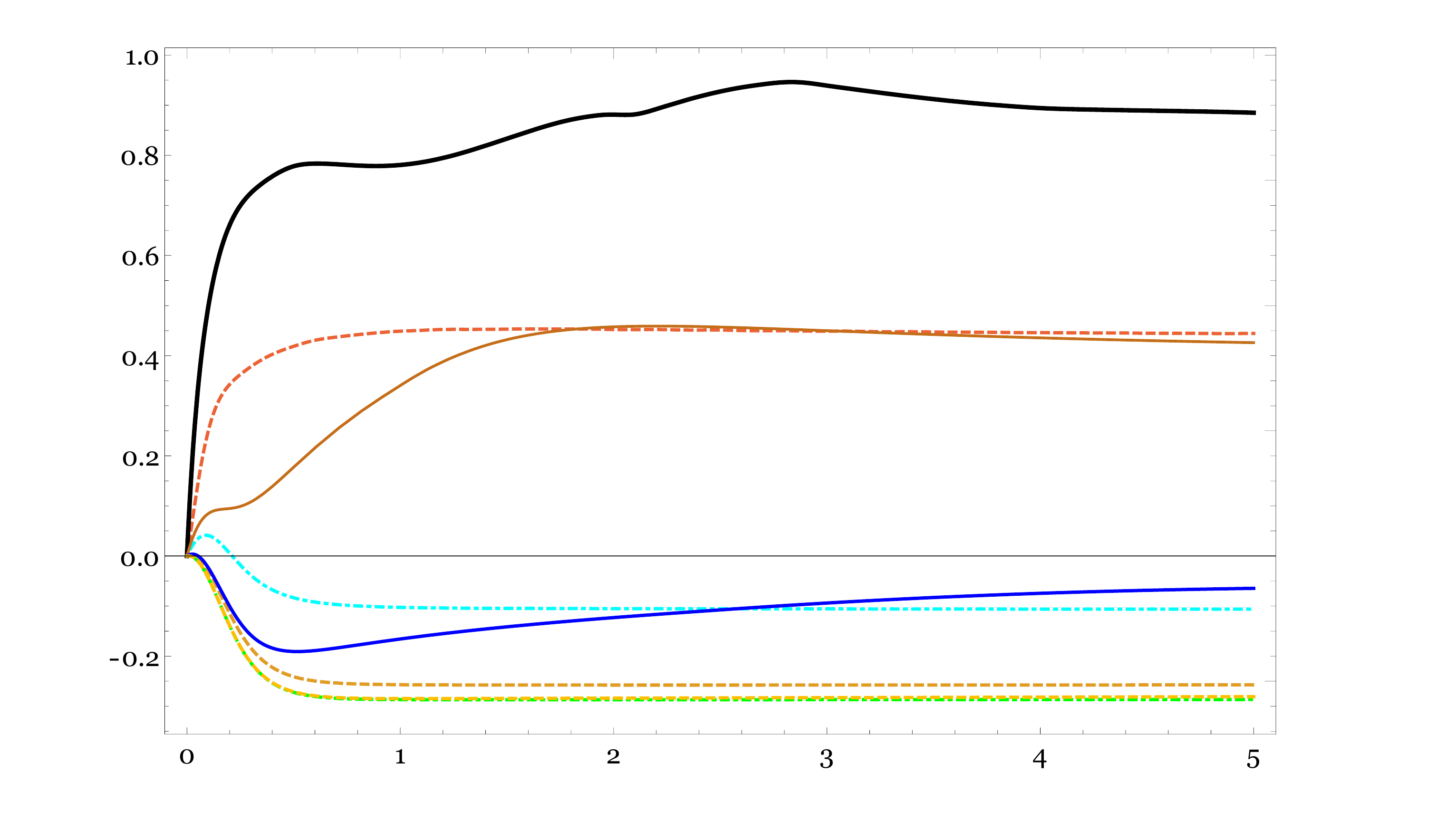}\put(-238,30){\rotatebox{-271}{\fontsize{14}{14}\selectfont $\frac{\Delta\mathcal{C}_{1}^{\text{LR}}(t)-\Delta\mathcal{C}_{1}^{\text{LR}}(0)}{S_{th}}$}}	\put(-125,-5){{\fontsize{13}{13}\selectfont$t\hspace{1mm} T$}}
		\put(-145,125){{\fontsize{11}{11}\selectfont$d=1+1$}}\hspace{-.1cm}\includegraphics[scale=.26]{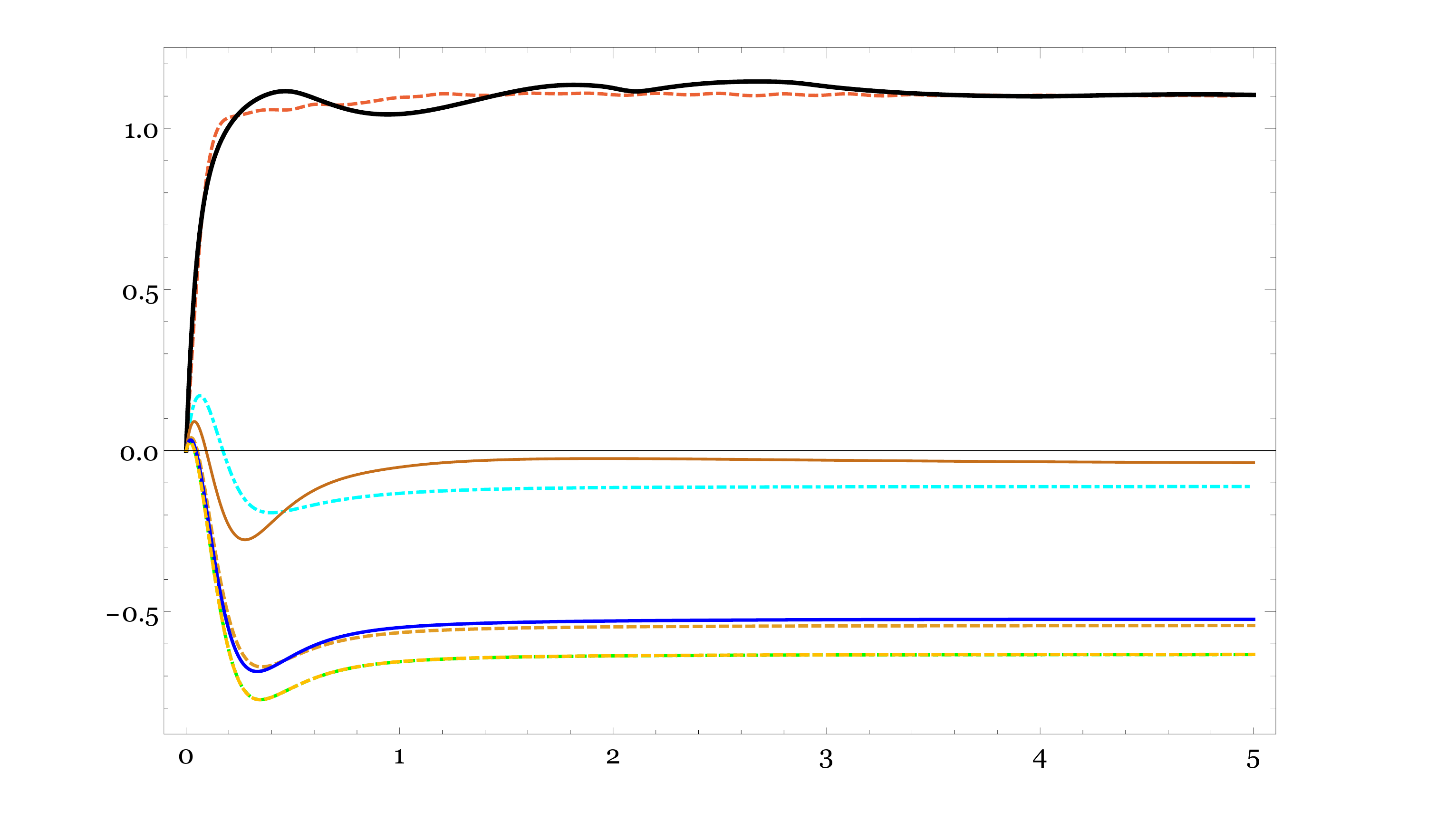}\put(-238,30){\rotatebox{-271}{\fontsize{14}{14}\selectfont $\frac{\Delta\mathcal{C}_{1}^{\text{LR}}(t)-\Delta\mathcal{C}_{1}^{\text{LR}}(0)}{S_{th}}$}}	\put(-125,-5){{\fontsize{13}{13}\selectfont$t\hspace{1mm} T$}}
		\put(-145,124){{\fontsize{11}{11}\selectfont$d=2+1$}}\put(-25,12){\includegraphics[scale=.55]{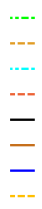}}\put(2,113){{\fontsize{10}{10}\selectfont$\tilde{\gamma}=10^{-5}$}}\put(2,100){{\fontsize{10}{10}\selectfont$\tilde{\gamma}=10^{-3}$}}\put(2,85){{\fontsize{10}{10}\selectfont$\tilde{\gamma}=10^{-2}$}}\put(2,70){{\fontsize{10}{10}\selectfont$\tilde{\gamma}=10^{-1}$}}\put(2,57){{\fontsize{10}{10}\selectfont$\tilde{\gamma}=1$}}\put(2,41){{\fontsize{10}{10}\selectfont$\tilde{\gamma}=10$}}\put(2,28){{\fontsize{10}{10}\selectfont$\tilde{\gamma}=10^{2}$}}\put(2,13){{\fontsize{10}{10}\selectfont$\tilde{\gamma}=10^{4}$}}
		\caption{The time evolution of $F_{1}$ complexity normalized by thermal entropy with varying reference scale for the massless complex scalar with $\tilde{Q} = 1/10$. $d=1+1$ (\textbf{left}) and $d=2+1$ (\textbf{right}).}\label{c1lrt2-7}	
	\end{figure}

\begin{figure}[H]
		\hspace{.5cm}\includegraphics[scale=.26]{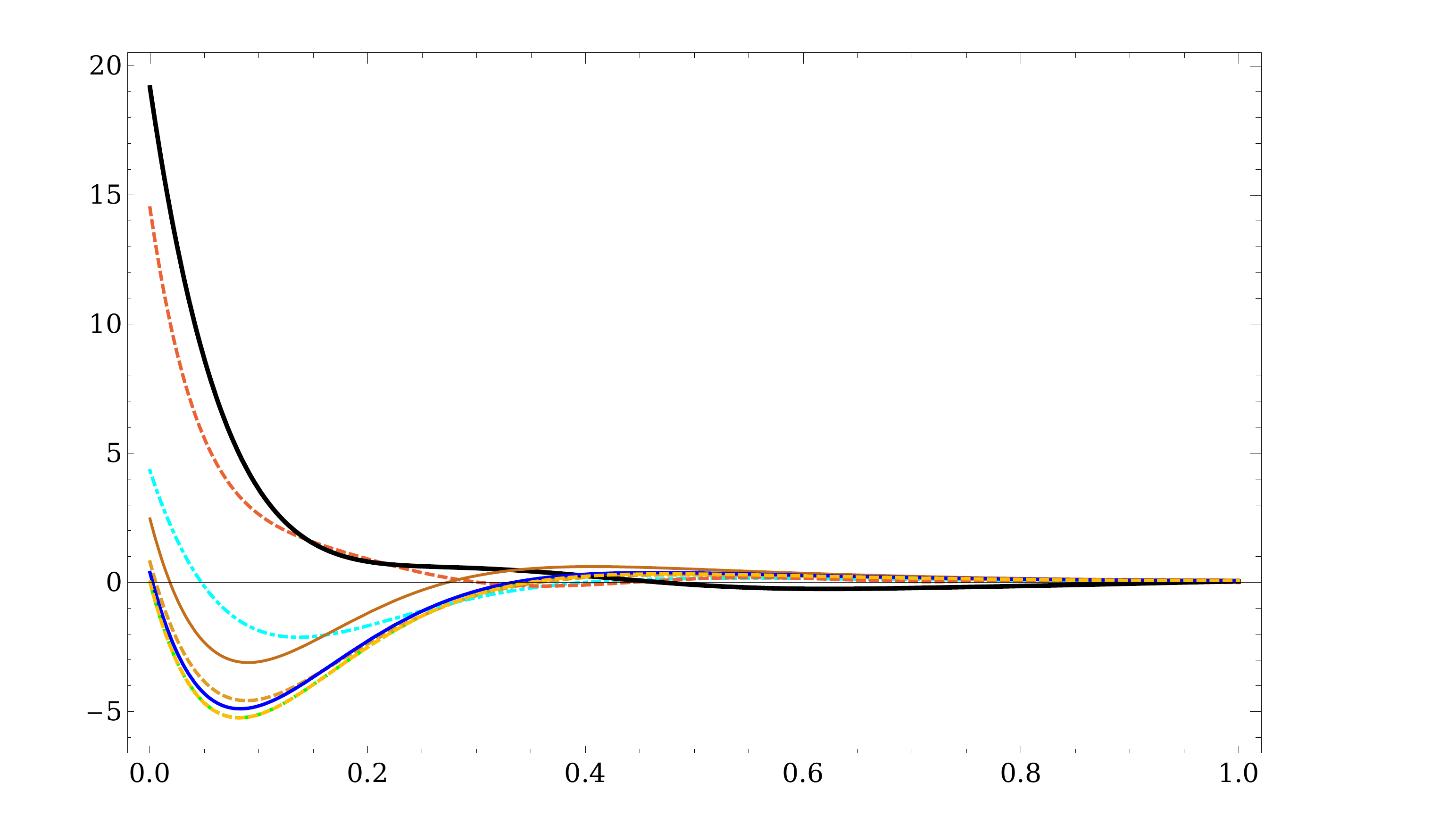}\put(-249,40){\rotatebox{-271}{\fontsize{17}{17}\selectfont $\frac{1}{T S_{th}}\frac{d\mathcal{C}_{1}^{(LR)}(t)}{dt}$}}	\put(-130,-5){{\fontsize{13}{13}\selectfont$t\hspace{1mm} T$}}
		\put(-145,129){{\fontsize{11}{11}\selectfont$d=2+1$}}\hspace{-0.2cm}
		\includegraphics[scale=.26]{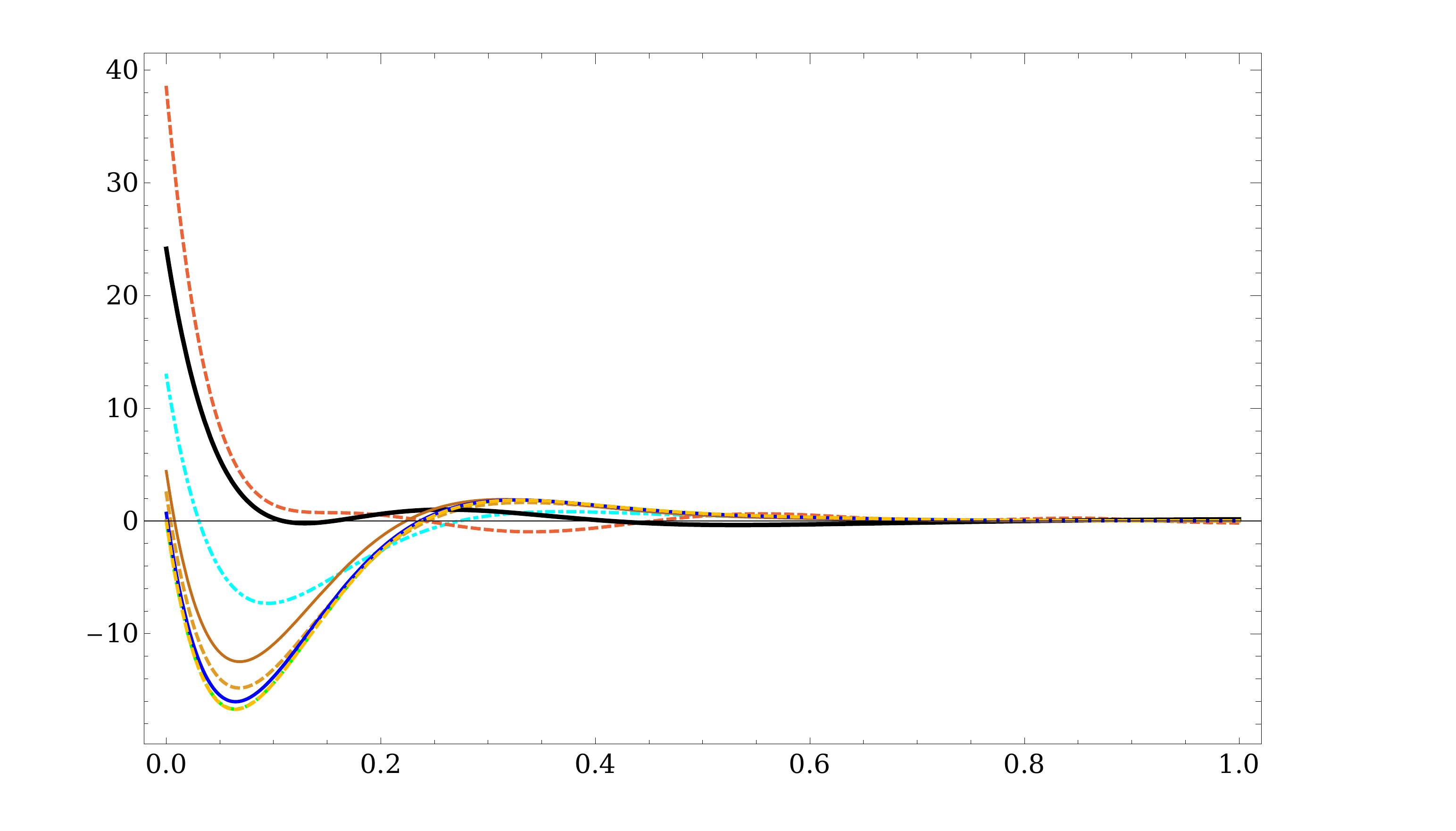}\put(-248,40){\rotatebox{-271}{\fontsize{17}{17}\selectfont $\frac{1}{T S_{th}}\frac{d\mathcal{C}_{1}^{(LR)}(t)}{dt}$}}	\put(-130,-5){{\fontsize{13}{13}\selectfont$t\hspace{1mm} T$}}
		\put(-145,126){{\fontsize{11}{11}\selectfont$d=3+1$}}\put(-28,12){\includegraphics[scale=.55]{crop2.png}}\put(2,113){{\fontsize{10}{10}\selectfont$\tilde{\gamma}=10^{-5}$}}\put(2,100){{\fontsize{10}{10}\selectfont$\tilde{\gamma}=10^{-3}$}}\put(2,85){{\fontsize{10}{10}\selectfont$\tilde{\gamma}=10^{-2}$}}\put(2,70){{\fontsize{10}{10}\selectfont$\tilde{\gamma}=10^{-1}$}}\put(2,57){{\fontsize{10}{10}\selectfont$\tilde{\gamma}=1$}}\put(2,41){{\fontsize{10}{10}\selectfont$\tilde{\gamma}=10$}}\put(2,28){{\fontsize{10}{10}\selectfont$\tilde{\gamma}=10^{2}$}}\put(2,13){{\fontsize{10}{10}\selectfont$\tilde{\gamma}=10^{4}$}}
		\caption{The $L^{1}$ norm complexity growth rate over temperature times thermal entropy for a cTFD state of massless complex scalar with  $\tilde{Q}=1/10$ and different values of reference state scale. $d=2+1$ (\textbf{left}), $d=3+1$ (\textbf{right}).}\label{c1lrt2-8}
	\end{figure}

	It will be instructive if we also study the effect of global charge in the presence of mass. According to figs.\ref{c1lrt2-5} and \ref{c1lrt2-6}, if the mass or chemical potential are larger than the temperature then complexity oscillates with damping amplitude. This oscillatory behavior is clear from (\ref{deltaC1LR}) and at late times we observe again a saturation to a constant value.
	\begin{figure}[H]	\hspace{0.7cm}\includegraphics[scale=.26]{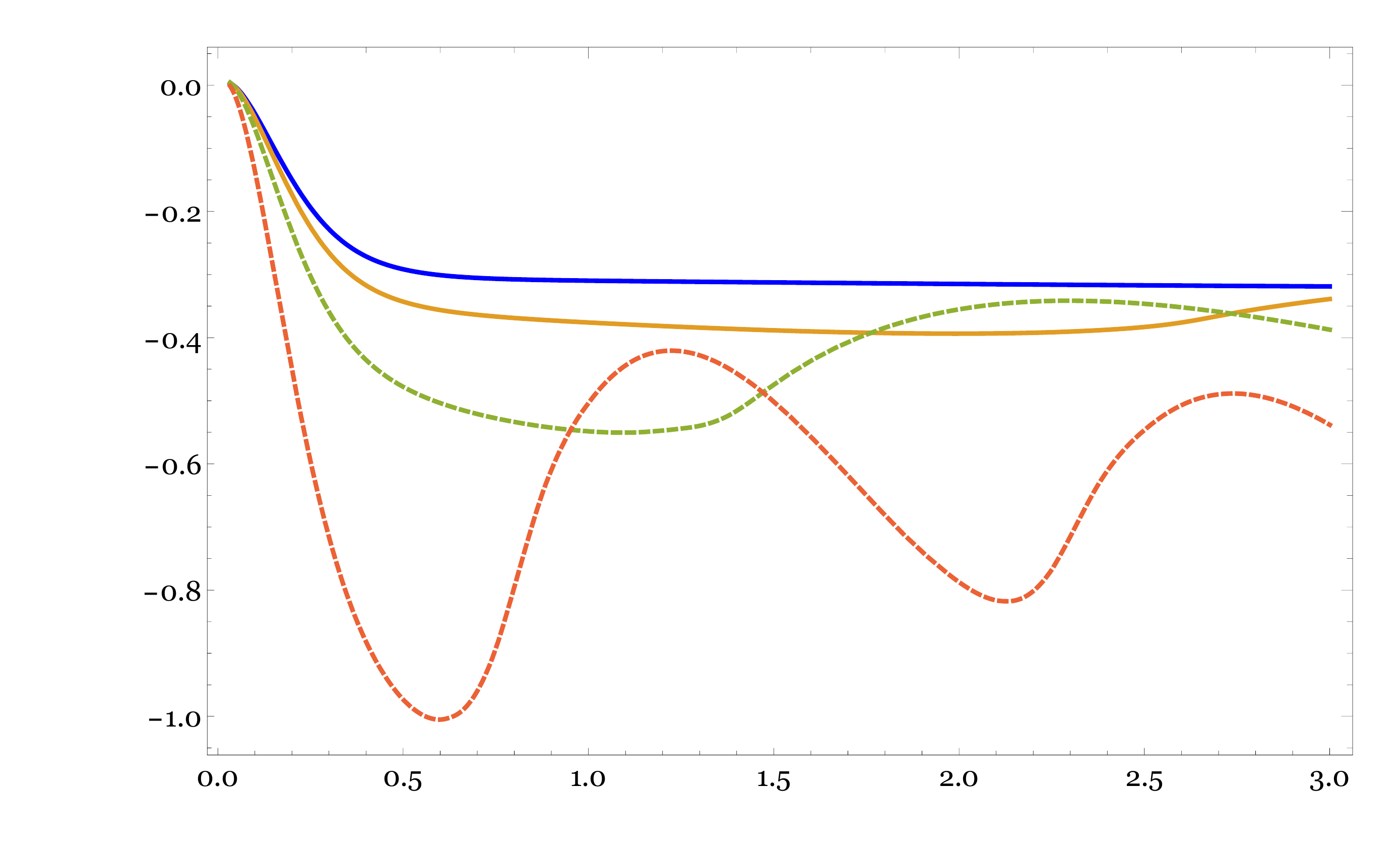}\put(-212,26){\rotatebox{-271}{\fontsize{14}{14}\selectfont $\frac{\Delta\mathcal{C}_{1}^{\text{LR}}(t)-\Delta\mathcal{C}_{1}^{\text{LR}}(0)}{S_{th}}$}}	\put(-105,-5){{\fontsize{13}{13}\selectfont$t\hspace{1mm} T$}}
		\put(-120,122){{\fontsize{11}{11}\selectfont$d=1+1$}}\hspace{.6cm}\includegraphics[scale=.25]{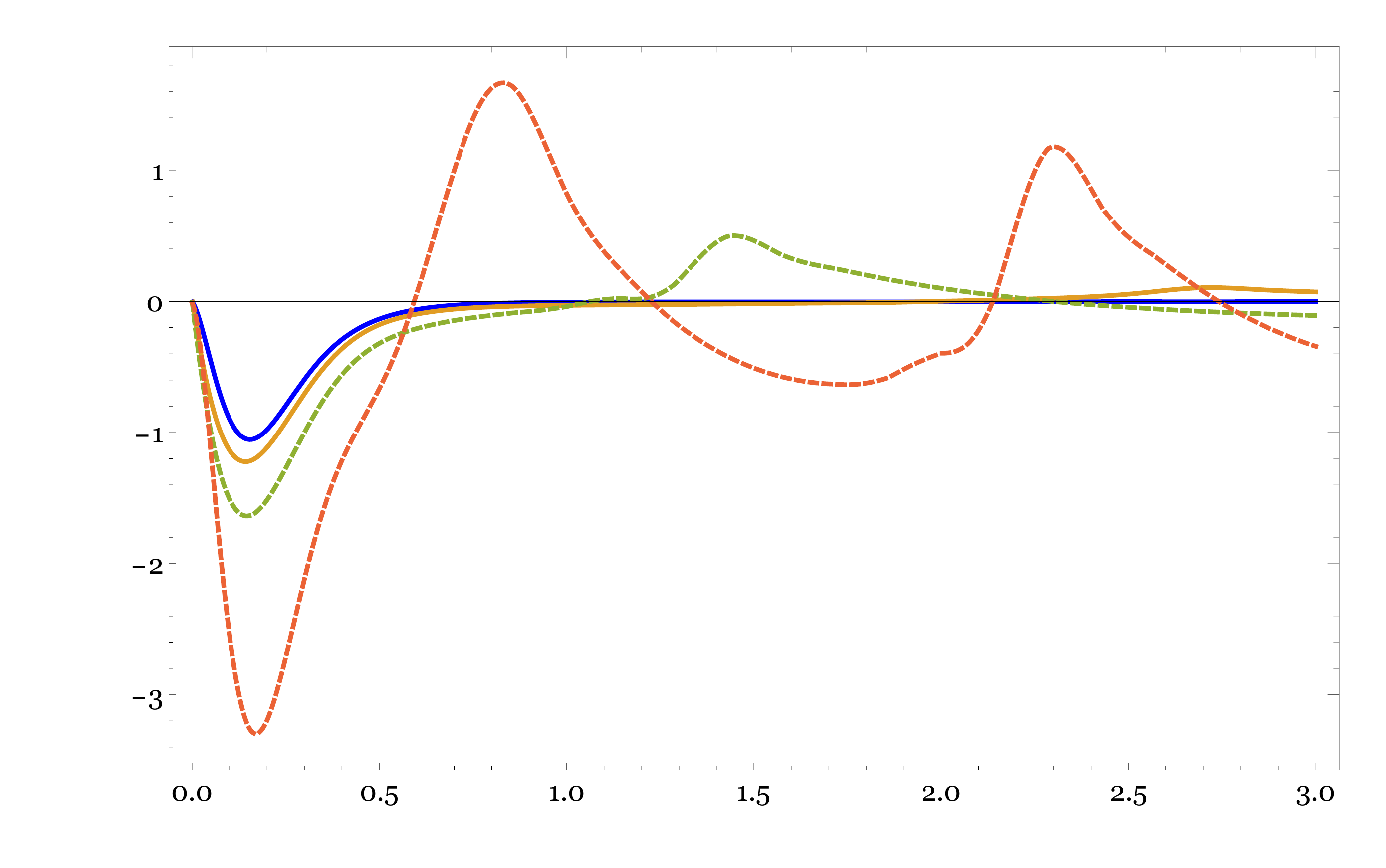}\put(-210,35){\rotatebox{-271}{\fontsize{17}{17}\selectfont $\frac{1}{T\hspace{.5mm}S_{th}}\frac{d\mathcal{C}_{1}^{\text{LR}}(t)}{dt}$}}	\put(-105,-5){{\fontsize{13}{13}\selectfont$t\hspace{1mm} T$}}
		\put(-120,125){{\fontsize{11}{11}\selectfont$d=1+1$}}
		\caption{The $L^{1}$ norm complexity and its time derivative in the LR basis for a massive cTFD state with $\tilde{Q} = 1/10$ in the simple limit ($\lambda_{p}\rightarrow 0$)  . $\tilde{m} = 1/5$ (blue), $\tilde{m}=1/2$ (brown), $\tilde{m} = 1$ (dashed green) and $\tilde{m}=2$ (dashed red). For large masses with respect to the thermal scale, there is an oscillatory behavior with period $\Delta t\sim \pi/ (m+\mu q)$. }\label{c1lrt2-5}	
	\end{figure}
	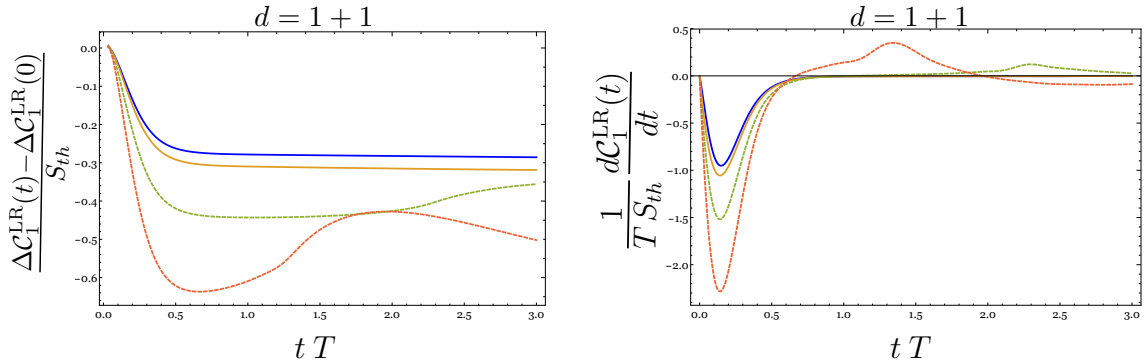
\begin{figure}[H]	\hspace{0.7cm}\includegraphics[scale=.26]{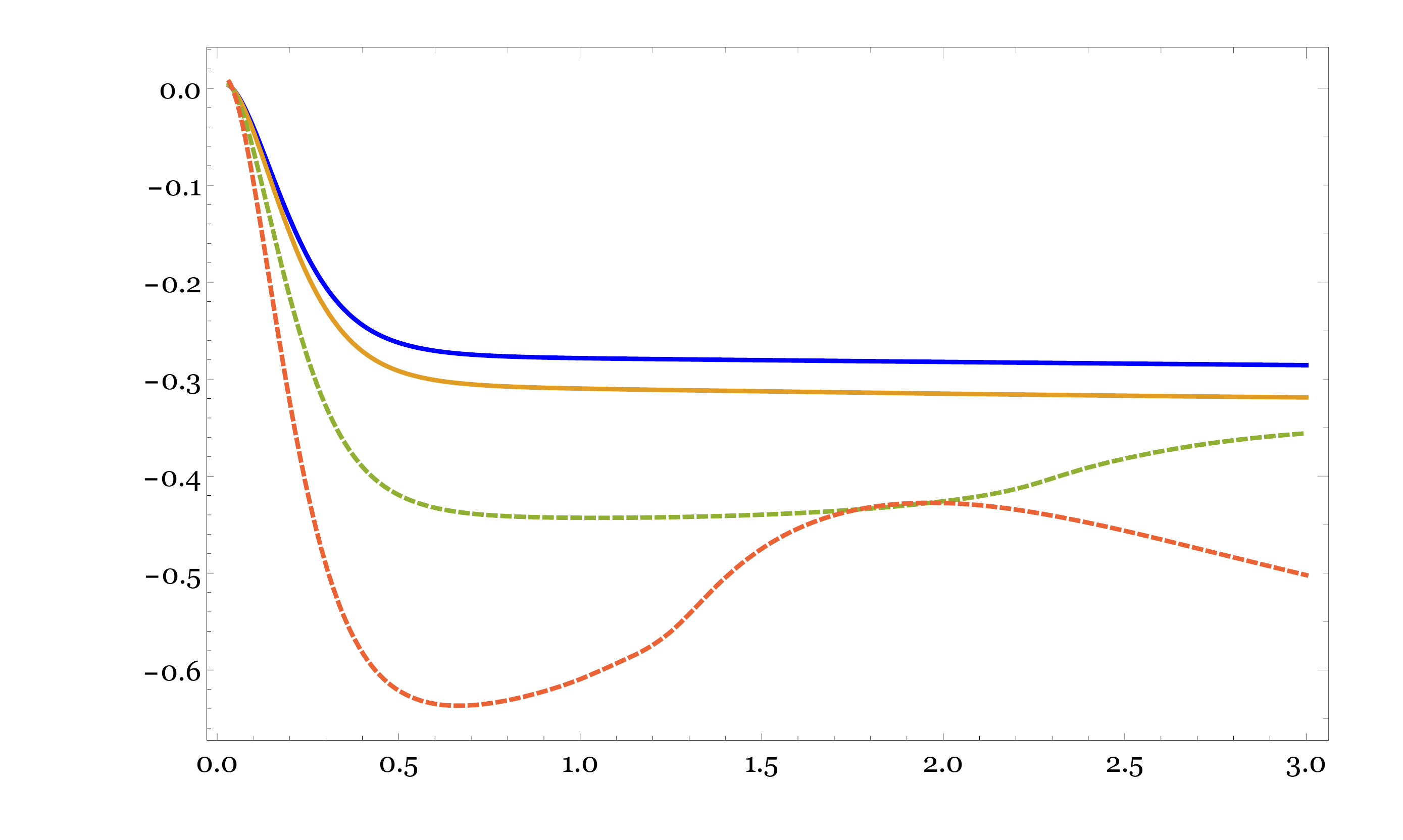}\put(-212,26){\rotatebox{-271}{\fontsize{14}{14}\selectfont $\frac{\Delta\mathcal{C}_{1}^{\text{LR}}(t)-\Delta\mathcal{C}_{1}^{\text{LR}}(0)}{S_{th}}$}}	\put(-105,-5){{\fontsize{13}{13}\selectfont$t\hspace{1mm} T$}}
		\put(-120,120){{\fontsize{11}{11}\selectfont$d=1+1$}}\hspace{.5cm}\includegraphics[scale=.26]{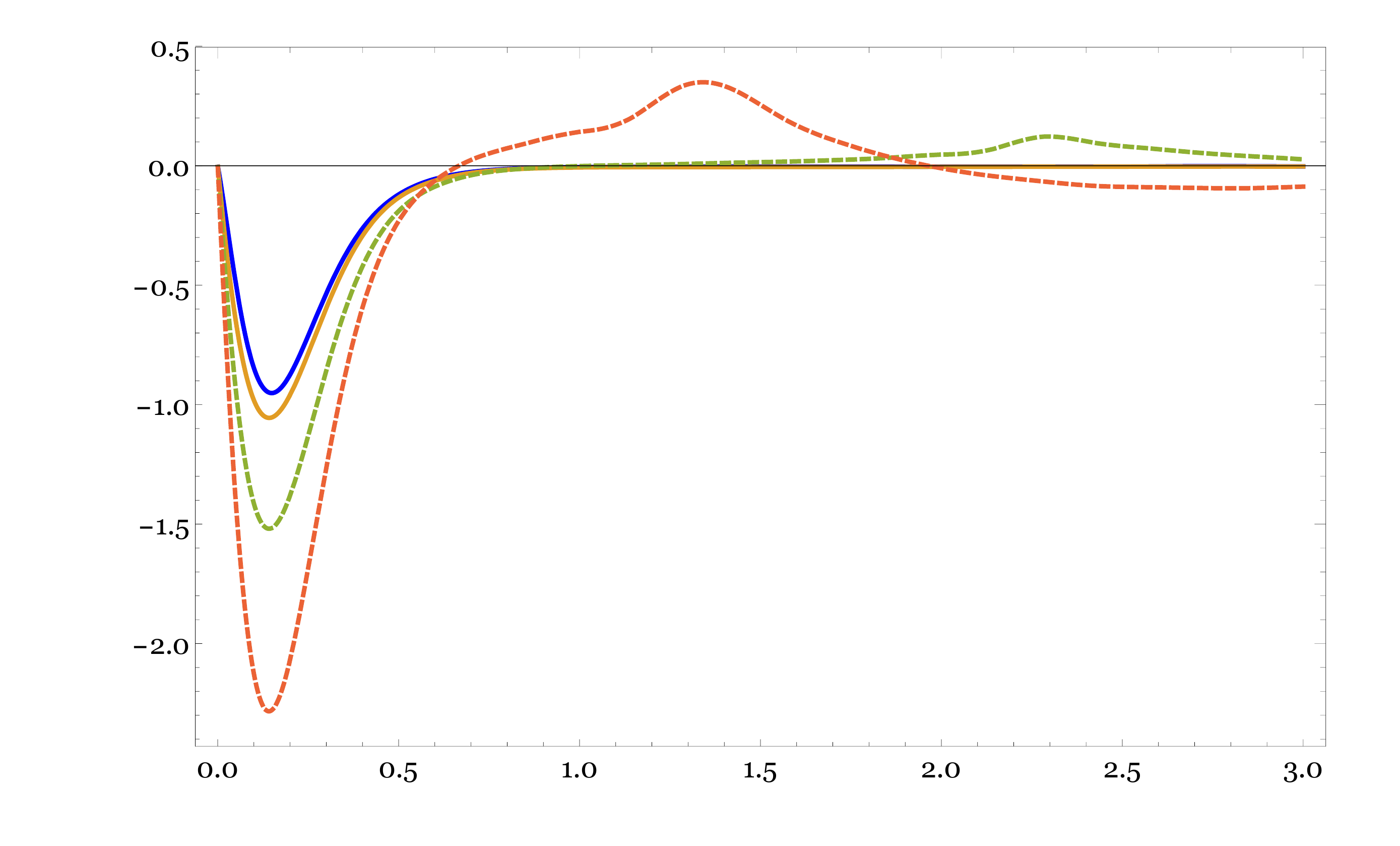}\put(-217,35){\rotatebox{-271}{\fontsize{17}{17}\selectfont $\frac{1}{T\hspace{.5mm}S_{th}}\frac{d\mathcal{C}_{1}^{\text{LR}}(t)}{dt}$}}	\put(-105,-5){{\fontsize{13}{13}\selectfont$t\hspace{1mm} T$}}
		\put(-120,120){{\fontsize{11}{11}\selectfont$d=1+1$}}
		\caption{The $L^{1}$ norm complexity and its time derivative in the LR basis for a massive cTFD state with   $\tilde{m} = 1/5$ in the simple limit ($\lambda_{p}\rightarrow 0$). $\tilde{Q} = 10^{-6}$ (blue), $\tilde{Q}=10^{-1}$ (brown), $\tilde{Q} = 1/2$ (dashed green) and $\tilde{Q}=1$ (dashed red). For large chemical potentials with respect to the thermal scale, there is an oscillatory behavior with period $\Delta t\sim \pi/ (m+\mu q)$.}\label{c1lrt2-6}	
	\end{figure}
	To close this section, in the following, we explore the effect of the reference scale on the time evolution of complexity of cTFD state by changing the dimensionless parameter $\tilde{\gamma}$. It is clear from fig.\ref{c1lrt2-7} that the decreasing of the complexity of formation to the minus values is an artifact of choosing the special value for reference scale and indeed it can grow very fast to positive values for intermediate scales, $\tilde{\gamma} \sim 1$. Of course, despite increasing at first for intermediate scale, it does not continue increasing for long times and it saturates fast. It is worth to emphasize that for both large and small values of $\tilde{\gamma}$ we see the same behavior, i.e. decreasing at first then saturating to a constant value. The constant value of saturation also depends on the value of $\tilde{\gamma}$, see figs.\ref{c1lrt2-7}-\ref{c1lrt2-8}. Moreover, it is worth noting that the dependency of complexity to reference scale at early transient times is the same as holographic results. Indeed, the time dependency of complexity for cTFD state dual to a charged eternal
	black hole exhibits non-universal behavior at early times due to the normalization of the null normals to boundary of WDW patch in the CA proposal which it is remained even after recovering affine parametrization. To be more precise, even after fixing reparametrization invariance by a boundary
	counterterm, the transient behavior is controlled by an arbitrary dimensionless parameter $l_{\text{ct}}/l$ which it can be dual to $\omega_{R}$ in the field theory side. The late time growth rate of complexity in holography was independent of these ambiguities in the null boundaries which
	also seems to be a property in figs.\ref{c1lrt2-8}.
	Similar analysis for the massive case is presented in figs.\ref{c1lrt2-9} and \ref{c1lrt2-10}. Same as the massless case, for the large and small values of $\tilde{\gamma}$ we see effectively the same behavior. This can be understood  easily by finding the complexity of formation in $LR$ bases for the simple limit in which $\omega_{R}$ is much smaller than any other scale, i.e. $\lambda_{p} \rightarrow \infty$. In this limit, using (\ref{upt}), we have
	\bea
	&& s_{1,\pm} = \frac{1}{2}\log\bigg(\cosh 2\alpha_{p}\mp \sinh 2\alpha_{p}\cos[(\omega_{p}+\mu q)t]\bigg) +\frac{1}{2}\log\lambda_{p},
	\cr\nonumber\\
	&&s_{2,\pm} \simeq \mp \text{sgn}\bigg(\sin(\omega_{p}+\mu q)t\bigg)\frac{\pi}{2}.  
	\eea
	Considering the above values for $s_{i,\pm}$ in (\ref{as}) implies that
	\bea
	&&a_{L,L}^{(1)} = -\frac{1}{2}\bigg(s_{1,-}\sin s_{2,-}+s_{1,+}\sin s_{2,+}\bigg),
	\hspace{1cm}a_{L,R}^{(1)} = \frac{1}{2}\bigg(s_{1,-}\sin s_{2,-}-s_{1,+}\sin s_{2,+}\bigg),
	\cr\nonumber\\
	&&\hspace{5cm}a_{L,R}^{(2)} \simeq 0,
	\hspace{1cm}a_{L,L}^{(2)} \simeq 0,
	\eea
	which by implementing them in (\ref{c1LRt}) we have
	\bea
	\label{lambdainfin}
	\mathcal{C}_1^{(LR)}(t) -\mathcal{C}_1^{(LR)}(\text{vac}) = \frac{\text{vol}}{(2\pi)^{d-1}}\int_{0}^{\infty} d^{d-1}p\hspace{1mm} \log\bigg(\cosh2\alpha_p+\bigg{|}\cos(\omega_{p}+\mu q)t\bigg{|}\sinh2\alpha_p\hspace{1mm}\bigg).
	\eea
	The result (\ref{lambdainfin}) exactly matches with (\ref{deltaC1LR}). Thus, even though we are considering the opposite limit here (i.e. $\lambda \rightarrow \infty$ rather than $\lambda \rightarrow 0$), the time dependency of complexity of formation remains unchanged. 
	\begin{figure}[H]
		\hspace{.7cm}\includegraphics[scale=.26]{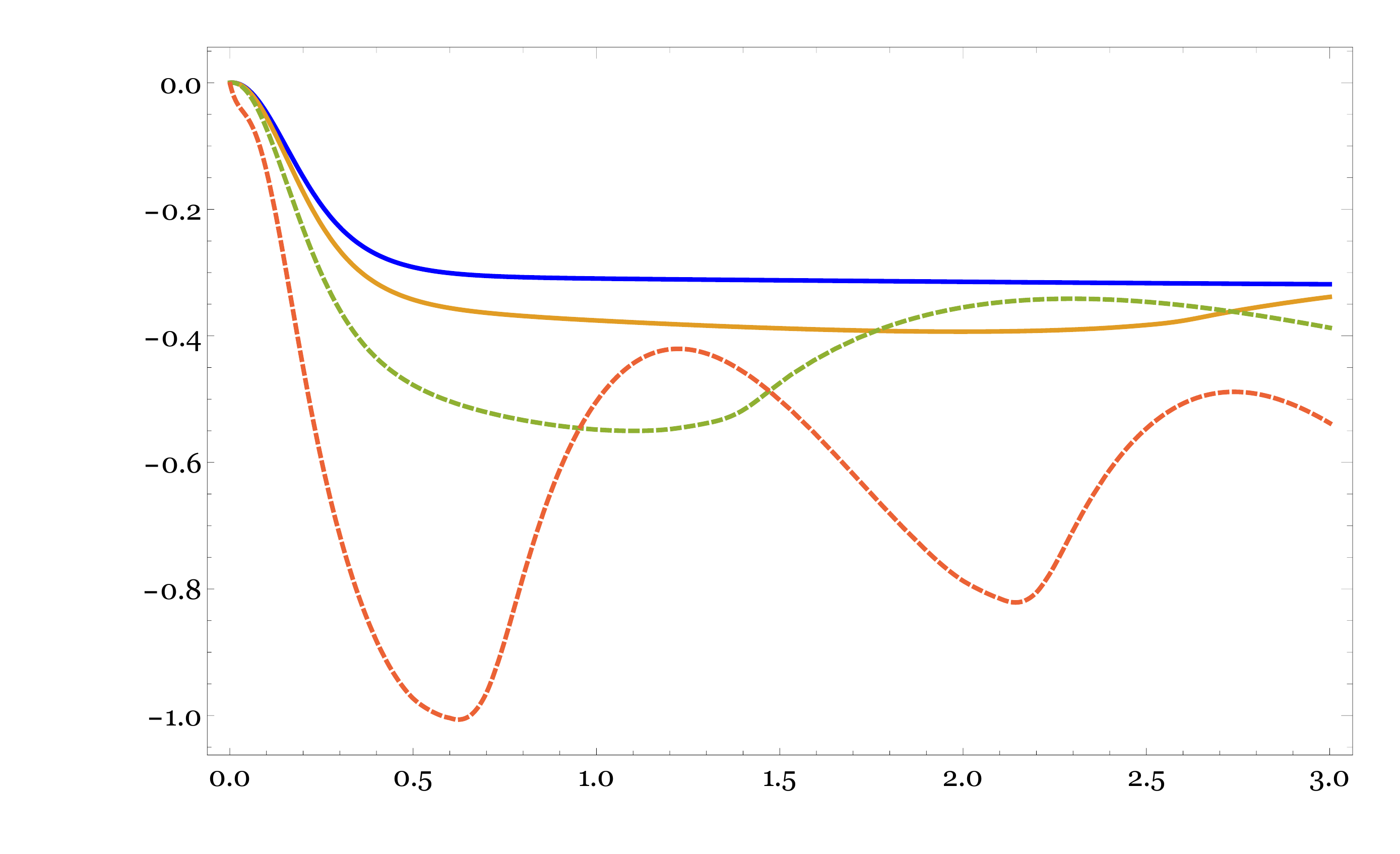}\put(-212,28){\rotatebox{-271}{\fontsize{14}{14}\selectfont $\frac{\Delta\mathcal{C}_{1}^{\text{LR}}(t)-\Delta\mathcal{C}_{1}^{\text{LR}}(0)}{S_{th}}$}}	\put(-100,-5){{\fontsize{13}{13}\selectfont$t\hspace{1mm} T$}}
		\put(-145,123){{\fontsize{11}{11}\selectfont$d=(1+1), \tilde{\gamma}= 10^{5}$}}\hspace{.5cm}\includegraphics[scale=.26]{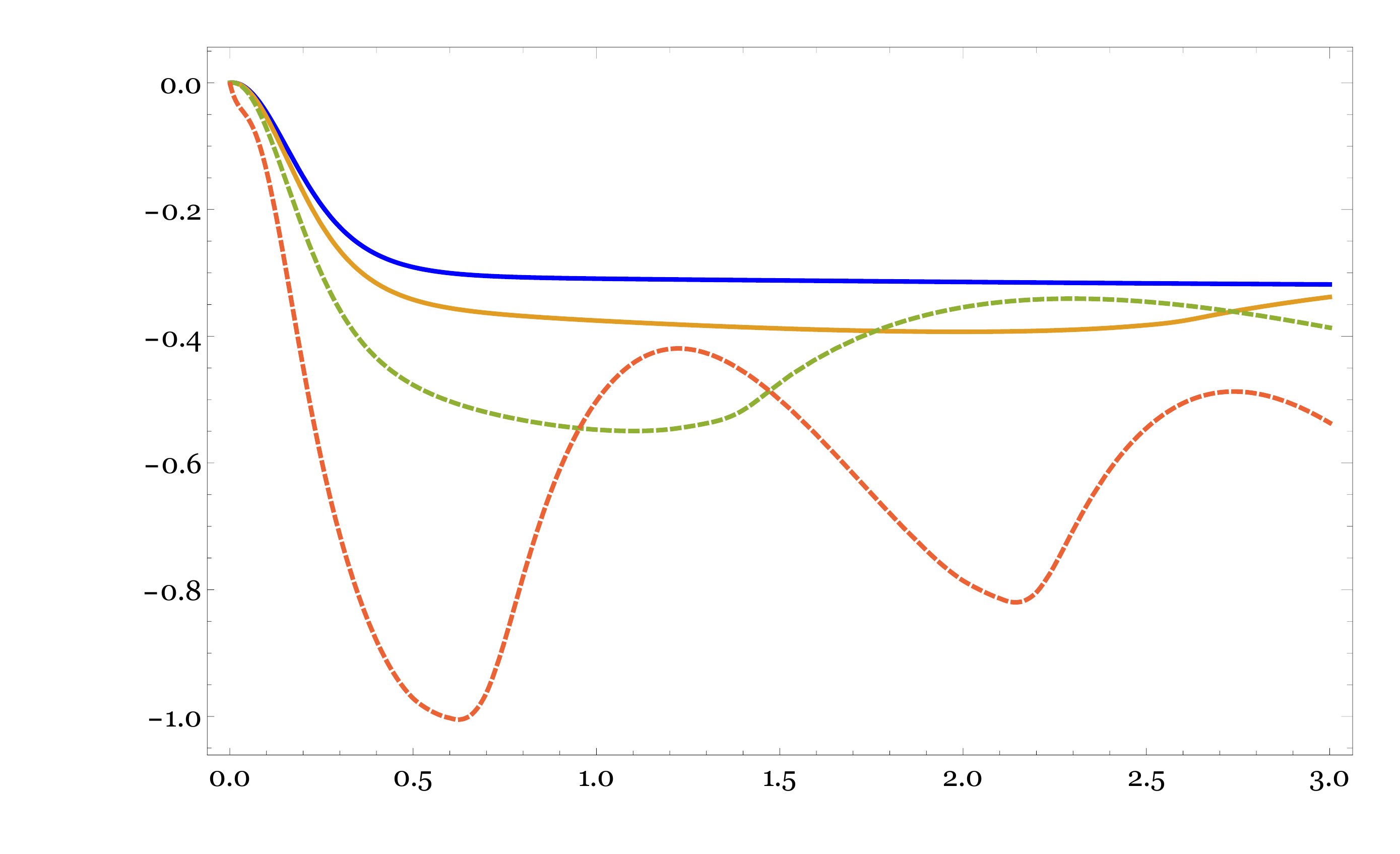}\put(-212,28){\rotatebox{-271}{\fontsize{14}{14}\selectfont $\frac{\Delta\mathcal{C}_{1}^{\text{LR}}(t)-\Delta\mathcal{C}_{1}^{\text{LR}}(0)}{S_{th}}$}}	\put(-100,-5){{\fontsize{13}{13}\selectfont$t\hspace{1mm} T$}}
		\put(-145,123){{\fontsize{11}{11}\selectfont$d=(1+1), \tilde{\gamma}= 10^{-5}$}}\vspace{.65cm}
		
		\hspace{.7cm}\includegraphics[scale=.26]{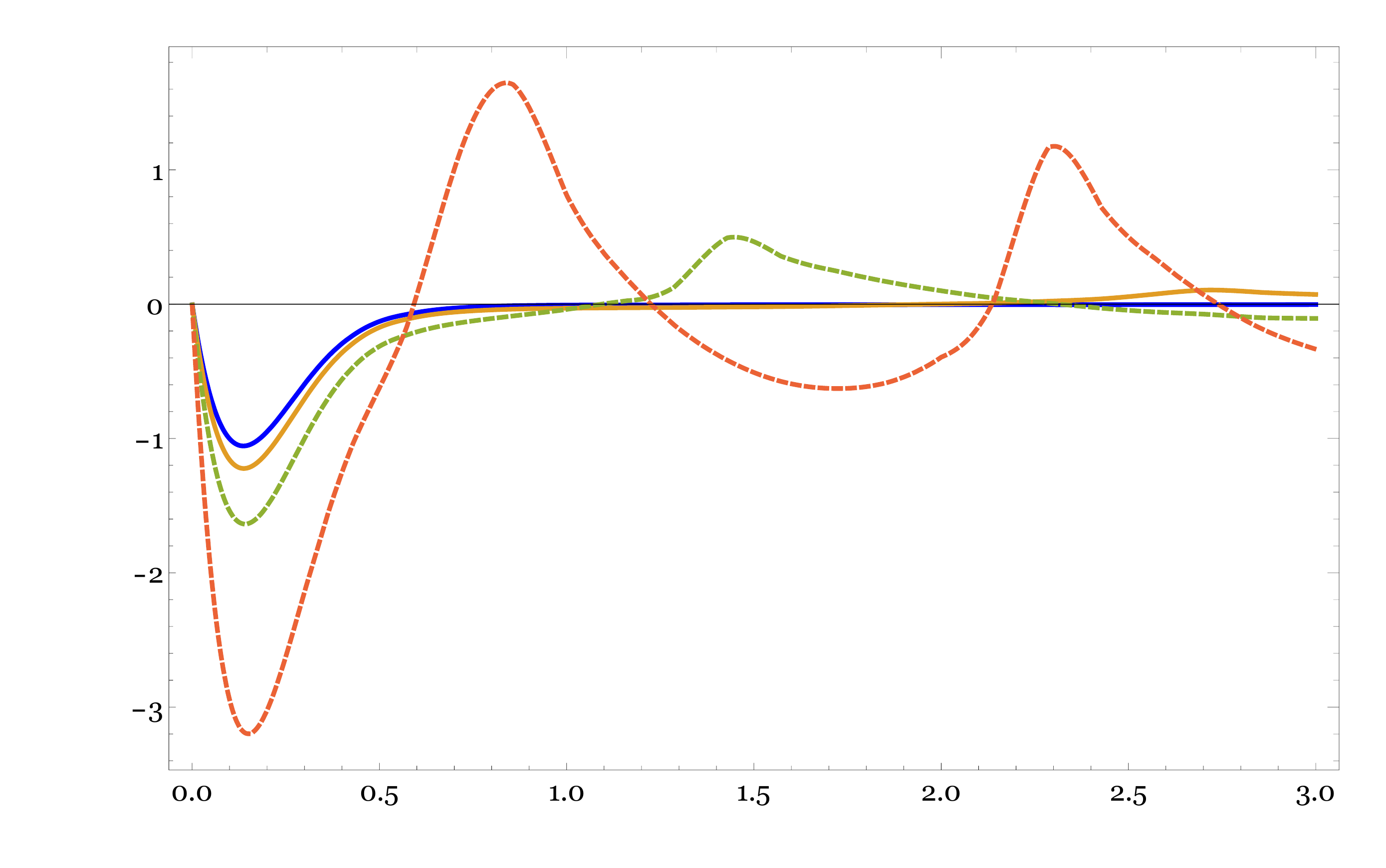}\put(-218,40){\rotatebox{-271}{\fontsize{17}{17}\selectfont $\frac{1}{T\hspace{.5mm}S_{th}}\frac{d\mathcal{C}_{1}^{\text{LR}}(t)}{dt}$}}	\put(-102,-5){{\fontsize{13}{13}\selectfont$t\hspace{1mm} T$}}
		\put(-145,128){{\fontsize{11}{11}\selectfont$d=(1+1), \tilde{\gamma}=10^{5}$}}\hspace{.7cm}\includegraphics[scale=.26]{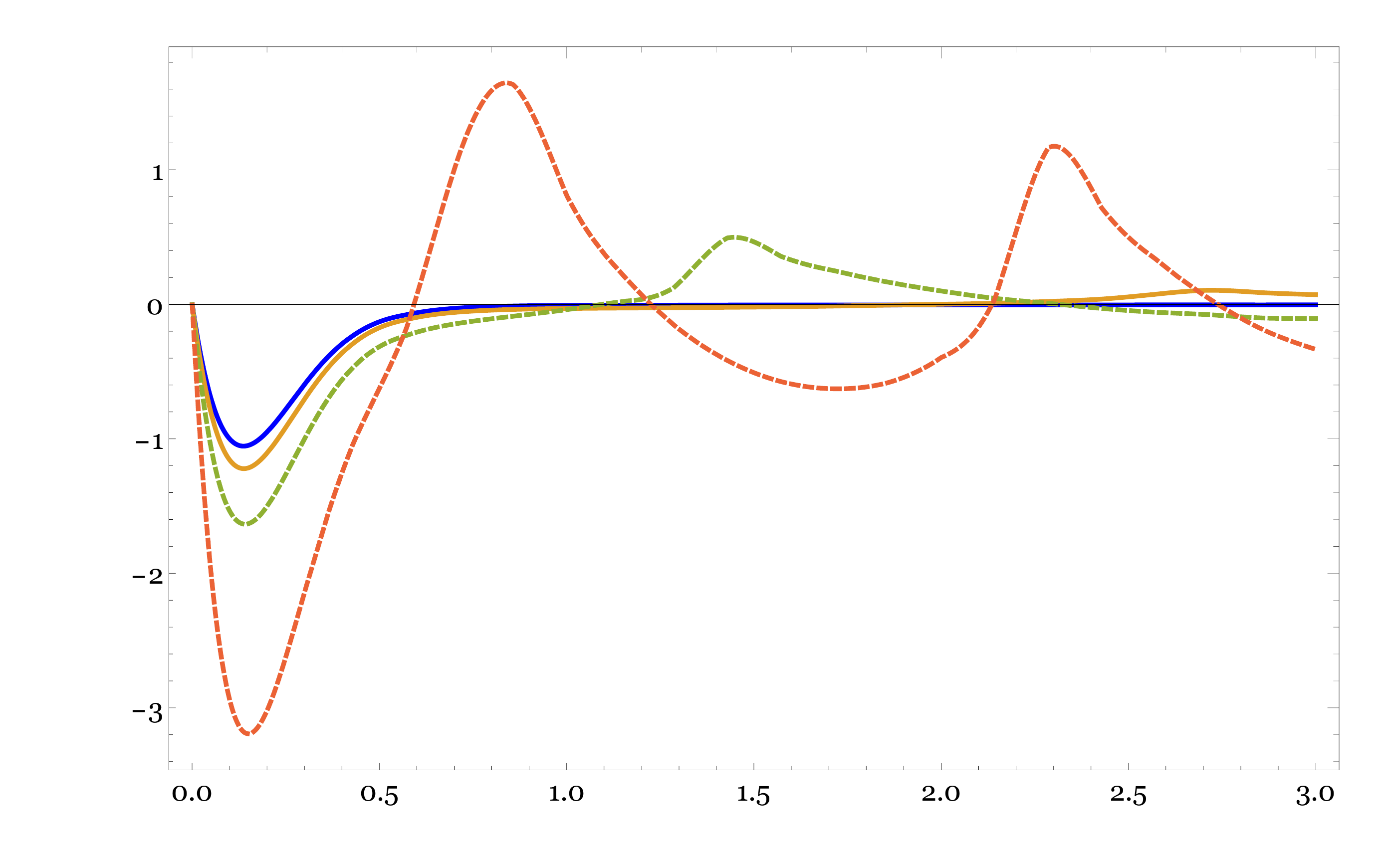}\put(-218,40){\rotatebox{-271}{\fontsize{17}{17}\selectfont $\frac{1}{T\hspace{.5mm}S_{th}}\frac{d\mathcal{C}_{1}^{\text{LR}}(t)}{dt}$}}	\put(-102,-5){{\fontsize{13}{13}\selectfont$t\hspace{1mm} T$}}
		\put(-145,128){{\fontsize{11}{11}\selectfont$d=(1+1), \tilde{\gamma}=10^{-5}$}}
		\caption{The time evolution of $L^{1}$ norm complexity of cTFD state in LR basis with $\tilde{Q} = 1/10$ in $d=1+1$ dimensions and different masses. $\tilde{m} = 1/5$ (blue), $\tilde{m}=1/2$ (brown), $\tilde{m} = 1$ (dashed green) and $\tilde{m}=2$ (dashed red). The left figure is for $\tilde{\gamma}= 10^{5}$ and the right one for $\tilde{\gamma}=10^{-5}$. It is clear that both the large and small values of reference scale led to the same time dependency for the complexity.}\label{c1lrt2-9}
	\end{figure}
	\begin{figure}[H]
		\hspace{.7cm}\includegraphics[scale=.26]{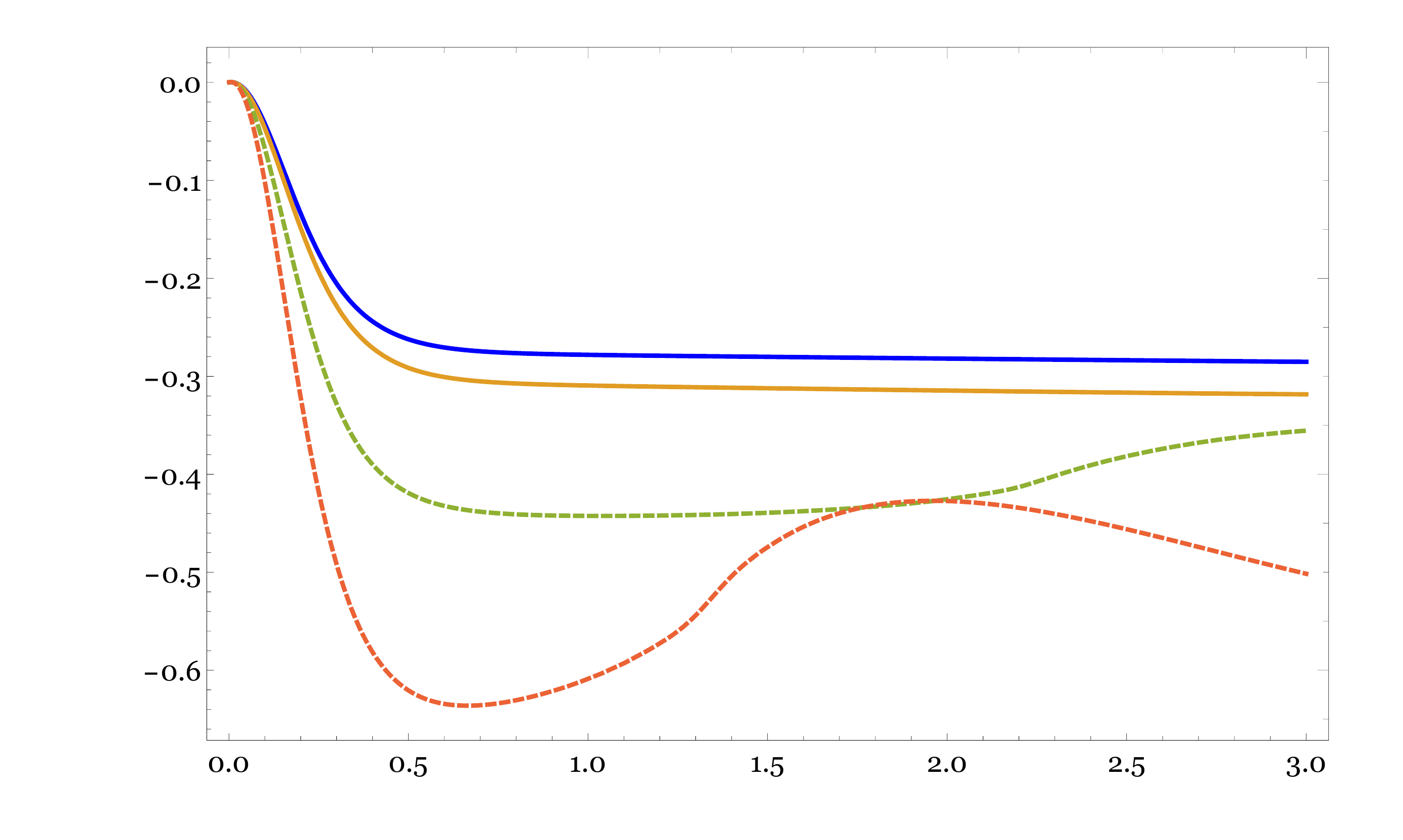}\put(-212,28){\rotatebox{-271}{\fontsize{14}{14}\selectfont $\frac{\Delta\mathcal{C}_{1}^{\text{LR}}(t)-\Delta\mathcal{C}_{1}^{\text{LR}}(0)}{S_{th}}$}}	\put(-103,-5){{\fontsize{13}{13}\selectfont$t\hspace{1mm} T$}}
		\put(-140,121){{\fontsize{11}{11}\selectfont$d=(1+1), \tilde{\gamma}= 10^{5}$}}\hspace{.5cm}\includegraphics[scale=.26]{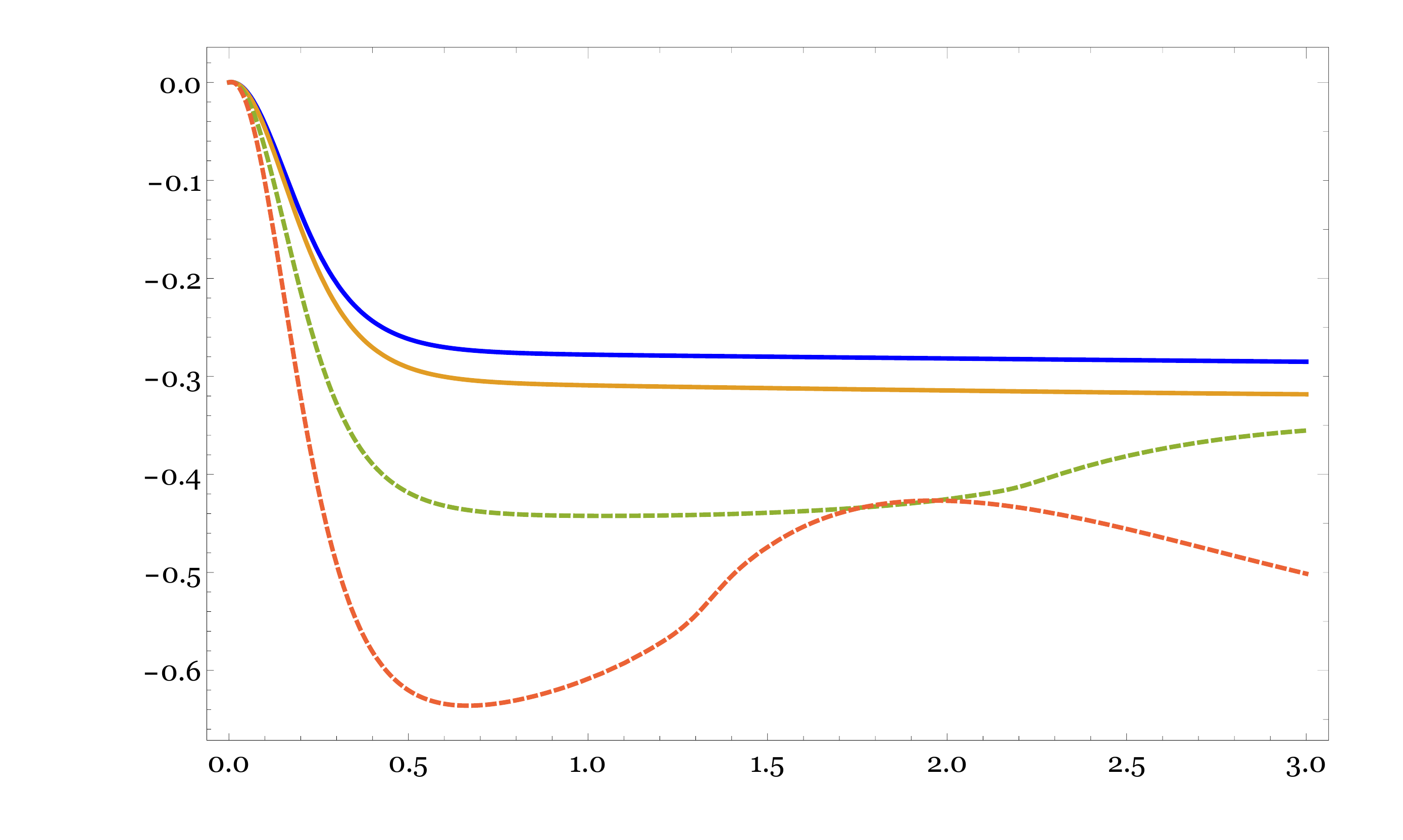}\put(-212,28){\rotatebox{-271}{\fontsize{14}{14}\selectfont $\frac{\Delta\mathcal{C}_{1}^{\text{LR}}(t)-\Delta\mathcal{C}_{1}^{\text{LR}}(0)}{S_{th}}$}}	\put(-103,-5){{\fontsize{13}{13}\selectfont$t\hspace{1mm} T$}}
		\put(-140,121){{\fontsize{11}{11}\selectfont$d=(1+1), \tilde{\gamma}= 10^{-5}$}}\vspace{.6cm}
		
		\hspace{.7cm}\includegraphics[scale=.26]{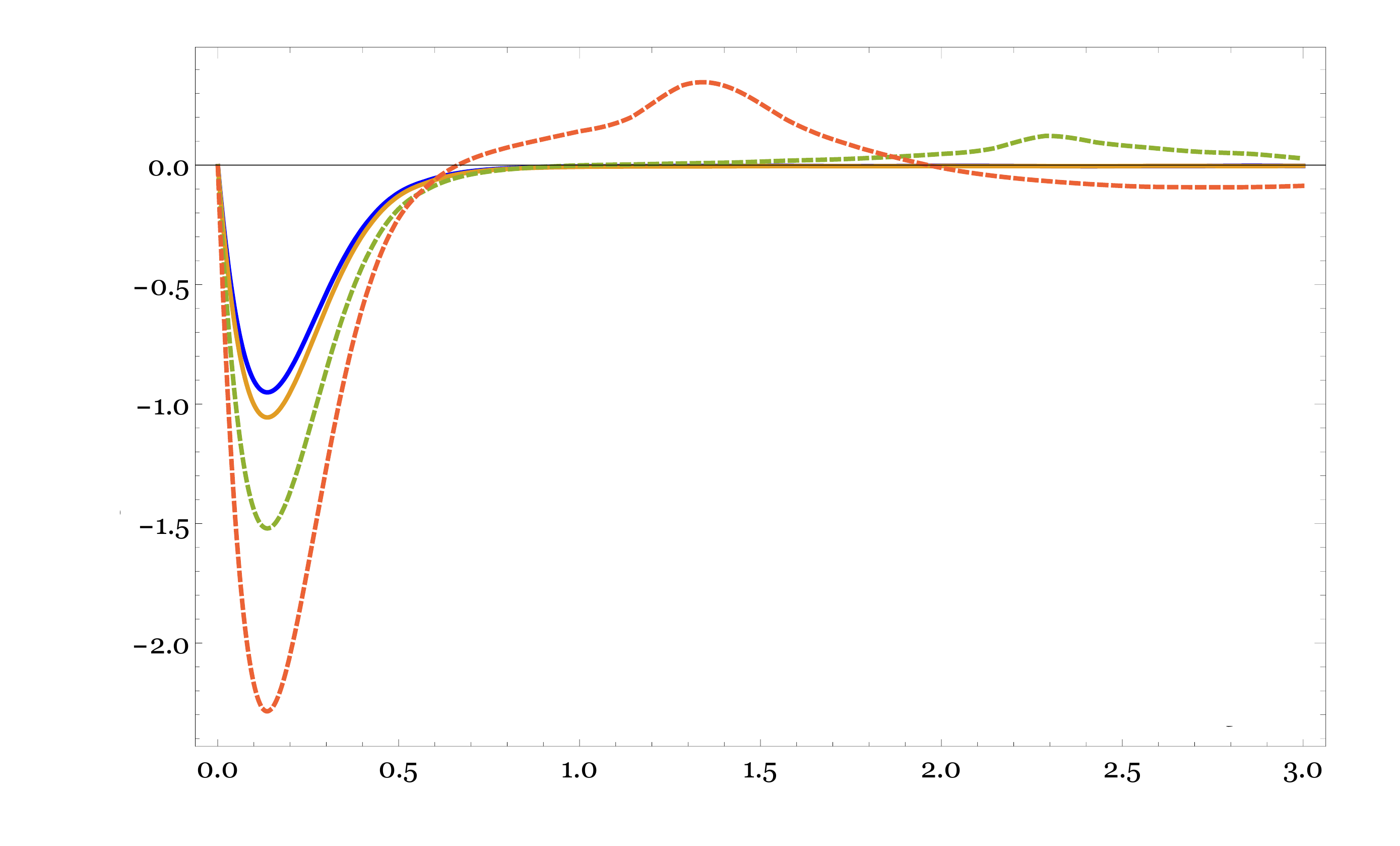}\put(-218,35){\rotatebox{-271}{\fontsize{17}{17}\selectfont $\frac{1}{T\hspace{.5mm}S_{th}}\frac{d\mathcal{C}_{1}^{\text{LR}}(t)}{dt}$}}	\put(-105,-5){{\fontsize{13}{13}\selectfont$t\hspace{1mm} T$}}
		\put(-145,125){{\fontsize{11}{11}\selectfont$d=(1+1), \tilde{\gamma}=10^{5}$}}\hspace{.5cm}\includegraphics[scale=.26]{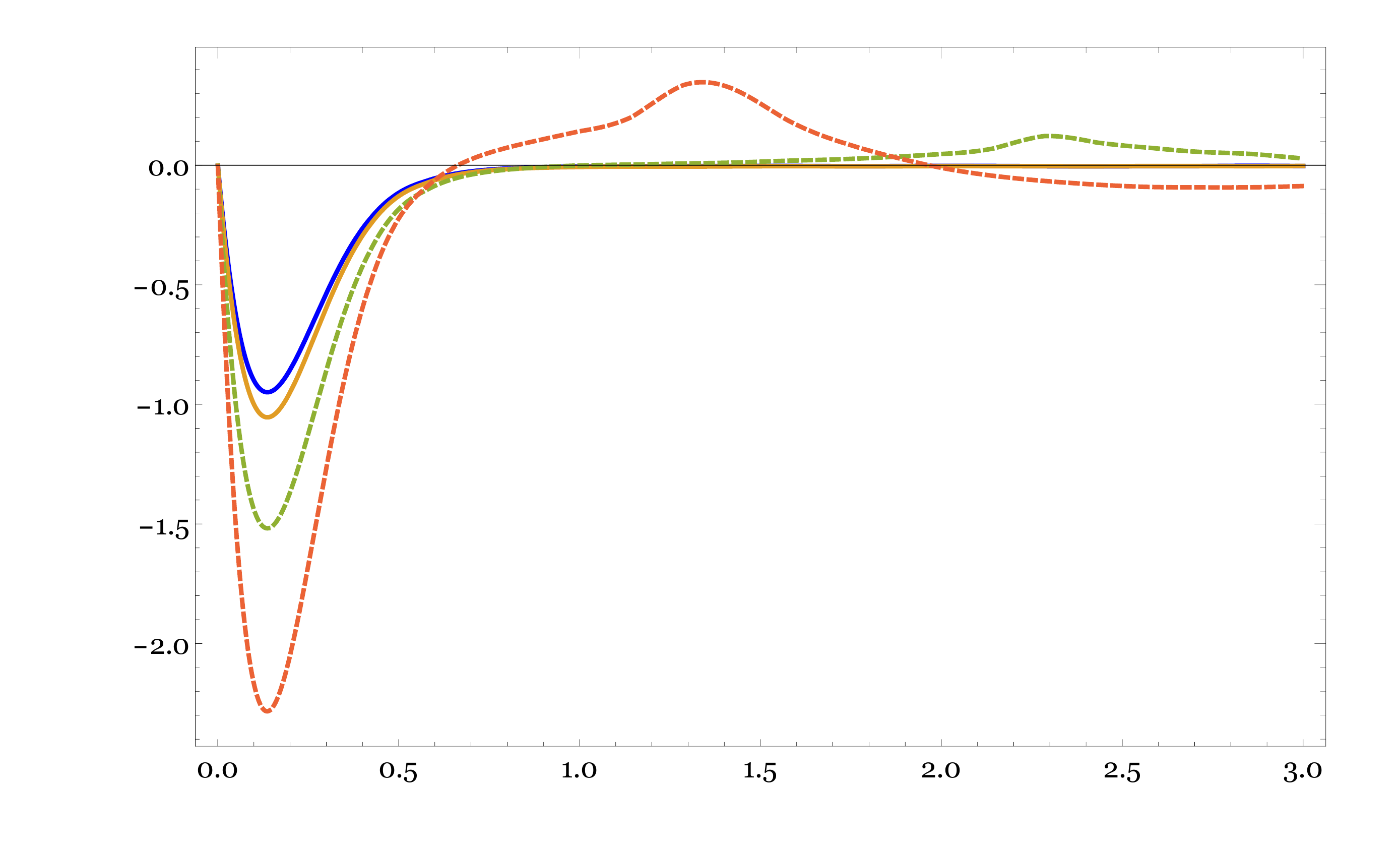}\put(-212,35){\rotatebox{-271}{\fontsize{17}{17}\selectfont $\frac{1}{T\hspace{.5mm}S_{th}}\frac{d\mathcal{C}_{1}^{\text{LR}}(t)}{dt}$}}	\put(-105,-5){{\fontsize{13}{13}\selectfont$t\hspace{1mm} T$}}
		\put(-145,123){{\fontsize{11}{11}\selectfont$d=(1+1), \tilde{\gamma}=10^{-5}$}}
		\caption{The time evolution of $L^{1}$ norm complexity of cTFD state in LR basis with $\tilde{m} = 1/5$ in $d=1+1$ dimensions and different charges. $\tilde{Q} = 10^{-6}$ (blue), $\tilde{Q}=10^{-1}$ (brown), $\tilde{Q} = 1/2$ (dashed green) and $\tilde{Q}=1$ (dashed red). The left figure is for $\tilde{\gamma}=10^{5}$ and the right one is for $\tilde{\gamma}=10^{-5}$. It is clear that both the large and small values of reference scale led to the same time dependency for the complexity.}\label{c1lrt2-10}
	\end{figure}
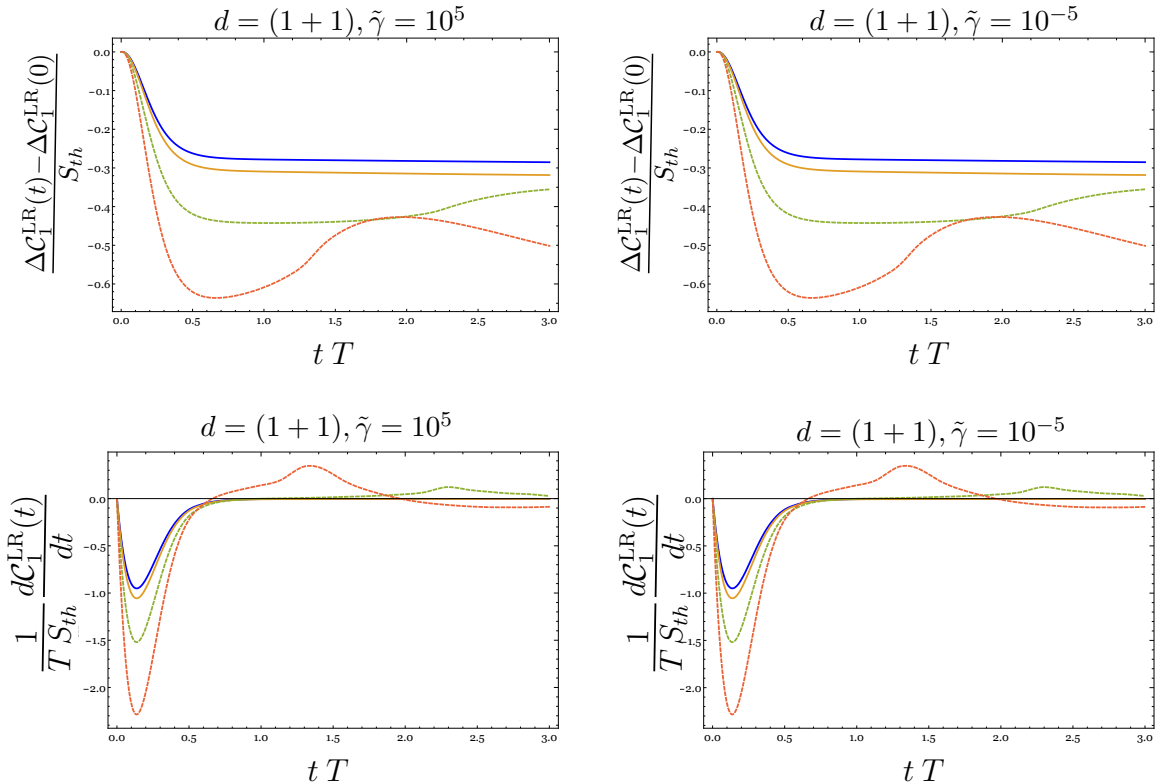
	\section{Conclusions}
	In this paper, we derived the complexity of formation over entropy for cTFD state of free complex  scalar theory with particles on one side and anti-particles on the other side. We chose Nielsen's geometric approach and used covariance matrix technique, since our states were Gaussian. We investigated $F_{\kappa=2}$, $F_2$ and $F_1$ cost functions to see which one is a better option to be the quantum field counterpart of holographic complexity. In general, $L^2$ norms are not basis-dependent and it makes them easier to work with. But their leading divergent term for $F_2$ and $F_{\kappa=2}$ cost functions are in conflict with the holographic results (\ref{Cholog}). On the other hand, if we take $\tilde{\alpha}=l \omega_R$, $\mathcal{C}_1$ would have the same leading term as holography. So after this observation, we mainly focused on $F_1$. As a matter of fact, we could just find an upper bound for this cost function and calculating the exact expression for $\mathcal{C}_1$ is very difficult. Also, this cost function is basis-dependent so we had to choose our basis. In the beginning, we used diagonal basis in which we can factorize our cTFD state. $\mathcal{C}_1$ in diagonal basis was in contrast with the third law of complexity, even though it had the correct leading terms. So we changed our basis to LR which uses physical degrees of freedom and it gave us the desired IR divergence that we see in holography.
	
	Then we investigated the time dependency of complexity and it's growth rate, using $F_1$ in LR basis. A difference with holographic complexity is saturation time, which is much less than the expected one in holography, because we are using free field theory so our state cannot probe a vast subspace of Hilbert space and complexity saturates in thermal time. Besides, we observed that the reference scale has the same rule that the ambiguities have in holography. For example, the dependency of complexity to reference scale at early transient times matches the holographic dependence on ambiguity.
	
	Our final observation regarding complexity was comparing each term of it with different holographic proposals. We derived complexity of formation in holography for different dimensions. In 3+1 dimensions, even though the neutral TFD was in agreement with holography, complexity of charged black hole had some terms missing. We discussed different ways to solve this problem. If we assume that the complexity we found in free cTFD does not change in strong coupling regime, we concluded that the most reasonable solution is adding boundary terms. One of these boundary terms were proposed in \cite{Brown:2018bms}. This term just recovers one of the missing terms and the remaining are still missing. But the other boundary term, proposed in \cite{Akhavan:2018wla} has all the missing terms and unlike the other one, smoothly approaches to neutral case. Moreover, in the finite chemical potential, the holographic complexity with the latter boundary term diverges similarly to the complexity of formation in cTFD.
	
	So $\mathcal{C}_1$ in LR basis not only is a very good candidate for neutral black hole, but also it is a very good match in the presence of $U(1)$ electric charge and complexity of formation in this basis matches very well with the theory that has cut off behind the horizon.
	\subsection*{Acknowledgements}
	We wish to thank Mohsen Alishahiha, Mostafa Ghasemi, S.Sedighe Hashemi, Ghadir Jafari, Ali Mollabashi, Behrad Taghavi, Farid Taghinavaz and Hamed Zolfi for fruitful discussions. We would also want to thank Michal Heller for useful comments on the early version of the paper and Shira Chapman for discussions and fruitful comments on the final draft. We especially would like to thank the IPM-Grid computing group for providing computing and storage facilities.
	
	\appendix
	
	\section{Thermal Entropy and Total Charge}
	\label{def}
	Entropy of any thermal state can be obtained by its partition function, using the equation
	\bea
	&& S_{th} = \frac{\partial}{\partial{T}}(T \log{Z})= \log{Z} - \beta \frac{\partial}{\partial \beta}\log{Z}.
	\eea
	For a single mode, the partition function becomes
	\bea
	Z = \frac{e^{-\beta(\omega+\mu q)/2}}{1-e^{-\beta(\omega+\mu q)}},
	\eea
	which it implies that the thermal entropy is given by
	\bea
	S_{th} = \mathrm{vol} \int \frac{\mathrm{d}^{d-1}k}{(2\pi)^{d-1}}\bigg[\frac{\beta(\omega_k+\mu q)}{e^{\beta(\omega_k+\mu q)}-1}-\log(1-e^{-\beta(\omega_k+\mu q)})\bigg].
	\eea
	Moreover, every particle has a charge $q$ so in order to find the total charge of the state, we need to find the average number of particles, using Bose-Einstein distribution. Accordingly,
	\bea
	&&Q = q\bigg[\frac{1}{Z}\sum_{n=0}^{\infty}ne^{-\beta\left(\omega(n+\frac{1}{2})+n\mu q\right)}\bigg]= \frac{q}{e^{\beta(\omega+\mu q)}-1}.
	\eea
	By integrating over every mode of the quantum field theory, we obtain
	\bea
	Q_{tot} =\mathrm{vol} \int \frac{\mathrm{d}^{d-1}k}{(2\pi)^{d-1}}\frac{q}{e^{\beta\left(\omega_k+\mu q\right)}-1}.
	\eea
	\section{Complexities in Diagonal Basis}\label{diagonal}
\begin{figure}[H]
		\hspace{1.5cm}\includegraphics[scale=.33]{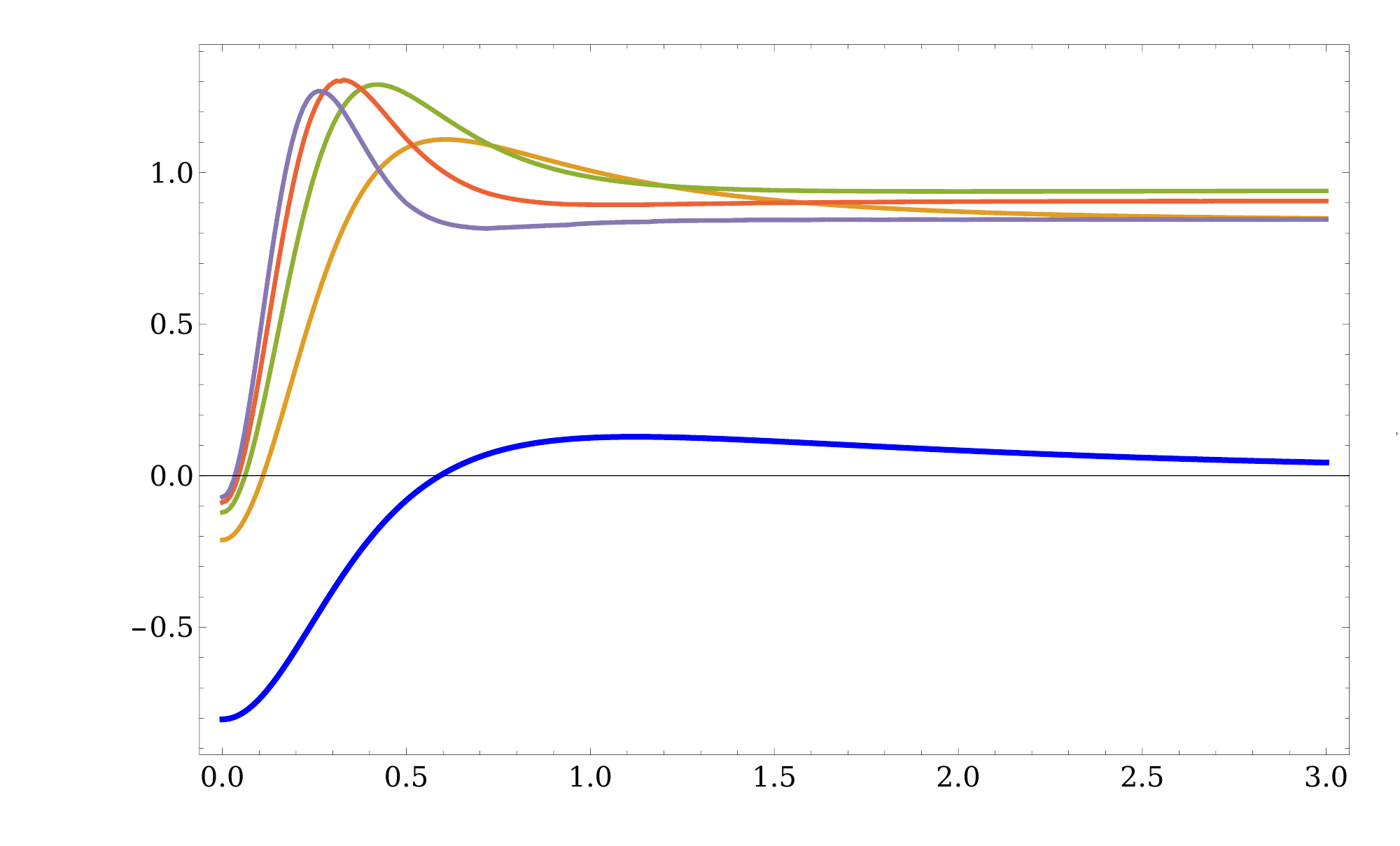}\put(-196 ,25){\rotatebox{-271}{\fontsize{14}{14}\selectfont $\frac{\mathcal{C}_{\kappa=2}(t)-\mathcal{C}^{\text{TFD}}_{\kappa=2}(0)}{S_{th}}$}}	\put(-95,-5){{\fontsize{13}{13}\selectfont$t\hspace{1mm} T$}}\hspace{.4cm}\includegraphics[scale=.325]{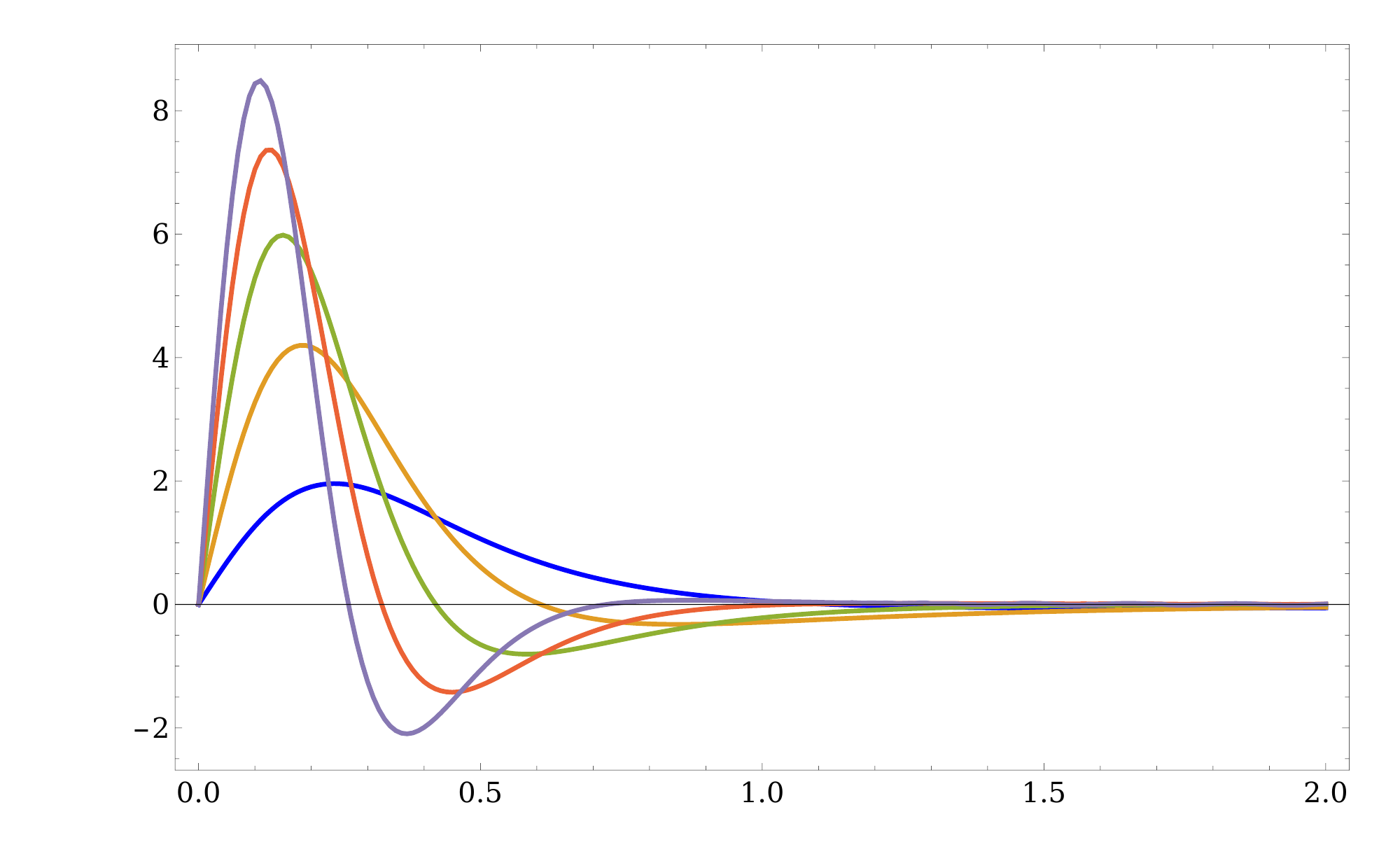}\put(-196,30){\rotatebox{-271}{\fontsize{14}{14}\selectfont $\frac{1}{T\hspace{.5mm}S_{th}}\frac{d\mathcal{C}_{\kappa=2}(t)}{dt}$}}	\put(-95,-5){{\fontsize{13}{13}\selectfont$t\hspace{1mm} T$}}\vspace{.3cm}
		
		\hspace{1.5cm}\includegraphics[scale=.33]{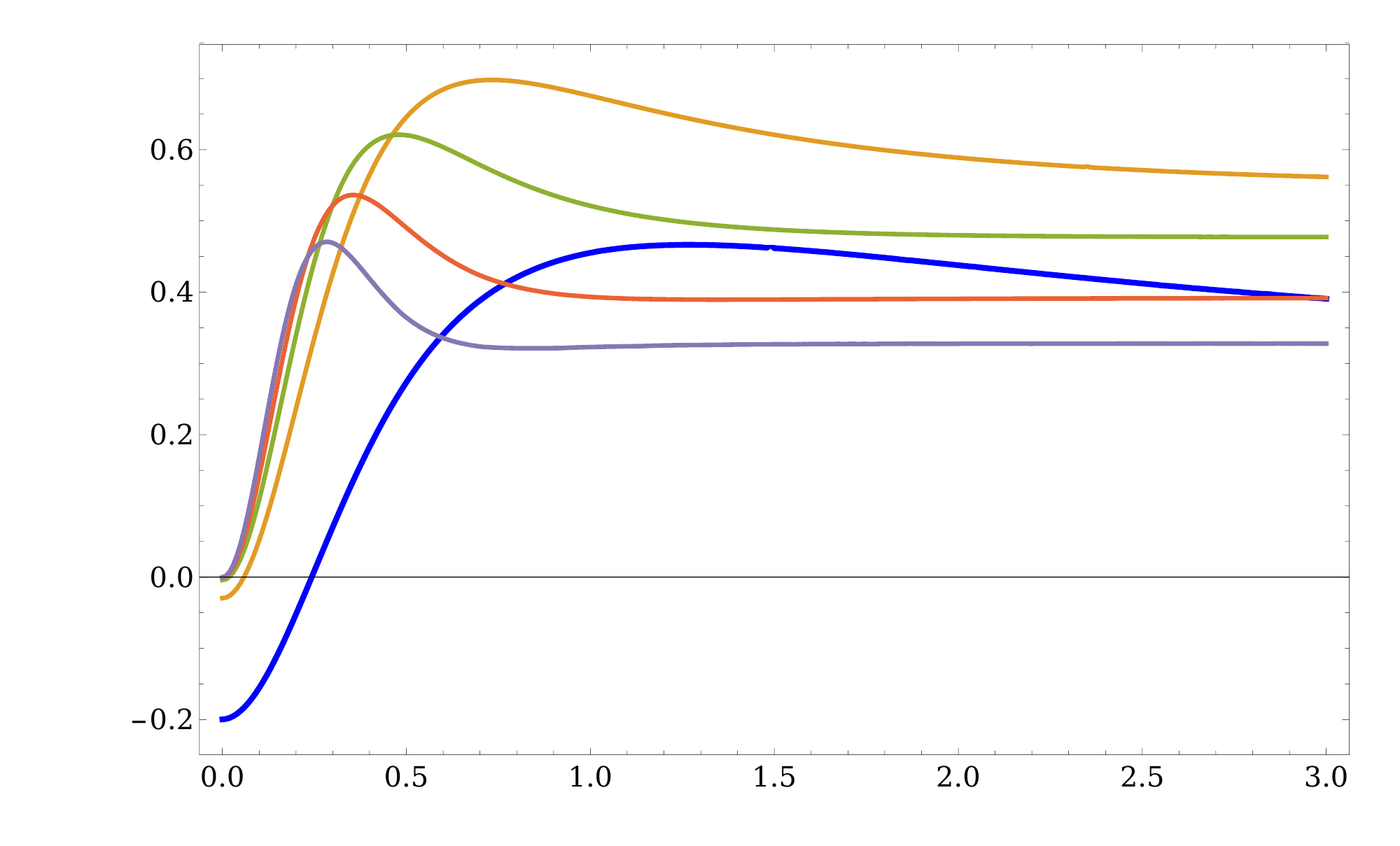}\put(-196 ,30){\rotatebox{-271}{\fontsize{14}{14}\selectfont $\frac{\mathcal{C}_{1}(t)-\mathcal{C}^{\text{TFD}}_{1}(0)}{S_{th}}$}}	\put(-95,-5){{\fontsize{13}{13}\selectfont$t\hspace{1mm} T$}}\hspace{.4cm}\includegraphics[scale=.33]{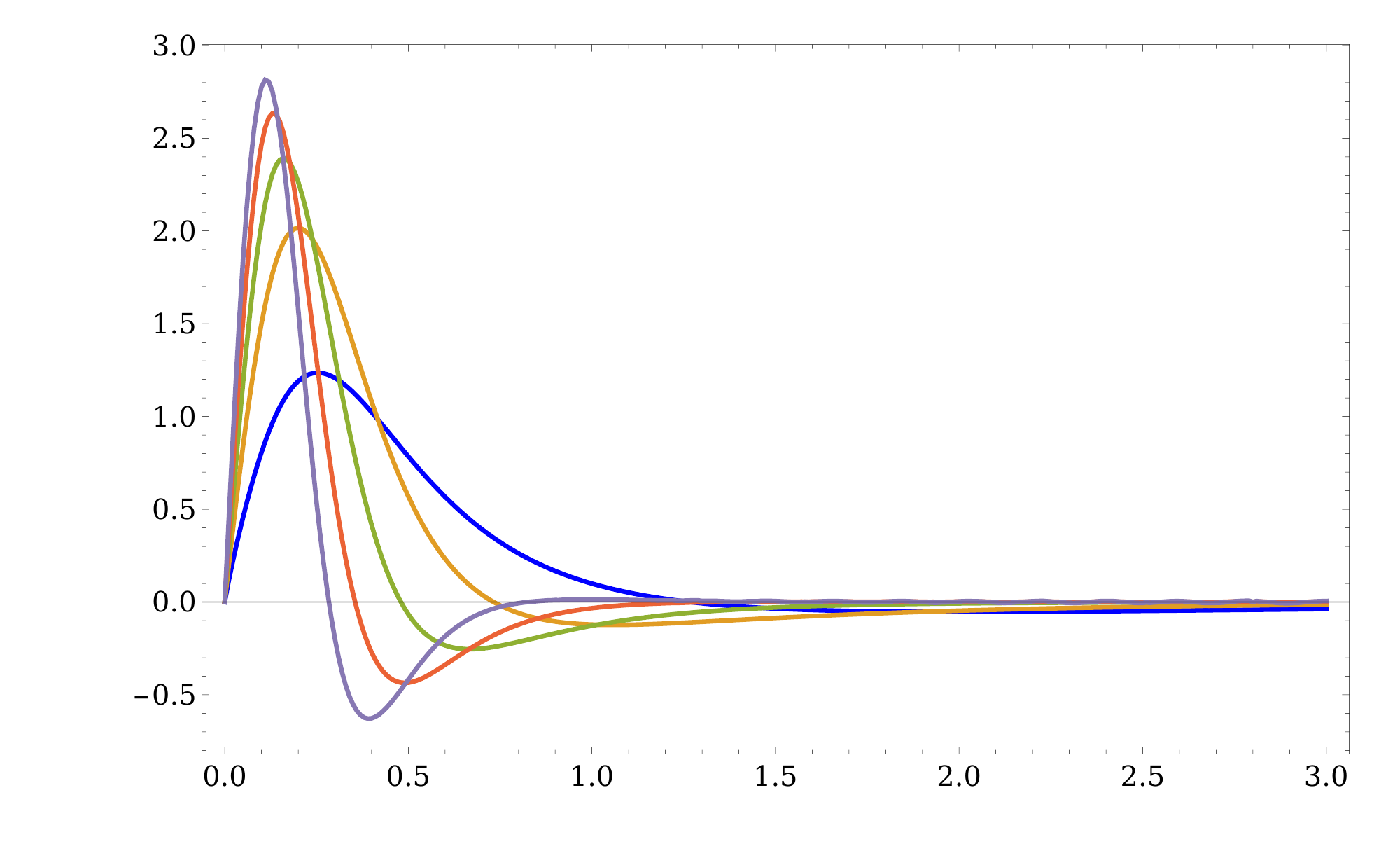}\put(-196,40){\rotatebox{-271}{\fontsize{14}{14}\selectfont $\frac{1}{T\hspace{.5mm}S_{th}}\frac{d\mathcal{C}_{1}(t)}{dt}$}}	\put(-95,-5){{\fontsize{13}{13}\selectfont$t\hspace{1mm} T$}}
		\caption{The time evolution of complexity of cTFD state in the diagonal basis for a massless complex scalar field  with $\tilde{Q}=1/10$ in different dimensions. $d=2$ (blue), $d=3$ (brown), $d=4$ (green), $d=5$ (red) and $d=6$ (purple). The reference scale is $\tilde{\gamma}=10$. \textbf{Up:} $\kappa=2$, \textbf{Down: $\kappa=1$}.}\label{diagonal1}	
	\end{figure}
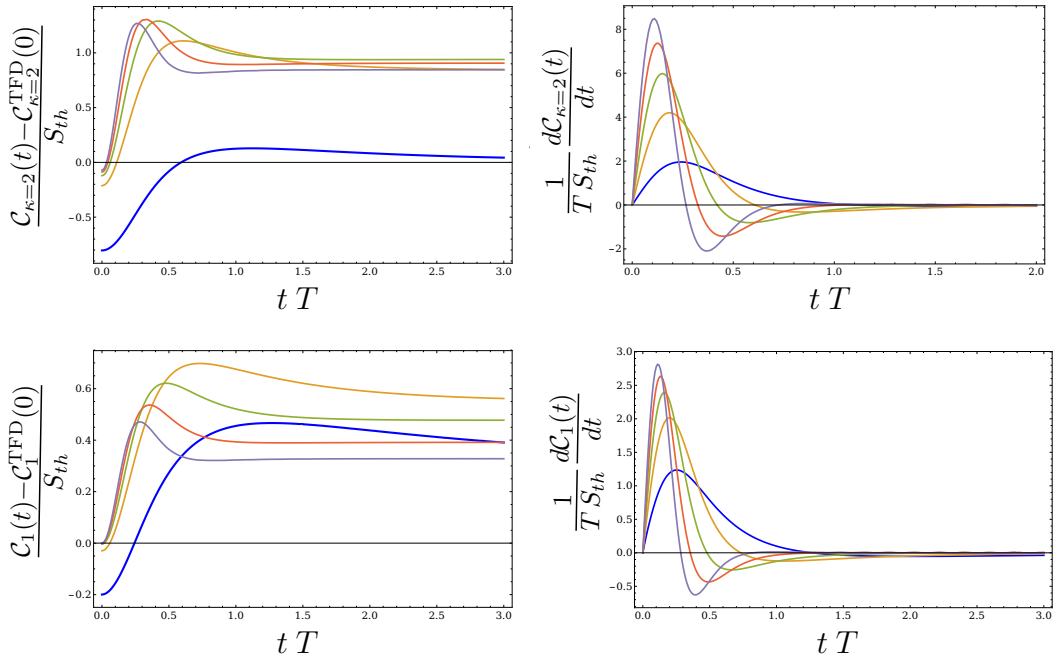

\begin{figure}[h]
		\hspace{1.5cm}\includegraphics[scale=.32]{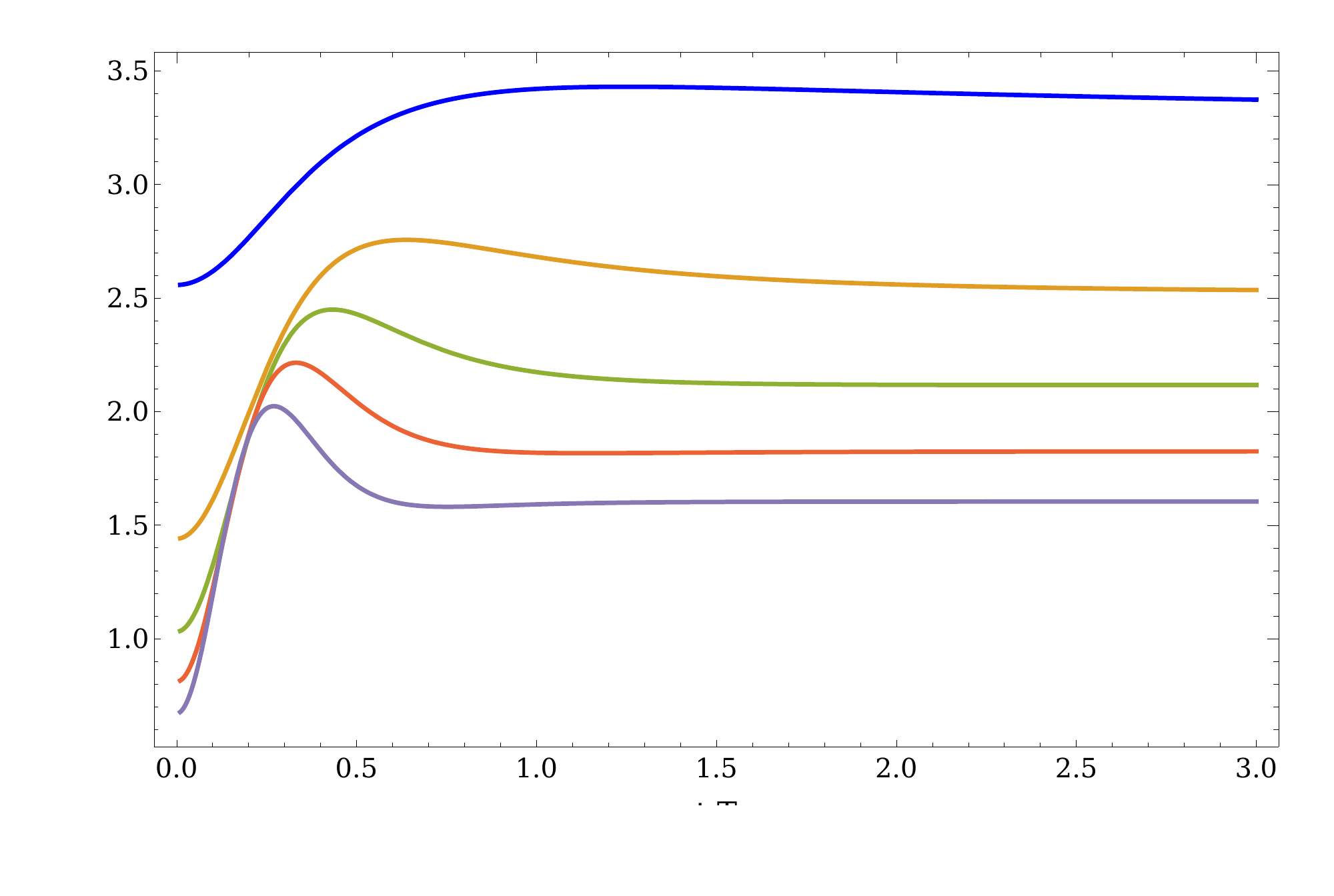}\put(-193 ,30){\rotatebox{-271}{\fontsize{13}{13}\selectfont $\frac{\mathcal{C}_{\kappa=2}(t)-\mathcal{C}_{\kappa=2}(\text{vac})}{S_{th}}$}}	\put(-95,0){{\fontsize{13}{13}\selectfont$t\hspace{1mm} T$}}\hspace{.5cm}\includegraphics[scale=.32]{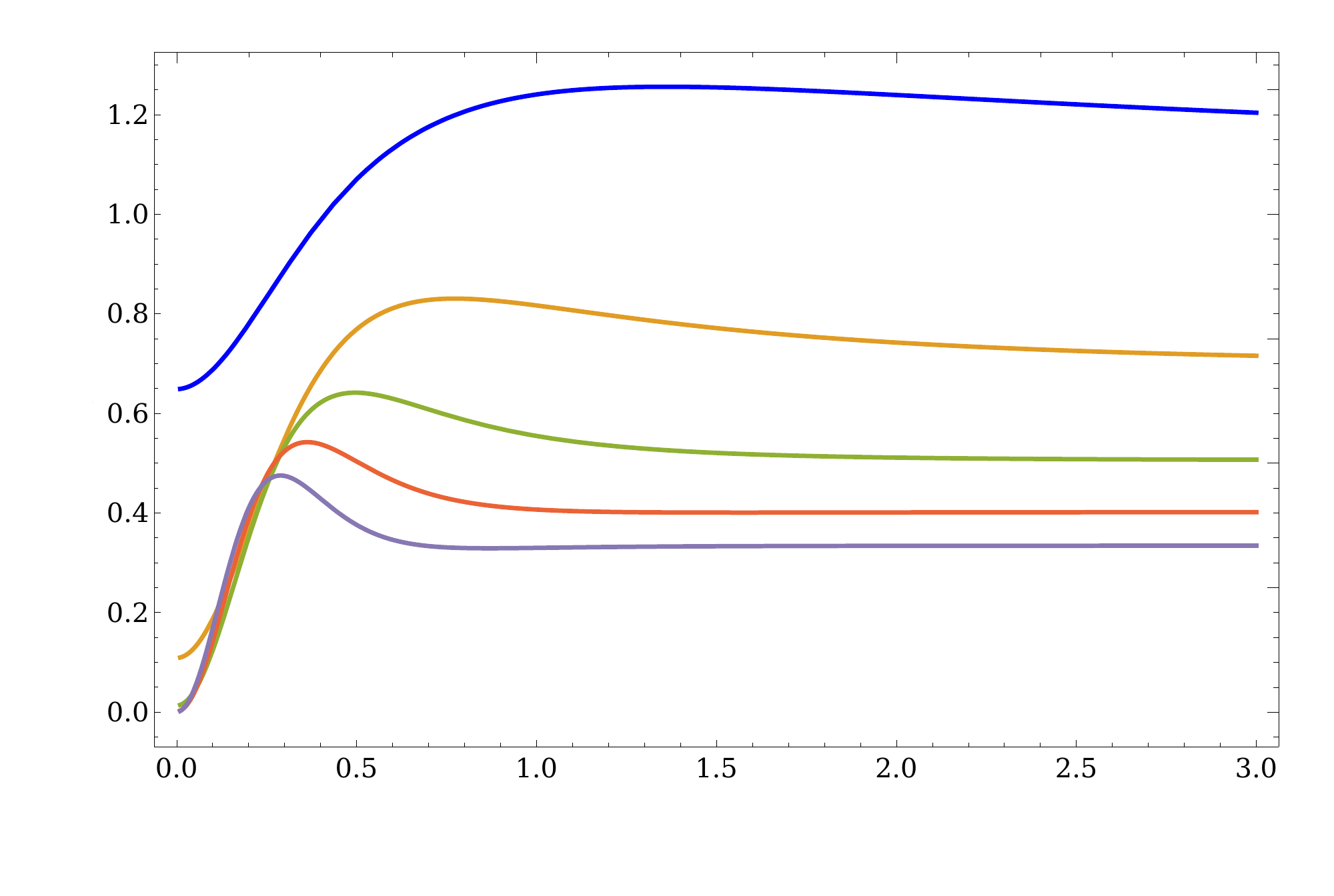}\put(-193 ,40){\rotatebox{-271}{\fontsize{13}{13}\selectfont $\frac{\mathcal{C}_{1}(t)-\mathcal{C}_{1}(\text{vac})}{S_{th}}$}}	\put(-95,0){{\fontsize{13}{13}\selectfont$t\hspace{1mm} T$}}\vspace{.5cm}
		
		\hspace{1.5cm}\includegraphics[scale=.32]{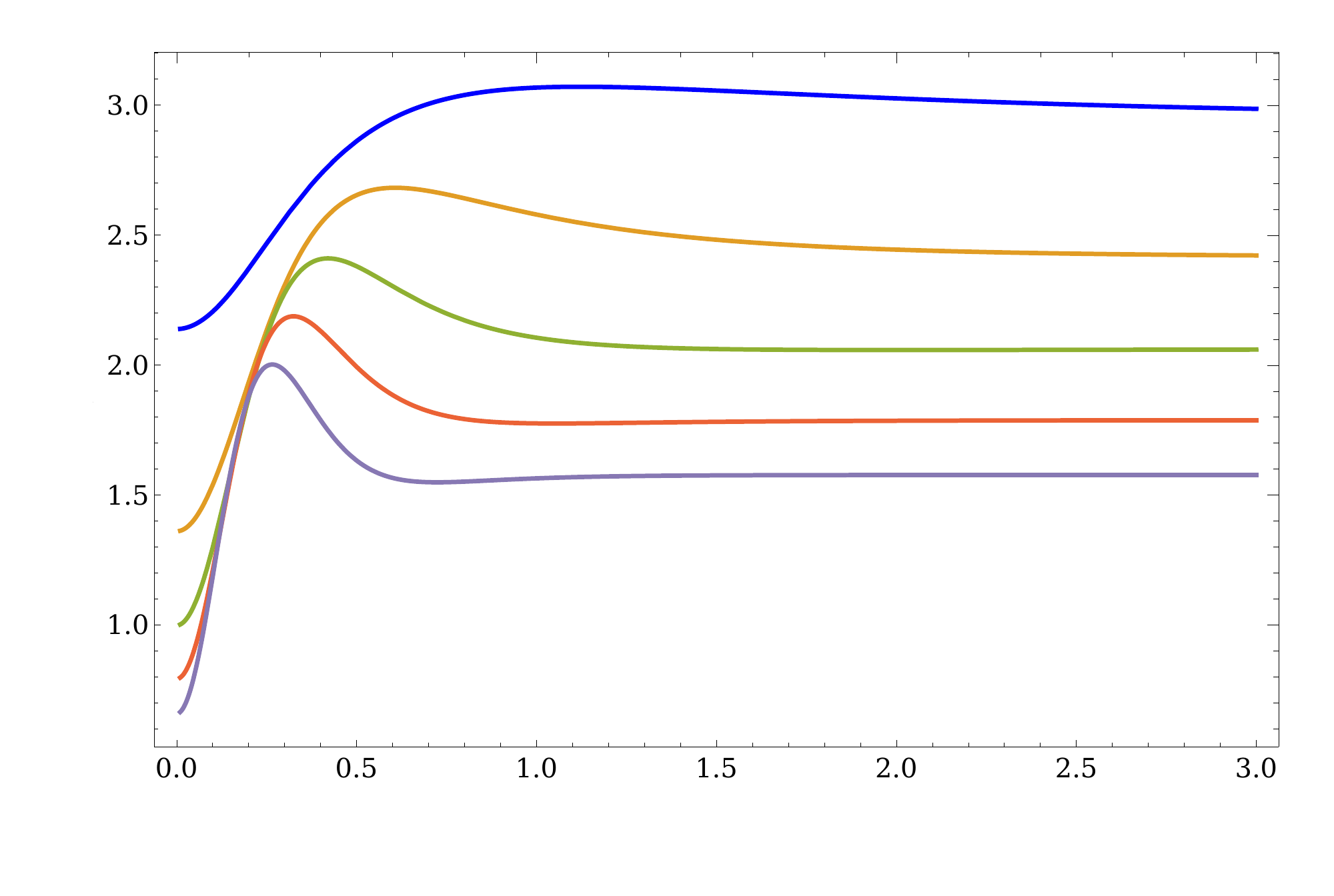}\put(-193 ,30){\rotatebox{-271}{\fontsize{13}{13}\selectfont $\frac{\mathcal{C}_{\kappa=2}(t)-\mathcal{C}_{\kappa=2}(\text{vac})}{S_{th}}$}}	\put(-95,0){{\fontsize{13}{13}\selectfont$t\hspace{1mm} T$}}\hspace{.5cm}\includegraphics[scale=.32]{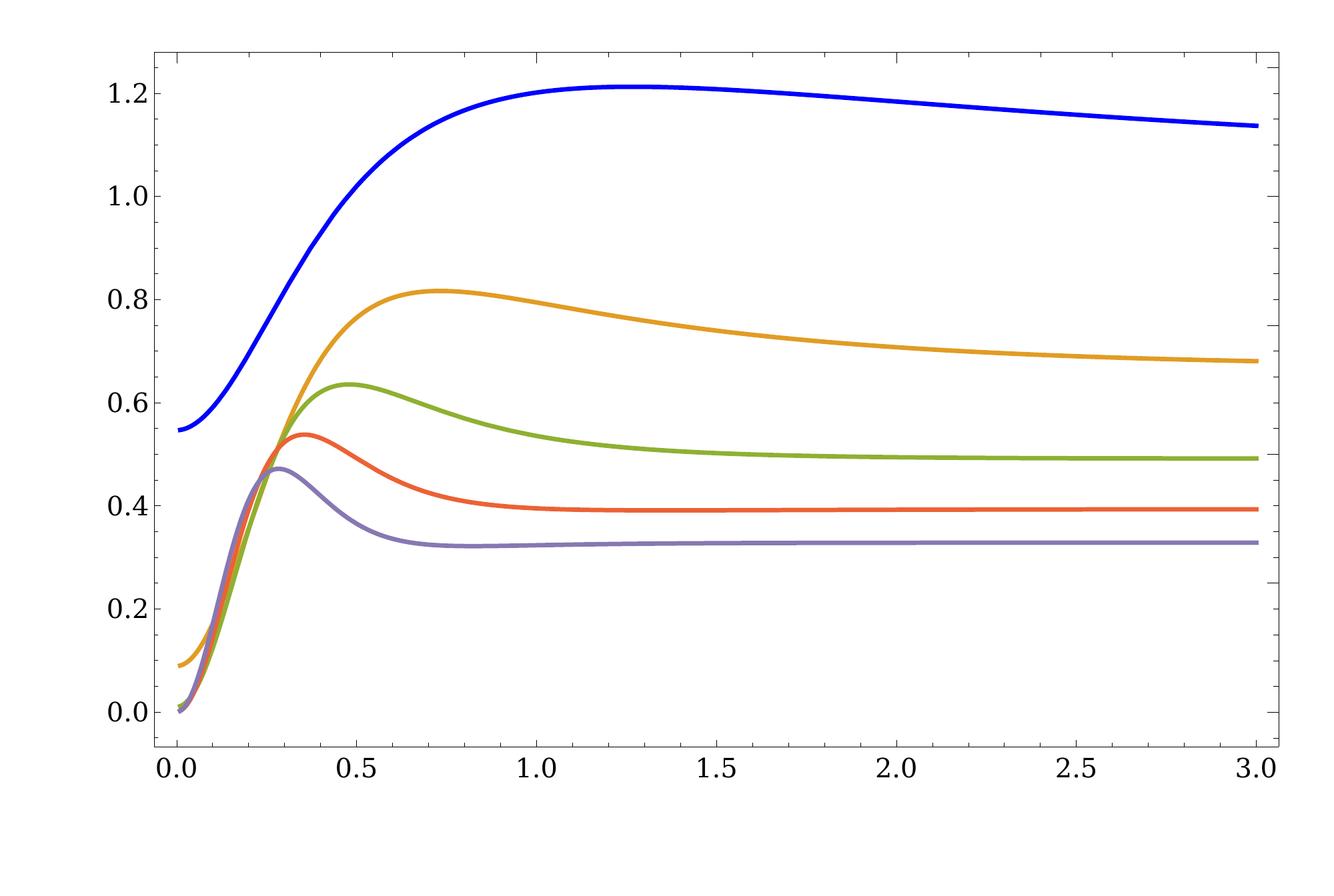}\put(-193 ,40){\rotatebox{-271}{\fontsize{13}{13}\selectfont $\frac{\mathcal{C}_{1}(t)-\mathcal{C}_{1}(\text{vac})}{S_{th}}$}}	\put(-95,0){{\fontsize{13}{13}\selectfont$t\hspace{1mm} T$}}
		\caption{\textbf{Up:} The time dependence of $\mathcal{C}_{\kappa=2}$ and $\mathcal{C}_{1}$ complexity of formation for TFD state of a massless real scalar theory with $\tilde{\gamma}=10$ in different dimensions. \textbf{Down:} The time dependence of $\mathcal{C}_{\kappa=2}$ and $\mathcal{C}_{1}$ complexity of formation for cTFD state of a massless complex scalar theory with $\tilde{Q}=1/10$ and $\tilde{\gamma}=10$ in different dimensions. $d=1+1$ (blue), $2+1$ (orange), $3+1$ (green) $4+1$ (red) and $5+1$ (purple).}\label{diagonal11}	
	\end{figure}

\begin{figure}[H]
		\hspace{1.5cm}\includegraphics[scale=.34]{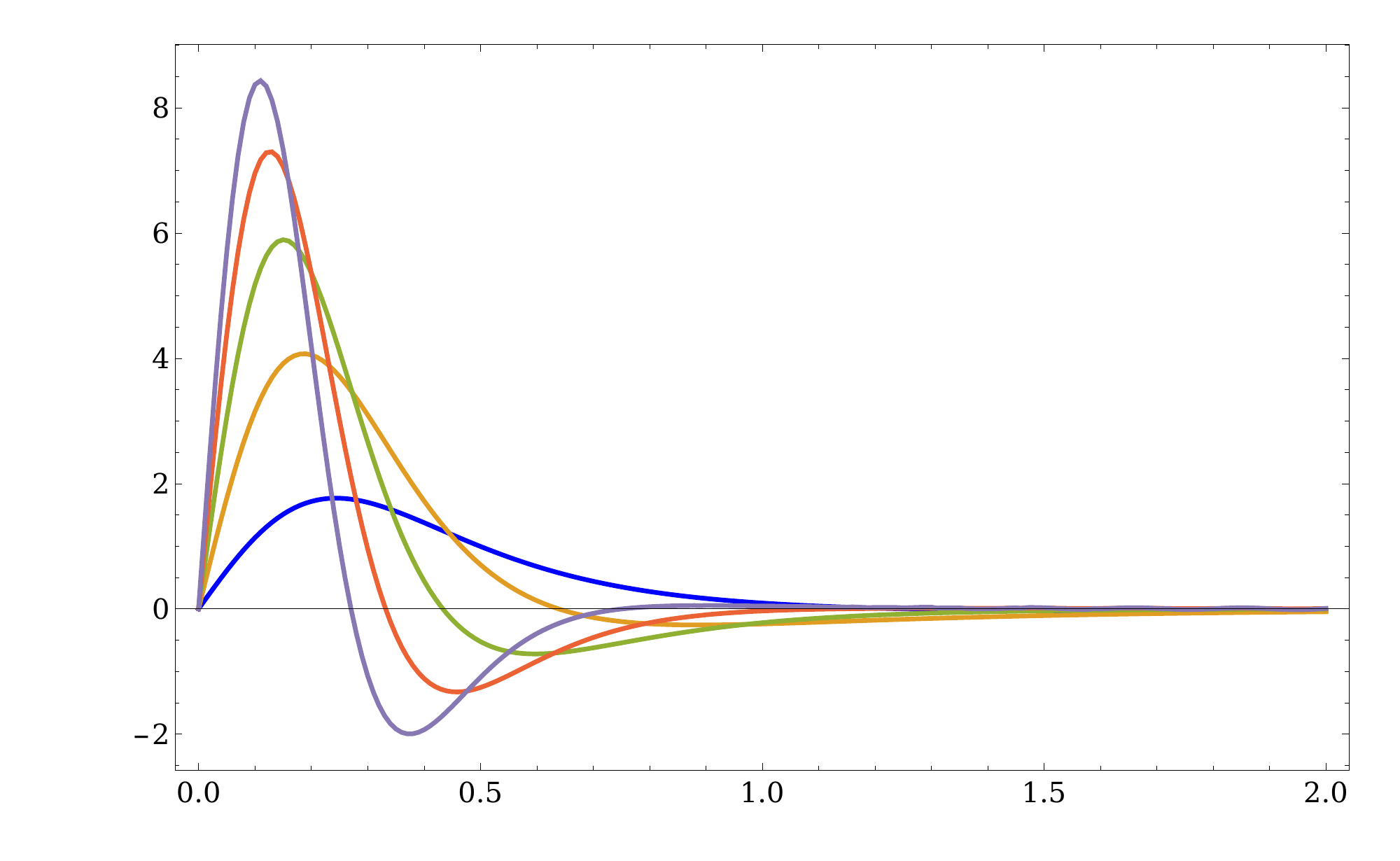}\put(-204 ,30){\rotatebox{-271}{\fontsize{15}{15}\selectfont $\frac{1}{T S_{th}}\frac{d \mathcal{C}_{\kappa=2}(t)}{dt}$}}	\put(-100,-5){{\fontsize{13}{13}\selectfont$t\hspace{1mm} T$}}\hspace{.5cm}\includegraphics[scale=.34]{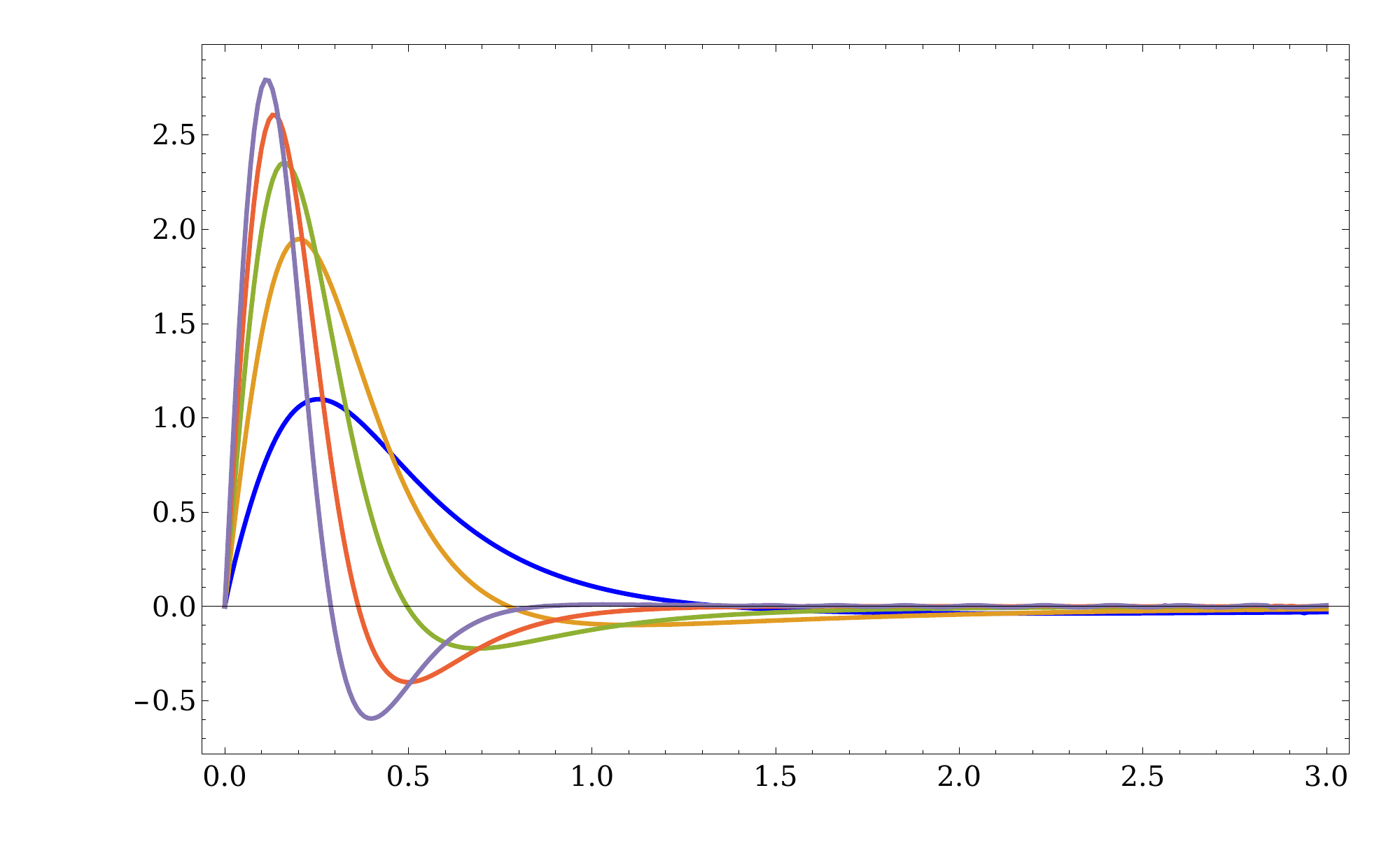}\put(-204 ,35){\rotatebox{-271}{\fontsize{15}{15}\selectfont $\frac{1}{T S_{th}}\frac{d \mathcal{C}_{1}(t)}{dt}$}}	\put(-100,-5){{\fontsize{13}{13}\selectfont$t\hspace{1mm} T$}}\vspace{.5cm}
		
		\hspace{1.2cm}
		\includegraphics[scale=.354]{contdsCk2dotTSthmt10m61qt10m1gt10}\put(-209 ,35){\rotatebox{-271}{\fontsize{15}{15}\selectfont $\frac{1}{T S_{th}}\frac{d \mathcal{C}_{\kappa=2}(t)}{dt}$}}	\put(-100,-5){{\fontsize{13}{13}\selectfont$t\hspace{1mm} T$}}\hspace{.3cm}\includegraphics[scale=.355]{contdsC1dotTSthmt10m61qt10m1gt10}\put(-208 ,44){\rotatebox{-271}{\fontsize{14}{14}\selectfont $\frac{1}{T S_{th}}\frac{d \mathcal{C}_{1}(t)}{dt}$}}	\put(-100,-5){{\fontsize{13}{13}\selectfont$t\hspace{1mm} T$}}
		\caption{\textbf{Up:} The $\mathcal{C}_{\kappa=2}$ and $\mathcal{C}_{1}$ complexity growth rate normalized by temperature times thermal entropy for TFD state of a massless real scalar theory with $\tilde{\gamma}=10$ in different dimensions. \textbf{Down:} The $\mathcal{C}_{\kappa=2}$ and $\mathcal{C}_{1}$ complexity growth rate normalized by temperature times thermal entropy for cTFD state of a massless complex scalar theory with $\tilde{Q}=1/10$, $\tilde{\gamma}=10$ in different dimensions. $d=1+1$ (blue), $2+1$ (orange), $3+1$ (green) $4+1$ (red) and $5+1$ (purple).}\label{diagonal12}	
	\end{figure}

	In this appendix we explore the time dependency of $\mathcal{C}_{\kappa=2}$ (or it's $\mathcal{C}_{2}$ counterpart) and  $\mathcal{C}_{1}$ complexities in diagonal basis. It is discussed in section.\ref{conti} that these complexities are not consistent with holographic results neither for neutral AdS black holes nor for RN-AdS ones. In fig.\ref{diagonal1}, time evolution of those complexities is presented for the massless complex scalar theory.
	Although these complexities are basis independent but they are scale dependent. In contrast with $\mathcal{C}_{1}$ complexity in LR basis, they decrease by increasing the dimension. The difference of complexity at time "t" with complexity of TFD state at zero time is presented in fig.\ref{diagonal1} and the difference with the value for vacuum state is presented in figs.\ref{diagonal11}-\ref{diagonal12}. Moreover, for both of them the complexity increases sharply at the beginning and reaches to its maximum value and then saturates very fast. This increasing in the beginning is another difference with $\mathcal{C}_{1}$ complexity in the LR basis but the fast saturation happens for both of the cases. The fast saturation is also in contrast with holographic results where one see a linear growth at late times. Since this linear growth is the effect of chaotic holographic CFTs this difference is not unexpected.
	
\begin{figure}[H]
		\hspace{1cm}\includegraphics[scale=.32]{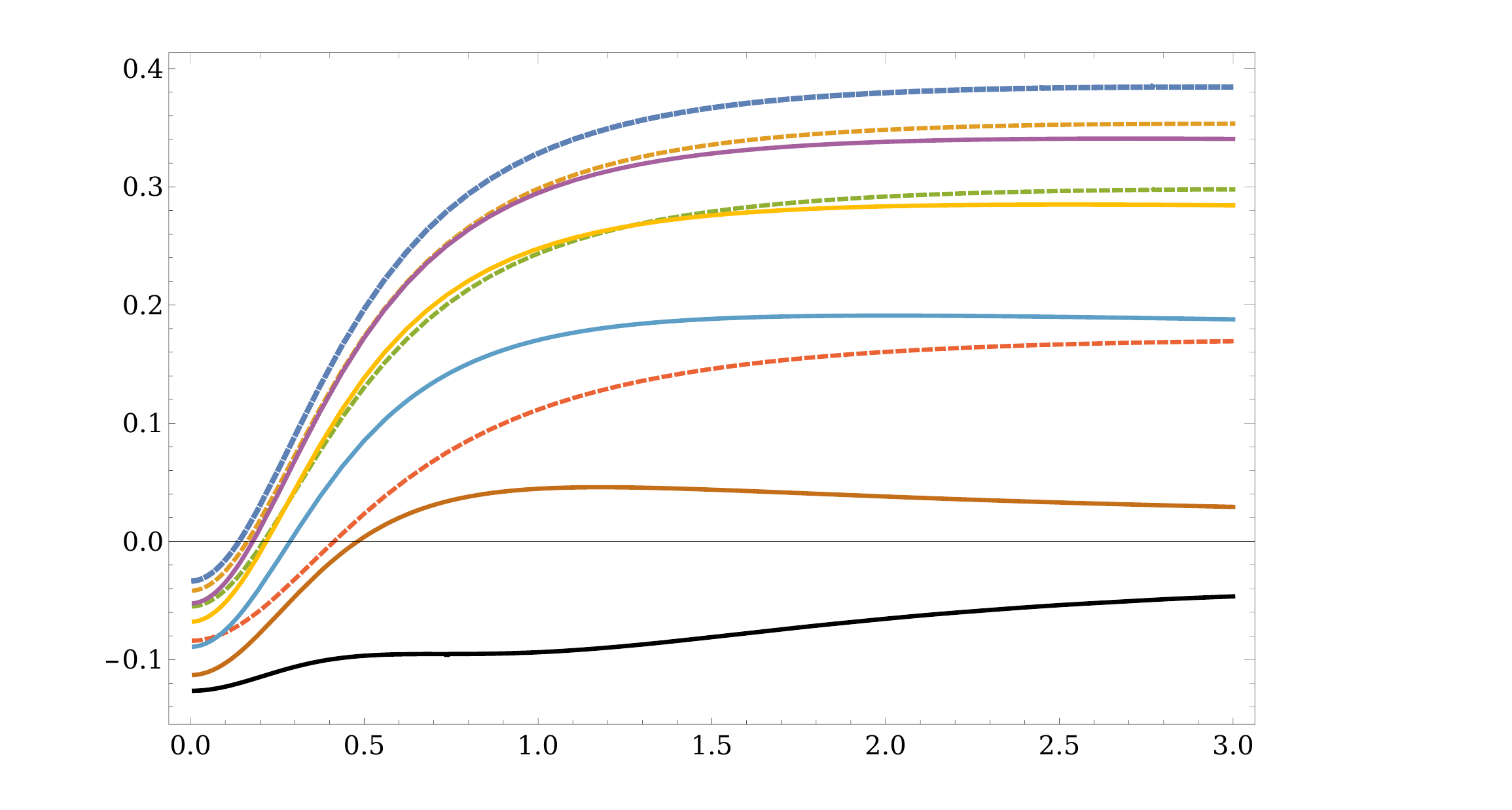}\put(-220 ,30){\rotatebox{-270}{\fontsize{14}{14}\selectfont $\frac{\mathcal{C}_{2}(t)-\mathcal{C}^{\text{TFD}}_{2}(0)}{S_{th}}$}}	\put(-120,-5){{\fontsize{13}{13}\selectfont$t\hspace{1mm} T$}}\put(-160,110){{\fontsize{10}{10}\selectfont$d=(1+1), \tilde{Q}=1/10$}}\hspace{-0.4cm}\includegraphics[scale=.32]{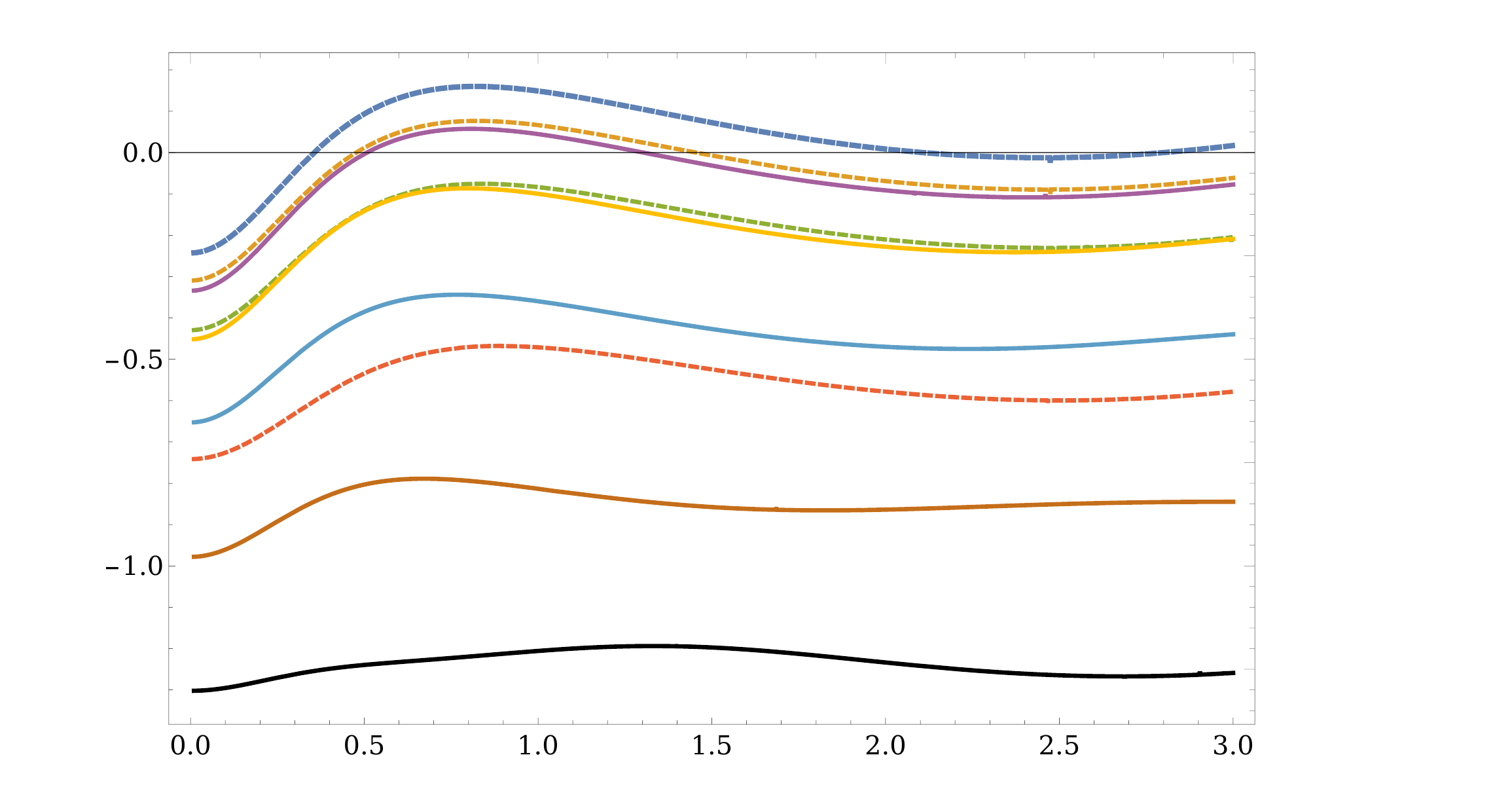}\put(-222 ,25){\rotatebox{-270}{\fontsize{14}{14}\selectfont $\frac{\mathcal{C}_{2}(t)-\mathcal{C}^{\text{TFD}}_{2}(0)}{S_{th}}$}}	\put(-120,-5){{\fontsize{13}{13}\selectfont$t\hspace{1mm} T$}}\put(-150,110){{\fontsize{10}{10}\selectfont$d=(1+1), \tilde{Q}=1$}}\put(-25,-38){\includegraphics[scale=.45]{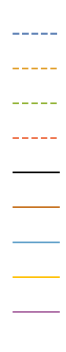}}\put(5,104){{\fontsize{10}{10}\selectfont$\tilde{\gamma}=10^{-4}$}}\put(5,88){{\fontsize{10}{10}\selectfont$\tilde{\gamma}=10^{-3}$}}\put(5,72){{\fontsize{10}{10}\selectfont$\tilde{\gamma}=10^{-2}$}}\put(5,57){{\fontsize{10}{10}\selectfont$\tilde{\gamma}=10^{-1}$}}\put(5,42){{\fontsize{10}{10}\selectfont$\tilde{\gamma}=1$}}\put(5,25){{\fontsize{10}{10}\selectfont$\tilde{\gamma}=10$}}\put(5,10){{\fontsize{10}{10}\selectfont$\tilde{\gamma}=10^{2}$}}\put(5,-6){{\fontsize{10}{10}\selectfont$\tilde{\gamma}=10^{3}$}}\put(5,-22){{\fontsize{10}{10}\selectfont$\tilde{\gamma}=10^{4}$}}\vspace{-.3cm}
		
		\hspace{1cm}\includegraphics[scale=.32]{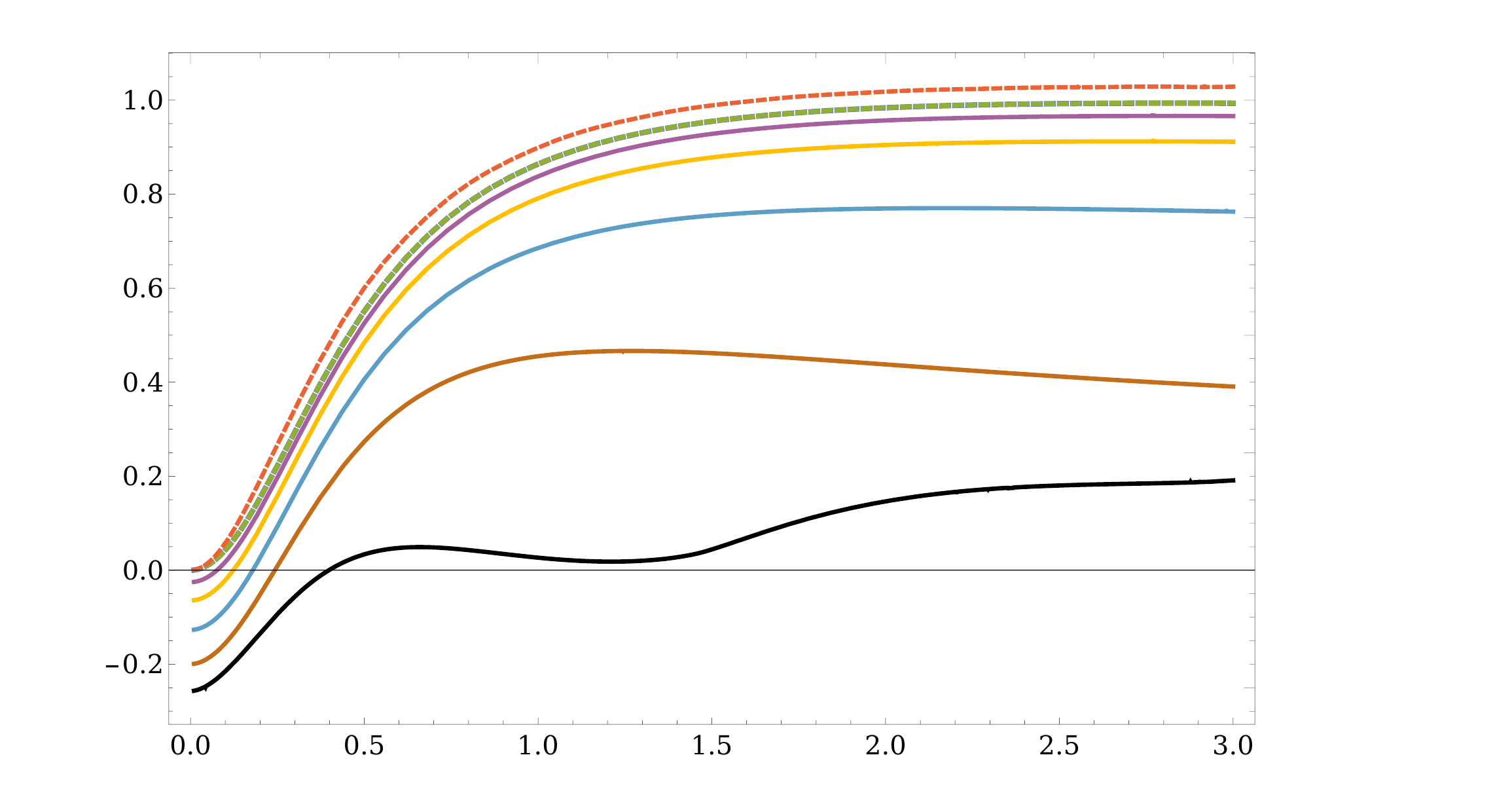}\put(-222 ,35){\rotatebox{-271}{\fontsize{14}{14}\selectfont $\frac{\mathcal{C}_{1}(t)-\mathcal{C}^{\text{TFD}}_{1}(0)}{S_{th}}$}}	\put(-120,-5){{\fontsize{13}{13}\selectfont$t\hspace{1mm} T$}}\put(-160,110){{\fontsize{10}{10}\selectfont$d=(1+1), \tilde{Q}=1/10$}}\hspace{-.4cm}\includegraphics[scale=.32]{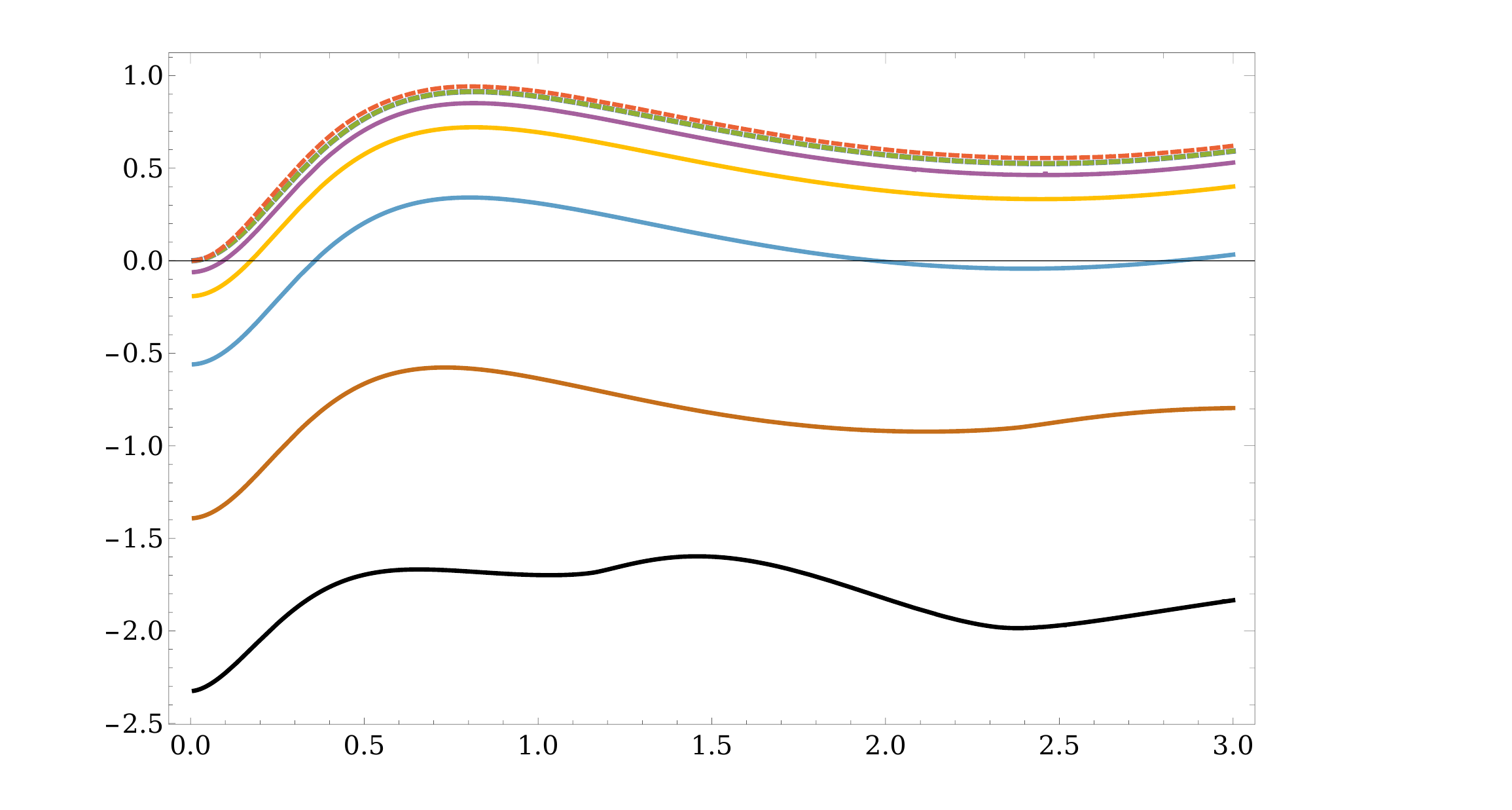}\put(-222 ,30){\rotatebox{-271}{\fontsize{14}{14}\selectfont $\frac{\mathcal{C}_{1}(t)-\mathcal{C}^{\text{TFD}}_{1}(0)}{S_{th}}$}}	\put(-125,-5){{\fontsize{13}{13}\selectfont$t\hspace{1mm} T$}}\put(-160,110){{\fontsize{10}{10}\selectfont$d=(1+1), \tilde{Q}=1$}}
		\caption{The time evolution of 
			complexity with varying reference scale for massless complex scalar in $d =1+1$. $\tilde{\gamma}=1$(solid black), $\tilde{\gamma} <1$ (dashed curves) and $\tilde{\gamma}>1$ (colored solid curves).
			For different values of $\tilde{\gamma}$ the complexity saturates to different constants at late times. We note that for large and small $\tilde{\gamma}$ we have a large time derivative during the transient period at early times. Increasing the charge actually smooth this transition. \textbf{Left:} $\tilde{Q}= 1/10$, \textbf{Right:} $\tilde{Q}= 1$, \textbf{Up:} $\mathcal{C}_2$ complexity and \textbf{Down:} $\mathcal{C}_1$ complexity.}\label{diagonal2}	
	\end{figure}
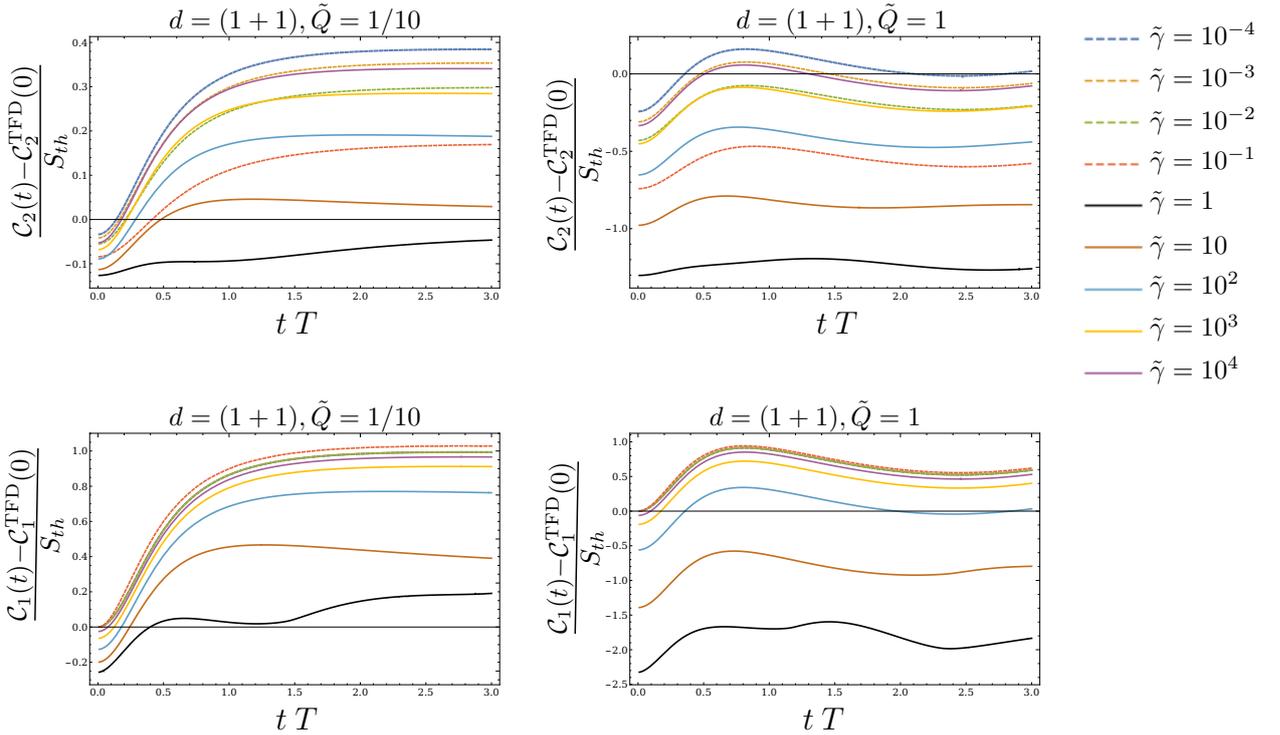

\begin{figure}[h]
		\hspace{1.5cm}\includegraphics[scale=.32]{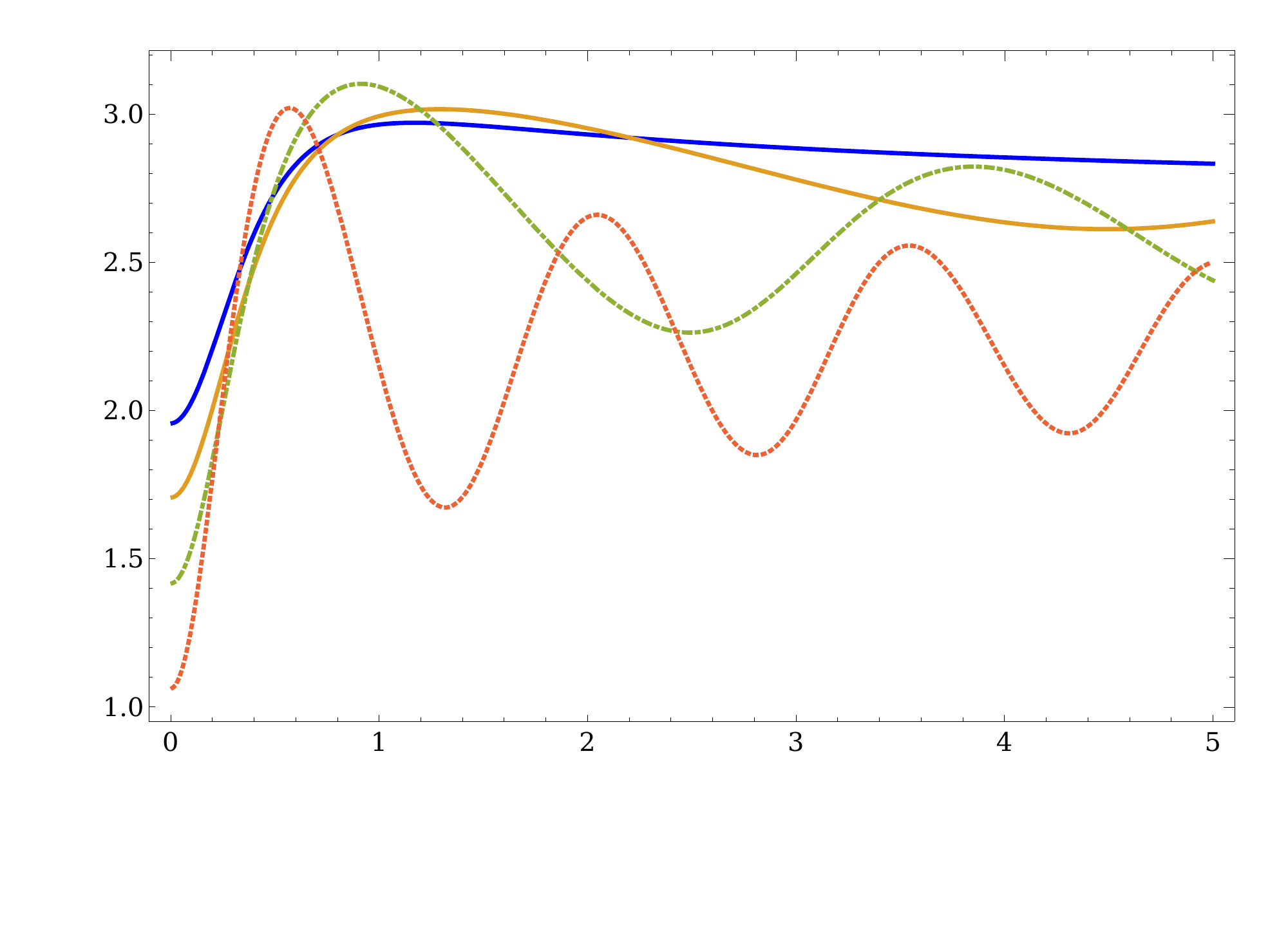}\put(-195 ,40){\rotatebox{-271}{\fontsize{14}{14}\selectfont $\frac{\mathcal{C}_{\kappa=2}(t)-\mathcal{C}^{\text{TFD}}_{\kappa=2}(0)}{S_{th}}$}}	\put(-95,10){{\fontsize{13}{13}\selectfont$t\hspace{1mm} T$}}\hspace{.6cm}\includegraphics[scale=.312]{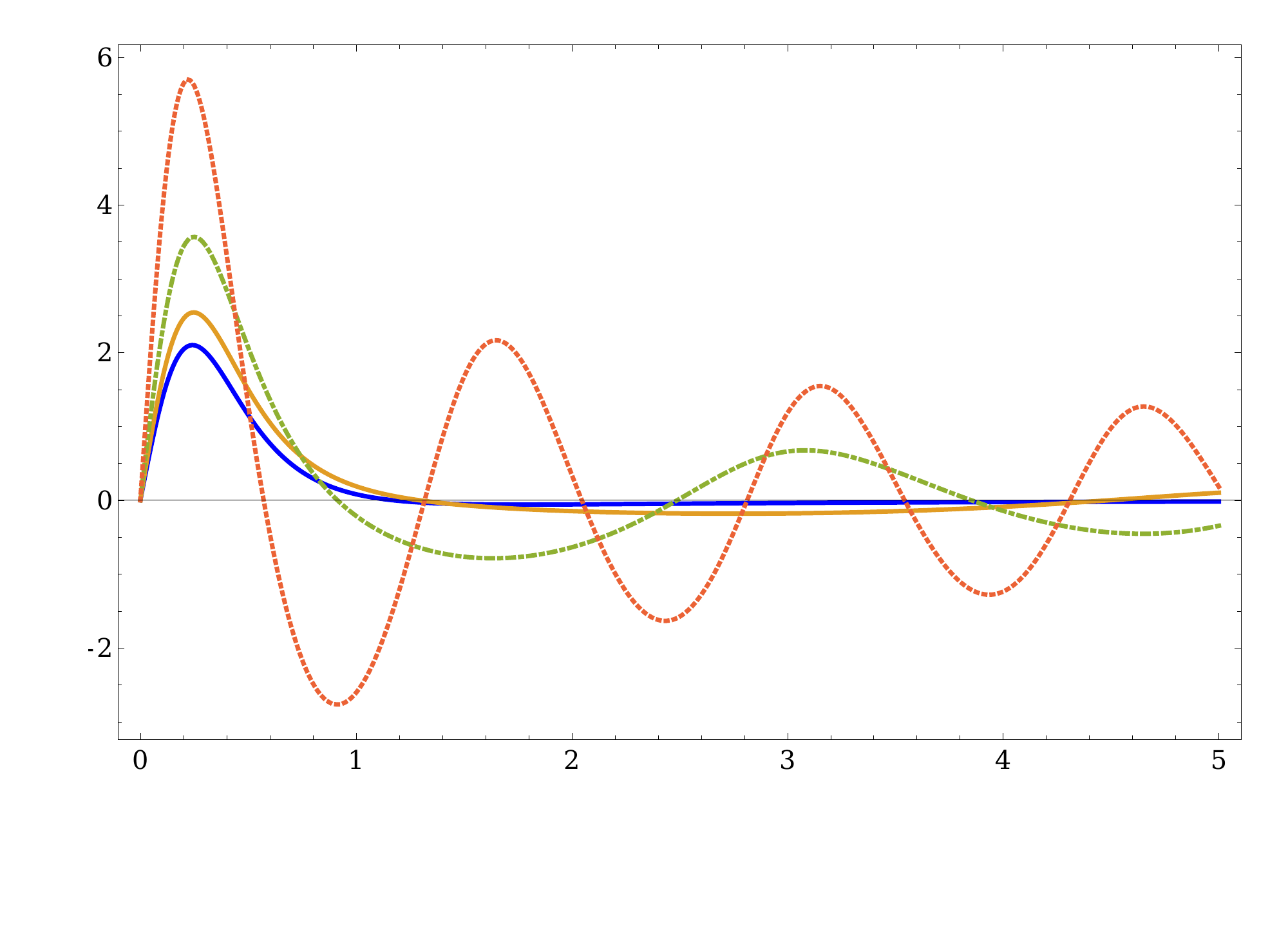}\put(-193 ,47){\rotatebox{-271}{\fontsize{14}{14}\selectfont $\frac{1}{T S_{th}}\frac{d \mathcal{C}_{\kappa=2}(t)}{dt}$}}	\put(-95,10){{\fontsize{13}{13}\selectfont$t\hspace{1mm} T$}}
		\vspace{-.2cm}
		
		\hspace{1.5cm}\includegraphics[scale=.32]{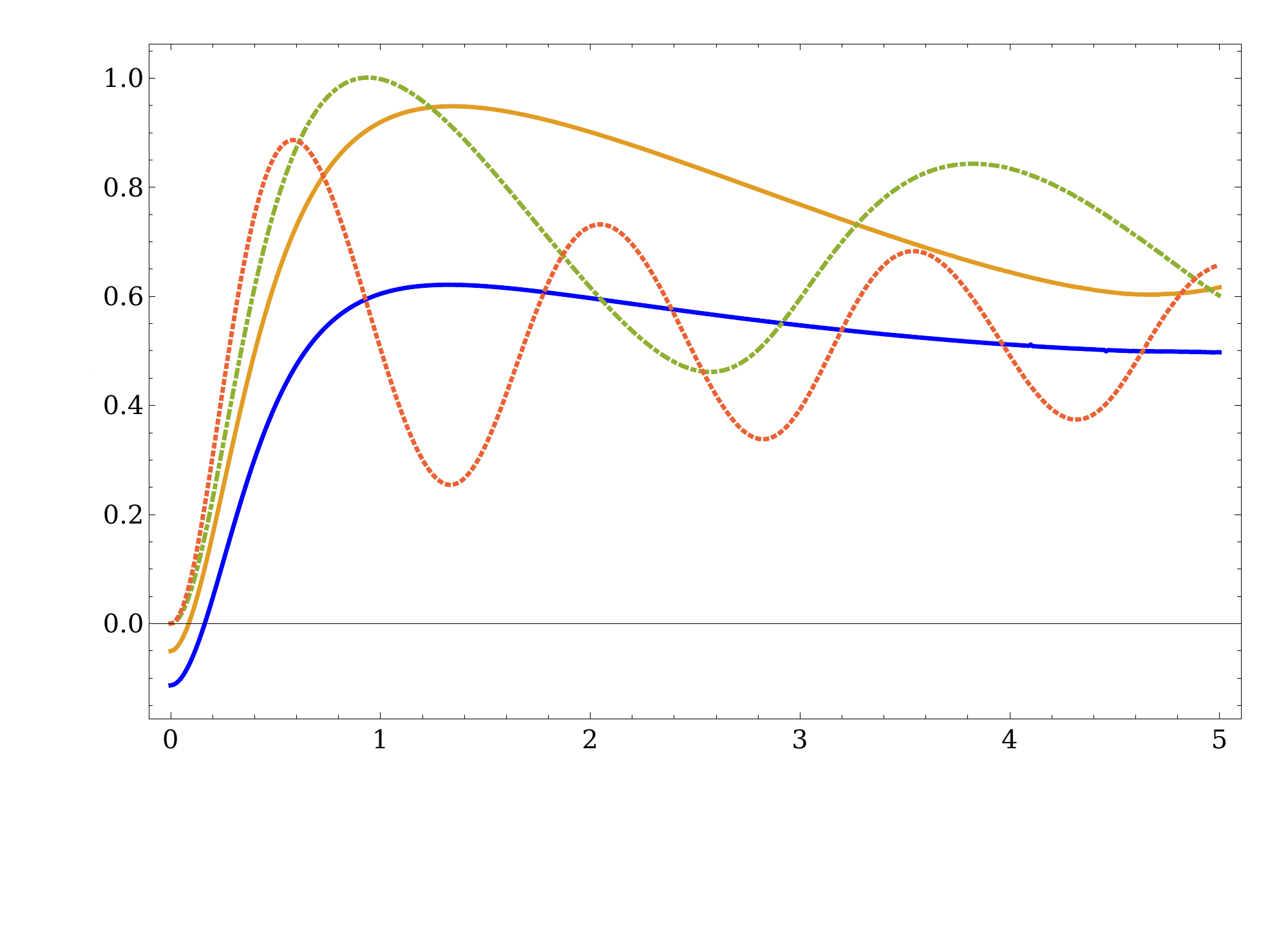}\put(-195 ,45){\rotatebox{-271}{\fontsize{14}{14}\selectfont $\frac{\mathcal{C}_{1}(t)-\mathcal{C}^{\text{TFD}}_{1}(0)}{S_{th}}$}}	\put(-95,10){{\fontsize{13}{13}\selectfont$t\hspace{1mm} T$}}\hspace{.5cm}\includegraphics[scale=.322]{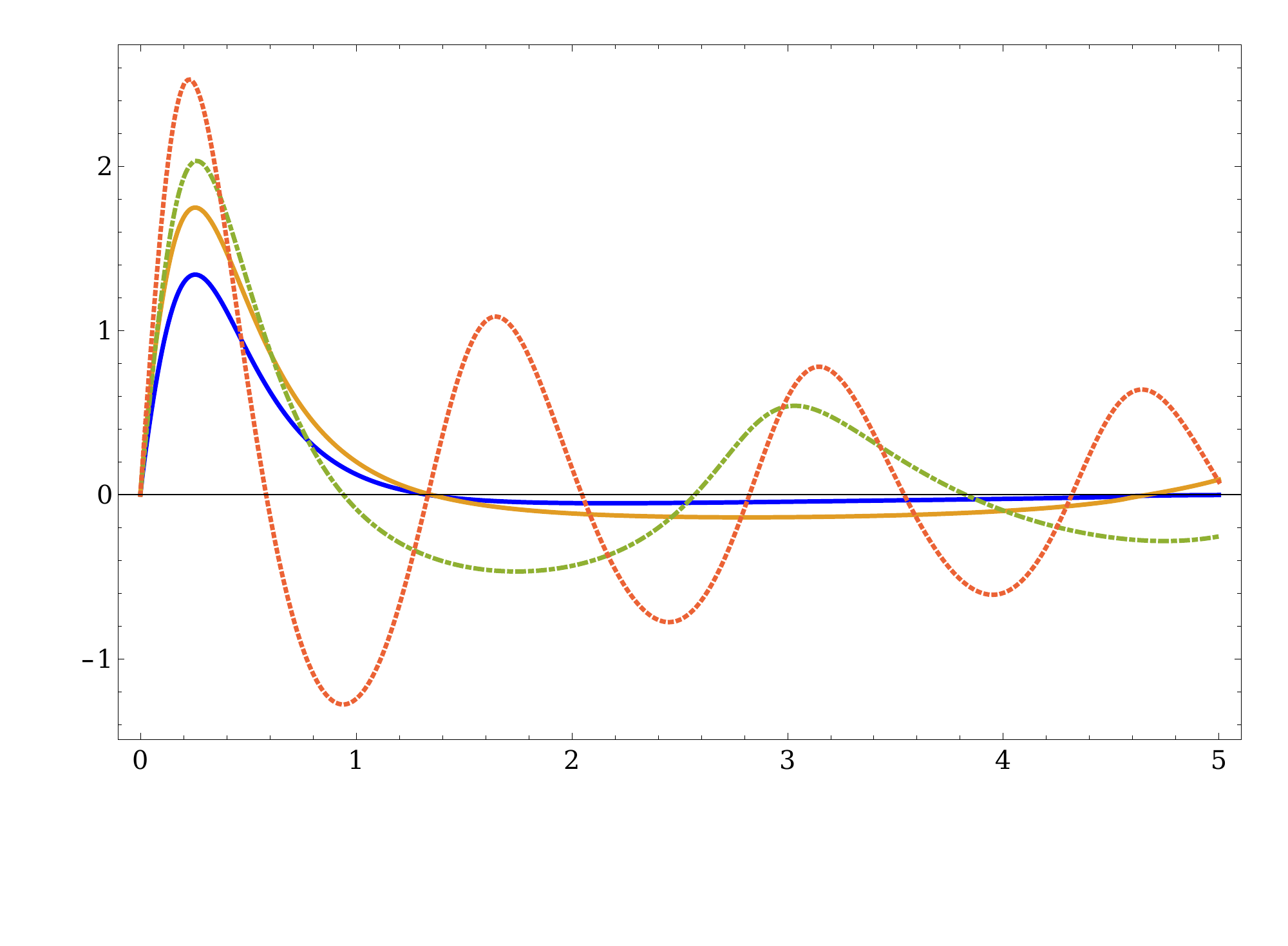}\put(-196 ,55){\rotatebox{-271}{\fontsize{14}{14}\selectfont $\frac{1}{T S_{th}}\frac{d \mathcal{C}_{1}(t)}{dt}$}}	\put(-95,10){{\fontsize{13}{13}\selectfont$t\hspace{1mm} T$}}
		\vspace{-.4cm}
		
		\caption{The time dependence of $\mathcal{C}_{\kappa=2}$ \textbf{(up)} and $\mathcal{C}_{1}$ \textbf{(down)} complexities for cTFD state of a massive complex scalar theory in $d=1+1$ dimensions with $\tilde{Q}=1/10$, $\tilde{\gamma}=10$ and different masses. $m = 0.2 T$ (blue), $m = 0.5 T$ (orange), $m = 1 T$ (dotted dashed green) and $m=2T$ (dotted red). For large masses with respect to the thermal scale, there is an oscillatory behavior with period $\Delta t = \pi/m$. At late times, we observe a saturation to a constant value.}\label{diagonal3}	
	\end{figure}

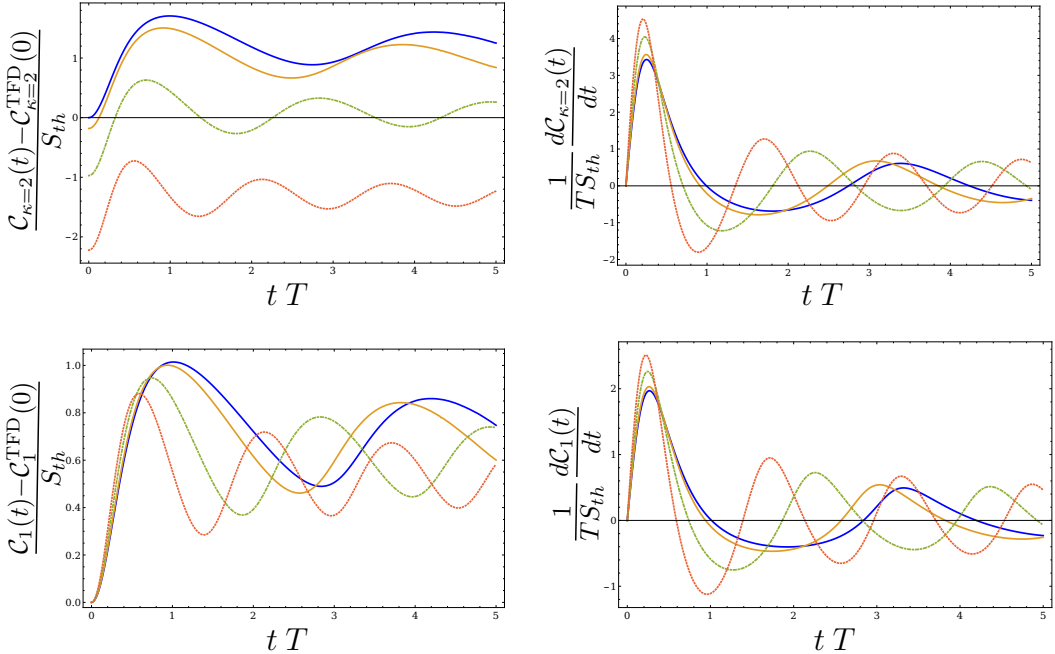
\begin{figure}[H]
		\hspace{1.5cm}\includegraphics[scale=.32]{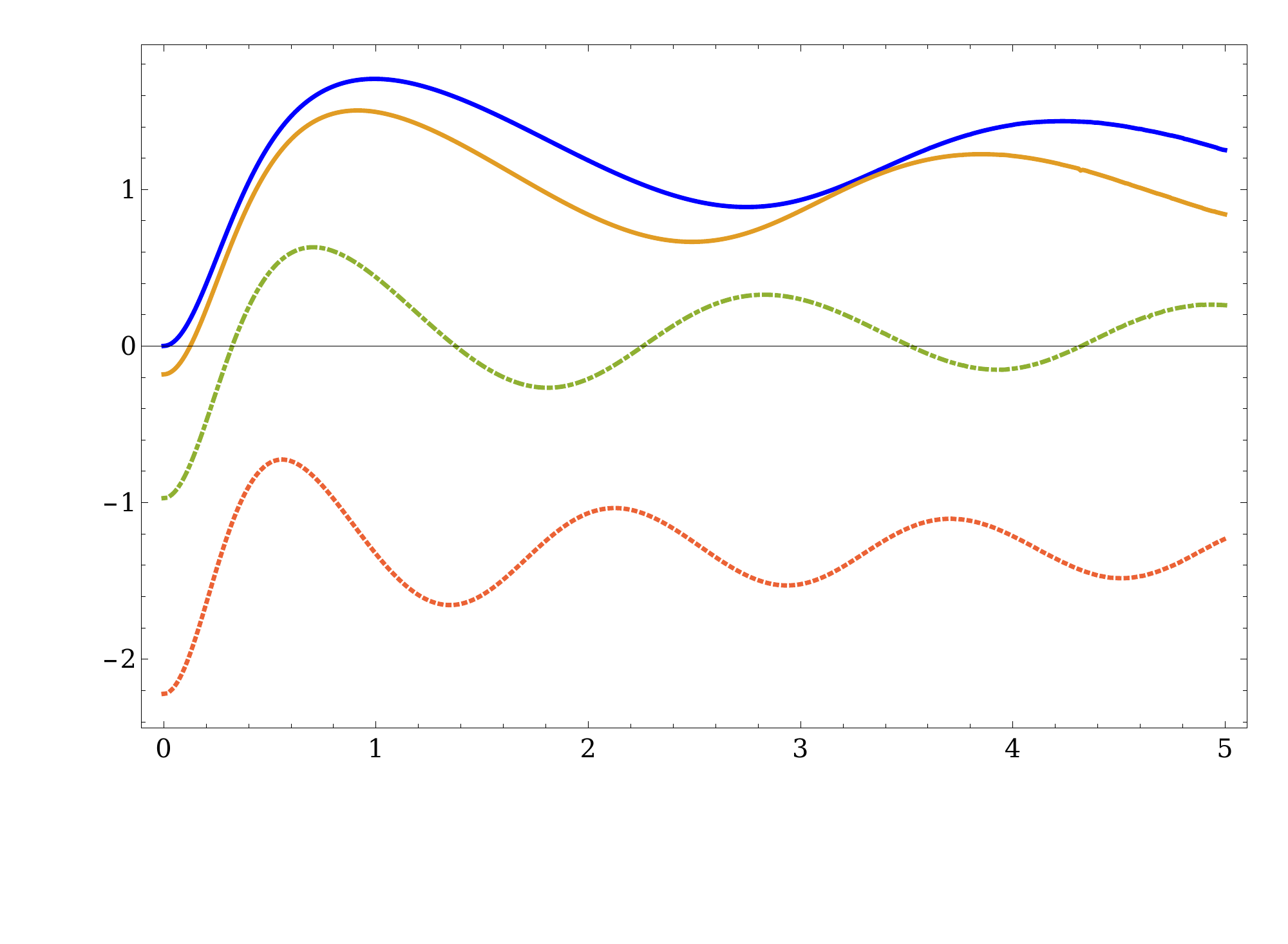}\put(-193 ,42){\rotatebox{-271}{\fontsize{14}{14}\selectfont $\frac{\mathcal{C}_{\kappa=2}(t)-\mathcal{C}^{\text{TFD}}_{\kappa=2}(0)}{S_{th}}$}}	\put(-95,15){{\fontsize{13}{13}\selectfont$t\hspace{1mm} T$}}\hspace{.7cm}\includegraphics[scale=.312]{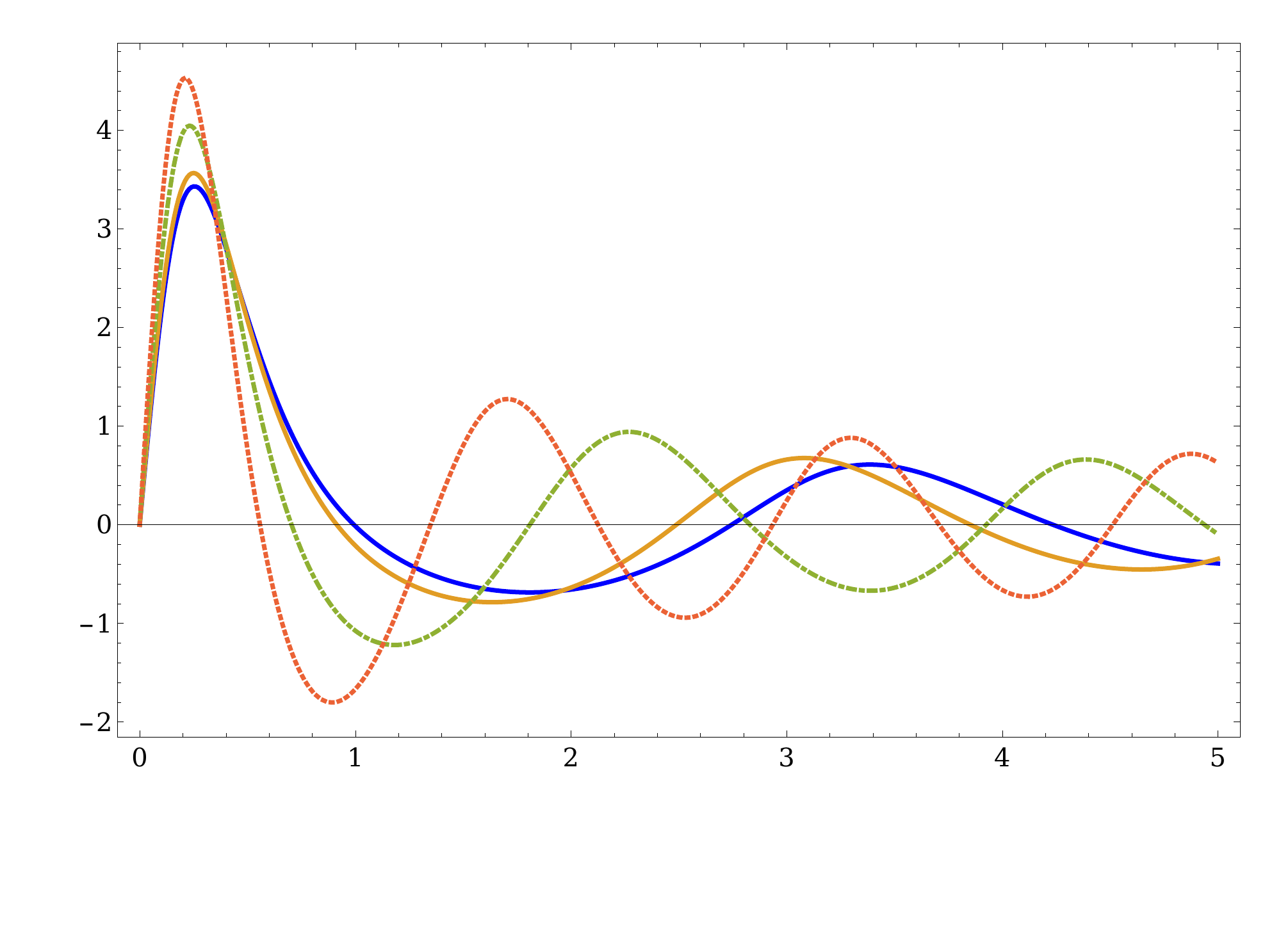}\put(-192 ,50){\rotatebox{-271}{\fontsize{14}{14}\selectfont $\frac{1}{T S_{th}}\frac{d \mathcal{C}_{\kappa=2}(t)}{dt}$}}	\put(-95,15){{\fontsize{13}{13}\selectfont$t\hspace{1mm} T$}}\vspace{-.3cm}
		
		\hspace{1.5cm}\includegraphics[scale=.32]{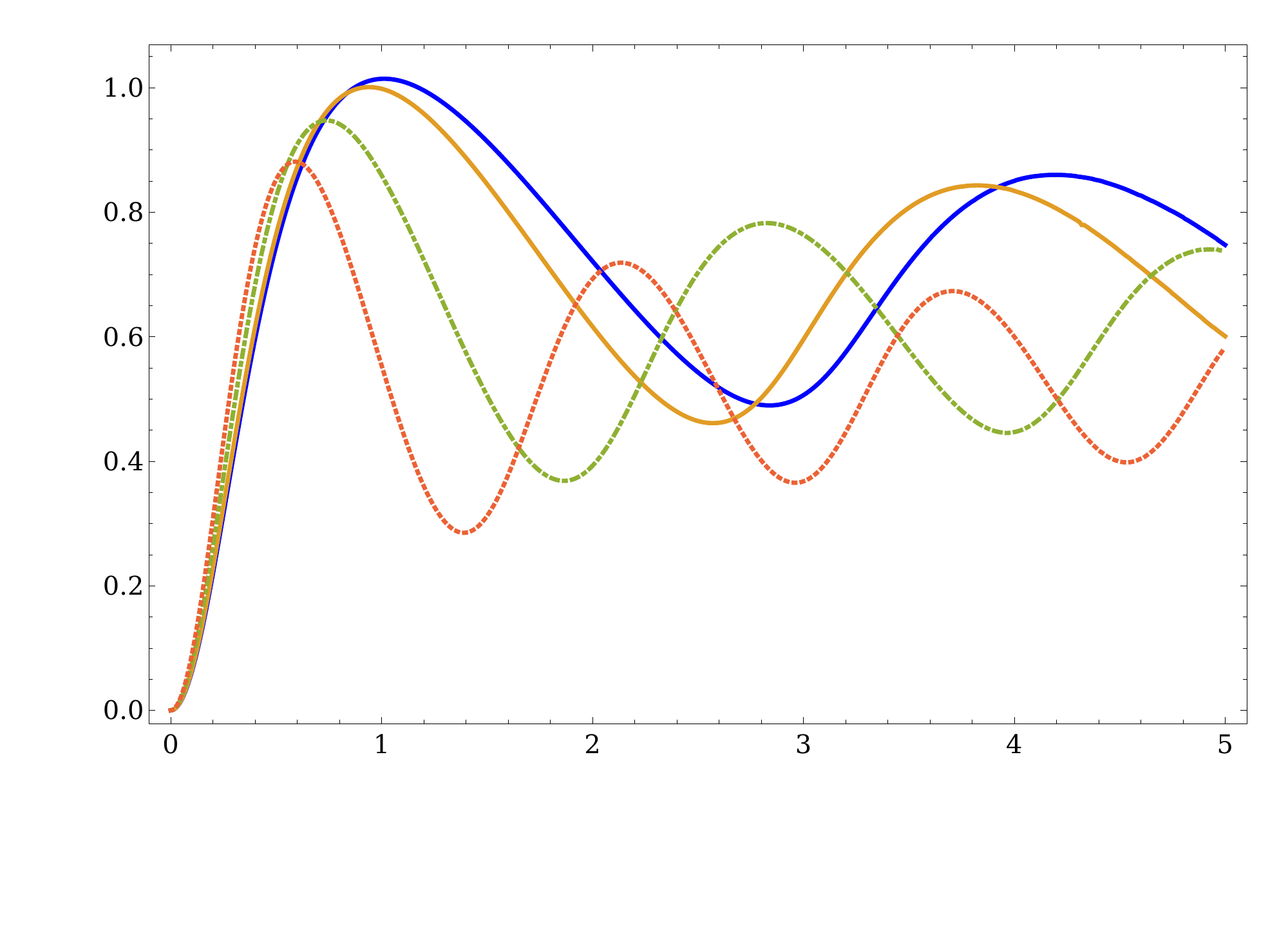}\put(-193 ,50){\rotatebox{-271}{\fontsize{14}{14}\selectfont $\frac{\mathcal{C}_{1}(t)-\mathcal{C}^{\text{TFD}}_{1}(0)}{S_{th}}$}}	\put(-95,15){{\fontsize{13}{13}\selectfont$t\hspace{1mm} T$}}\hspace{.7cm}\includegraphics[scale=.32]{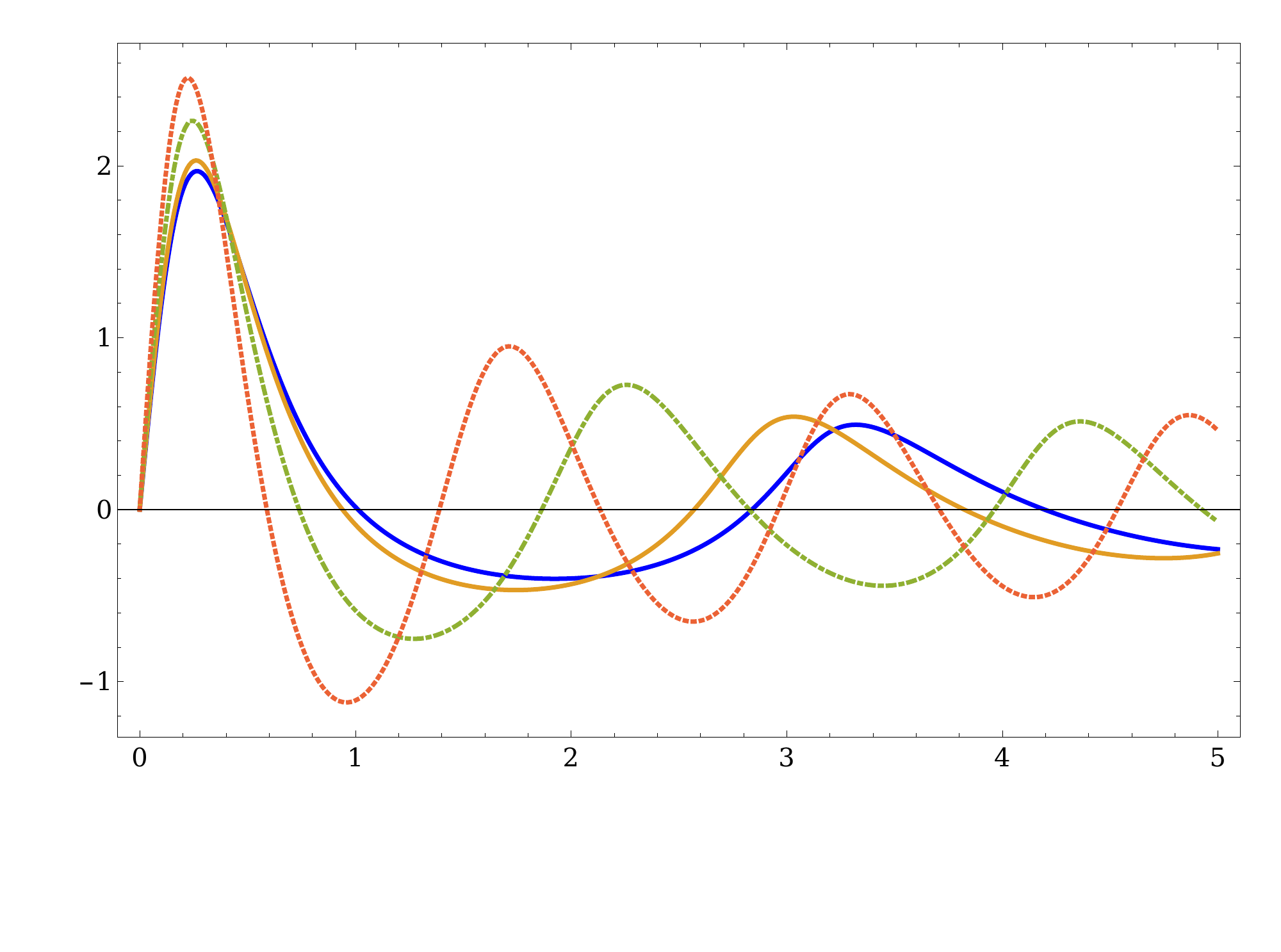}\put(-195 ,55){\rotatebox{-271}{\fontsize{14}{14}\selectfont $\frac{1}{T S_{th}}\frac{d \mathcal{C}_{1}(t)}{dt}$}}	\put(-95,15){{\fontsize{13}{13}\selectfont$t\hspace{1mm} T$}}
		\vspace{-.6cm}
		
		\caption{The time dependence of $\mathcal{C}_{\kappa=2}$ \textbf{(up)} and $\mathcal{C}_{1}$ \textbf{(down)} complexities for cTFD state of a massive complex scalar theory in $d=1+1$ dimensions with $\tilde{m}=1$, $\tilde{\gamma}=10$ and different charges. $\tilde{Q} = 10^{-6}$ (blue), $\tilde{Q} = 10^{-1}$ (orange), $\tilde{Q} = 1/2 $ (dotted dashed green) and $\tilde{Q}=1$ (dotted red). For large charges with respect to the thermal scale, there is an oscillatory behavior with period $\Delta t = \pi\beta/\tilde{Q}$. At late times, we observe a saturation to a constant value.}\label{diagonal4}	
	\end{figure}

	From fig.\ref{diagonal12}, it is clear that the complexity growth rate for both $\mathcal{C}_{\kappa=2}$ and $\mathcal{C}_{1}$ measures  is independent from the chemical potential which it is another difference with holographic proposals. Beside previously indicated reasons, this observation also implies that these complexities are not well suited to compare with holography. In fig.\ref{diagonal2}, we study the effect of changing reference scale and moreover we see that sharply increasing value of complexity can become smooth by increasing value of $\tilde{Q}$. These complexities also saturate to different values by changing the reference scale and moreover for large and small $\tilde{\gamma}$ we have a large time derivative during the transient period at early times.  This dependency to reference scale in early times is similar to the $\mathcal{C}_{1}$ complexity in LR basis and moreover is the same as holography and ambiguity in the normalization of null surfaces in CA conjecture.
	To complete the study, we do the same analysis for massive theory and results are presented in figs.\ref{diagonal3} -\ref{diagonal4}. Again we see a sharp increasing in the value of complexity but in contrast to massless case, for large masses (or large values of charges) with respect to the thermal scale there is an oscillatory behavior with period $\Delta t = \pi/m$ (or $\Delta t=\pi/\mu q$). The complexity becomes smaller by increasing the mass or chemical potential and moreover at late times, we observe a saturation for complexity to a constant value. Albeit in all above cases we have chosen the UV cut-off as infinity, but  complexity of formation is finite. This indicates that the UV divergences in the complexity of the time-dependent cTFD
	state is exactly those of (two copies of) the vacuum not only in massless theory but also in massive one. Moreover, the UV divergences for the time-dependent cTFD state are the same with the ones for TFD state. This means that in presence of the U(1) global charge no new UV divergence appears which is similar to holography. In holographic dual, we do not need to add any new gravitational counterterm in presence of $U(1)$ bulk field. The reason for the absence of new counterterms comes from the special fall of $U(1)$ gauge field near the boundary of spacetime. 
	\vspace{-.5cm}
	
\begin{figure}[H]
		\hspace{.9cm}\includegraphics[scale=.34]{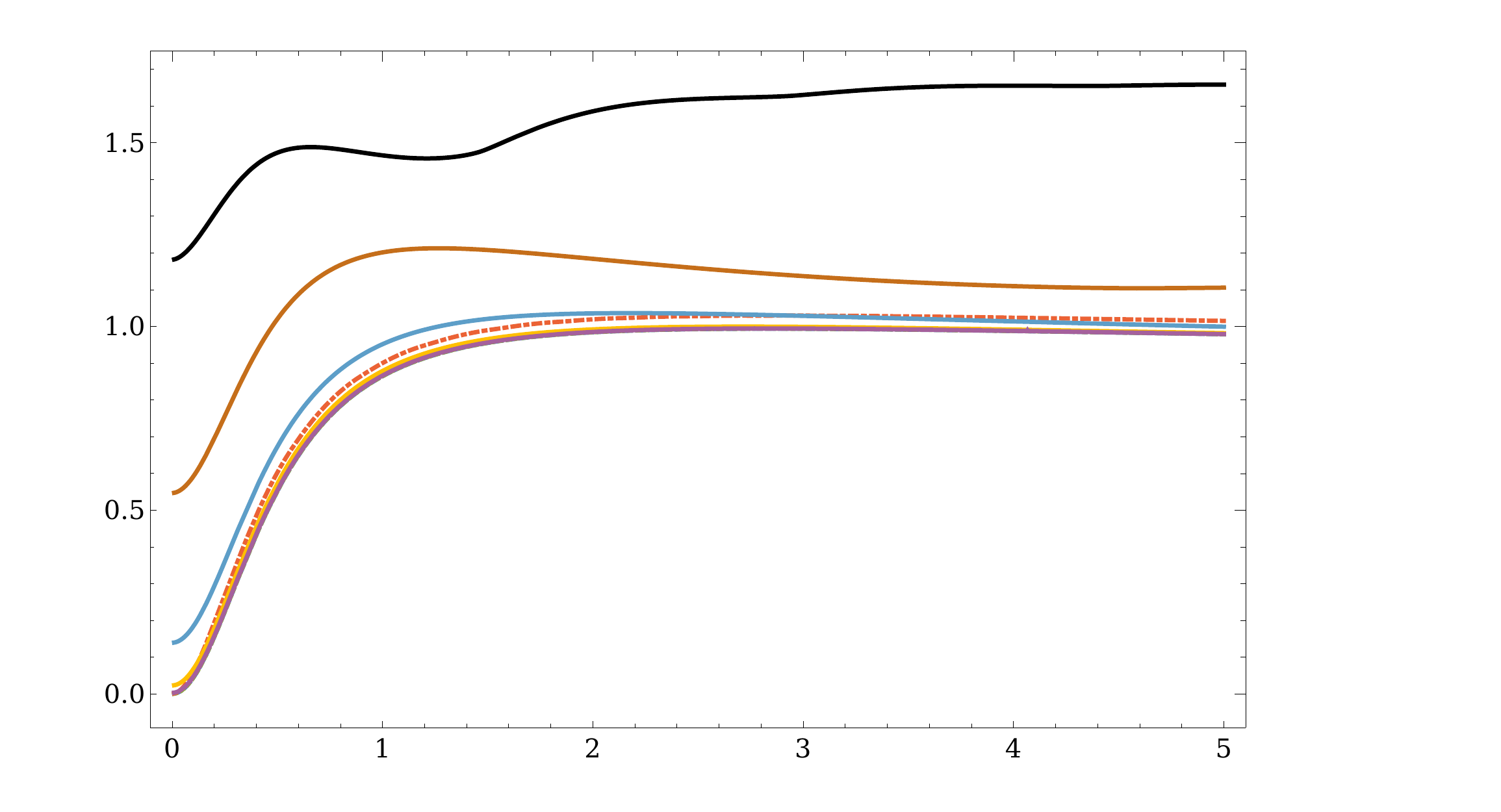}\put(-235 ,35){\rotatebox{-271}{\fontsize{14}{14}\selectfont $\frac{\mathcal{C}_{1}(t)-\mathcal{C}_{1}(\text{vac})}{S_{th}}$}}	\put(-130,-5){{\fontsize{13}{13}\selectfont$t\hspace{1mm} T$}}\hspace{-.7cm}
		\includegraphics[scale=.34]{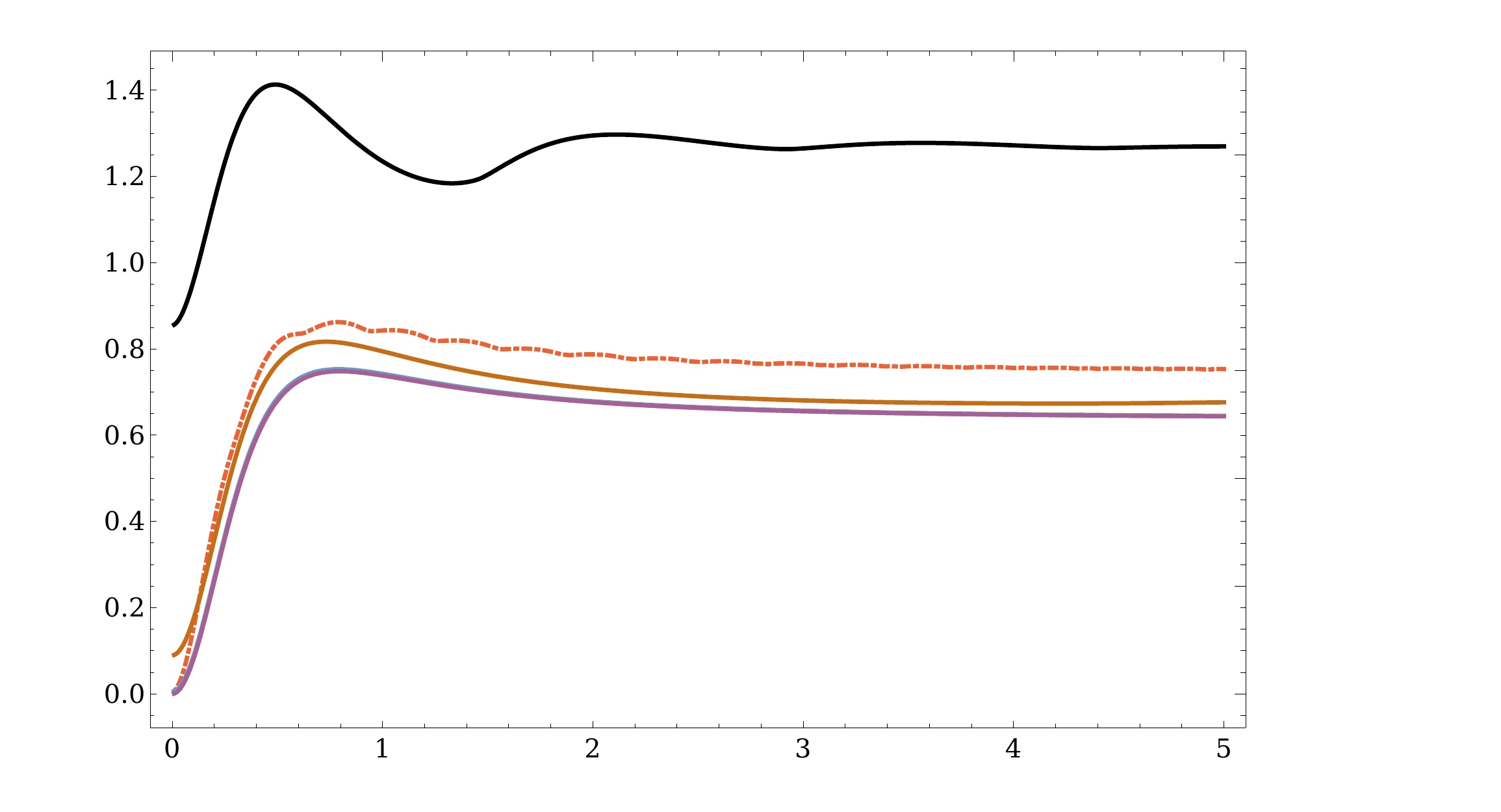}\put(-235 ,35){\rotatebox{-271}{\fontsize{13}{13}\selectfont $\frac{\mathcal{C}_{1}(t)-\mathcal{C}_{1}(\text{vac})}{S_{th}}$}}	\put(-130,-5){{\fontsize{13}{13}\selectfont$t\hspace{1mm} T$}}\put(-33,-30){\includegraphics[scale=.45]{crop3.png}}\put(5,112){{\fontsize{10}{10}\selectfont$\tilde{\gamma}=10^{-4}$}}\put(5,95){{\fontsize{10}{10}\selectfont$\tilde{\gamma}=10^{-3}$}}\put(5,80){{\fontsize{10}{10}\selectfont$\tilde{\gamma}=10^{-2}$}}\put(5,65){{\fontsize{10}{10}\selectfont$\tilde{\gamma}=10^{-1}$}}\put(5,49){{\fontsize{10}{10}\selectfont$\tilde{\gamma}=1$}}\put(5,33){{\fontsize{10}{10}\selectfont$\tilde{\gamma}=10$}}\put(5,17){{\fontsize{10}{10}\selectfont$\tilde{\gamma}=10^{2}$}}\put(5,0){{\fontsize{10}{10}\selectfont$\tilde{\gamma}=10^{3}$}}\put(5,-14){{\fontsize{10}{10}\selectfont$\tilde{\gamma}=10^{4}$}}
		\caption{Time dependency of $\mathcal{C}_{1}$ complexity of formation in diagonal basis for a cTFD state of massless complex scalar theory with $\tilde{Q}=1/10$ and different values of reference scale. \textbf{Left:} $d=1+1$ and \textbf{Right:} d=2+1}\label{diagonal5}	
	\end{figure}

\begin{figure}[H]
		\hspace{.9cm}\includegraphics[scale=.34]{formtd2Ck2thmtsqt10m1gt10}\put(-205 ,45){\rotatebox{-271}{\fontsize{14}{14}\selectfont $\frac{\mathcal{C}_{\kappa=2}(t)-\mathcal{C}_{\kappa=2}(\text{vac})}{S_{th}}$}}	\put(-100,15){{\fontsize{13}{13}\selectfont$t\hspace{1mm} T$}}\hspace{.7cm}\includegraphics[scale=.34]{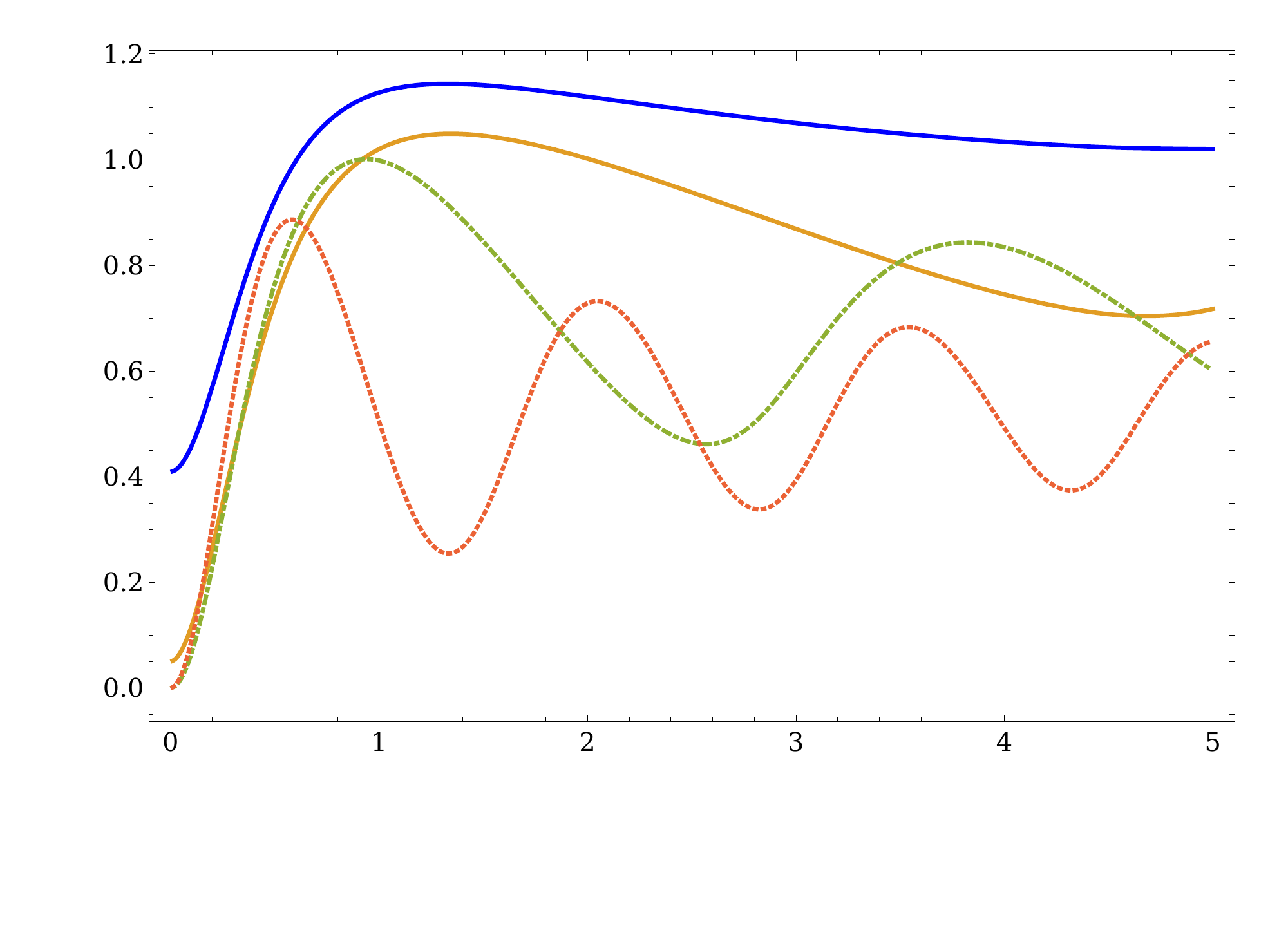}\hspace{.5cm}\put(-205 ,55){\rotatebox{-271}{\fontsize{14}{14}\selectfont $\frac{\mathcal{C}_{1}(t)-\mathcal{C}_{1}(\text{vac})}{S_{th}}$}}	\put(-100,15){{\fontsize{13}{13}\selectfont$t\hspace{1mm} T$}}\vspace{-.6cm}
		
		\caption{The time dependence of $\mathcal{C}_{\kappa=2}$ and $\mathcal{C}_{1}$ complexity of formation for cTFD state of a massive complex scalar theory in $d=1+1$ dimensions with $\tilde{Q}=1/10$, $\tilde{\gamma}=10$ and different masses. $\tilde{m} = 1/5$ (blue), $\tilde{m} = 1/2$ (orange), $\tilde{m} = 1$ (dotted dashed green) and $\tilde{m}=2$ (dotted red). For large masses with respect to the thermal scale, there is an oscillatory behavior with period $\Delta t = \pi/\tilde{m}$. At late times, we observe a saturation to a constant value.}\label{diagonal6}	
	\end{figure}

\begin{figure}[H]
		\hspace{1.5cm}\includegraphics[scale=.34]{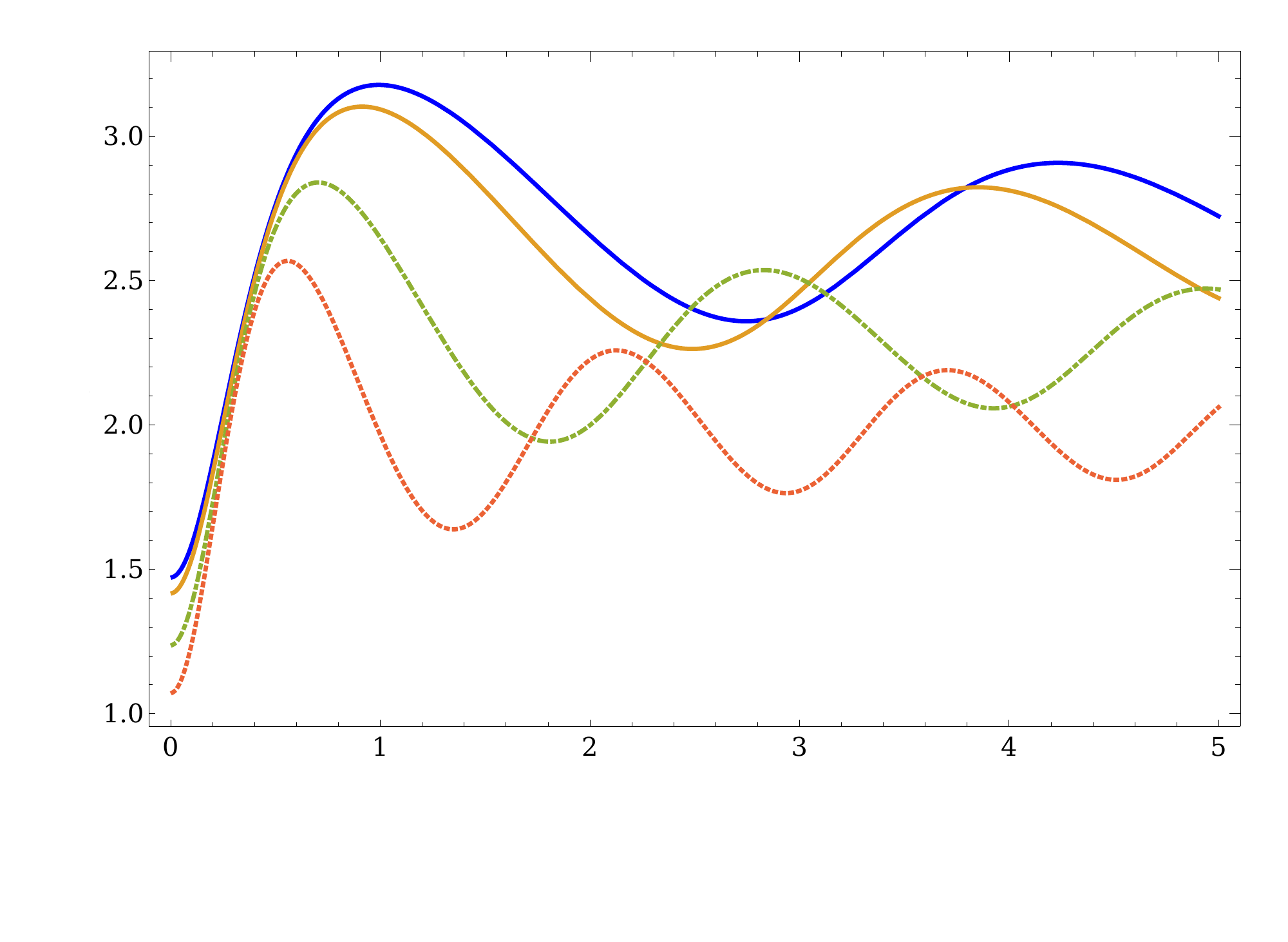}\put(-205 ,45){\rotatebox{-271}{\fontsize{14}{14}\selectfont $\frac{\mathcal{C}_{\kappa=2}(t)-\mathcal{C}_{\kappa=2}(\text{vac})}{S_{th}}$}}	\put(-100,15){{\fontsize{13}{13}\selectfont$t\hspace{1mm} T$}}\hspace{.6cm}\includegraphics[scale=.34]{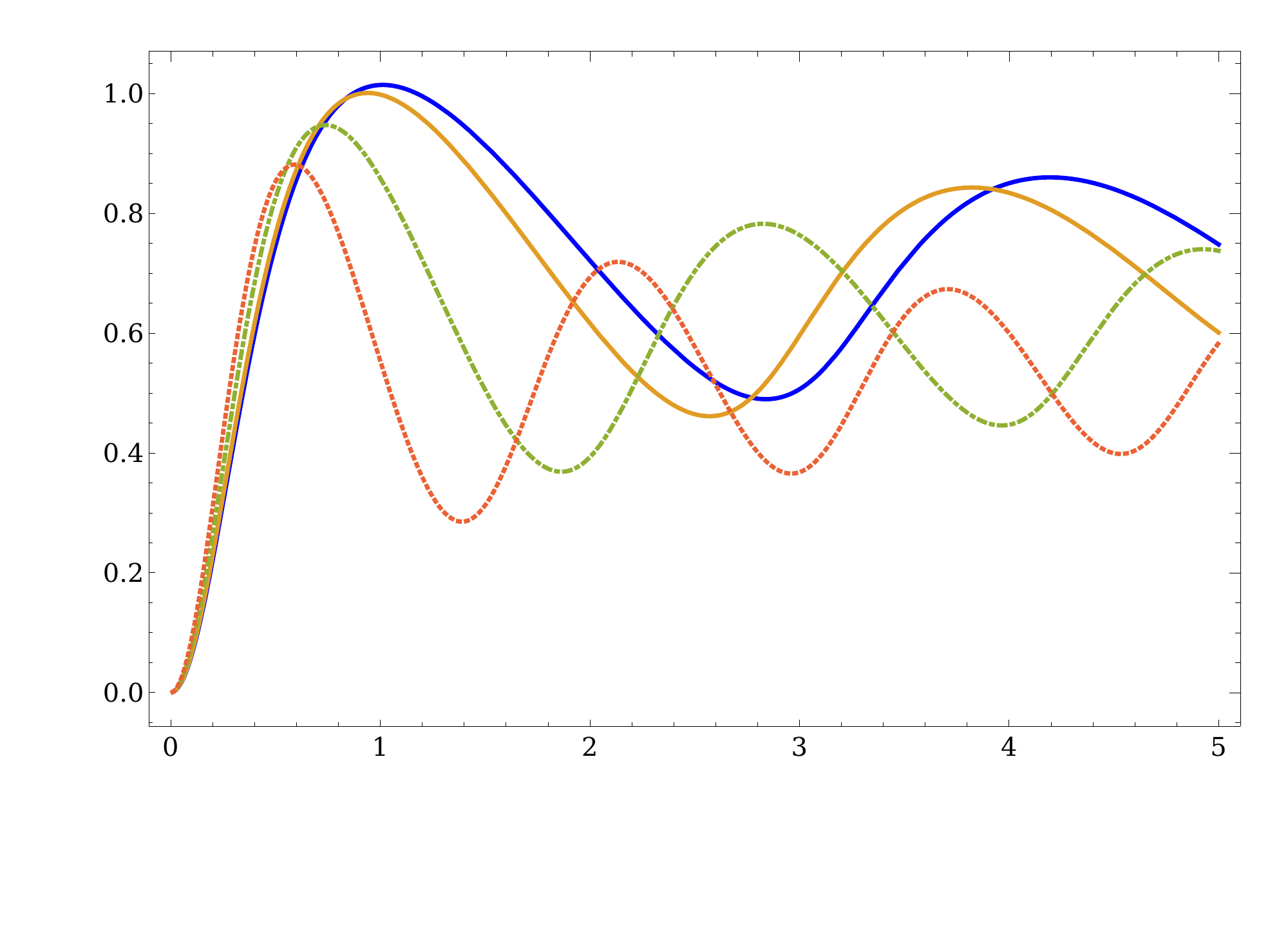}\put(-205 ,55){\rotatebox{-271}{\fontsize{14}{14}\selectfont $\frac{\mathcal{C}_{1}(t)-\mathcal{C}_{1}(\text{vac})}{S_{th}}$}}	\put(-100,15){{\fontsize{13}{13}\selectfont$t\hspace{1mm} T$}}\vspace{-.6cm}
		
		\caption{The time dependence of $\mathcal{C}_{\kappa=2}$ and $\mathcal{C}_{1}$ complexity of formation for cTFD state of a massive complex scalar theory in $d=1+1$ dimensions with $\tilde{m}=1$, $\tilde{\gamma}=10$ and different charges. $\tilde{Q} = 10^{-6}$ (blue), $\tilde{Q} = 10^{-1}$ (orange), $\tilde{Q} = 1/2 $ (dotted dashed green) and $\tilde{Q}=1$ (dotted red). For large charges with respect to the thermal scale, there is an oscillatory behavior with period $\Delta t = \pi\beta/\tilde{Q}$. At late times, we observe a saturation to a constant value.}\label{diagonal7}	
	\end{figure}

	The effect of changing the reference scale on time dependence of complexity is explored above, fig.\ref{diagonal2}. Similarly, the results for time dependency of complexity of formation is presented in fig.\ref{diagonal5}. Figs.\ref{diagonal2} and \ref{diagonal5} show that by changing the reference scale, the general structure of time dependency of complexity of formation, which is increasing in the beginning and saturating after short time,  in this basis does not change. This is in contrast with the result of  $\mathcal{C}_{1}$ complexity in LR basis, fig.\ref{c1lrt2-7}, where by changing the reference scale, the decreasing of complexity in the beginning changes to increasing behavior and vice versa. But similar to the $\mathcal{C}_{1}$ in LR basis, for the large and small values of reference scale we see the same behavior. The effects of changing mass and charge on complexity of formation in this basis are provided respectively in fig.\ref{diagonal6} and fig.\ref{diagonal7}.

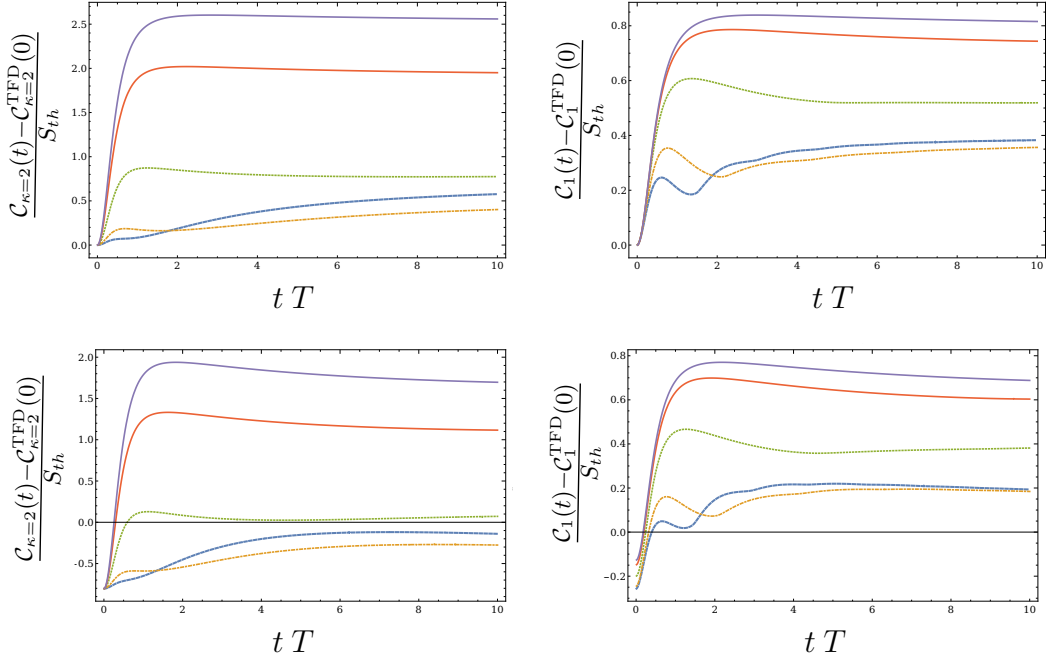
\begin{figure}[H]
		\hspace{1.5cm}\includegraphics[scale=.32]{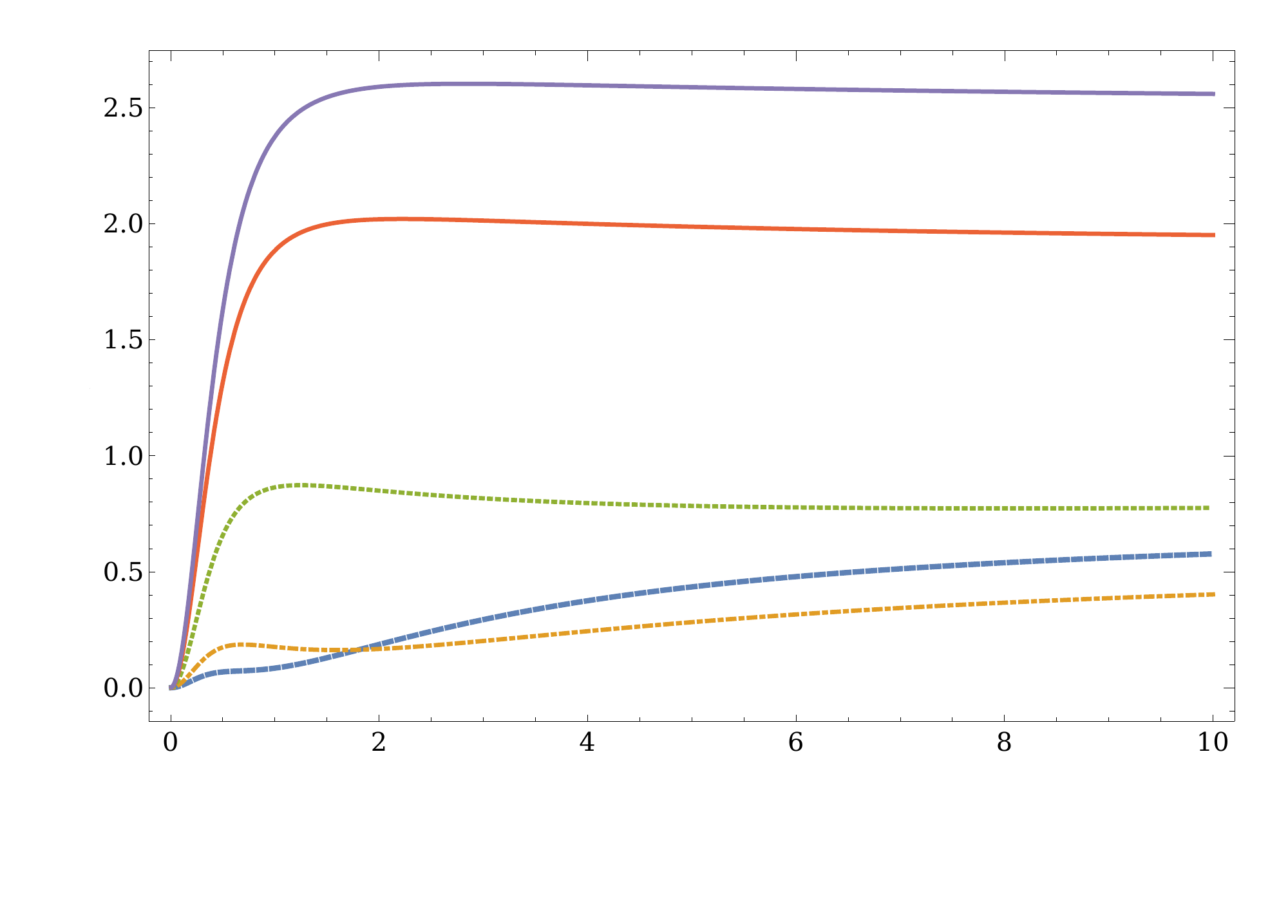}\put(-195 ,43){\rotatebox{-271}{\fontsize{13}{13}\selectfont $\frac{\mathcal{C}_{\kappa=2}(t)-\mathcal{C}^{\text{TFD}}_{\kappa=2}(0)}{S_{th}}$}}	\put(-95,10){{\fontsize{13}{13}\selectfont$t\hspace{1mm} T$}}\hspace{.6cm}\includegraphics[scale=.32]{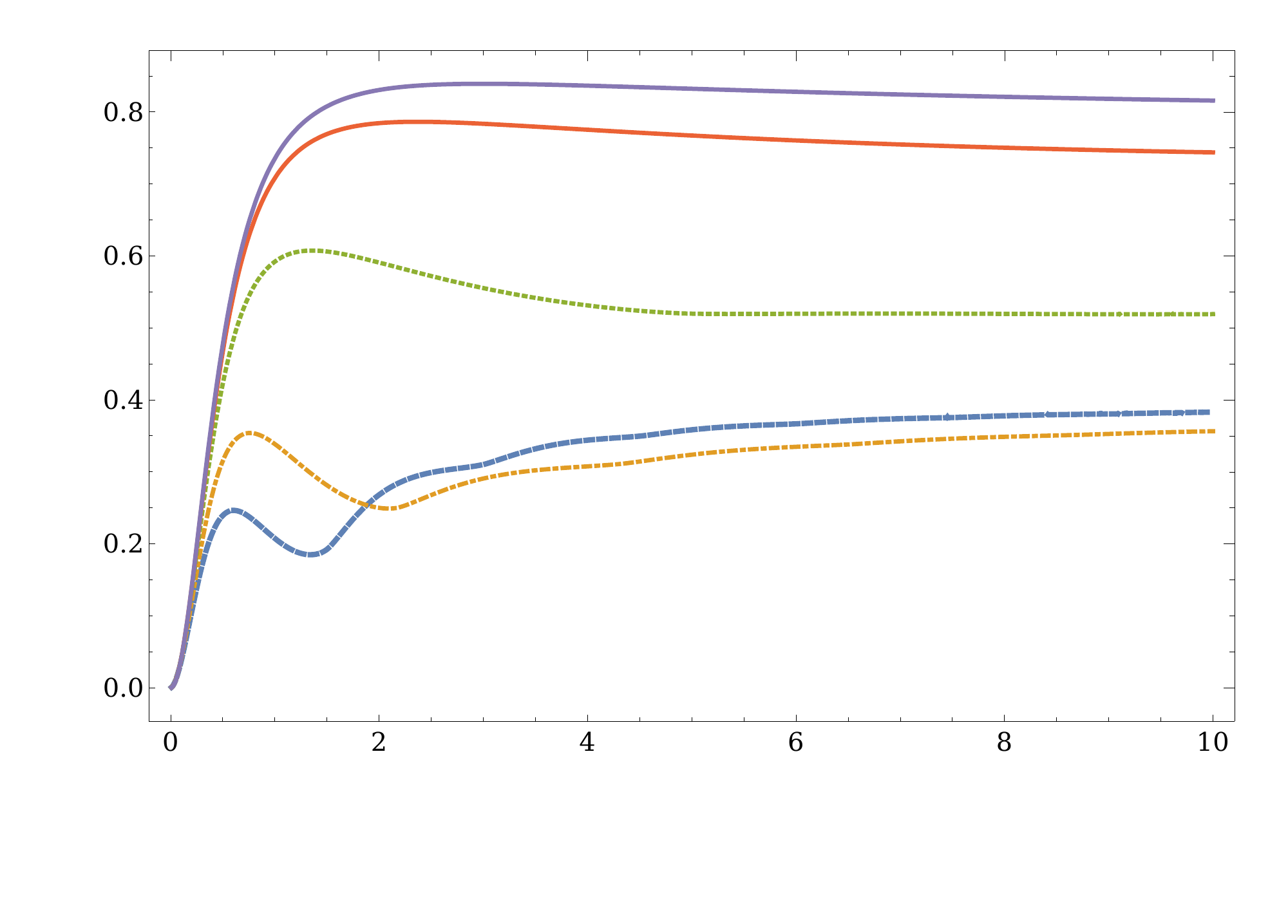}\put(-193 ,50){\rotatebox{-271}{\fontsize{13}{13}\selectfont $\frac{\mathcal{C}_{1}(t)-\mathcal{C}^{\text{TFD}}_{1}(0)}{S_{th}}$}}	\put(-95,10){{\fontsize{13}{13}\selectfont$t\hspace{1mm} T$}}\vspace{-.1cm}
		
		\hspace{1.5cm}\includegraphics[scale=.32]{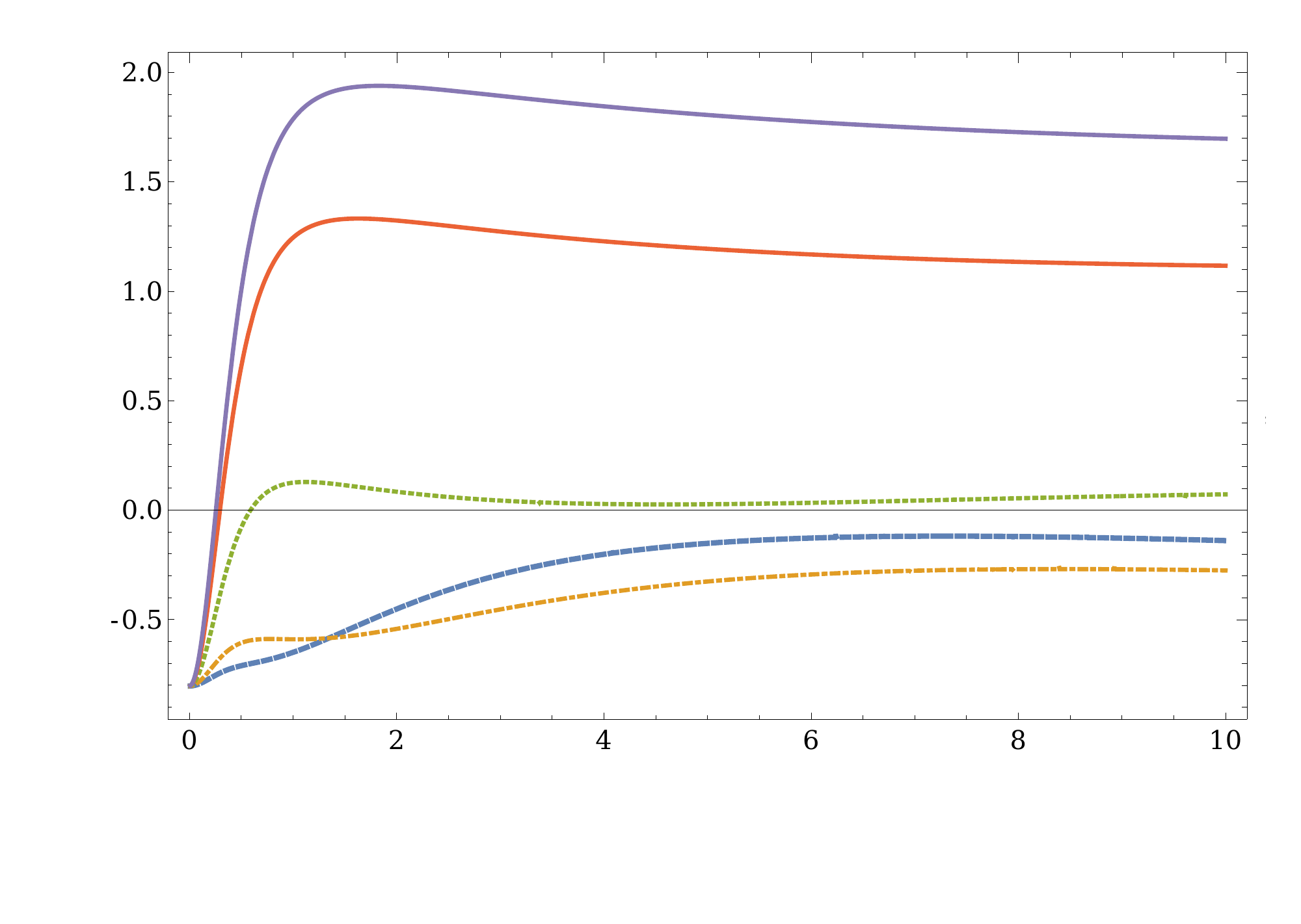}\put(-193 ,43){\rotatebox{-271}{\fontsize{13}{13}\selectfont $\frac{\mathcal{C}_{\kappa=2}(t)-\mathcal{C}^{\text{TFD}}_{\kappa=2}(0)}{S_{th}}$}}	\put(-95,10){{\fontsize{13}{13}\selectfont$t\hspace{1mm} T$}}\hspace{.5cm}\includegraphics[scale=.32]{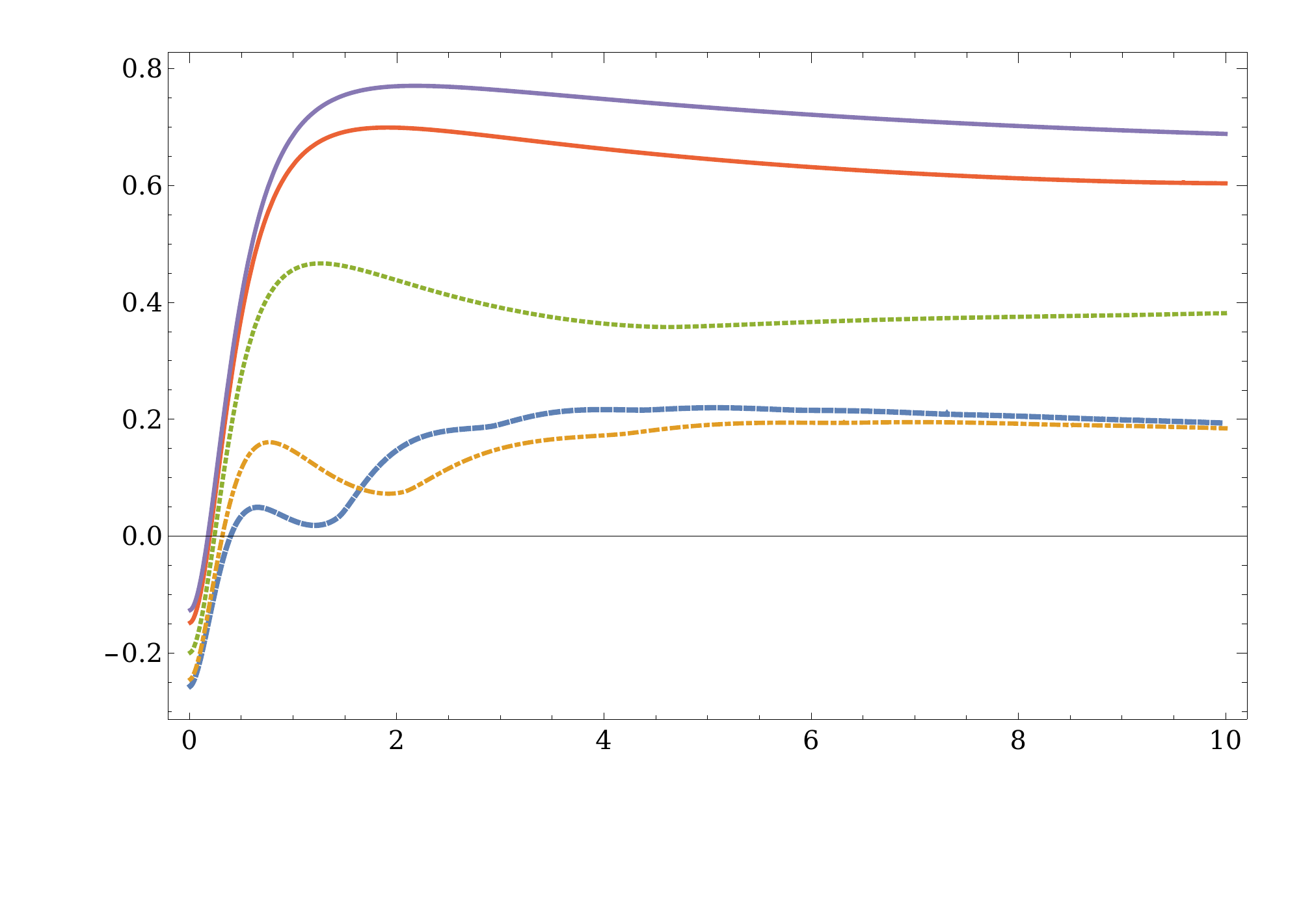}\put(-191 ,50){\rotatebox{-271}{\fontsize{13}{13}\selectfont $\frac{\mathcal{C}_{1}(t)-\mathcal{C}^{\text{TFD}}_{1}(0)}{S_{th}}$}}	\put(-95,10){{\fontsize{13}{13}\selectfont$t\hspace{1mm} T$}}\vspace{-.4cm}
		
		\caption{\textbf{Up:} The time dependence of $\mathcal{C}_{\kappa=2}$ and $\mathcal{C}_{1}$ complexities for TFD state of a massless real scalar theory in $d=1+1$ dimensions with  $\tilde{\gamma}=10$ and different $\lambda_{R}$. \textbf{Down:} The time dependence of $\mathcal{C}_{\kappa=2}$ and $\mathcal{C}_{1}$ complexities for cTFD state of a massless complex scalar theory in $d=1+1$ dimensions with $\tilde{Q}=10^{-1}$ and $\tilde{\gamma}=10$ and different $\lambda_{R}$. $\lambda_{R}=10$ (dashed blue), $\lambda_{R}=5$ (dotted dashed orange), $\lambda_{R}=1$ (dotted green), $\lambda_{R}=1/5$ (red) and $\lambda_{R}=1/10$ (purple).}\label{diagonal8}	
	\end{figure}

	To close this section, we explore the effect of changing the $\lambda_{R}$. Changing the ratio of the reference state scale to the gate scale i.e. $\lambda_{R}$ does not change the conclusions above significantly. The complexity still saturates at times of the order of $\beta$. From fig.\ref{diagonal8} we see that sharply changes of  complexity becomes more and more smooth by increasing $\lambda_{R}$. This is consistent with this fact that by increasing $\lambda_{R}$, target state approaches to the reference state and therefore the complexity becomes smaller. For these norms the net effect of charge is just decreasing the value of complexity in each time without altering the general structure of graph specially its first time derivative. This is in contrast with the holographic results \cite{Carmi:2017jqz} which again, implies that these norms are not well suited to compare with large $N$ CFTs.
	\vspace{-.35cm}
	
	\section{More concrete comparison with Holography}\label{compareholo}
\begin{figure}
		\centering\includegraphics[scale=.39]{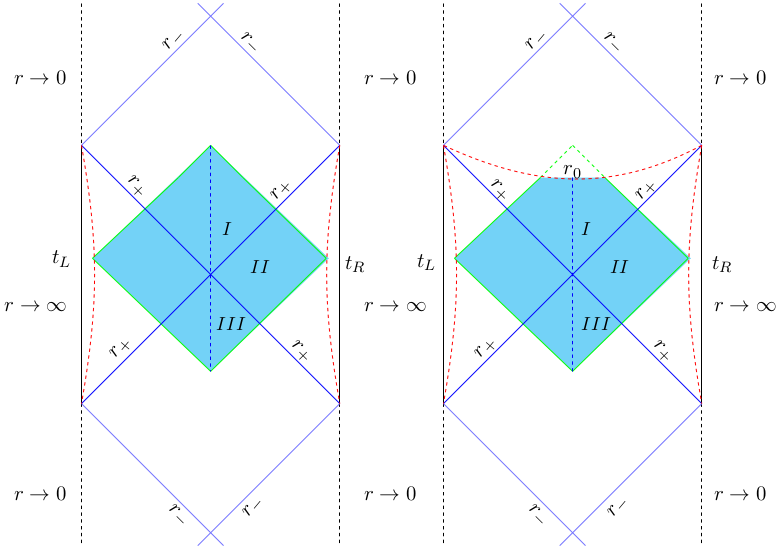}
		\caption{\textbf{Left:} Penrose diagram of charged AdS-BH and WDW patch is shown by shaded region. \textbf{Right:} Penrose diagram of charged AdS-BH and modified WDW patch is shown by shaded region. In comparison with left panel, one part of WDW patch is removed where $r_-<r_0<r_+$. Existence of this extra surface, $r=r_0$, implies that one also needs to consider GH term, reparametrization recovery counterterm together with two joint terms and bulk counterterm on that.}\label{penrose}
	\end{figure}

	To explore the relation between QFT calculations in subsection.\ref{prefereC} with holographic results more concretely, it would be better to focus on the complexity of formation. The $\mathcal{C}_{1}$ function (\ref{c1lr2}) in LR basis together with (\ref{Sthq}) imply that the ratio of complexity of formation to thermal entropy for  massless theory has the following expansion in $\tilde{Q}$,
	\bea
	\label{QFTC1}
	\frac{\mathcal{C}_{1}^{(LR)}-\mathcal{C}_{1}^{(LR)}(\text{vac})}{S_{\text{th}}} =\frac{2^{d}-1}{d}+\sum_{n=1}^{d-1} a(n,d)\hspace{1mm}\tilde{Q}^{n}+b(d)\hspace{1mm}\tilde{Q}^{d-1}\log\tilde{Q} +\mathcal{O}(\tilde{Q}^d),
	\eea
	where $a(n,d)$ and $b(d)$ are constant numbers that depend on the dimensions, for example
	\bea
	&&a(1,2) = \frac{6}{\pi^{2}}(1-\log2),\hspace{.5cm}a(1,3) = \frac{5\pi^2}{54\hspace{1mm}\zeta(3)},\hspace{.5cm}a(1,4) = \frac{765 \hspace{1mm}\zeta(3)}{8\pi^4},\cr \nonumber\\
	&& a(2,3) = \frac{1}{486\hspace{1mm}\zeta(3)^2}\big(5\pi^4+27(-13+6\log2)\hspace{1mm}\zeta(3)\big),
	\hspace{.5cm} a(2,4)= -
	\frac{135}{16\pi^8}\big(\pi^6-765 \hspace{1mm}\zeta(3)^2\big),
	\cr\nonumber\\
	&& a(3,4)= -\frac{5}{32\pi^{12}}\big(-211\pi^8+48\pi^8\log{2}+5940\pi^6 \hspace{1mm}\zeta(3)-2788425 \hspace{1mm}\zeta(3)^3\big),
	\eea
	and
	\bea
	b(2) =-\frac{3}{2\pi^2},\hspace{.5cm}b(3) = \frac{2}{9\hspace{.5mm} \zeta(3)},\hspace{.5cm}b(4)=-\frac{165}{16\pi^{4}}.
	\eea
	According to CA proposal, to compute the complexity of cTFD state dual to  charged eternal AdS-BH, one may find on-shell bulk action in a special part of geometry which is known as WDW patch.
	The background geometry is 
	\bea
	\label{metric}
	ds^{2} = -f(r) dt^2+ \frac{dr^2}{f(r)} +r^{2} d\Omega^2,\hspace{.7cm}f(r) = \frac{r^2}{l^2}-\frac{\omega^{d-2}}{r^{d-2}}+\frac{q^2}{r^{2(d-2)}}
	\eea
	and the WDW patch is shown as shaded region in left panel of  fig.\ref{penrose}. Temperature and entropy of this solution is given by
	\bea
	\label{Sth}
	T=\frac{1}{4\pi}\frac{\partial f}{\partial r}\bigg{|}_{r=r_+} ,\hspace{.5cm} S_{th} = \frac{V_{d-1}}{4G_{N}}\hspace{1mm}r_{+}^{d-1}.
	\eea
	Furthermore,  bulk gravitational action is as follows
	\bea
	\label{BulkAction}
	S_{\text{Bulk}} = \frac{1}{16\pi G_{N}} \int d^{d+1}x\hspace{1mm}\sqrt{-G} \left(R-2\Lambda\right) -\frac{1}{4g^{2}} \int d^{d+1}x\hspace{1mm}\sqrt{-G}\hspace{1mm}F_{\mu\nu}F^{\mu\nu}.
	\eea
	To have an action which has well defined variational principle and its on-shell value also has reparamitrization invariance, one needs to add respectively  Gibbons-Hawking term and proper counterterm. According to fig.\ref{penrose} for zero time at the boundary ($t_{L}=t_{R}=0$), the final expression for each of those parts is as following 
	\bea
	&& I_{\text{EH}} = 2\times \frac{V_{d-1}}{8\pi G_{N}} \left(-\frac{2d}{l^2}\right) \int _{r_{m}}^{r_{\infty}} dr\hspace{1mm} r^{d-1} \left(r^{*}_{\infty}-r^{*}(r)\right),
	\cr\nonumber\\
	&& I_{\text{Maxwell}} = 2\times \frac{V_{d-1}}{8\pi G_{N}} (1-2\gamma) \big(2(d-2)q^2\big) \int _{r_{m}}^{r_{\infty}} dr\hspace{1mm} \frac{1}{r^{d-1}} \big(r^{*}_{\infty}-r^{*}(r)\big),
	\cr\nonumber\\
	&& I_{\text{joint}} = -2\times\frac{V_{d-1}}{8\pi G_{N}} \hspace{1mm}r_{m}^{d-1} \log \bigg(\frac{l^2 |f(r_{m})|}{\tilde{\alpha}^{2}R^{2}}\bigg) -2\times\frac{V_{d-1}}{8\pi G} \hspace{1mm}r_{\infty}^{d-1} \log \bigg(\frac{l^2 |f(r_{\infty})|}{\tilde{\alpha}^{2}R^{2}}\bigg),
	\cr\nonumber\\
	&& I_{\text{ct}} = -2\times\frac{V_{d-1}}{8\pi G_{N}} \hspace{1mm} r_{m}^{d-1} \bigg[\log\bigg(\frac{(d-1)^{2}\hspace{.5mm}l^{2}_{\text{ct}} \hspace{.5mm}\tilde{\alpha}^2 \hspace{.5mm}R^{2}}{r_m^2 \hspace{.5mm}l^2}\bigg)+\frac{2}{d-1}\bigg]
	\cr\nonumber\\
	&&\hspace{2.2cm}-2\times\frac{V_{d-1}}{8\pi G_{N}} \hspace{1mm} r_{\infty}^{d-1} \bigg[\log\bigg(\frac{(d-1)^2 \hspace{.5mm}l^{2}_{\text{ct}} \hspace{.5mm}\tilde{\alpha}^2 \hspace{.5mm}R^{2}}{r_{\infty}^2 \hspace{.5mm}l^2}\bigg)+\frac{2}{d-1}\bigg],\label{bulk}
	\eea
	where $r_{\infty}$ indicates the conformal boundary, $r_m$ is chosen for the joint points between two horizons, $\tilde{\alpha}$ refers to normalization of null boundaries of WDW patch\footnote{One can choose different normalization for each null boundary of WDW patch but the final results do not alter.} and moreover total action is $I_{\text{tot}} =I_{\text{EH}}+I_{\text{Maxwee}}+I_{\text{joint}}+I_{\text{ct}}$. To find the complexity of formation, we also need on-shell action for the vacuum state (pure AdS spacetime), $I^{\text{vac}} = I_{\text{Bulk}}^{\text{vac}}+I_{\text{joint}}^{\text{vac}}+I_{\text{ct}}^{\text{vac}}$, which its different parts are given by 
	\bea
	&& I_{\text{Bulk}}^{\text{vac}} = 2\times\frac{V_{d-1}}{8\pi G_{N}} \left(-\frac{2d}{l^2}\right) \int _{0}^{r_{\infty}} dr\hspace{1mm} r^{d-1} \left(r^{*\hspace{.5mm}\text{AdS}}_{\infty}-r^{*\hspace{.5mm}\text{AdS}}(r)\right),
	\cr\nonumber\\
	&& I_{\text{joint}}^{\text{vac}} = -2\times\frac{V_{d-1}}{8\pi G_{N}} \hspace{1mm}r_{\infty}^{d-1} \log \bigg(\frac{l^2 |f^{\text{vac}}(r_{\infty})|}{\tilde{\alpha}^{2}R^{2}}\bigg),
	\cr\nonumber\\
	&& I_{\text{ct}}^{\text{vac}} = -2\times\frac{V_{d-1}}{8\pi G_{N}} \hspace{1mm} r_{\infty}^{d-1} \bigg[\log\bigg(\frac{4 l^{2}_{\text{ct}} \hspace{.5mm}\tilde{\alpha}^2 \hspace{.5mm}R^{2}}{r_{\infty}^2 \hspace{.5mm}l^2}\bigg)+\frac{2}{d-1}\bigg].\label{Hvac}
	\eea
	It is easy to check that IR bulk divergences of total action in  limit $r\rightarrow r_{\infty}$ are exactly the same as IR divergences for vacuum state. Therefore, holographic complexity of formation  
	\bea
	\label{holocomplexform}
	\Delta\mathcal{C}_{\text{H}} = \frac{1}{\pi} \left(I_{\text{tot}}- I^{\text{vac}}\right),
	\eea
	is finite and this cancellation of IR bulk divergences is consistent with the cancellation of UV divergences in the field theory side, (\ref{QFTC1}). Apart from cancellation of IR divergences, there is another important reason for choosing AdS spacetime for reference state. It is shown \cite{Carmi:2017jqz} that the complexity of extremal BH diverges and therefore it can not be used as a meaningful tool for comparison. To compare with field theory results it will be useful to define following dimensionless parameters
	\bea
	z =\frac{l}{r_+},\hspace{1cm} y= \frac{r_-}{r_+},\hspace{1cm} x= \frac{r}{r_{+}},
	\eea
	which by using them for small charges we have\footnote{The ‘non-normalizable’ mode of the gauge potential is identified with $\tilde{\mu}$, i.e. $\tilde{\mu}= \lim_{r\rightarrow \text{bdy}} A_{t}$.}
	\bea
	\label{def1}
	&& R\hspace{.5mm}T \sim \frac{r_+}{4\pi L}\bigg(d-(d-2)y^{d-2}\bigg)+\mathcal{O}(y^{2(d-2)}),
	\cr\nonumber\\
	&&\hspace{.2cm} \nu \equiv \sqrt{\frac{C_{J}}{C_{T}}}\hspace{1mm}\tilde{Q} \sim \frac{2\sqrt{2}\pi(d-1)}{d\sqrt{d(d+1)}}\left[y^{\frac{d}{2}-1}+\left(\frac{3}{2}-\frac{2}{d}\right)y^{\frac{3d}{2}-3}+\mathcal{O}(y^{\frac{3d}{2}-1})\right],
	\eea
	where
	\bea
	\label{centralcharges}
	C_{J} = (d-2)\hspace{1mm} \frac{\Gamma[d]}{2\pi^{d/2}\Gamma[d/2]}\hspace{1mm}\frac{l^{d-3}}{g^{2}}, \hspace{1cm}C_{T} =\frac{d+1}{d-1}\hspace{1mm}\frac{\Gamma[d+1]}{8\pi^{(d+2)/2}\Gamma[d/2]}\frac{l^{d-1}}{G_{N}}.
	\eea
	The $C_{J}$ and $C_{T}$ are central charges associated with the two-point functions of the boundary currents and stress tensor
	\bea
	\langle J_{\mu}(x) J_{\nu}(0)\rangle = \frac{C_{J}}{x^{2(d-1)}}\hspace{1mm} \mathcal{A}_{\mu\nu}(x)+...,\hspace{1cm}\langle T_{\mu\nu}(x)T_{\alpha\beta}(0)\rangle =\
	\frac{C_{T}}{x^{2d}}\hspace{1mm} \mathcal{B}_{\mu\nu,\alpha\beta}+...,
	\eea
	with
	\bea
	&&\mathcal{A}_{\mu\nu} = \eta_{\mu\nu} -2 \frac{x_{\mu}x_{\nu}}{x^{2}},
	\hspace{.5cm}\mathcal{B}_{\mu\nu,\alpha\beta} = \frac{1}{2}\bigg(\mathcal{A}_{\mu\nu}(x)\mathcal{A}_{\alpha\beta}(x)+\mathcal{A}_{\mu\beta}(x)\mathcal{A}_{\nu\alpha}(x)\bigg)-\frac{1}{d}\hspace{.5mm}\eta_{\mu\nu}\hspace{.5mm}\eta_{\alpha\beta}.
	\eea
	Let us calculate the different contributions in (\ref{bulk}) and (\ref{Hvac}) for $d=2+1$ and $d=3+1$ separately in the following two subsections.
	\subsection{$d=2+1$ dimensions}
	By defining 
	\bea
	\tilde{f}(x,y) = z^2 f(r), \hspace{.7cm} x^{*}(x,y) = r^{*}(r)/z^2 r_+,
	\eea
	in $d=2+1$ dimensions we have
	\bea
	\label{xstar}
	&&\tilde{f}(x,y) = \frac{1}{x^2} (x-1) (x-y) \left(1+x+x^2 +y + y^2 +x y\right),
	\cr \nonumber\\
	&& x^{*} (x,y) = \frac{(3+6y+10 y^2 +6 y^3 +3 y^4)}{(3+2y+y^2)(1+2y+3y^2)\sqrt{3+2y+3y^2}} \arctan [\frac{1+2x+y}{\sqrt{3+2y+3y^2}}]
	\cr\nonumber\\
	&&\hspace{1.5cm} -\frac{1}{(-3+y+y^2+y^3)} \log|x-1| -\frac{y^2}{(1+y+y^2-3y^3)} \log(x-y)
	\cr\nonumber\\
	&&\hspace{1.5cm}-\frac{(1+y)^3}{2(3+2y+y^2)(1+2y+3y^2)}\log \bigg(1+x+x^2+y+y^2+x y\bigg),
	\eea
	where $r^{*}(r)= \int dr/f(r)$. According to (\ref{xstar}) for boundary and joint surfaces one can see
	\bea
	&&x^{*}_{\infty} = \frac{\pi \left(3+6y+10y^2 +6y^3 +3y^4\right)}{2(3+2y+y^2)(1+2y+3y^2)\sqrt{3+2y+3y^2}},
	\cr \nonumber\\
	&&x_{m} \simeq y \left(1 +\exp\bigg(-\frac{\pi}{3\sqrt{3} y^2}+ \mathcal{O}(\frac{1}{y})\bigg)\right),
	\eea
	which they all together imply that
	\bea
	&& I_{\text{AdS}} = - \frac{V_{2}\hspace{1mm} r_+^2}{4 G_{N}}\bigg(\frac{4\hspace{1mm} x_{\infty}^2}{\pi}+\frac{ x_{\infty}^2}{\pi}\log\left(\frac{4\hspace{.5mm} l_{\text{ct}}^2}{l^2}\right)\bigg),
	\cr\nonumber\\
	&& \text{I}_{\text{EH}} = -\frac{V_{2}\hspace{1mm}r_+^2}{4G_{N}}\bigg(\frac{3\hspace{1mm}x_{\infty}^2 }{\pi}+\frac{2\sqrt{3}}{9}+\frac{2y}{3\sqrt{3}}+\frac{(9+4\sqrt{3}\pi)y^2}{27\pi}+\mathcal{O}(y^3)\bigg),
	\cr\nonumber\\
	&& I_{\text{Maxwell}} =  \frac{V_{2}\hspace{1mm} r_+^2}{4 G_{N}} (1-2\gamma)\bigg(\frac{2}{3\sqrt{3}} +\frac{2y}{3\sqrt{3}}+(\frac{4}{9\sqrt{3}}+\frac{4}{3\pi})y^2+\mathcal{O}(y^3)\bigg),
	\cr\nonumber\\
	&& I_{\text{joint}} = \frac{V_{2}\hspace{1mm} r_+^2}{4 G_{N}} \bigg(-\frac{x_{\infty}^2}{\pi}\log\left(\frac{l^2 x_{\infty}^2}{R^2\hspace{.5mm}z^2\hspace{.5mm}\tilde{\alpha}^2}\right)+\frac{1}{3\sqrt{3}} +\frac{y^2}{\pi} \log \bigg(\frac{R^2 z^2 \tilde{\alpha}^2 y}{l^2}\bigg)\hspace{1mm}\bigg),
	\cr\nonumber\\
	&& I_{\text{ct}} = -\frac{V_{2}\hspace{1mm} r_+^2}{4 G_{N}} \bigg( \frac{x_{\infty}^2}{\pi}+\frac{x_{\infty}^2}{\pi}\log\left(\frac{4 R^2\hspace{.5mm}z^2\hspace{.5mm}\tilde{\alpha}^2\hspace{.5mm}l_{\text{ct}}^2}{x_{\infty}^2\hspace{.5mm}l^4}\right)+\frac{y^2}{\pi}+\frac{y^2}{\pi}\log\bigg(\frac{4 R^2 z^2 \tilde{\alpha}^2 l_{ct}^2}{l^4 y^2}\bigg)\bigg).\label{3d}
	\eea
	By using final expressions in (\ref{3d}) and noting to (\ref{holocomplexform}), (\ref{def1}) and (\ref{Sth}), it is easy to see that
	\bea
	\label{d3h}
	\frac{\Delta\mathcal{C}_{\text{H}}}{S_{\text{th}}} = \frac{1}{3\sqrt{3}\pi} +\frac{27}{64\pi^6}\left(162\log\left(\frac{3\sqrt{3}\hspace{.5mm}l^{\frac{1}{3}}\hspace{.5mm}}{2^{\frac{11}{6}}\hspace{.5mm}l_{ct}^{\frac{1}{3}} \hspace{.5mm}\pi}\hspace{1mm}\nu\right)\right) \nu^4+\mathcal{O}({\nu^6}).
	\eea
	\subsection{$d=3+1$ dimensions}
	In $d=3+1$ dimensions we have
	\bea
	&&\tilde{f} (x,y)= \frac{(x^2-1)(x^2-y^2)(x^2+y^2+1)}{x^4},\hspace{.7cm}x_{m} \simeq y \left(1+\text{exp}\left(-\frac{\pi}{2y^{3}}+\mathcal{O}(1/y)\right)\right),
	\cr \nonumber\\
	&&x^{*}(x,y) = \frac{(1+y^2)^{\frac{3}{2}}}{(2+y^2)(1+2y^2)} \arctan (\frac{x}{\sqrt{1+y^2}})+\frac{1}{2(1-y^2)(2+y^2)} \log\frac{|x-1|}{x+1} \cr \nonumber\\
	&&\hspace{1.5cm}-\frac{y^{3}}{2(1-y^2)(1+2y^2)}\log \frac{x-y}{x+y},
	\eea
	which they all together imply that
	\bea
	\label{detail4}
	&& I_{\text{AdS}} = -\frac{V_{3}\hspace{1mm}r_{+}^3}{4 G_{N}}\hspace{1mm}\left( \frac{10\hspace{.5mm}x^{3}_{\infty}}{3\pi}+\frac{x_{\infty}^3}{\pi}\log\left(\frac{9\hspace{.5mm}l_{\text{ct}}^2}{l^2}\right)\right),
	\cr\nonumber\\
	&& I_{\text{EH}} = -\frac{V_{3}\hspace{1mm}r_{+}^3}{4 G_{N}}\hspace{1mm}\left(\frac{8\hspace{.5mm}x^{3}_{\infty}}{3\pi}+\frac{1}{2}+\frac{y^2}{2}+\mathcal{O}(y^4)\right),
	\cr\nonumber\\
	&& I_{\text{Maxwell}} = \frac{V_{3}\hspace{1mm}r_{+}^3}{4 G_{N}}\hspace{1mm}\left(\frac{1}{2}+\frac{y^2}{2}+\frac{2y^3}{3\pi}+\mathcal{O}(y^4)\right),
	\cr\nonumber\\
	&&I_{\text{joint}} = \frac{V_{3}\hspace{1mm}r_{+}^3}{4 G_{N}} \left(-\frac{x_{\infty}^3}{\pi}\log\left(\frac{l^2\hspace{.5mm}x_{\infty}^2}{R^2\hspace{.5mm}z^2\hspace{.5mm}\tilde{\alpha}^2}\right)+\frac{1}{2} +\frac{y^3}{\pi}\log \bigg(\frac{R^2\hspace{.5mm}z^{2}\hspace{.5mm}\tilde{\alpha}^2 y^2}{2l^2}\bigg)\right),
	\cr\nonumber\\
	&& I_{\text{ct}} = -\frac{V_{3}\hspace{1mm}r_{+}^3}{4 G_{N}} \left(+\frac{2\hspace{.5mm}x_{\infty}^3}{3\pi}+\frac{x_{\infty}^3}{\pi}\log\left(\frac{9 R^2\hspace{.5mm}z^{2}\hspace{.5mm}\tilde{\alpha}^2\hspace{.5mm}l_{\text{ct}}^2}{x_{\infty}^2\hspace{.5mm}l^4}\right)+\frac{2y^3}{3\pi} +\frac{y^3}{\pi}\log \bigg(\frac{9 R^2\hspace{.5mm}z^{2}\hspace{.5mm}\tilde{\alpha}^2  \hspace{1mm}l_{\text{ct}}^2}{l^4\hspace{.5mm}y^2}\bigg)\right).
	\eea
	Having above results and noting to (\ref{holocomplexform}), (\ref{def1}) and (\ref{Sth}) after short computation one can see
	\bea
	\label{d4h}
	\frac{\Delta \mathcal{C}_{\text{H}}}{S_{\text{th}}} = \frac{1}{2\pi}  +\frac{320\sqrt{10}}{27\pi^{5}}\bigg( \log\left(\frac{2^{\frac{5}{4}} \sqrt{5}}{3\sqrt{3}\pi}\sqrt{\frac{l}{l_{\text{ct}}}} \hspace{1mm}\nu\right)\bigg)\hspace{1mm}\nu^{3} +\mathcal{O}(\nu)^{4}.
	\eea
	For the neutral case, $\tilde{Q}=0$, the complexity of formation is proportional to thermal entropy which is completely in agreement with the result of $\mathcal{C}_{1}$ complexity in LR basis. According to this agreement, it is  claimed that \cite{Chapman:2018hou} this norm of complexity is a corner stone of holographic complexity. Unlike the neutral case, QFT result (\ref{QFTC1}) is independent of reference state scale $\omega_{R}$ but holographic results (\ref{d3h}) and (\ref{d4h}) depend to $l_\text{ct}$. Apart from that, in comparison with (\ref{QFTC1}) also some terms in $\tilde{Q}$ expansion are missed in holographic results (\ref{d3h}) and (\ref{d4h}). Now we have at least four options. We can change the complexity in the QFT side by using the charged dependent penalty factors to set these extra terms zero or we can conclude that the CA proposal needs more boundary terms or we can say that these missing terms are actually zero in the strong coupling regime or we can conclude that $l_{\text{ct}}$ is proportional to AdS radius $l$. If one chooses the first option, the new UV charged dependent divergences appear which they do not have the same counterpart in the holographic side because the on-shell action for the Maxwell term is finite and so the holographic UV divergent part is independent of chemical potential. According to the second option, recently two different boundary terms are proposed in \cite{Brown:2018bms} and \cite{Akhavan:2018wla}. In \cite{Brown:2018bms} a new boundary term for the Maxwell field is added to the action of Einstein-Hilbert- Maxwell theory (\ref{BulkAction}) in $D =d+1 = 3+1$ dimensions\footnote{Adding this term puts the electric and magnetic charges on an equal footing in the holographic complexity.}
	\bea
	\label{new.b.1}
	I_{\tilde{\mu}} = \frac{\gamma}{g^{2}} \int_{\partial{\mathcal{M}}} \hspace{1mm} d\Sigma_{\mu}\hspace{1mm}F^{\mu\nu}A_{\nu}. 
	\eea
	While introducing this boundary term does not change the equations of motion, it does change the nature of variational principle of the Maxwell field. That is, it changes the boundary conditions that must be imposed for consistency of the variational principle. If we use the Maxwell equations, $\nabla_{\mu}F^{\mu\nu}=0$, then this boundary term can be converted into a bulk term via stokes's theorem as 
	\bea
	I_{\tilde{\mu},\text{on-shell}} =\frac{\gamma}{2 g^{2}} \int_{\mathcal{M}}\hspace{1mm} d^{4}x \sqrt{-G}\hspace{.5mm}F_{\mu\nu}F^{\mu\nu}.
	\eea
	According to (\ref{detail4}), this new boundary term changes (\ref{d4h}) to the following expression 
	\bea
	&&\frac{\Delta\mathcal{C}_{\text{H}}}{S_{th}} = \frac{(1-4\gamma)}{3\sqrt{3}\pi}-\frac{\sqrt{3}\hspace{.5mm}\gamma}{\pi^2}\left(\frac{G_{N}}{g^{2}\hspace{.5mm}l^2}\right)\tilde{Q}^{2} +\mathcal{O}(\tilde{Q}^{4},l,l_{\text{ct}}),
	\eea
	where we have used the (\ref{def1}). For $\gamma =0$ and $\tilde{Q} =0$ we recover the known result \cite{Carmi:2017jqz}. In comparison with the QFT result (\ref{QFTC1}), it is clear that adding boundary term (\ref{new.b.1}) just recovers $\tilde{Q}^{2}$ term but $\tilde{Q}$ and $\tilde{Q}^{2}\log{\tilde{Q}}$ ones are remained absent. Apart from that, in the zero charge limit, the complexity of formation changes in comparison with neutral case, since this boundary term actually changes the boundary condition and therefore ensemble of system under consideration. Furthermore, by using this extra term, one can not see the dependence of complexity to the reference scale in general time, (\ref{c1LRt}), since by comparison with QFT result at zero time we may choose $l_{\text{ct}}$ as a function of the AdS scale $l$ and so no other arbitrary scale remains for another times.
	
	Instead of changing the variational principle in \cite{Brown:2018bms,Goto:2018iay}, authors \cite{Akhavan:2018wla,Alishahiha:2018swh} suggest a very different understanding of the holographic complexity for JT gravity. This approach relies on defining a new cut-off surface, $r=r_{0}$, behind the outer horizon according to the left panel in fig.\ref{penrose}. Based on this proposal, one needs to remove one part of the WDW patch and instead, add the following contributions to the action
	\bea
	&& I_{\text{GH}} = \frac{V_{d-1}}{8\pi G_{N}}\hspace{1mm} r_{0}^{d-1} \left(\partial_{r}f(r) +\frac{2(d-1)}{r}f(r)\right)\big(r^{*}(r_{0})-r^{*}_{\infty}\big),\cr\nonumber\\
	&& I_{\text{joint}} = -\frac{V_{d-1}}{8\pi G_{N}} \hspace{1mm}r_{0}^{d-1} \log \bigg(\frac{l^2 |f(r_{0})|}{\tilde{\alpha}^{2}R^{2}}\bigg),
	\cr\nonumber\\
	&&I_{\text{ct}} = -\frac{V_{d-1}}{8\pi G_{N}} \hspace{1mm} r_{0}^{d-1} \bigg[\log\bigg(\frac{(d-1)^{2}\hspace{.5mm}l^{2}_{\text{ct}} \hspace{.5mm}\tilde{\alpha}^2 \hspace{.5mm}R^{2}}{r_0^2 \hspace{.5mm}l^2}\bigg)+\frac{2}{d-1}\bigg].
	\cr\nonumber\\
	&& I_{\text{ct}}^{\text{new}} = \frac{V_{d-1}}{8\pi G}\hspace{1mm} r_{0}^{d-1}\left( \frac{2(d-1)}{l}\hspace{1mm} \sqrt{f(r_{0})}\hspace{2mm}\big(r^{*}_{\infty}-r^{*}(r_0)\big)\right),
	\eea
	where $I_{\text{ct}}^{\text{new}}$ is standard volume counterterm which is calculated on the $r=r_0$ hypersurface. By choosing the behind the horizon cut-off $r_0$ as following
	\bea\label{x0}
	x_{0} \equiv \frac{r_0}{r_+} = y^{\frac{(d-2)}{2}} \left(1 + \exp\left[-\frac{c_{3}}{y^{3}}-\frac{c_{2}}{y^{2}}-\frac{c_{1}}{y} + c_{0} + \tilde{c}_{1}y +...\right]\right),
	\eea
	for $d=2+1$ we have
	\bea
	&&\hspace{-.5cm}I_{\text{EH}}^{r_0} =  \frac{V_{2}\hspace{1mm}r_{+}^{2}}{4G_{N}}\big(-\frac{3\hspace{1mm}x_{\infty}^2}{2\pi}-\frac{1}{3\sqrt{3}}-\frac{y}{3\sqrt{3}}\big)+\mathcal{O}(y^{\frac{3}{2}}),\hspace{.2cm}I_{\text{Maxwell}}^{r_0} =  \frac{V_{2}\hspace{1mm}r_{+}^{2}}{4G_{N}}\big(\frac{(1-2\gamma)\hspace{.5mm}y^{\frac{1}{2}}}{3\sqrt{3}}+\frac{(1-2\gamma)\hspace{.5mm}y}{3\sqrt{3}}\big) +\mathcal{O} (y^{\frac{3}{2}}),
	\cr\nonumber\\
	&&\hspace{-.5cm}I_{\text{GH}}^{r_{0}} = \frac{V_{2}\hspace{1mm}r_{+}^{2}}{4G_{N}}\left(\frac{1}{2\sqrt{3}}-\frac{y^{\frac{1}{2}}}{3\sqrt{3}}+\frac{3 y}{4\pi}\right)
	+\mathcal{O}(y^{\frac{3}{2}}),
	\hspace{.3cm}I_{\text{joint}}^{r_{0}} = \frac{V_{2} \hspace{.5mm}r_{+}^{2}}{4G_{N}} \left(\frac{y}{4\pi}\hspace{.5mm} \log\left(\frac{ z^{4}\hspace{.5mm}R^{4}\tilde{\alpha}^{4}y}{l^{4}}\right)
	\right)+\mathcal{O}(y^{\frac{3}{2}}),
	\cr\nonumber\\
	&&\hspace{-.5cm}I_{\text{ct}}^{r_{0}} = -\frac{V_{2}\hspace{1mm}r_{+}^2}{4G_{N}}\frac{y}{2\pi} \left(1- \log\left(\frac{y\hspace{.5mm}l^{4}}{4\hspace{.5mm}z^2\hspace{.5mm}l_{\text{ct}}^2\hspace{.5mm}R^{2}\hspace{.5mm} \tilde{\alpha}^{2}}\right)\right),
	\hspace{.3cm}I_{\text{ct}}^{\text{new}} \sim \mathcal{O}(y^{\frac{3}{4}}).\label{Ir03}
	\eea
	Also for $d=3+1$ one can check that
	\bea
	&&\hspace{-.5cm}I_{\text{EH}}^{r_0} = \frac{V_{3}\hspace{1mm}r_{+}^{3}}{4G_{N}}\left(-\frac{4\hspace{.5mm}x_{\infty}^{3}}{3\pi}-\frac{1}{4} +\frac{y}{4\pi} -\frac{1+2\pi}{8\pi}y^{2}+\frac{5 y^3}{24\pi}\right)+\mathcal{O}(y^4),
	\cr\nonumber\\
	&&\hspace{-.5cm}I_{\text{Maxwell}}^{r_{0}} = \frac{V_{3}\hspace{1mm}r_{+}^{3}}{4G_{N}}\left(\frac{1}{4} -\frac{y}{4\pi} +\frac{(1+2\pi)}{8\pi}y^{2}+\frac{y^{3}}{8\pi}\right)+\mathcal{O}(y^{4}),
	\cr\nonumber\\
	&&\hspace{-.5cm}I_{\text{GH}}^{r_{0}} = \frac{V_{3}\hspace{1mm}r_{+}^{3}}{4G_{N}}\left( \frac{(\pi-2c_{3})}{4\pi} -\frac{(1+2c_{2})}{4\pi}\hspace{1mm} y -\frac{(-1+4c_{1})}{8\pi}\hspace{1mm} y^{2} -\frac{(-7-4c_{0}+4\log2)}{8\pi} \hspace{1mm}y^{3} \right)+\mathcal{O}(y^{4}),
	\cr\nonumber\\
	&&\hspace{-.5cm}I_{\text{joint}}^{r_{0}} =\frac{V_{3}\hspace{1mm}r_{+}^{3}}{4G_{N}}\big(\frac{c_{3}}{2\pi} +\frac{c_{2}}{2\pi}\hspace{1mm}y +\frac{c_{1}}{2\pi}\hspace{1mm}y^{2}+\frac{1}{2\pi}\big(-c_{0}+2\log\left[\frac{z\hspace{.5mm}R\hspace{.5mm}y\hspace{.5mm}\tilde{\alpha}}{\sqrt{2}\hspace{.5mm} l}\right]\big)y^{3}\big),
	\cr\nonumber\\
	&&\hspace{-.5cm}I_{\text{ct}}^{r_{0}} =-\frac{V_{3}\hspace{1mm}r_{+}^{3}}{4G_{N}}\hspace{1mm}\frac{y^{3}}{2\pi}\big(\frac{2}{3}-2\log\left[\frac{l^2\hspace{.5mm} y}{3\hspace{.5mm}z\hspace{.5mm} l_{\text{ct}}\hspace{.5mm}R\hspace{.5mm} \tilde{\alpha}}\right]\big),
	\hspace{.3cm}I_{\text{ct}}^{\text{new}} = 0.\label{Ir04}
	\eea
	It is also easy to see that by choosing the behind horizon cut off according to (\ref{x0}) (for small charges), not only the complexity of formation smoothly approach to the nuetral case in the limit $y\rightarrow 0$ but also the complexity growth rate matches with the Lloyd's bound in the late time in the same limit. Putting all ingredients (\ref{Ir03}) and (\ref{Ir04}) together the complexity for formation for $d=2+1$ and $d=3+1$ respectively becomes 
	\bea
	\label{deltacholo1}
	&&\hspace{-.5cm}\frac{\Delta \mathcal{C}_{\text{holo}}}{S_{\text{th}}} = \frac{(1-2\gamma)}{3\sqrt{3}\pi} - \left(\frac{\gamma}{\sqrt{3}\pi^{\frac{3}{2}}}\frac{G_{N}^{\frac{1}{2}}}{g\hspace{.5mm}l}\right) \tilde{Q}
	\cr\nonumber\\
	&&\hspace{-.3cm} + \frac{G_{N}}{16\pi^{3} g^{2}l^{2}}\left(9-16\sqrt{3}\pi\hspace{.5mm}\gamma-54\log\left(\frac{2^{\frac{5}{3}}\pi^{\frac{1}{2}}}{3}\hspace{.5mm}\frac{g\hspace{.5mm}l^{\frac{1}{3}}\hspace{.5mm}{l_{\text{ct}}^{\frac{2}{3}}}}{G_{N}^{\frac{1}{2}}}\right)+54\log\tilde{Q}\right)\hspace{-.5mm}\tilde{Q}^{2} +\mathcal{O}(\tilde{Q}^{3}),
	\eea
	and
	\bea
	\label{deltacholo2}
	&&\frac{\Delta \mathcal{C}_{\text{holo}}}{S_{\text{th}}} = \frac{1}{2\pi} - \left(\frac{1}{\sqrt{3}\pi^{\frac{5}{2}}}\frac{G_{N}^{\frac{1}{2}}}{g\hspace{.5mm}l}\right)\tilde{Q}+ \left(\frac{2G_{N}}{3\pi^{3}g^2 \hspace{.5mm}l^2}\right)\tilde{Q}^{2}+
	\cr\nonumber\\
	&&\hspace{1.7cm} + \frac{8}{3\sqrt{3}\hspace{.5mm}\pi^{\frac{7}{2}}}\left(\frac{G_{N}}{g^{2}\hspace{.5mm}l^{2}}\right)^{\frac{3}{2}}\left(7-16\log\left(\frac{9\pi}{2^{\frac{13}{4}}}\frac{g^{2}\hspace{.5mm}l\hspace{.5mm}l_{\text{ct}}}{G_{N}}\right)+32\log{\tilde{Q}}\right)\tilde{Q}^3+\mathcal{O}(\tilde{Q}^{4}).\nonumber\\
	\eea
	Intriguingly, holographic results (\ref{deltacholo1})-(\ref{deltacholo2}) which are obtained in presence of behind the horizon cut-off are compatible with field theory result (\ref{QFTC1}) since we could recover all the $\tilde{Q}$ terms. Moreover, the QFT result (\ref{QFTC1}) does not depend on the reference scale and its consistency with holographic results (\ref{deltacholo1})-(\ref{deltacholo2}) implies that $l_{\text{ct}}$ is actually proportional to AdS scale. It is worth noting that the identification between $l_{\text{ct}}$ and $l$ implies that the complexity of formation as a function of  $G_{N}, g, l$ can be completely expressed as a ratio of two boundary central charges, $C_{J}/C_{T}$\footnote{ Another choice can be $l_{\text{ct}} \sim l\times \mathbf{L}/\delta$ which implies that the leading divergence of complexity remains as $1/\delta^{d-1}\log{\delta}$. But this choice implies that the complexity of formation in (\ref{deltacholo1})-(\ref{deltacholo2}) not only depends on the arbitrary length scale $\mathbf{L}$ but also it becomes UV divergent which is clearly in contradiction with the QFT result (\ref{QFTC1}).}. Moreover, at finite chemical potential and zero temperature $\tilde{Q}\equiv\tilde{\mu}/T \rightarrow \infty$, the complexity of formation of holographic states (\ref{deltacholo1})-(\ref{deltacholo2}) diverges similarly to the complexity of formation of free field theory (\ref{QFTC1}). According to the above analysis, we see that the $F_1$ complexity (\ref{C1LR}) not only is compatible with holographic complexity for neutral BH but also works for  $U(1)$ electric charged ones. 
	
	To close this section let us emphasize that we have chosen an arbitrary length scale $l_{\text{ct}} \sim l$ by comparing with the QFT results for complexity of formation. This identification is obtained for zero time $t_{L}=t_{R} =0$ and for another times, apart from geometrical variables $G_{N}, l ,g$ we also have the time $\tau = t_{L}+t_{R}$ itself. This extra dimensionful variable may also appear in the expression of length scale $l_{\text{ct}}$. This time dependency of $l_{\text{ct}}$ can cause the complexity itself becomes a function of a completely new scale, similar to the appearance of $\omega_{R}$ in non-zero times in QFT side, (\ref{c1LRt}). But this is not a good way to reconstruct the dependence on $\omega_{R}$ for nonzero times because of two reasons. Firstly, the time dependency of $l_{\text{ct}}$ causes the counterterterm action,  $I_{\text{ct}}$, breaks bulk general covariance. Secondly, this time dependency implies that UV divergences of holographic complexity for AdS geometry become time-dependent which by comparison with (\ref{Hvac}) it is clear that this can not be correct. This observation needs to be considered more carefully and we hope to address a resolution for that in near future.

\end{document}